\documentclass[twocolumn,aps,superscriptaddress,showpacs,floatfix,prd,noshowpacs]{revtex4-2}
\usepackage[body={18.cm,23.5cm}]{geometry}     
\usepackage{mathrsfs}
\usepackage{amssymb}
\usepackage{amsmath}
\usepackage{graphicx}
\usepackage[normalem]{ulem}
\usepackage[dvips]{color}
\usepackage{bm}
\usepackage{longtable}
\usepackage{slashed}
\usepackage{enumitem}
\usepackage{empheq}
\usepackage{booktabs}
\usepackage{multirow}

\usepackage{titlesec}
\renewcommand{\thesection}{\Roman{section}}
\titleformat{\section}{\small\bfseries\centering}{\thesection.}{0.5em}{}

\usepackage[T1]{fontenc}
\usepackage[utopia]{mathdesign}

\usepackage[breaklinks=true]{hyperref}
\hypersetup{
  colorlinks=true,
  citecolor=magenta,
  linkcolor=black,
  urlcolor=teal,
}

\renewcommand\sout{\bgroup \color{red} \ULdepth=-.5ex \ULset}

\renewcommand{\v}[1]{\textbf{#1}}
\renewcommand{\rm}[1]{\textrm{#1}}
\renewcommand{\d}{\mathrm{d}}

\usepackage{tikz,xcolor,hyperref}
\DeclareSymbolFont{cmsymbols}{OMS}{cmsy}{m}{n}
\DeclareSymbolFontAlphabet{\mathcal}{cmsymbols}

\usepackage{xcolor}
\usepackage{academicons}
\usepackage{hyperref}

\definecolor{orcidgreen}{HTML}{A6CE39}

\newcommand{\orcid}[1]{%
  \href{https://orcid.org/#1}{%
    \textcolor{orcidgreen}{\aiOrcid}}%
}

\usepackage{setspace}

\renewcommand{\thesection}{\arabic{section}}

\begin{document}

\title{Neutron Skin Effects on Particle Emission in Heavy-Ion Collisions:
A Topic Review with Astrophysical and Nuclear Structure Connections}

\author{Bao-Jun Cai\,\orcid{0000-0002-8150-1020}}
\email{bjcai@fudan.edu.cn}
\affiliation{Key Laboratory of Nuclear Physics and Ion-beam Application (MOE), Institute of Modern Physics, Fudan University, Shanghai 200433, China}
\affiliation{Shanghai Research Center for Theoretical Nuclear Physics, NSFC and Fudan University, Shanghai 200438, China}
\author{De-Qing Fang\,\orcid{0000-0002-6123-3014}}
\email{dqfang@fudan.edu.cn}
\affiliation{Key Laboratory of Nuclear Physics and Ion-beam Application (MOE), Institute of Modern Physics, Fudan University, Shanghai 200433, China}
\affiliation{Shanghai Research Center for Theoretical Nuclear Physics, NSFC and Fudan University, Shanghai 200438, China}
\author{Yu-Gang Ma\,\orcid{0000-0002-0233-9900}}
\email{mayugang@fudan.edu.cn}
\affiliation{Key Laboratory of Nuclear Physics and Ion-beam Application (MOE), Institute of Modern Physics, Fudan University, Shanghai 200433, China}
\affiliation{Shanghai Research Center for Theoretical Nuclear Physics, NSFC and Fudan University, Shanghai 200438, China}
\affiliation{School of Physics, East China Normal University, Shanghai 200241, China}

\date{\today}

\begin{abstract}
\footnotesize

The neutron skin, defined by the difference between the neutron and proton root-mean-square radii of a nucleus, is a characteristic manifestation of isospin asymmetry and an important probe of the isovector nuclear interaction. This focused review examines how neutron skins influence particle emission and collective dynamics in heavy-ion collisions over a broad energy range, from reactions near the Fermi energy to ultra-relativistic nuclear collisions. By modifying the initial neutron and proton density profiles, the neutron skin affects the isospin composition and geometry of the participant region, pre-equilibrium emission, nucleon transport, particle production, fragment formation, and collective flow. We review the sensitivity of neutron-to-proton and $\rm{t}/^3\rm{He}$ yield ratios, light clusters, pion ratios, bremsstrahlung photons, isoscaling parameters, fragment parallel-momentum distributions, and neutron-proton differential flow and momentum observables. Particular attention is given to the density regions and reaction stages probed by these observables, as well as to their dependence on the nuclear symmetry energy and transport dynamics. The review also discusses the extension of neutron-skin studies from intermediate-energy reactions to isobar and heavy-nucleus collisions at RHIC and other high-energy facilities, where the neutron density profile modifies the initial geometry, eccentricity fluctuations, multiplicities, and anisotropic flows. The associated challenge of disentangling neutron-skin effects from nuclear deformation, surface diffuseness, shell structure, clustering, and model-dependent reaction dynamics is emphasized. Beyond heavy-ion collisions, broader nuclear-structure and astrophysical connections of neutron skins are explored. These include parity-violating electron scattering, dipole responses, coherent elastic neutrino-nucleus scattering, and the connection between coordinate-space neutron skins and SRC-induced proton skins in momentum space, as well as neutron star radii, tidal deformabilities, and multimessenger constraints on neutron-rich matter. Finally, we outline future opportunities offered by radioactive-beam facilities, improved intermediate- and high-energy collision experiments, microscopic many-body calculations, transport-model comparisons, and Bayesian inference. A coordinated analysis of multiple reaction systems and observables, together with complementary nuclear-structure and astrophysical information, will be essential for establishing particle emission in heavy-ion collisions as a quantitative probe of neutron skins and the density dependence of the symmetry energy.

\end{abstract}

\maketitle

\fontdimen2\font=1.5pt

\makeatletter
\renewcommand{\l@subsubsection}[2]{}
\makeatother

{
\footnotesize
\begin{spacing}{1.}
\tableofcontents
\end{spacing}
}

\section{Introduction}
\label{sec:introduction}

Understanding the properties of neutron-rich nuclei and dense matter is a central objective of contemporary nuclear physics and nuclear astrophysics\,\cite{MaYG_arxiv,Steiner2005PhysRep,Yang2020ARNPS,Tsang2024NatAstron,Drischler2021ARNPS,Lovato2022LRP,Fukushima2011RPP,Ghiglieri2020PhysRep,Haque2025PPNP,Laine2016Book,Romatschke2019Book}. Its Equation of State (EOS) governs phenomena across an enormous range of length, time, density, and temperature scales: it influences the structure and stability of neutron-rich nuclei, the non-equilibrium dynamics and particle emission of heavy-ion collisions (HICs)\,\cite{Li2008PhysRep,Sorensen2024PPNP,BraunMunzinger2016PhysRep,Shuryak2017RMP,Busza2018ARNPS,Heinz2013ARNPS,Gale2013IJMPA,Florkowski2018RPP,Elfner2023JPG,Chen2018PhysRep,Ma2018PPNP,Deng2024PPNP,Chen2024NST,Shou2024NST,Zhao2024NSTEM,Xu2025CPL,Ma2025NST}, the composition and evolution of supernova matter\,\cite{Oertel2017RMP,Vidana2018PRSA,Furusawa2023PPNP,Mezzacappa2020LRCA,Fischer2024PPNP}, the synthesis of heavy elements, and the masses, radii, tidal deformabilities, cooling, and merger dynamics of neutron stars (NSs)\,\cite{Li2021Universe,Alford2008RMP,Watts2016RMP,Burgio2021PPNP,Baym2018RPP,Baiotti2019PPNP,Orsaria2019JPG,Li2019EPJA,Dexheimer2021JPG,Lattimer2021ARNPS}. These systems probe different thermodynamic conditions and density regimes\,\cite{Chen2025SCPMA,Shen2025Research,He2023NST,Ma2023CPL,He2026SCPMA,Wang2025NST,Fang2026NST}, but they are connected by the poorly constrained isovector sector of the strong interaction.
The neutron skin of a finite nucleus provides a particularly important laboratory manifestation of this physics. It encodes how neutron excess is distributed between the nuclear interior and surface and thereby offers experimentally accessible information on the density dependence of the symmetry energy\,\cite{Brown2000PRL,Typel2001PRC,
Horowitz2001PRL,Centelles2009PRL}. At the same time, the neutron-rich surface constitutes part of the initial state of a heavy-ion collision and can leave measurable signatures in nucleon and cluster emission, particle production, fragmentation, and collective flow. { Neutron-skin therefore provides a bridge connecting nuclear structure, reaction dynamics, the EOS of neutron-rich matter, and astrophysical phenomena extending from supernovae and nucleosynthesis to NSs and gravitational waves (GWs)\,\cite{Thiel2019JPG,Horowitz2014JPG}.}

The spatial distributions of neutrons and protons constitute the microscopic starting point for this connection. In a neutron-rich nucleus, the competition among the nuclear symmetry energy, surface tension, Coulomb interaction, shell structure, and many-body correlations generally causes the neutron distribution to extend beyond the proton distribution, producing a neutron-rich surface commonly referred to as the \textit{neutron skin}. Its thickness is defined as
\begin{equation}
R_{\rm{skin}}
\equiv
R_{\rm n}-R_{\rm p}
=
\Delta r_{\rm{np}},
\label{eq:Rskin_definition}
\end{equation}
where the point- and point-proton rms radii are\,\cite{BohrMottelson1998Book,RingSchuck1980Book}
\begin{equation}
R_J
=
\sqrt{\left\langle r_J^2\right\rangle}
=
\left[
N_J^{-1}
\int r^2\rho_J(\vec{r})\d\vec{r}
\right]^{1/2},
~~
J=\rm{n,p}.
\label{eq:rms_radius}
\end{equation}
Here, $N_{\rm n}=N$ and $N_{\rm p}=Z$; the point-nucleon densities satisfy
$\int\rho_J(\vec{r})\d\vec{r}=N_J$. The point-proton radius should be distinguished from the experimentally measured charge radius, which also contains contributions from the finite sizes of the nucleons and other relatively small corrections.

\begin{figure}[h!]
\centering
\includegraphics[width=0.48\textwidth]{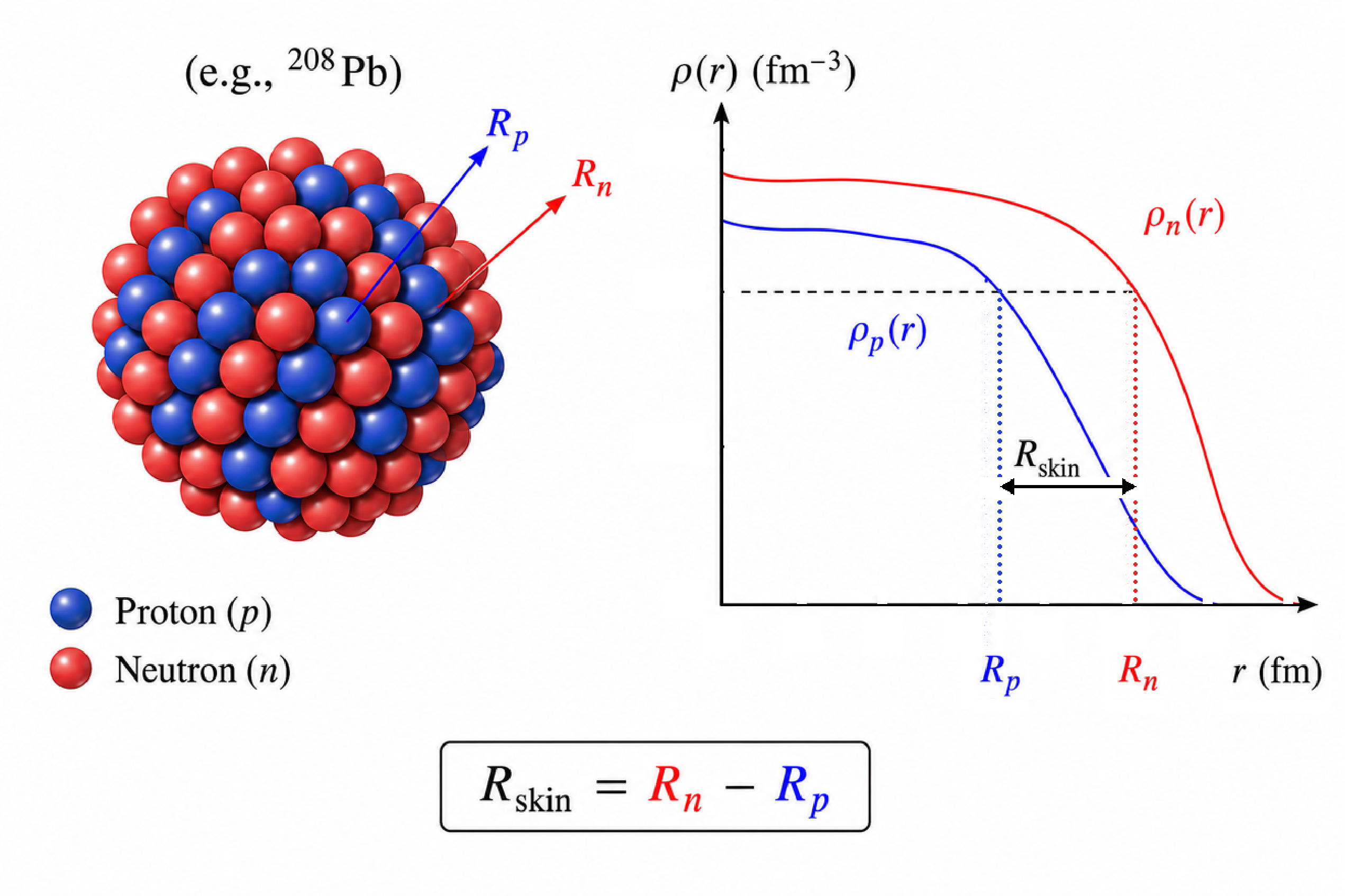}
\caption{(Color Online). Schematic illustration of a neutron-rich nucleus and its neutron and proton density distributions. Red and blue spheres denote neutrons and protons, respectively. The neutron distribution $\rho_{\rm n}(r)$ extends beyond the proton distribution $\rho_{\rm p}(r)$, producing a neutron skin characterized by $R_{\rm{skin}}=R_{\rm n}-R_{\rm p}$, where $R_J$ denotes the point rms radius for $J=\rm{n,p}$. The skin may reflect differences in both the radial extension and surface diffuseness of the two distributions.}
\label{fig:skin_schematic}
\end{figure}

For a spherical nucleus, neutron and proton density profiles are frequently characterized by two-parameter Fermi distributions\,\cite{DeVries1987ADNDT,Warda2009PRC},
\begin{equation}
\rho_J(r)
=
\frac{\rho_{0,J}}
{1+\exp[(r-C_J)/a_J]},
~~
J=\rm{n,p},
\label{eq:2pf_density}
\end{equation}
where $C_J$ is the half-density radius and $a_J$ characterizes the surface diffuseness. The neutron-skin thickness is therefore an integrated measure of the difference between the neutron and proton distributions. A given $R_{\rm{skin}}$ can originate from different combinations of $C_{\rm n}-C_{\rm p}$ and $a_{\rm n}-a_{\rm p}$. This distinction is especially relevant for peripheral nuclear reactions, which may respond differently to the radial displacement and diffuseness of the neutron-rich surface even when the corresponding rms neutron-skin thicknesses are similar. The basic coordinate-space picture is illustrated in FIG.\,\ref{fig:skin_schematic}.

\renewcommand*\figurename{\small FIG.}

\begin{figure}[h!]
\centering
\hspace{0.1cm}
\includegraphics[width=0.47\textwidth]{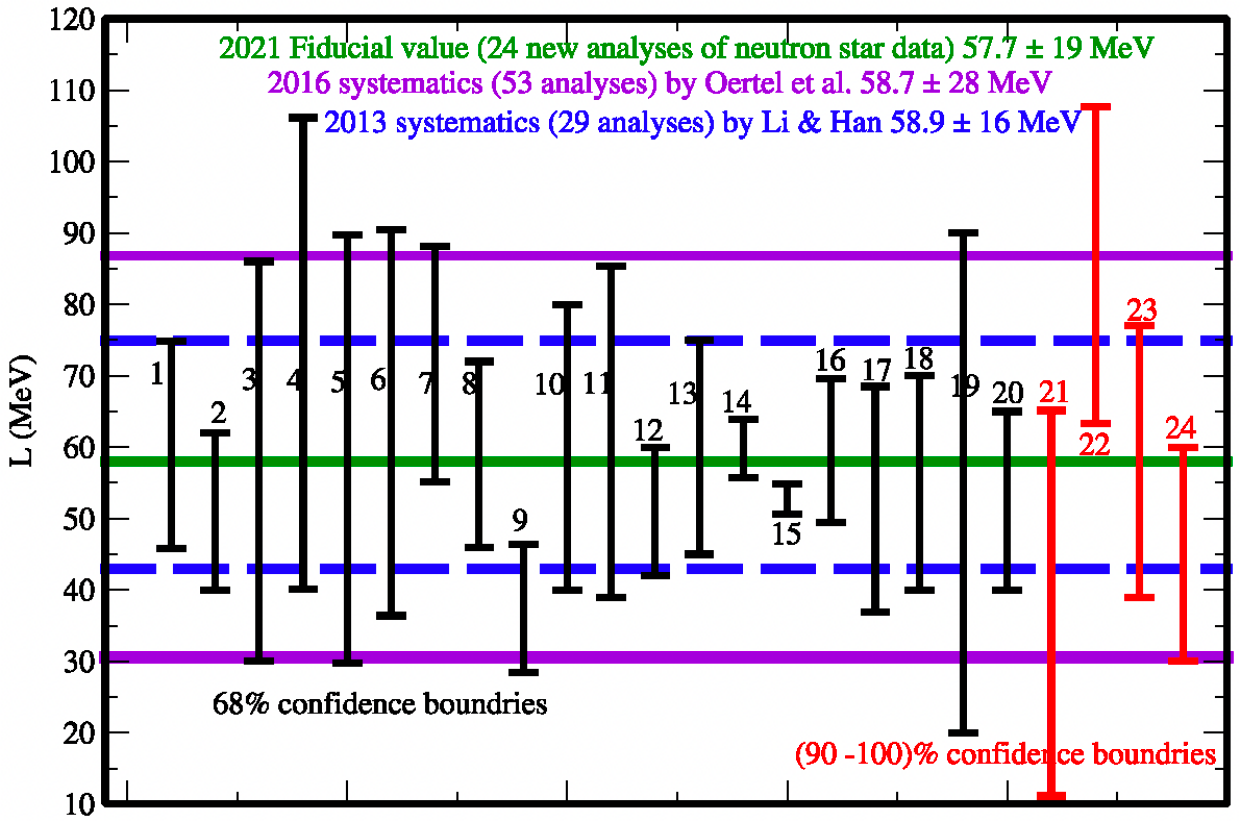}\\[0.0cm]
\includegraphics[width=0.48\textwidth]{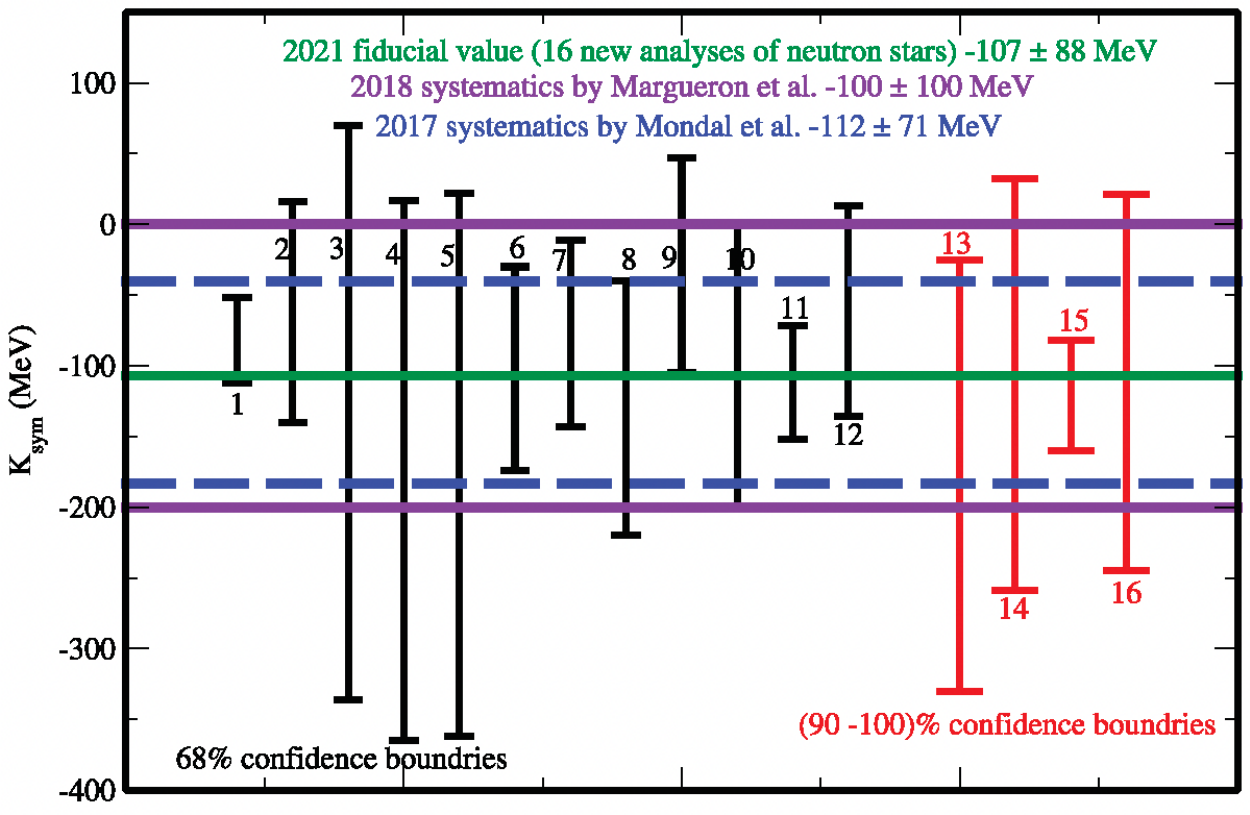}
\caption{(Color Online). Summary of representative constraints on the slope parameter $L$ (upper) and curvature parameter $K_{\rm{sym}}$ (lower) of the symmetry energy at saturation density. Upper: values of $L$ extracted from 24 analyses of NS observables after GW170817, compared with the ranges obtained from earlier surveys of terrestrial nuclear experiments, HIC observables, nuclear theory, and astrophysical information. The green line denotes their average, $L\approx57.7\pm19$\,MeV. Lower: values of $K_{\rm{sym}}$ extracted from 16 analyses of post-GW170817 NS observations, compared with earlier estimates from constrained energy-density functionals and nuclear EOS meta modeling. Their average is $K_{\rm{sym}}\approx-107\pm88$\,MeV. The comparison shows that $L$ is relatively well constrained around saturation density, whereas substantial uncertainty remains in the curvature and hence in the extrapolation of $E_{\rm{sym}}(\rho)$ toward higher densities.
Adapted from Ref.\,\cite{Li2021Universe}.}
\label{fig:L_Ksym_summary}
\end{figure}

The connection between the neutron skin and neutron-rich matter is conventionally expressed through the EOS of uniform nuclear matter. At zero temperature, the energy per nucleon can be expanded in the isospin asymmetry as\,\cite{Li2008PhysRep}
\begin{equation}
E(\rho,\delta)
\approx
E_0(\rho)
+
E_{\rm{sym}}(\rho)\delta^2
+
\mathcal{O}(\delta^4),
~~
\delta=\frac{\rho_{\rm n}-\rho_{\rm p}}{\rho},
\label{eq:parabolic_EOS}
\end{equation}
where $\rho=\rho_{\rm n}+\rho_{\rm p}$, $E_0(\rho)$ is the energy per nucleon of symmetric nuclear matter, and $E_{\rm{sym}}(\rho)$ is the nuclear symmetry energy. Around the saturation density $\rho_0$, the latter may be expanded around as\,\cite{Li2008PhysRep}
\begin{equation}
E_{\rm{sym}}(\rho)
\approx
S+L\chi+2^{-1}K_{\rm{sym}}\chi^2
+\mathcal{O}(\chi^3),
~~
\chi=\frac{\rho-\rho_0}{3\rho_0},
\label{eq:Esym_expansion}
\end{equation}
where $S\equiv E_{\rm{sym}}(\rho_0)$, $L$ (slope) and $K_{\rm{sym}}$ (curvature) are the corresponding expansion coefficients.
As summarized in
FIG.\,\ref{fig:L_Ksym_summary}, the slope parameter is relatively well
constrained by terrestrial nuclear experiments, HIC observables,
nuclear-structure measurements, microscopic calculations, and NS
observations. In particular, a survey of 29 analyses available by 2013 yielded
$L\approx58.9\pm16$\,MeV, while 24 analyses of NS observables published
after GW170817 gave the closely consistent estimate
$L\approx57.7\pm19$\,MeV, see also Ref.\,\cite{Sotani2022PTEP}. In contrast, the curvature parameter remains considerably less certain. A survey of more than 500 energy density functionals constrained by the available terrestrial and astrophysical information suggested
$K_{\rm{sym}}\approx-112\pm71$\,MeV, whereas 16 analyses of NS
observables after GW170817 yielded $K_{\rm{sym}}\approx-107\pm88$\,MeV\,\cite{Li2021Universe}. The agreement of the central values is encouraging, but the broad uncertainty in $K_{\rm{sym}}$ illustrates why a determination of $L$ alone cannot reliably fix the symmetry energy at suprasaturation densities. Moreover, the quoted constraints are generally correlated and depend on the adopted EOS parametrization, priors, and density
sensitivity of the observables.

Within the parabolic approximation, the pressure of pure neutron matter at saturation density is approximately\,\cite{Brown2000PRL}
\begin{equation}
P_{\rm{PNM}}(\rho_0)
\approx
3^{-1}\rho_0L.
\label{eq:symmetry_pressure}
\end{equation}
A larger symmetry pressure generally increases the energetic preference for placing excess neutrons toward the nuclear surface and hence favors a thicker neutron skin. This produces the familiar correlation between $R_{\rm{skin}}$ and $L$ in many nuclear models. The underlying physics should not, however, be interpreted simply as a local pressure gradient pushing neutrons outward. Rather, the skin results from a balance between the symmetry energy of the nuclear interior and the energetic cost of accommodating neutron excess in the dilute surface. Because finite nuclei sample a range of sub-saturation densities, $R_{\rm{skin}}$ also depends on the surface symmetry energy, $K_{\rm{sym}}$, shell structure, deformation, and Coulomb effects. The $R_{\rm{skin}}$-$L$ correlation can therefore be strong within a restricted class of models without representing an exact or universal one-to-one relation. See relevant disucssion of Subsection \ref{subs:EsymNSkin}.

Heavy-ion collisions provide a dynamical approach to accessing this
nuclear-structure information. The evolution of the neutron and proton phase-space distributions may be represented schematically by a
transport equation\,\cite{Bertsch1988PhysRep,Aichelin1991PhysRep,
Botermans1990PhysRep,Danielewicz1984AnnPhys,
Bonasera1994PhysRep,Ko1996JPhysG,
Bass1998PPNP,Cassing1999PhysRep,Li1998IJMPE,
Buss2012PhysRep}
\begin{equation}
\frac{\partial f_J}{\partial t}
+
\nabla_{\vec{p}}H_J\cdot\nabla_{\vec{r}}f_J
-
\nabla_{\vec{r}}H_J\cdot\nabla_{\vec{p}}f_J
=
\mathcal{C}_J[f_{\rm n},f_{\rm p}],
~~
J=\rm{n,p},
\label{eq:transport_schematic}
\end{equation}
where $f_J(\vec{r},\vec{p},t)$ is the one-body phase-space distribution,
$H_J$ is the corresponding single-particle Hamiltonian, and
$\mathcal{C}_J$ denotes the collision integral. The neutron skin enters primarily through the initial neutron and proton distributions, whereas the symmetry energy affects their subsequent evolution through the isovector mean field. Experimentally measured observables consequently
reflect both the initial neutron skin and the dynamical processes that transmit, modify, or partially erase its signal\,\cite{Xiao2009PRL,Tsang2009PRL,Li2016NST,Ma2013PRC,Fang2010PRC}.
The microscopic derivation and numerical solution of such transport
equations are important subjects in their own right. Within the
Schwinger-Keldysh closed-time-path formalism\,\cite{Schwinger1961JMP,Keldysh1965JETP,Chou1985PhysRep}, nonequilibrium Green
functions obey the Kadanoff-Baym equations\,\cite{KadanoffBaym1962Book}, from which semiclassical transport equations can be obtained through a Wigner transformation, gradient expansion, and suitable quasiparticle or off-shell approximations; controlling these approximations while preserving conservation laws and incorporating memory and correlation effects remains a central challenge in transport theory\,\cite{Buss2012PhysRep,Danielewicz1984AnnPhys,Botermans1990PhysRep,
Ivanov1999NPA,Rios2011AnnPhys}.

\begin{figure}[h!]
\centering
\includegraphics[width=0.45\textwidth]{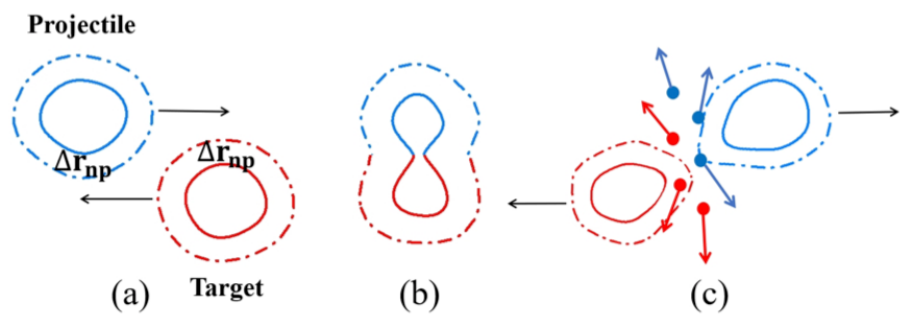}
\caption{(Color Online). Schematic illustration of the evolution of a peripheral heavy-ion collision from the initial configuration, through the reaction stage, to the formation of projectile-like and target-like residues and the emission of light particles from the neck region. The neutron-rich nuclear surface directly affects the composition of
the participant region and the emitted particles and fragments.
Adapated from Ref.\,\cite{Yang2023Universe}.} \label{fig:skin_intermediate_HIC}
\end{figure}

At Fermi and intermediate energies, the dilute surfaces of the
projectile and target play an especially important role in peripheral
and semi-peripheral collisions. Variations in $R_{\rm{skin}}$ modify
both the spatial extent and the neutron-to-proton composition of the
first-contact and overlap regions. During the subsequent evolution,
nucleon exchange and isospin transport occur through the low-density
neck connecting the projectile-like and target-like residues, while
pre-equilibrium nucleons and light clusters are emitted from the
participant and neck regions. Consequently, the collision geometry
provides a direct mechanism through which information carried by the
initial neutron-rich surface can be transferred to measurable
final-state particles and fragments, as illustrated schematically in
FIG.\,\ref{fig:skin_intermediate_HIC}.

The influence of the neutron skin is not restricted to a single
reaction stage or class of observables. It can affect isospin diffusion and fractionation, pre-equilibrium emission, cluster and fragment formation, and particle production thresholds. The resulting signatures may appear in neutron-to-proton and $\rm{t}/^3\rm{He}$ yield ratios, light-particle spectra, fragment momentum distributions, isoscaling parameters, pion ratios, hard-photon emission, and neutron-proton differential flow. For example, transport calculations of peripheral $^{124,132}\rm{Sn}+{}^{124}\rm{Sn}$ collisions at $200\,\rm{MeV}/$nucleon indicate that projectile-like-residue cross
sections and transverse neutron-to-proton yield ratios retain
sensitivity to the neutron skin of the incident nucleus\,\cite{Yang2023Universe}. Since particles and fragments are produced
or emitted at different stages of the collision, their sensitivities to the initial surface, compressed participant matter, neck dynamics, and subsequent expansion are complementary. Their quantitative interpretation nevertheless requires transport calculations that consistently describe the initial nuclear density and momentum distributions, isoscalar and isovector mean fields, in-medium scattering, Pauli blocking, resonance dynamics, cluster formation, and final-state interactions\,\cite{Colonna2020PPNP,Wolter2022PPNP}.

\begin{figure}[h!]
\centering
\includegraphics[width=0.45\textwidth]{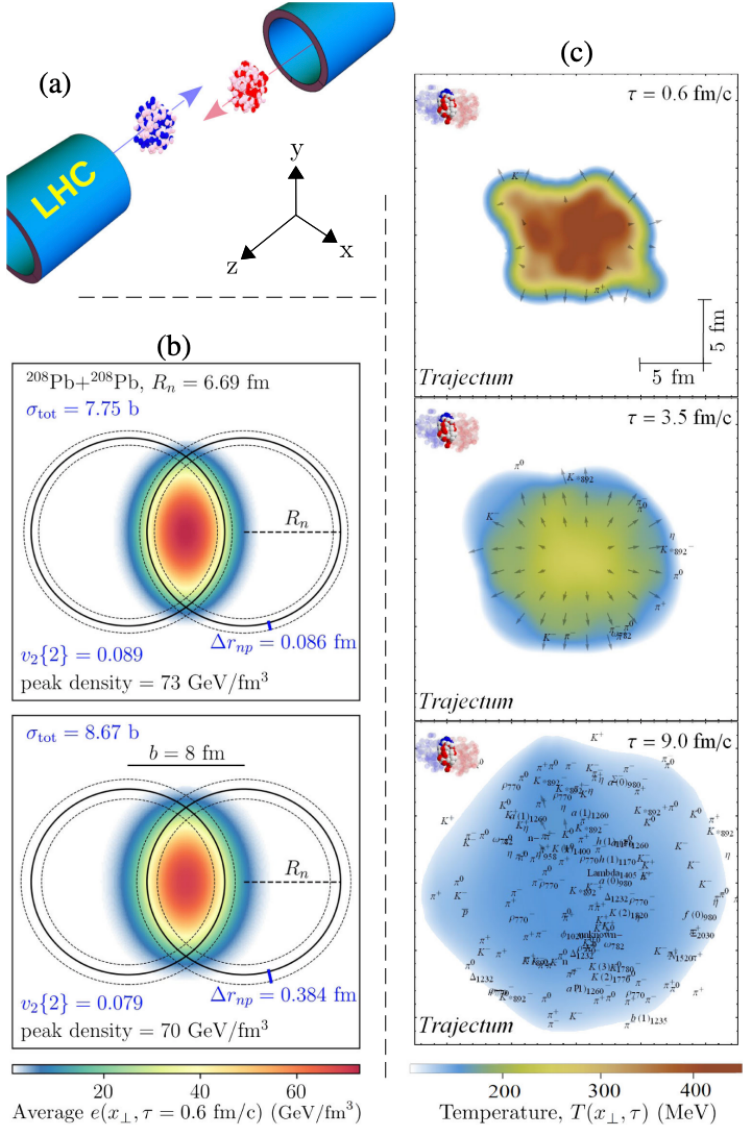}
\caption{(Color Online). Connection between the neutron skin of $^{208}\rm{Pb}$ and collective observables in ultra-relativistic Pb+Pb collisions. The Lorentz-contracted nuclei deposit energy in the transverse overlap region, whose size, diffuseness, and eccentricity depend on the neutron density profile. Hydrodynamic evolution converts this initial geometry into final-state particle spectra and collective flow. In the examples shown, a larger neutron diffuseness increases the total hadronic cross section and produces a more diffuse and less elliptical QGP profile. Adapted from Ref.\,\cite{Giacalone2023PRL}.}
\label{fig:skin_HIC_schematic}
\end{figure}

At ultra-relativistic energies, the microscopic mechanism changes, but the initial nuclear density distribution remains relevant. The colliding nuclei are strongly Lorentz contracted, and their transverse density profiles determine the geometry and energy deposition of the produced quark-gluon plasma (QGP). Hydrodynamic expansion subsequently converts the initial size, eccentricity, and density fluctuations into final-state multiplicities, transverse-momentum spectra, and anisotropic flows. As illustrated in FIG.\,\ref{fig:skin_HIC_schematic}, changing the neutron diffuseness of $^{208}\rm{Pb}$ modifies the total hadronic cross section as well as the size, diffuseness, and ellipticity of the initially produced medium. Within the calculation shown, a larger neutron skin produces a more diffuse QGP profile, weaker pressure gradients, and reduced elliptic flow. A global Bayesian analysis of Pb+Pb data at the LHC obtained
$R_{\rm{skin}}(^{208}\rm{Pb})\approx 0.217\pm0.058$\,fm, demonstrating that ultra-relativistic collisions can provide an independent probe of ground-state neutron distributions\,\cite{Giacalone2023PRL}. 
Although this extraction remains conditional on the descriptions of initial energy deposition and hydrodynamic response, it establishes a striking connection between low-energy nuclear structure and high-energy QGP observables.
More related discussion on this example is given in Subsection \ref{subs:Nk208-Ultra}.

\begin{figure}[h!]
\centering
\includegraphics[width=0.42\textwidth]{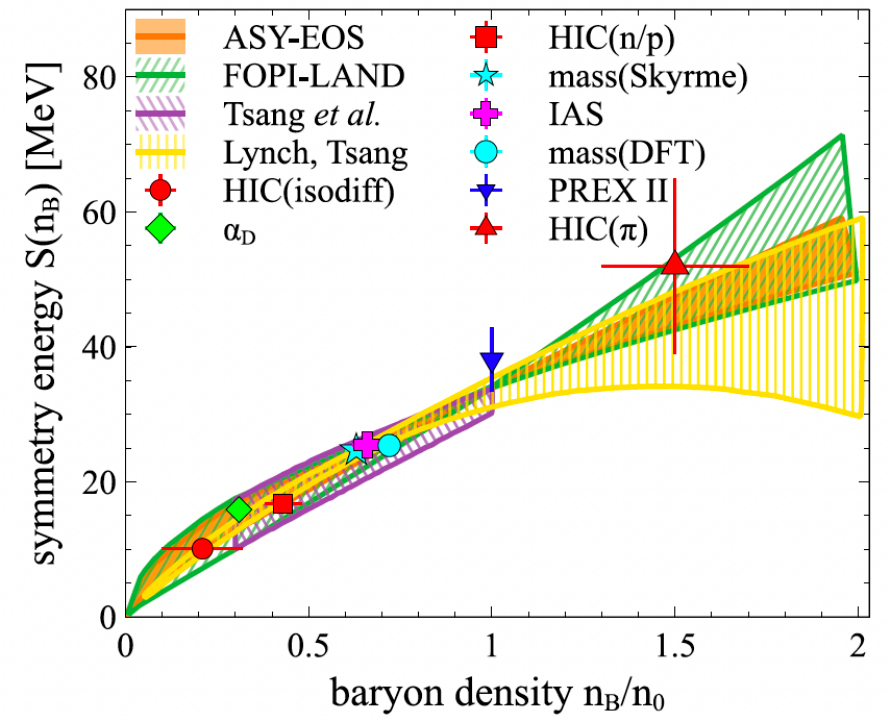}
\caption{(Color Online). Representative constraints on the density dependence of the nuclear symmetry energy obtained from heavy-ion collisions. Different observables and beam energies probe different sub- and supra-saturation density regions, illustrating both the broad reach of heavy-ion experiments and the importance of identifying the effective density sensitivity of each extraction. Adapted from Ref.\,\cite{Sorensen2024PPNP}.}
\label{fig:symmetry_energy_constraints}
\end{figure}

The broader importance of heavy-ion collisions lies in their ability to probe the EOS over density and temperature regions inaccessible to static nuclear measurements. By changing the collision energy, system size, isospin asymmetry, and impact parameter, one can vary the maximum density reached and the relative importance of the participant, spectator, and surface regions. FIG.\,\ref{fig:symmetry_energy_constraints} summarizes representative constraints on the density dependence of the nuclear symmetry energy derived from comparisons between heavy-ion collision data and transport calculations. The different constraints probe sub- and supra-saturation densities through observables such as isospin diffusion, particle-yield ratios, and collective flows\,\cite{Sorensen2024PPNP}. The spread among the extracted regions reflects differences in the observables, collision systems and energies, effective density sensitivities, symmetry-energy parametrizations, and transport treatments. A parameter inferred from a particular reaction should therefore be interpreted mainly as a constraint on $E_{\rm{sym}}(\rho)$ over the density interval sampled by that observable, rather than as a universal determination of its behavior at all densities\,\cite{Sorensen2024PPNP}.

\begin{figure}[h!]
\centering
\includegraphics[width=0.32\textwidth]{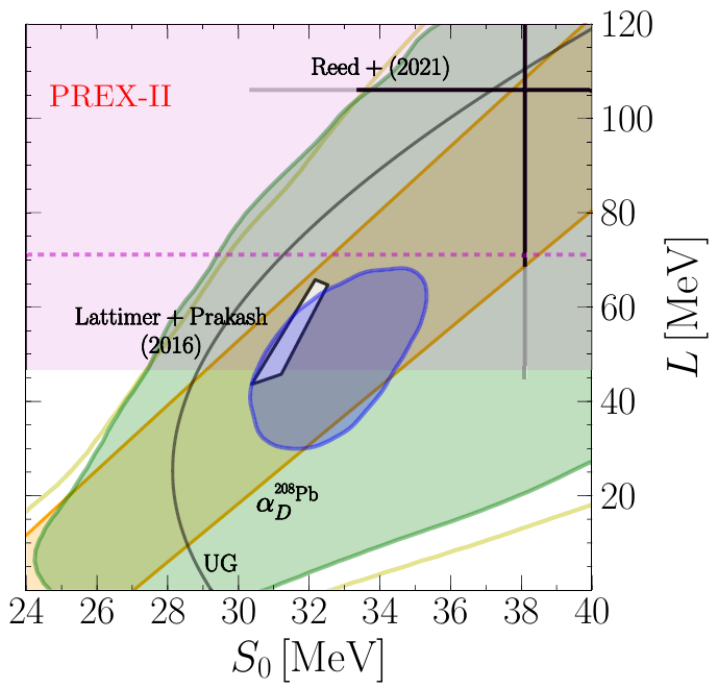}
\caption{(Color Online). Representative constraints in the $S$-$L$ plane from PREX-II (here $S_0\equiv S$), other nuclear experiments, microscopic nuclear theory, and astrophysical observations. The orientations and widths of the allowed regions reflect the different parameter combinations and density intervals probed by each source of information. Their comparison illustrates the complementarity of finite-nucleus, microscopic, and astrophysical constraints, as well as the importance of retaining their correlations and theoretical uncertainties. Adapted from Ref.\,\cite{Essick2021PRC}.}
\label{fig:S0_L_constraints}
\end{figure}

Neutron-skin studies in heavy-ion collisions are complemented by other terrestrial and astrophysical measurements. Parity-violating electron scattering (PVES) provides a comparatively clean electroweak probe of neutron distributions, with PREX-II and CREX constraining the weak form factors of $^{208}\rm{Pb}$ and $^{48}\rm{Ca}$, respectively\,\cite{PREX2021PRL,CREX}. Nuclear masses, dipole polarizabilities, collective excitations, nucleon scattering, and microscopic many-body calculations probe overlapping but nonidentical combinations of $S$, $L$, $K_{\rm{sym}}$, and nuclear surface properties. At higher densities, NS radii, tidal deformabilities, cooling, and merger dynamics provide additional information on the EOS of strongly neutron-rich matter. Finite nuclei, heavy-ion collisions\,\cite{Li2008PhysRep}, and NSs nevertheless\,\cite{Chatziioannou2025RMP,Li2025eXTP} probe different density ranges and thermodynamic conditions, and no unique mapping among their observables should be assumed.
FIG.\,\ref{fig:S0_L_constraints} illustrates representative constraints in the $S$-$L$ plane obtained from PREX-II, other nuclear experiments, microscopic theory, and astrophysical observations. The different orientations and widths of the allowed regions show that these probes constrain different combinations of the symmetry-energy parameters. Their comparison demonstrates the value of combining complementary information, but also emphasizes that the inferred $S$, $L$, and $R_{\rm{skin}}$ depend on the adopted EOS representation and on the treatment of theoretical correlations and uncertainties\,\cite{Essick2021PRC}.

This focused review examines how neutron skins affect particle emission and collective dynamics in heavy-ion collisions from the Fermi-energy regime to ultra-relativistic collider energies. Particular attention is devoted to neutron-to-proton and $\rm{t}/^3\rm{He}$ yield ratios, light clusters, pions, hard photons, fragments, pre-equilibrium nucleons, and collective-flow observables. Their sensitivities to the initial neutron and proton distributions are discussed together with their dependence on the symmetry energy, collision geometry, and reaction dynamics. Isobar and Pb+Pb collisions are considered as extensions of neutron-skin studies to relativistic energies, where ground-state nuclear structure is encoded in the initial collision geometry and subsequently reflected in QGP observables.
The discussion is further placed in the broader context of SRC-induced momentum distributions, the possible coexistence of a coordinate-space neutron skin and a momentum-space proton skin, and the connection between finite nuclei and the EOS of NS matter. We also address the principal challenges in separating neutron-skin effects from deformation, surface diffuseness, shell and clustering effects, and transport-model uncertainties. The central perspective of this review is that heavy-ion observables do not measure $R_{\rm{skin}}$ directly; rather, they probe its dynamical consequences. Establishing particle emission and collective flow as quantitative neutron-skin probes therefore requires coordinated analyses of multiple collision systems and observables, realistic nuclear initial states, controlled reaction models, and comparisons with complementary nuclear and astrophysical information.

The remainder of this review is organized as follows.
Section \ref{sec:foundations} introduces the theoretical foundations of neutron skins, their experimental determination, their connection to the nuclear symmetry energy, and the basic mechanisms through which they enter heavy-ion collision dynamics. Section \ref{sec:observables} reviews neutron-skin effects on particle-emission observables, including nucleons, light clusters, pions, hard photons, and fragments, while Section \ref{sec:collective} discusses collective-flow observables and
neutron-skin constraints from intermediate-energy reactions, isobar
collisions, and ultra-relativistic collisions.
Section \ref{SEC_Pskin} extends the discussion to SRCs and the possible coexistence of a neutron skin in coordinate space and a proton skin in momentum space. Section \ref{sec:astrophysics} places neutron skins and heavy-ion constraints in the broader context of NS radii, tidal deformabilities, and multi-messenger observations. Section \ref{sec:challenges} summarizes the principal challenges and future directions, including nuclear deformation and surface diffuseness, transport-model dependence, radioactive-beam and collider opportunities, ab initio and Bayesian approaches, the PREX-CREX puzzle, spectral functions, and neutrino probes. Finally, Section \ref{sec:conclusion} presents the main conclusions.

This review is not intended to provide an exhaustive account of all
aspects of neutron-skin physics. Instead, it offers a selective and
deliberately focused perspective on particle emission and collective dynamics in heavy-ion collisions, their connections to nuclear structure and astrophysics, and several emerging questions that may become particularly important in future studies. For broader reviews of the nuclear symmetry energy, neutron distributions, and EOS constraints obtained from nuclear experiments, heavy-ion collisions, and astrophysical observations, we refer the reader to
Refs.\,\cite{Lattimer2023,Mammei2024ARNPS,
Sammarruca2024Symmetry,Atkinson2024FrontPhys,
Tanaka2024FrontPhys,Ding2024NST,
Miyagi2025FrontPhys,NeumannCosel2025FrontPhys,Jia2025RPP,LimHolt22}.
For a comprehensive overview of nuclear science encompassing heavy-ion
collisions, nuclear structure, and nuclear astrophysics, we refer the reader to Ref.\,\cite{MaYG_arxiv}.

\section{Theoretical Foundations}
\label{sec:foundations}

The neutron skin provides an important connection between nuclear
structure, the isovector nuclear interaction, and the dynamics of
neutron-rich systems. Its thickness, defined by the difference
between neutron and proton rms radii, can be constrained by a variety of electroweak and hadronic probes and is closely related to the density dependence of the nuclear symmetry energy. In heavy-ion collisions, the neutron skin further acts as an initial-state
property that influences the isospin composition and subsequent
dynamical evolution of the reaction, particularly in peripheral
collisions where the nuclear surface is preferentially sampled.
This section summarizes these theoretical foundations, from the
definition and experimental determination of the neutron skin and
its connection to the symmetry energy, to the transport description
that links the initial neutron distribution to measurable
heavy ion collision observables.

\subsection{Neutron skin: experimental determination}

The neutron skin is a characteristic feature of neutron-rich nuclei,
arising from the different spatial distributions of neutrons and
protons. A positive $R_{\rm{skin}}$ indicates that the neutron distribution extends, on average, farther than the proton distribution. The formation of a neutron skin reflects the competition among the nuclear symmetry energy, surface tension, Coulomb interaction, and shell effects, and its magnitude therefore provides important information on the isovector sector of the nuclear interaction.

\begin{figure}[h!]
\centering
\includegraphics[width=0.45\textwidth]{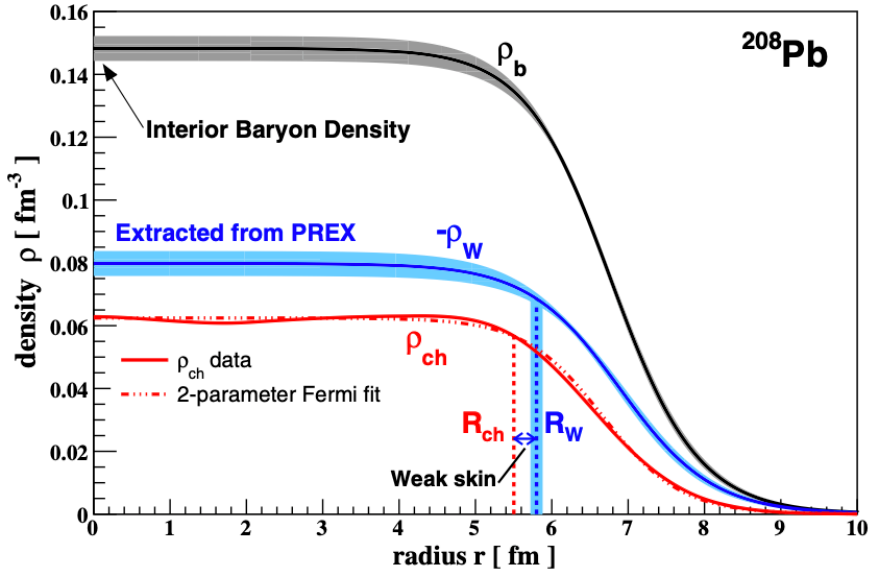}
\caption{(Color Online). Radial charge, weak-charge, and baryon density
distributions of $^{208}\rm{Pb}$ inferred from the combined PREX
datasets. The accurately known charge density $\rho_{\rm{ch}}(r)$ is
shown together with its two-parameter Fermi fit, while
$-\rho_{\rm W}(r)$ denotes the magnitude of the extracted weak-charge
density, whose uncertainties are indicated by the shaded band. The
weak radius $R_{\rm W}$ extends beyond the charge radius $R_{\rm{ch}}$,
providing direct evidence for a weak skin and, after accounting for
finite nucleon form factors and electroweak corrections, a neutron
skin. The corresponding baryon density $\rho_{\rm b}(r)$ is also
shown. Adapted from Ref.\,\cite{PREX2021PRL}.}
\label{fig:PREX_weak_density}
\end{figure}

Experimentally, the proton distribution is relatively well
constrained because it can be accessed through electromagnetic
interactions. Nuclear charge radii have been measured with high
precision using elastic electron scattering, muonic atoms, and
isotope-shift measurements in atomic spectroscopy. After accounting
for the finite sizes of the proton and neutron and other small
corrections, these measurements provide accurate information on
$R_{\rm{p}}$. In contrast, determining $R_{\rm{n}}$ is considerably
more difficult because neutrons carry no electric charge. A variety
of electroweak and hadronic probes have therefore been developed to
constrain neutron density distributions and neutron-skin
thicknesses. Representative approaches include:
\begin{enumerate}[label=(\alph*),leftmargin=*]
\item \textbf{Parity-violating electron scattering (PVES):}
PVES exploits the interference between electromagnetic and weak neutral-current amplitudes in the elastic scattering of longitudinally polarized electrons. Because the neutron weak charge is much larger in magnitude than that of the proton, the measured helicity-dependent asymmetry is predominantly sensitive to the nuclear weak form factor and hence to the neutron distribution, whereas the charge distribution is already accurately known from conventional electron scattering. PREX and CREX have applied this method to $^{208}\rm{Pb}$ and $^{48}\rm{Ca}$, respectively\,\cite{PREX2021PRL,CREX}. As illustrated in FIG.\,\ref{fig:PREX_weak_density}, the PREX weak-charge distribution extends beyond the charge distribution, providing direct evidence for a weak skin and, after accounting for finite nucleon form factors and electroweak corrections, a neutron skin. Owing to its well-controlled electroweak reaction mechanism, PVES is one of the least model-dependent probes of neutron distributions, although extracting the point-neutron radius from a measurement at finite momentum transfer retains some dependence on the assumed density profile and on Coulomb-distortion and radiative corrections\,\cite{Horowitz2001PRC-a}.

\item \textbf{Proton elastic scattering:}
The strong interaction between incident protons and target nucleons
makes proton elastic scattering sensitive to both neutron and proton density distributions. Measurements of differential cross sections, analyzing powers, and isotope-dependent cross-section ratios can therefore be used to extract neutron density distributions and $R_{\rm{skin}}$\,\cite{Clark2003PRC,Terashima2008PRC,
Zenihiro2010PRC,KanadaEnyo2021arXiv}. Compared with PVES, however,
the extraction depends more strongly on the reaction model and on the effective nucleon-nucleus interaction employed in the
analysis\,\cite{Clark2003PRC,Terashima2008PRC,Zenihiro2010PRC}.

\item \textbf{Pion photoproduction:}
Coherent neutral-pion photoproduction provides another hadronic probe of nuclear density distributions. When combined with the accurately known proton density from electron scattering, the measured $\left(\gamma,\pi^0\right)$ cross section has been used to extract the neutron density and $R_{\rm{skin}}$ of $^{208}\rm{Pb}$\,\cite{Tarbert2014PRL}. The reliability and precision of this extraction, however, remain sensitive to the description of the reaction mechanism, including pion final-state interactions, charge-exchange processes, medium modifications of the elementary production amplitude, and the predominantly isoscalar character of coherent $\pi^0$
photoproduction\,\cite{Miller2019PRC,Colomer2022PRC}.
More generally, pion dynamics in the nuclear medium is intrinsically complex because pion propagation is coupled to Pauli blocking, $\Delta(1232)$ excitation and its in-medium self-energy, pion absorption, multiple scattering, and charge-exchange processes\,\cite{Ericson1988Book}. These effects are not restricted to pion photoproduction but also play an important role in heavy-ion collisions, where pion production and propagation occur within a dynamically evolving, compressed, and isospin-asymmetric
medium.

\item \textbf{Heavy-ion collisions:}
Heavy-ion reactions provide a complementary dynamical approach to
probing neutron skins\,\cite{Ding2024NST,XuWang2024SCPMA}. In peripheral and semi-peripheral collisions, the overlap region preferentially samples the nuclear surface, allowing the initial difference between neutron and proton density distributions to influence the subsequent reaction dynamics. Neutron-skin information may consequently be encoded in particle yields, momentum distributions, collective flow, and projectile-like fragments. As reviewed in this article, proposed observables at intermediate energies include neutron-to-proton and $\rm{t}/^3\rm{He}$ yield ratios, $\pi^-/\pi^+$ production, hard-photon emission, fragment isoscaling and momentum distributions, and neutron-proton momentum differences\,\cite{Ding2024NST}. In particular, the interpretation of pion observables must account for the complex in-medium dynamics discussed above, together with their coupling to the density and isospin evolution of the collision system. Although the extraction of neutron-skin information from these observables is intrinsically more model dependent than that from PVES, heavy-ion collisions offer the distinctive possibility of tracing how an initial neutron-rich surface is transformed into dynamical signals at different reaction stages and across different density regimes\,\cite{Xu2016PRC,Sorensen2024PPNP}. At ultra-relativistic energies, neutron skins can additionally modify the initial collision geometry, and energy deposition, thereby affecting charged-particle multiplicities, collective-flow observables, and their fluctuations. Comparisons of isobaric systems and precision Pb+Pb data thus provide complementary opportunities to constrain neutron distributions while
disentangling skin effects from nuclear deformation\,\cite{Li2020PRL,Jia2023PRL,Giacalone2023PRL}.

\item \textbf{Astrophysical constraints:}
Neutron-star observations provide an astrophysical connection to
neutron-skin properties through their common dependence on the
isovector EOS of neutron-rich matter\,\cite{Lattimer2014NPA}. In particular,
the density dependence of the nuclear symmetry energy influences
both the neutron pressure that drives the formation of neutron skins in finite nuclei and the pressure of neutron-rich matter relevant to NS structure\,\cite{Lattimer2001ApJ}. Measurements of NS radii and
masses, together with constraints on tidal deformability from
gravitational-wave observations, can therefore provide complementary information on the symmetry energy and, within a specified nuclear model or EOS framework, on neutron-skin thicknesses\,\cite{Tsang2019PLB,Huth2022Nature}. Representative examples include radius constraints from NICER\,\cite{Riley19,Fonseca21,Miller19,Riley21,Miller21,Salmi22,Salmi24,Ditt24,Choud24,Mauviard26,Mauviard25,Miller26} and tidal-deformability constraints from binary NS mergers such as GW170817\,\cite{Abbott2017PRL,Abbott2018PRL}. Unlike electroweak or hadronic measurements on finite nuclei, however, these observations do not probe the neutron skin directly. Their connection to $R_{\rm{skin}}$ is inferred through correlations with the underlying symmetry energy and is consequently sensitive to assumptions concerning the EOS, particularly its behavior beyond densities characteristic of finite nuclei\,\cite{Lattimer2014NPA,Li2021Universe,Chatziioannou2025RMP,Li2025eXTP,Tang2021PRD,Kumar2024LRR,Burgio2024FASS,Miyatsu2025FrontPhys,Xie2026SCPMA,Li2026PRL,Sedrakian2023PPNP,
Sammarruca2025FASS,Tong2025FASS,Yunes2022NRP,Cui2025NST,Xu2026SCPMA}.
\end{enumerate}

These approaches are complementary because they probe neutron-rich matter through different interactions, momentum transfers, density regimes, and physical systems. Electroweak measurements, particularly PVES, provide comparatively clean and direct constraints on neutron radii, whereas hadronic probes generally require a more detailed description of the reaction mechanism. Heavy-ion collisions extend this program into the dynamical domain, connecting the initial neutron-rich surface with particle emission, collective motion, and fragment production during the reaction. Astrophysical observations provide a further, indirect connection by constraining the EOS of neutron-rich matter over densities extending well beyond those encountered in finite nuclei. Comparing and combining these different sources of information is therefore valuable not only for constraining neutron-skin thicknesses, but also for testing the consistency of our description of the nuclear symmetry energy and EOS of neutron-rich matter across finite nuclei, nuclear reactions, and NSs\,\cite{CaiLi2025EPJA,Li2025eXTP}.

\subsection{Symmetry energy and the neutron skin}\label{subs:EsymNSkin}

The formation of a neutron skin reflects the competition between the tendency of the symmetry energy to distribute the neutron excess throughout the nuclear volume and the energetic cost of separating the neutron and proton surfaces. In many nuclear energy-density functionals, the neutron-skin thickness of a heavy nucleus exhibits an approximately linear correlation with the symmetry-energy slope parameter $L$.
A larger $L$ generally corresponds to a larger pressure in neutron-rich matter around saturation density, which energetically favors moving excess neutrons toward the nuclear surface and hence produces a thicker neutron skin\,\cite{Brown2000PRL,Typel2001PRC}. This correlation makes $R_{\rm{skin}}$ an important experimental probe of the isovector nuclear interaction. Its quantitative interpretation, however, depends on the surface symmetry energy, Coulomb interaction, shell structure, and the density dependence of $E_{\rm{sym}}(\rho)$ below saturation.

\begin{figure}[h!]
\centering
\includegraphics[width=0.48\textwidth]{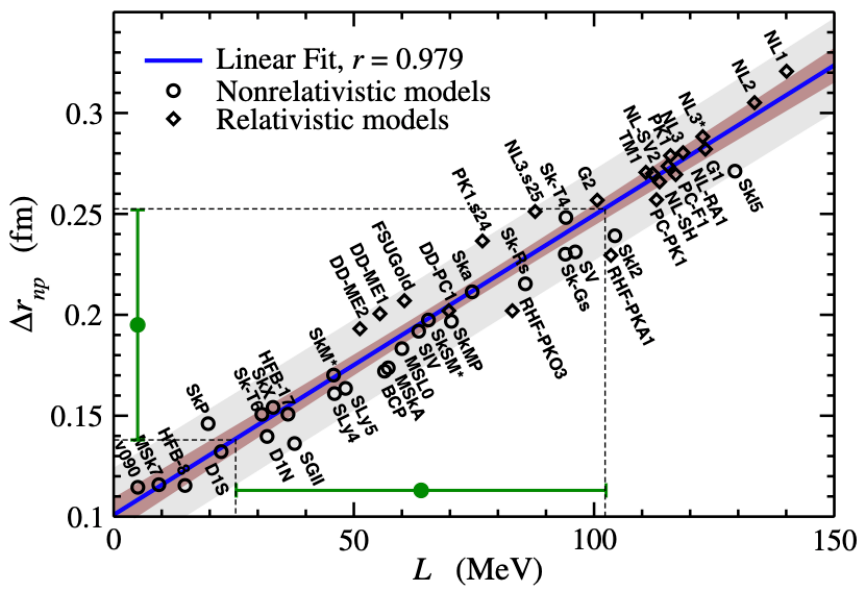}
\caption{(Color Online). Correlation between the neutron-skin thickness $R_{\rm{skin}}(^{208}\rm{Pb})$ and the symmetry-energy slope parameter $L$ predicted by non-relativistic and relativistic
mean-field models. The solid line represents the linear fit
$R_{\rm{skin}}\approx0.101+0.00147L$, with $R_{\rm{skin}}$ in fm and $L$ in MeV. The shaded region illustrates a representative constraint derived by assuming a 3\% measurement of the parity-violating asymmetry in PREX. Taken from Ref.\,\cite{RocaMaza2011PRL}.}
\label{fig:Rskin-L-correlation}
\end{figure}

The role of these finite-size effects can be illustrated in the droplet model. The neutron-skin thickness may be expressed as\,\cite{Centelles2009PRL}
\begin{equation}
R_{\rm{skin}}
\approx
\sqrt{\frac{3}{5}}
\left[
t-\frac{e^2Z}{70S}
+\frac{5}{2R}\left(b_{\rm n}^2-b_{\rm p}^2\right)
\right],
\label{eq:droplet_skin}
\end{equation}
where $R=r_0A^{1/3}$, $b_{\rm n}$ and $b_{\rm p}$ characterize the neutron and proton surface widths, and $t$ is the separation between their mean surface locations. The latter is approximately\,\cite{Centelles2009PRL}
\begin{equation}
t
\approx
\frac{3r_0}{2}
\frac{S/Q}{1+x_A}
\left(I_A-I_C\right),
~~
x_A\approx
\frac{9S}{4Q}A^{-1/3},
~~
I_C\approx
\frac{e^2Z}{20SR},
\label{eq:droplet_t}
\end{equation}
with $I_A=(N-Z)/A\approx\delta$ and $Q$ denoting the surface-stiffness coefficient. These expressions show that the skin is governed directly by the competition between volume and surface symmetry energies, together with Coulomb and surface-width corrections\,\cite{Centelles2009PRL}. The familiar correlation with $L$ (as shown in FIG.\,\ref{fig:Rskin-L-correlation}) emerges because the surface symmetry properties of calibrated nuclear interactions are themselves correlated with the density dependence of the symmetry energy. It should therefore not be regarded as an exact or model-independent relation.

\begin{figure}[h!]
\centering
\hspace{-0.5cm}
\includegraphics[width=0.45\textwidth]{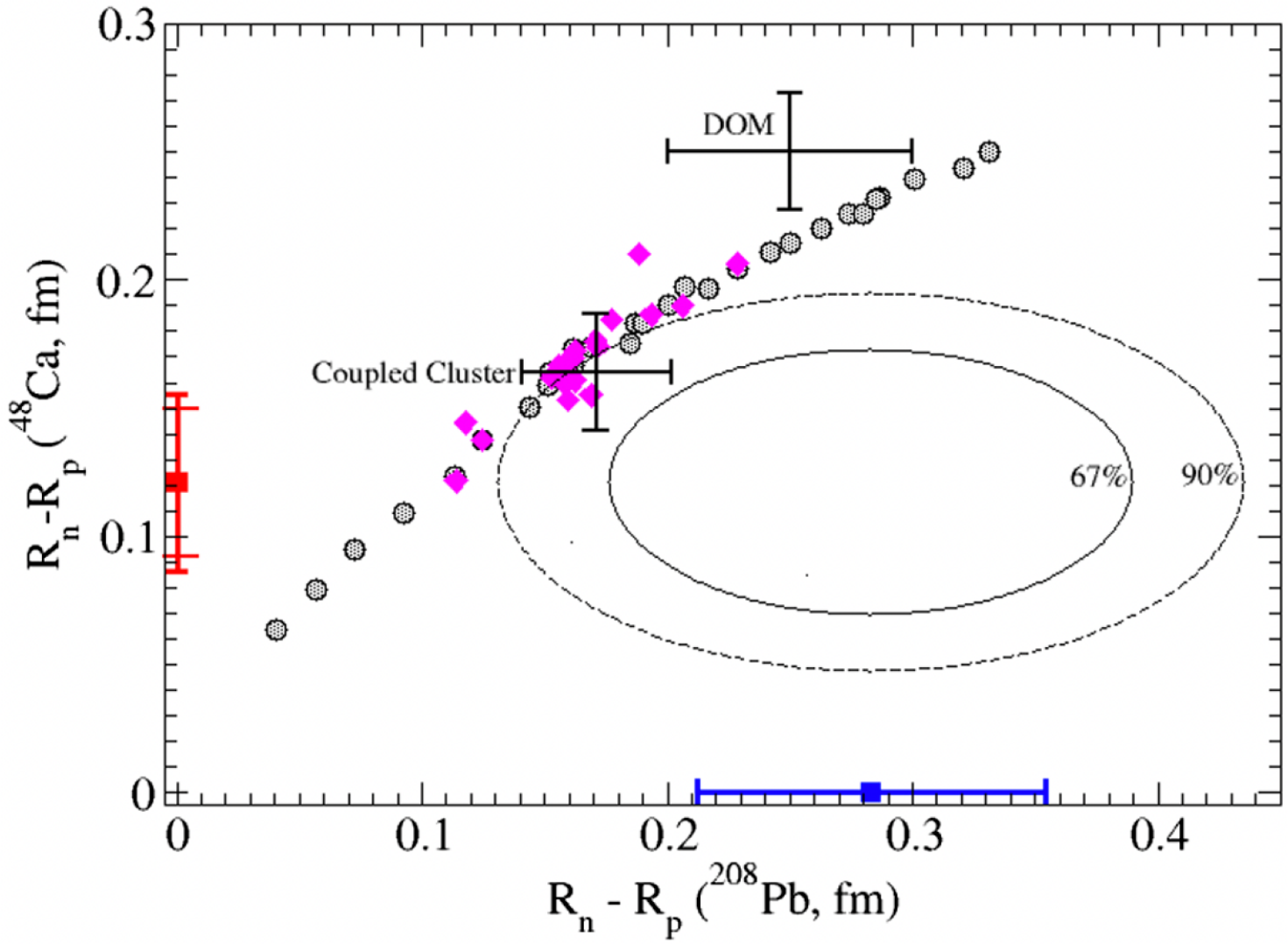}\\[0.25cm]
\includegraphics[width=0.48\textwidth]{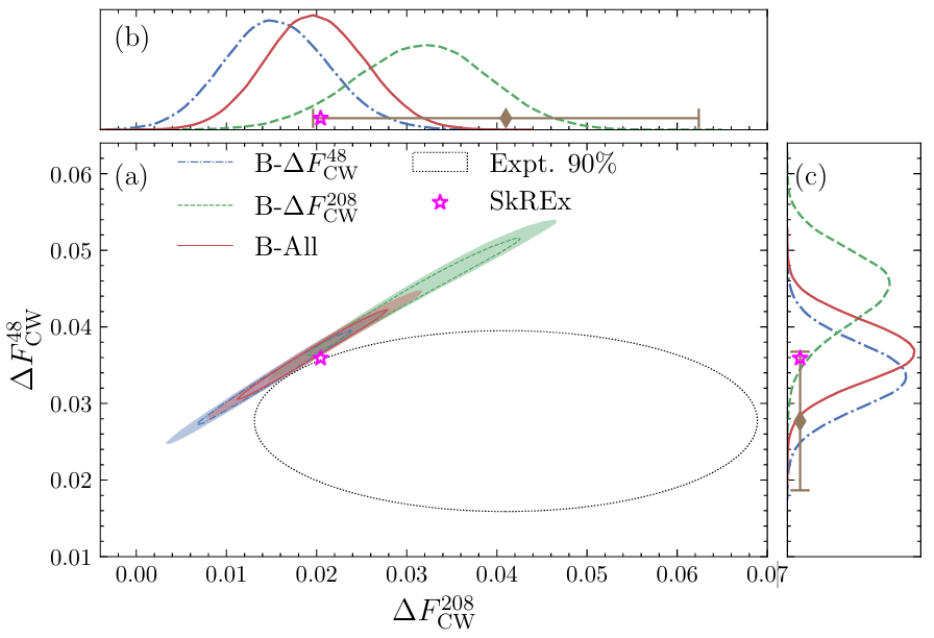}
\caption{(Color Online). Comparison of the PREX-II and CREX
constraints on neutron-rich nuclei.
Upper: neutron-skin thicknesses of $^{208}\rm{Pb}$ and
$^{48}\rm{Ca}$ from PREX-II and CREX, respectively, compared with predictions from representative nuclear models and many-body calculations. The ellipses indicate the joint PREX-II/CREX probability regions.
Lower: Bayesian analysis of the charge-weak form factor
differences measured by PREX-II and CREX within a Skyrme
energy-density-functional framework. The comparison illustrates the tension between the two measurements and the extent to which they can be accommodated within a common theoretical description. The upper and lower panels are adapted from
Refs.\,\cite{CREX,Zhang2023PRC}, respectively.}
\label{fig:prex_crex}
\end{figure}

The PREX-II measurement of $^{208}\rm{Pb}$ yielded
$R_{\rm{skin}}\approx0.283\pm0.071$\,fm, which, when interpreted using conventional $R_{\rm{skin}}$-$L$ correlations, favors a relatively stiff symmetry energy and a large value of $L$\,\cite{PREX2021PRL}. In contrast, CREX obtained
$R_{\rm{skin}}\approx0.121\pm0.026\,(\rm{exp})\pm0.024\,(\rm{model})$\,fm for $^{48}\rm{Ca}$, generally favoring a softer symmetry energy\,\cite{CREX}. As shown in the upper panel of FIG.\,\ref{fig:prex_crex}, simultaneously reproducing the central values of the two measurements remains difficult for many conventional nuclear models, giving rise to the {\it PREX-CREX puzzle.} The Bayesian analysis displayed in the lower panel shows that the two measurements are incompatible at the 68.3\% credible level but become compatible at the 90\% level within the adopted Skyrme framework\,\cite{Zhang2023PRC}. Their combined analysis favors a relatively soft symmetry energy around saturation density, closer to the constraint inferred from CREX alone. This comparison illustrates that translating neutron-skin measurements into $L$ involves nontrivial correlations with other nuclear-matter and finite-nucleus parameters.

\begin{figure}[h!]
\centering
\includegraphics[width=0.48\textwidth]{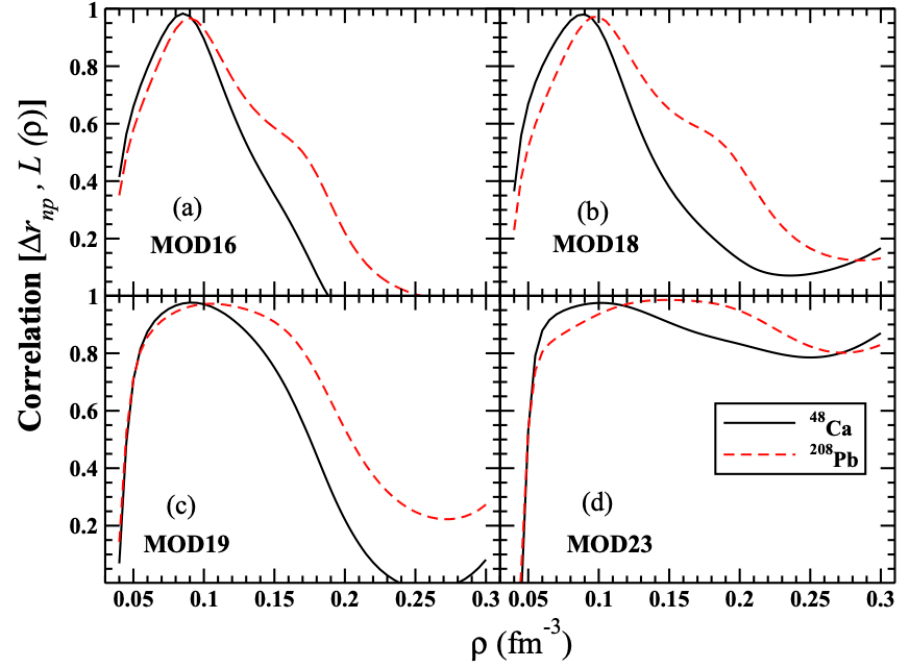}
\caption{(Color Online). Correlation coefficient between the
neutron-skin thicknesses of $^{48}\rm{Ca}$ and $^{208}\rm{Pb}$ and
the symmetry energy slope $L(\rho)$ as a function
of density. Results are shown for four relativistic mean field
parametrizations calibrated to different values of
$R_{\rm{skin}}(^{208}\rm{Pb})$. For the parametrizations predicting
relatively thin neutron skins, the correlation exhibits a pronounced maximum at a sub-saturation density around
$\rho\approx0.10$-$0.12\,\rm{fm}^{-3}$. For larger assumed neutron
skins, the maximum becomes broader, illustrating the model dependence of the density region probed by $R_{\rm{skin}}$. Adapted from
Ref.\,\cite{Mondal2022PRC}.}
\label{fig:skin_Lrho_correlation}
\end{figure}

The interpretation in terms of $L\equiv L(\rho_0)$ also raises a more fundamental question: at which density does the neutron skin most directly constrain the symmetry energy? A finite nucleus is spatially inhomogeneous, and a significant part of its isospin asymmetry is carried by the surface at densities below $\rho_0$. It is therefore useful to define a density-dependent local slope $L(\rho_{\rm r})\equiv3\rho_{\rm r}\partial E_{\rm{sym}}(\rho)/\partial\rho|_{\rho_{\rm{r}}}$.
Droplet-model and energy-density-functional analyses indicate that the symmetry-energy coefficient of a heavy finite nucleus approximately corresponds to $E_{\rm{sym}}(\rho)$ at a characteristic sub-saturation density around $0.10$-$0.11\,\rm{fm}^{-3}$\,\cite{Centelles2009PRL}. In particular, the neutron-skin thickness of heavy nuclei has been found to correlate especially strongly with $L(\rho)$ at a cross density
$\rho_{\rm c}\approx0.11\,\rm{fm}^{-3}$, rather than directly with the slope at $\rho_0$\,\cite{Zhang2013PLB}. The frequently quoted $R_{\rm{skin}}$-$L(\rho_0)$ correlation arises partly because many calibrated interactions correlate the behavior of $E_{\rm{sym}}(\rho)$ at $\rho_{\rm c}$ with its extrapolation to saturation density.
This distinction also helps explain part of the model dependence of the $R_{\rm{skin}}$-$L$ relation. Models having similar values of $L$ may predict different neutron skins if their symmetry-energy slopes differ over the sub-saturation densities sampled by the nuclear surface. An effective slope averaged over the finite-nucleus density profile can therefore describe the neutron-skin systematics more directly than $L$ alone\,\cite{Mondal2016PRC}. Conversely, extracting $L$ from a measured neutron skin requires extrapolating from the sub-saturation region toward $\rho_0$, and the result inevitably depends on the assumed functional form of $E_{\rm{sym}}(\rho)$ and its correlations with higher-order parameters such as $K_{\rm{sym}}$. An example is shown in FIG.\,\ref{fig:skin_Lrho_correlation}\,\cite{Mondal2022PRC}.

This density perspective sharpens the broader connection among neutron skins, HICs, and NS physics. Neutron-skin measurements primarily constrain the isovector interaction at sub-saturation densities, whereas particle-emission and collective-flow observables in HICs probe density windows that depend on the beam energy, collision geometry, and emission stage. NS radii and tidal deformabilities, in turn, depend on the EOS over a wider range extending to supra-saturation densities. Combining these probes is therefore valuable not because they all determine the same parameter $L$, but because their complementary density sensitivities can trace the evolution of $E_{\rm{sym}}(\rho)$ from the dilute nuclear surface toward dense neutron-rich matter. Such comparisons require sufficiently flexible EOS parametrizations and consistent uncertainty propagation so that information obtained near $\rho_{\rm c}$ is not artificially converted into a tight constraint at $\rho_0$ or higher densities.

\subsection{Heavy-ion collisions as neutron skin probes}

The initial stage of a heavy-ion collision is strongly influenced by the neutron and proton density distributions of the colliding nuclei\,\cite{Ma_2023}. In neutron rich nuclei, the spatial extension of the neutron distribution beyond the proton distribution modifies the local isospin asymmetry, particularly in the low-density surface region\,\cite{Sun2010PLB,Dai2014PRC,Ding2024PRC}. Consequently, a thicker neutron skin generally implies:
\begin{enumerate}[label=(\alph*),leftmargin=*]
\item a more extended neutron tail, which increases the neutron
content of the nuclear surface and can enhance neutron emission from the surface region\,\cite{Sun2010PLB,Ding2024PRC};
\item a modified neutron excess in the participant region, thereby changing the isospin composition of the interacting nuclear matter;
and
\item different mean field propagation and nucleon-nucleon collision histories for neutrons and protons, through the interplay between the initial density distributions and the isospin-dependent nuclear interaction\,\cite{Li2008PhysRep}.
\end{enumerate}

The sensitivity to these initial-state differences depends strongly on the collision geometry\,\cite{Ma_2017,Ma_2023,GiacaloneNST,SchenkeeNST,Jia2024NST}.
In central collisions, a large fraction of the two nuclei participates in the reaction and the dynamics can reach densities at or above nuclear saturation density. Observables from such collisions are therefore generally more sensitive to the bulk nuclear EOS and, in particular, to the symmetry energy at relatively high densities\,\cite{Li2002PRL,Xiao2009PRL}. In contrast,
peripheral and semi-peripheral collisions involve substantial overlap of the nuclear surface regions. They preferentially sample the low-density neutron and proton distributions and are consequently more sensitive to the neutron-skin thickness\,\cite{Sun2010PLB,Dai2014PRC,Dai2015PRC}. This geometrical surface sensitivity is one of the principal reasons why peripheral collisions have been widely employed in searches for reaction observables correlated with the neutron skin.

The initial neutron-skin information is not measured directly in a heavy-ion collision. Rather, it is propagated through the dynamical evolution of the reaction and encoded in the phase-space distributions of the emitted particles and residual fragments. An extended neutron distribution can modify the relative numbers and spatial locations of neutron-proton and neutron-neutron collisions, the isospin asymmetry of the participant and spectator matter, and the subsequent emission
and fragmentation processes. As a result, neutron-skin effects can appear in a broad range of observables, including neutron-to-proton and $\rm{t}/^3\rm{He}$ yield ratios\,\cite{Sun2010PLB,Dai2014PRC,
Ding2024PRC}, charged-pion ratios\,\cite{Wei2014PRC}, hard-photon
production\,\cite{Wei2015PRC,Wang2022PRC,Guo2023PRC}, neutron-proton momentum differences\,\cite{Ding2024PRC}, and the isotopic and momentum distributions of projectile-like fragments\,\cite{Ma2013PRC,Dai2015PRC,Ma2024NST}. Since these observables are generated at different stages of the collision and probe different density regions, their sensitivities to the neutron skin need not be identical and can also be entangled with the density dependence of the symmetry energy and other ingredients of the reaction dynamics\,\cite{Tsang2009PRL,Sorensen2024PPNP}.

Transport models at intermediate energies provide the essential
theoretical framework for establishing the connection between the
initial neutron and proton density distributions and experimentally accessible final-state observables\,\cite{Bertsch1988PhysRep,
Aichelin1991PhysRep,Li2008PhysRep,Buss2012PhysRep}. The most commonly employed approaches include Boltzmann-Uehling-Uhlenbeck (BUU) and related Boltzmann transport models, quantum molecular dynamics (QMD) and its isospin-dependent extensions, and antisymmetrized molecular dynamics (AMD). Although their microscopic implementations differ, these approaches evolve the nuclear phase-space distributions under the combined effects of mean field propagation, nucleon-nucleon scattering, Pauli blocking, particle production, and, where appropriate, cluster formation and secondary de-excitation. Neutron-skin
effects can then be studied by systematically varying the initial
neutron density distribution, while keeping other properties of the projectile and target as controlled as possible, and examining the resulting changes in reaction observables\,\cite{Sun2010PLB,
Dai2014PRC,Dai2015PRC,Ding2024PRC}.

An important issue in such analyses is that neutron-skin sensitivity does not by itself imply a model-independent extraction of $R_{\rm{skin}}$. Many of the same observables are also affected by the isovector mean field, the density dependence of the symmetry energy, in-medium nucleon-nucleon cross sections, cluster-production mechanisms, and secondary decay\,\cite{Li2008PhysRep,Buss2012PhysRep,
Xu2016PRC,Ono2019PRC}. A useful neutron-skin observable should
therefore exhibit not only a sizable response to $R_{\rm{skin}}$,
but also a sufficiently robust correlation against these competing dynamical uncertainties. Ratios, double ratios, and comparisons between different centrality classes are particularly useful in this respect because common systematic and model dependences can be partially reduced\,\cite{Tsang2009PRL}. The complementary use of several observables probing different stages of the reaction provides a further means of separating genuine initial-state neutron-skin effects from uncertainties associated with the subsequent collision dynamics\,\cite{Ding2024NST,Sorensen2024PPNP}.

\section{Particle Emission Observables Sensitive to Neutron Skin}
\label{sec:observables}

Particle emission in heavy-ion collisions provides a variety of
complementary observables for probing the neutron-rich surface of
atomic nuclei and hence the neutron-skin thickness. The underlying
sensitivity arises because the initial neutron and proton density
distributions influence the composition, geometry, and isospin
content of the participant and spectator regions, which are
subsequently reflected in the yields, momentum distributions, and
correlations of emitted particles and fragments. Different
observables probe different stages and density regions of the
reaction. Neutron-to-proton and $\rm{t}/^3\rm{He}$ yield ratios are
particularly sensitive to the isospin composition of the low-density surface and generally exhibit enhanced neutron-skin sensitivity in peripheral collisions. Charged-pion production provides a connection between neutron-skin effects and the symmetry energy, with central collisions predominantly probing the high-density participant region and peripheral collisions becoming increasingly sensitive to the initial neutron-rich surface. Hard photons offer a complementary electromagnetic probe through proton-neutron bremsstrahlung and are especially sensitive to the collision geometry and surface neutron distribution in peripheral reactions. Projectile-like fragments can retain information on the initial neutron distribution through their isotopic composition and parallel-momentum distributions, while pre-equilibrium nucleon emission and neutron-proton momentum differences provide access to the early isospin-dependent reaction dynamics. Taken together, these observables demonstrate that neutron-skin information can be encoded in several distinct stages of heavy-ion collisions, from early nucleon and photon emission to pion production and the formation of final-state fragments. Their different sensitivities to collision centrality, density, and reaction dynamics make them complementary rather than interchangeable probes of the neutron-skin thickness.
Beyond collision-induced particle emission, coherent neutral-pion
photoproduction provides a conceptually distinct photonuclear probe of the neutron density and its surface profile. Its quantitative sensitivity to $R_{\rm{skin}}$, however, remains strongly dependent on the reaction model, particularly the predominantly isoscalar production amplitude
and the treatment of pion final-state interactions.

\subsection{Light particles and clusters: neutron/proton and t/$^3$He ratios}

Light particles — neutrons, protons, deuterons, tritons, and $^3$He — are copiously emitted in heavy-ion collisions across all energy regimes. The neutron-to-proton yield ratio is among the most direct probes of the isospin asymmetry of the collision system.

\subsubsection{\it Neutron-to-proton yield ratio $R_{\rm{np}}$}

The neutron-to-proton yield ratio, 
\begin{equation}
R_{\rm{np}}=Y_{\rm n}/Y_{\rm p},
\end{equation}
provides a direct probe of the isospin content of the emitting source and is sensitive to the neutron distribution in the nuclear surface. In neutron-rich nuclei, an increased neutron-skin thickness enhances the neutron abundance in the low-density surface region, leading to preferential neutron emission, particularly in peripheral collisions where the reaction dynamics is more strongly influenced by the nuclear surface. Transport-model calculations have demonstrated an approximately linear correlation between $R_{\rm{np}}$ and the neutron-skin thickness $R_{\rm{skin}}$, with the sensitivity generally becoming stronger toward peripheral collisions and for particles emitted from the projectile-like region.

\begin{figure}[h!]
\centering
\includegraphics[width=0.4\textwidth]{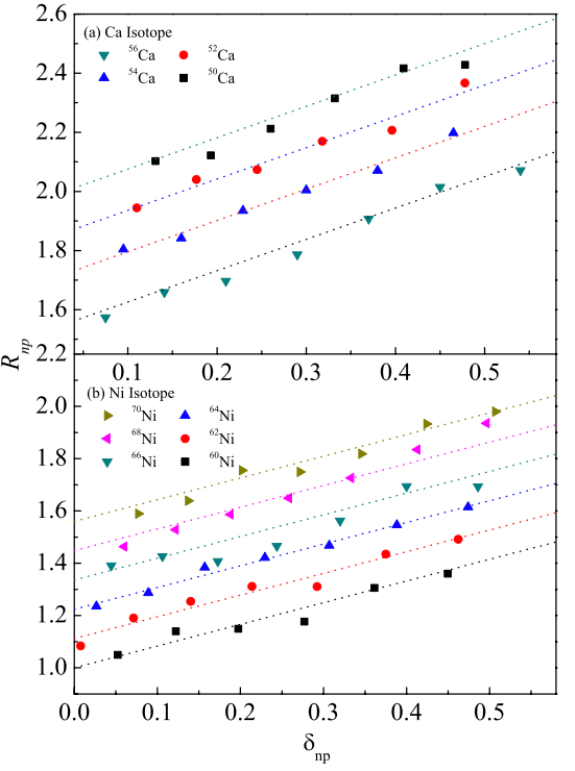}
\caption{(Color Online). Neutron-to-proton yield ratio $R_{\rm{np}}$ as a function of
the neutron-skin thickness $R_{\rm{skin}}$ (denoted as $\delta_{\rm{np}}$ in the figure) for
$^{50,52,54,56}$Ca (upper panel) and $^{60,62,64,66,70}$Ni
(lower panel) projectiles incident on a $^{12}$C target at
50$A$\,MeV, calculated with the IQMD model. Peripheral collisions
with $0.8<b/b_{\max}<1.0$ and positive rapidity $Y>0$ are selected.
The approximately linear increase of $R_{\rm{np}}$ with
$R_{\rm{skin}}$ demonstrates the sensitivity of projectile-side
nucleon emission to the initial neutron distribution.
Taken from Ref.\,\cite{Sun2010PLB}.}
\label{fig_Sun2010PLB}
\end{figure}

Such a correlation was investigated within the isospin-dependent
quantum molecular dynamics (IQMD) model for neutron-rich Ca and Ni
isotopes incident on a $^{12}$C target at 50$A$\,MeV\,\cite{Sun2010PLB}. By varying the
diffuseness of the neutron density distribution while keeping the proton
distribution fixed, different neutron-skin thicknesses were generated
for each projectile. FIG.\,\ref{fig_Sun2010PLB} shows the resulting
$R_{\rm{np}}$ as a function of $R_{\rm{skin}}$ for several Ca and Ni
isotopes. Peripheral collisions with $0.8<b/b_{\max}<1.0$ were selected, together with a positive-rapidity cut $Y>0$ to enhance the contribution from projectile nucleons.
A clear approximately linear increase of $R_{\rm{np}}$ with
$R_{\rm{skin}}$ is observed for all the isotopes considered. The
correlation can be parametrized as $
R_{\rm{np}}\approx a+bR_{\rm{skin}}$, with average slopes of about $1.06\,\rm{fm}^{-1}$ and 0.83\,fm$^{-1}$ for the Ca and Ni
isotope chains, respectively. The systematic behavior indicates that
the neutron-to-proton yield ratio retains direct information on the
initial neutron distribution of the projectile and may therefore serve
as a reaction observable for constraining neutron-skin thickness.

\subsubsection{\it Triton-to-$^3$He yield ratio $R_{\rm{t}/^3\text{He}}$}

The triton-to-$^3\rm{He}$ yield ratio,
$R_{\rm{t}/^3\rm{He}}=Y_{\rm t}/Y_{^3\rm{He}}$, provides a complementary
observable to the neutron-to-proton yield ratio for probing the
isospin composition of the emitting source. Since tritons contain
two neutrons and one proton, whereas $^3\rm{He}$ nuclei contain one
neutron and two protons, their relative production is sensitive to
the local neutron-to-proton composition during cluster formation.
Within a simple coalescence picture, one expects approximately
$Y_{\rm t}\sim \rm{n}^2\rm{p}$ and $Y_{^3\rm{He}}\sim\rm{np}^2$, and hence
$R_{\rm{t}/^3\rm{He}}\sim \rm{n}/\rm{p}$. The $\rm{t}/^3\rm{He}$ ratio can therefore
retain information similar to that contained in the free
neutron-to-proton yield ratio, while involving only charged
particles that are experimentally easier to detect.

\begin{figure}[h!]
\centering
\includegraphics[width=0.41\textwidth]{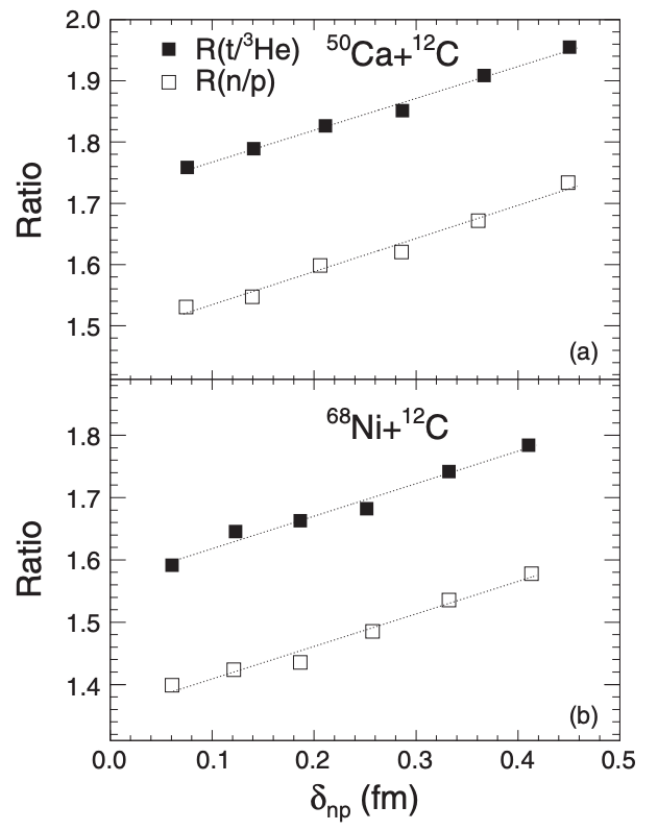}\\
\includegraphics[width=0.41\textwidth]{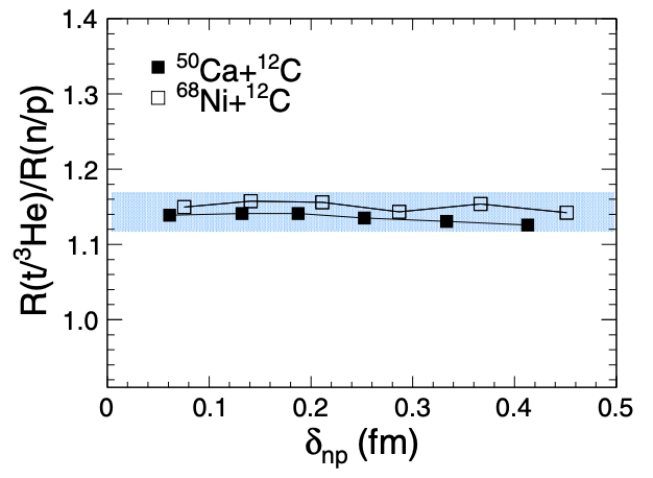}
\caption{(Color Online). Upper panels: Neutron-to-proton yield ratio $R_{\rm{np}}$ and
triton-to-$^3\rm{He}$ yield ratio $R_{\rm{t}/^3\rm{He}}$ as functions of the neutron-skin thickness $R_{\rm{skin}}$ (denoted as $\delta_{\rm{np}}$ in the figure) for
$^{50}$Ca+$^{12}$C (panel (a)) and $^{68}$Ni+$^{12}$C (panel (b)) collisions at
50\,MeV/nucleon. Both observables exhibit an approximately linear
increase with $R_{\rm{skin}}$. Lower panel: The corresponding double ratio
$R_{\rm{t}/^3\rm{He}}/R_{\rm{np}}$, which remains nearly constant with
increasing $R_{\rm{skin}}$, indicating an approximate proportionality
between $R_{\rm{t}/^3\rm{He}}$ and $R_{\rm{np}}$. The calculations are
performed for $0.6<b/b_{\max}<1.0$ and $y>0$ within the IQMD model.
Adapted from Ref.\,\cite{Dai2014PRC}.}
\label{fig_Dai2014PRC}
\end{figure}

The sensitivity of $R_{\rm{t}/^3\rm{He}}$ to the neutron skin was
investigated within the IQMD model for semiperipheral
$^{50}$Ca+$^{12}$C and $^{68}$Ni+$^{12}$C collisions at
50\,MeV/nucleon. Upper two panels of FIG.\,\ref{fig_Dai2014PRC} show both
$R_{\rm{t}/^3\rm{He}}$ and $R_{\rm{np}}$ as functions of the
neutron-skin thickness $R_{\rm{skin}}$. A positive-rapidity cut,
$y>0$, is imposed to enhance the contribution from the neutron-rich
projectile, while $0.6<b/b_{\max}<1.0$ selects semiperipheral and
peripheral collisions that are particularly sensitive to the nuclear
surface. Both ratios exhibit a strong, approximately linear increase
with $R_{\rm{skin}}$ for the $^{50}$Ca and $^{68}$Ni projectiles.
Moreover, $R_{\rm{t}/^3\rm{He}}$ follows essentially the same trend as
$R_{\rm{np}}$, although its magnitude is systematically larger.
This close correspondence is further illustrated in
the lower panel of FIG.\,\ref{fig_Dai2014PRC}, where the double ratio
$R_{\rm{t}/^3\rm{He}}/R_{\rm{np}}$ remains nearly independent of
$R_{\rm{skin}}$ for both reaction systems. This behavior indicates
an approximate proportionality between the two observables, $
R_{\rm{t}/^3\rm{He}}\approx C R_{\rm{np}}$, with $C$ being nearly independent of $R_{\rm{skin}}$, consistent
with the coalescence picture of light-cluster formation. Consequently,
the neutron-skin dependence carried by the emitted neutron-to-proton
ratio is also reflected in the relative production of the mirror
nuclei t and $^3\rm{He}$. Since both t and $^3\rm{He}$ are
charged particles, $R_{\rm{t}/^3\rm{He}}$ provides an experimentally
attractive alternative to direct neutron measurements for probing
the neutron-rich nuclear surface.

A more systematic investigation was recently performed by Ding
et al. in Ref.\,\cite{Ding2024PRC}, who extended the IQMD calculations to a broader set of neutron-rich Ca, Mg, and Ne isotopes incident on a $^{12}$C target at 50\,MeV/nucleon. As shown in their systematic comparison, $R_{\rm{np}}$ exhibits a robust, approximately linear increase with $R_{\rm{skin}}$ for all the isotopes considered. For the most neutron-rich systems, such as $^{60}$Ca, $^{37}$Mg, and $^{31}$Ne, a similarly strong correlation is also observed for $R_{\rm{t}/^3\rm{He}}$, supporting the basic conclusion of Ref.\,\cite{Dai2014PRC} that the $\rm{t}/^3\rm{He}$ ratio can carry information on the neutron-rich surface of the projectile. The extended isotope-systematics, however, also reveals an important limitation of the cluster observable. In contrast to
$R_{\rm{np}}$, whose correlation with $R_{\rm{skin}}$ remains
robust over the systems studied, the dependence of $R_{\rm{t}/^3\rm{He}}$ on $R_{\rm{skin}}$ becomes considerably weaker
for projectiles closer to the $\beta$-stability line. The
$\rm{t}/^3\rm{He}$ observable therefore appears to be most effective
for very neutron-rich nuclei with large $N/Z$, rather than being a universal substitute for $R_{\rm{np}}$. This systematic study thus both supports and qualifies the earlier result of
Ref.\,\cite{Dai2014PRC}: light-cluster ratios provide an experimentally attractive probe of neutron-skin effects, but their sensitivity depends more strongly on the isospin asymmetry of the projectile than does the free neutron-to-proton ratio.

\subsection{Pion production: $\pi^-/\pi^+$ ratio as a high-density probe}

At intermediate beam energies, charged pions are produced predominantly through the excitation and subsequent decay of $\Delta(1232)$ resonances in nucleon-nucleon collisions. Owing to its sensitivity to the isospin asymmetry of the participant matter, the $\pi^-/\pi^+$ ratio has long been proposed as a probe of the density dependence of the nuclear symmetry energy, particularly near the pion-production threshold\,\cite{Li2002PRL,Li2003PRC,Yong2006PRC,
Xiao2009PRL,Xiao2014EPJA,Li2015PRC}. Early transport studies by Li and collaborators demonstrated that the high-density symmetry energy controls the degree of isospin fractionation in the compressed participant matter and thereby modifies the relative production of the different $\Delta$ charge states and their daughter pions\,\cite{Li2002PRL,Li2003PRC}. These studies further showed that the sensitivity is generally enhanced for heavy neutron-rich systems and beam energies close to the pion-production threshold, while double $\pi^-/\pi^+$ ratios between isotopic reactions may reduce part of the systematic uncertainty\,\cite{Yong2006PRC}. Comparisons of IBUU
calculations with the FOPI data initially favored a relatively soft symmetry energy above saturation density\,\cite{Xiao2009PRL}. However, subsequent investigations emphasized that this inference depends appreciably on the treatment of pion production, absorption, and rescattering, as well as the mean-field potentials of pions and $\Delta$ resonances\,\cite{Xiao2014EPJA,Li2015PRC}. The $\pi^-/\pi^+$ ratio should therefore be regarded as a potentially powerful but intrinsically model dependent probe of the high-density isovector interaction.

\begin{figure}[h!]
\centering
\includegraphics[width=0.48\textwidth]{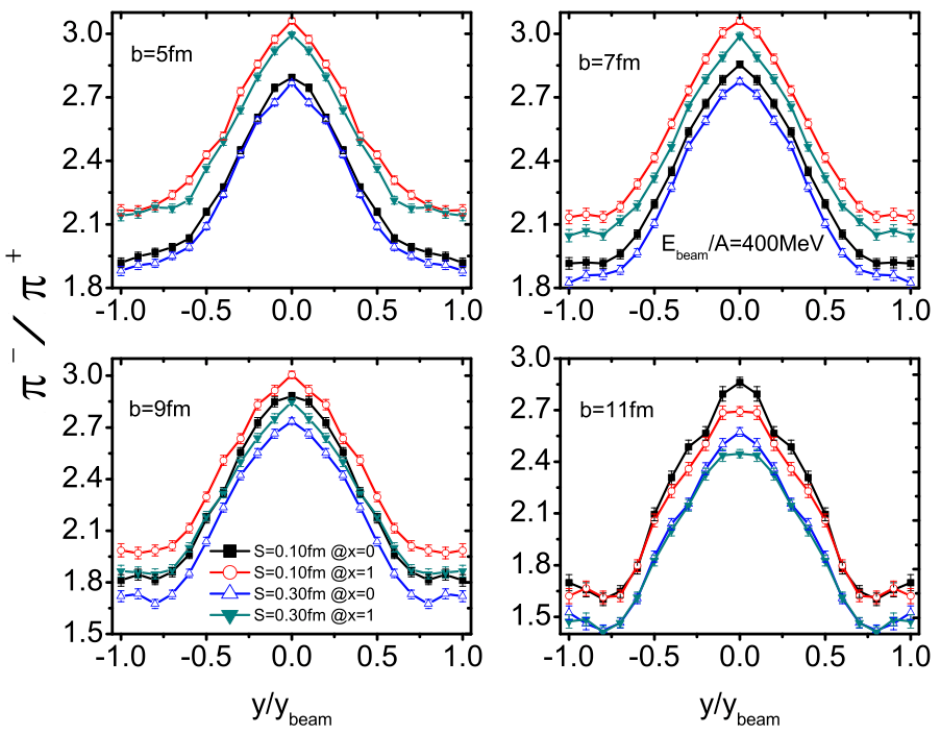}
\caption{(Color Online). Influence of the neutron-skin thickness on the
$\pi^-/\pi^+$ ratio in $^{208}$Pb+$^{208}$Pb collisions at
400\,MeV/nucleon. The neutron-skin effect is relatively small in
central collisions but increases substantially toward peripheral
collisions, where it becomes comparable to or even larger than the symmetry-energy effect. The results demonstrate the different roles played by the symmetry energy and the neutron-rich surface of the colliding nuclei as the impact parameter increases. Adapted from Ref.\,\cite{Wei2014PRC}.}
\label{fig:pi_ratio}
\end{figure}

An important distinction between central and peripheral collisions was demonstrated in Ref.\,\cite{Wei2014PRC} using the IBUU11 transport
model. In central and midcentral $^{208}$Pb+$^{208}$Pb collisions,
the participant region reaches suprasaturation densities and the
$\pi^-/\pi^+$ ratio is primarily sensitive to the high-density
symmetry energy, with only a weak dependence on the uncertainty of
the initial neutron-skin thickness. In contrast, as the impact
parameter increases, the neutron-rich surface plays an increasingly
important role. The neutron-skin effect on the charged-pion ratio
then grows rapidly and can become comparable to, or even larger than, the symmetry-energy effect in peripheral collisions, particularly at
400\,MeV/nucleon, as shown in FIG.\,\ref{fig:pi_ratio}, here the $x$ characterizes the density dependence of $E_{\rm{sym}}(\rho)$.

This centrality dependence also changes the density region probed by
the observable. While central collisions provide access to the
symmetry energy at suprasaturation densities, the participant matter
in sufficiently peripheral collisions is dominated by densities below $\rho_0$. Consequently, the ordering of the $\pi^-/\pi^+$ ratio between soft and stiff symmetry energies may even be reversed from central to peripheral collisions. Peripheral pion production can therefore provide complementary information on the neutron-rich
surface and neutron-skin thickness of the colliding nuclei, whereas
central collisions are better suited for probing the high-density
symmetry energy. A well-defined centrality or impact-parameter
selection is thus essential when interpreting the $\pi^-/\pi^+$
ratio in terms of either the symmetry energy or neutron-skin effects.

\subsection{Photon emission: bremsstrahlung and hard photon production}

Hard photons ($E_\gamma \gtrsim 30$\,MeV) in intermediate-energy
heavy-ion collisions are emitted primarily from incoherent
proton-neutron bremsstrahlung, $\rm{p}+\rm{n}\rightarrow \rm{p}+\rm{n}+\gamma$.
In particular, direct hard photons originate mainly from the early
stage of the reaction and constitute the dominant component of
hard-photon emission. Owing to their weak final-state interactions
with the surrounding nuclear medium, they can retain information on
the initial neutron and proton density distributions. The neutron-skin
thickness can therefore affect hard-photon emission by modifying the
spatial distribution of neutrons and, consequently, the number and
geometry of effective pn collisions in the projectile-target
overlap region.
The possibility of probing neutron-skin thickness through hard-photon
production was investigated in Ref.\,\cite{Wei2015PRC} using
intermediate-energy proton-induced reactions. Since hard photons are
predominantly produced through pn bremsstrahlung, the incident
protons provide a means of sampling the neutron distribution,
particularly in the nuclear surface. This study provided an early
indication that hard-photon observables could carry information on
the neutron-skin thickness.

The neutron-skin effect on direct hard-photon emission in heavy-ion
collisions was subsequently investigated systematically by Wang
et al.\,\cite{Wang2022PRC} within the IQMD model for
$^{50}$Ca+$^{12}$C and $^{50}$Ca+$^{40}$Ca reactions. Different
neutron-skin thicknesses of $^{50}$Ca were generated by varying the
diffuseness of its neutron density distribution while keeping the
proton distribution fixed. The calculations show that the neutron-skin
effect is particularly pronounced in peripheral collisions, where the
reaction preferentially samples the neutron-rich surface. A thicker
neutron skin enhances the opportunity for incoherent pn
bremsstrahlung in such collisions, resulting in increased direct
hard-photon production.

\begin{figure}[h!]
\centering
\includegraphics[width=0.48\textwidth]{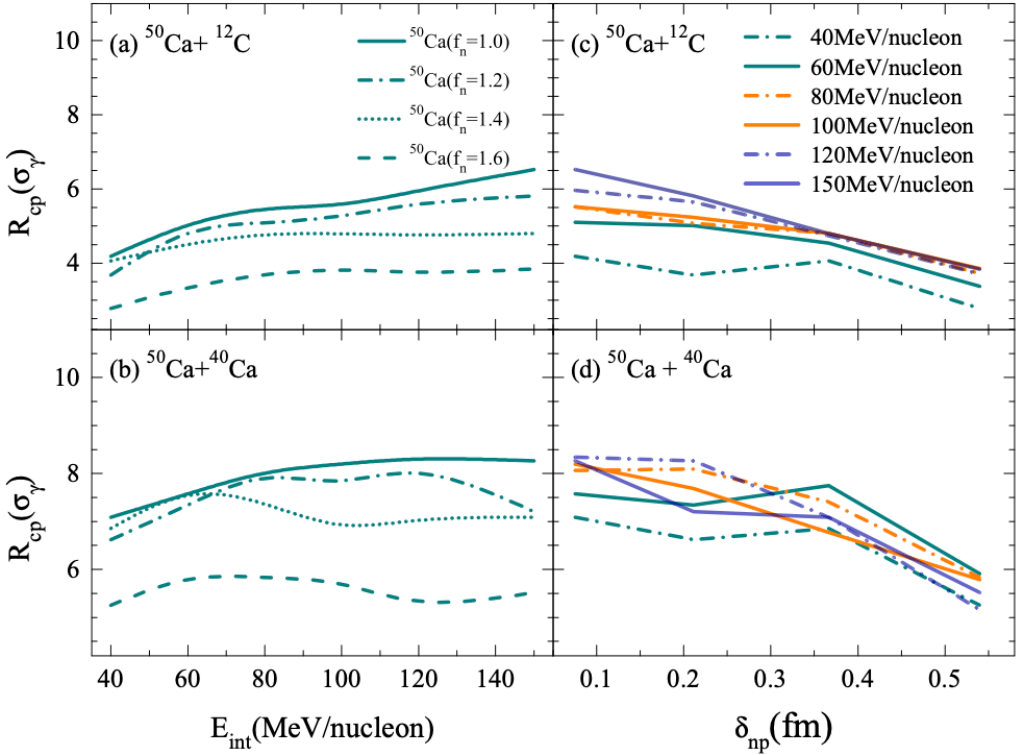}
\caption{(Color Online). Neutron-skin effect on the central-to-peripheral yield ratio
$R_{\rm{cp}}(\sigma_\gamma)$ of direct hard photons in
$^{50}$Ca+$^{12}$C (upper panels) and $^{50}$Ca+$^{40}$Ca
(lower panels) reactions. The left panels show the incident-energy
dependence for different neutron-density diffuseness parameters of
$^{50}$Ca, while the right panels show the dependence on the
neutron-skin thickness $R_{\rm{skin}}\equiv\delta_{\rm{np}}$ at different incident
energies. The overall decrease of $R_{\rm{cp}}(\sigma_\gamma)$ with
increasing $\delta_{\rm{np}}$ reflects the enhanced hard-photon
production from the neutron-rich surface in peripheral collisions.
The sensitivity to the neutron skin is more pronounced for the
$^{50}$Ca+$^{12}$C system. Adapted from
Ref.\,\cite{Wang2022PRC}.}
\label{fig:hard-photon-skin}
\end{figure}

A particularly sensitive measure of the neutron-skin effect is the
central-to-peripheral yield ratio\,\cite{Wang2022PRC}
\begin{equation}
R_{\rm{cp}}(\sigma_\gamma)
=
\frac{\sigma_\gamma(\mathrm{central})}
     {\sigma_\gamma(\mathrm{peripheral})},
\end{equation}
where central and peripheral collisions correspond to centralities
of 0-10\% and 80-100\%, respectively. As shown in
FIG.\,\ref{fig:hard-photon-skin}, $R_{\rm{cp}}(\sigma_\gamma)$
generally decreases with increasing neutron-skin thickness. This
behavior reflects the different sensitivities of central and
peripheral collisions to the initial neutron distribution. Whereas
central collisions involve a substantial fraction of the projectile,
peripheral collisions preferentially probe its surface and are
therefore more strongly affected by an extended neutron distribution.
Increasing the neutron-skin thickness consequently enhances peripheral
hard-photon production more strongly than central production, thereby
reducing $R_{\rm{cp}}(\sigma_\gamma)$. Among the hard-photon
observables considered in Ref.\,\cite{Wang2022PRC},
$R_{\rm{cp}}(\sigma_\gamma)$ exhibits a stronger sensitivity to
neutron-skin thickness than either the absolute photon yield or the
yield ratio between two similar reactions.

The interpretation of hard photons as neutron-skin probes can,
however, be influenced by additional aspects of the nucleon
phase-space distribution. Guo et al.\,\cite{Guo2023PRC} investigated the interplay between the coordinate-space neutron skin and the momentum-space proton skin associated with nucleon short-range correlations (SRCs) based on the earlier work of Cai et al.\,\cite{Cai16b}. Their study indicates that momentum-space modifications can significantly affect hard-photon production and should therefore be considered when extracting neutron-skin information quantitatively, see relevant discussions of Section \ref{SEC_Pskin}.
Nevertheless, from the perspective of coordinate-space neutron-skin effects, the results of Refs.\,\cite{Wei2015PRC,Wang2022PRC} support
direct hard-photon emission as a complementary electromagnetic probe of the neutron-rich nuclear surface. 

A recent CSHINE measurement demonstrates that the similar hard-photon channel is also sensitive to the initial nucleon momentum
distribution. In $^{124}\rm{Sn}+^{124}\rm{Sn}$ collisions at
25\,MeV/nucleon, Xu et al.\,\cite{JHXu25PRR,JHXu2026PRC} measured the neutron-proton
bremsstrahlung spectrum and compared the detector-filtered data with IBUU calculations employing different fractions of SRC-induced high-momentum nucleons. Nucleons in the high-momentum tail (HMT) possess larger relative momenta and therefore generate a harder photon spectrum through
$\rm{n}+\rm{p}\rightarrow\rm{n}+\rm{p}+\gamma$. From a likelihood
analysis of the measured spectrum, an HMT fraction of
$R_{\rm{HMT}}\approx(20\pm3)\%$ was extracted for $^{124}\rm{Sn}$,
providing quantitative evidence that bremsstrahlung photons from
low-energy heavy-ion collisions can probe SRC-induced high-momentum components.
This result is directly relevant to neutron-skin studies because
hard-photon production reflects both the coordinate-space distributions of neutrons and protons and their momentum-space distributions. An experimentally constrained HMT fraction can therefore reduce an important uncertainty in the initialization of transport calculations and help disentangle an enhancement caused by a spatially extended neutron-rich surface from that generated by energetic SRC pairs. Earlier CSHINE measurements of
$^{86}\rm{Kr}+^{124}\rm{Sn}$ and subsequent photon-spectrum
reconstruction analyses had already established the feasibility of this approach\,\cite{YHQin24PLB-a,JHXu24PLB}; the higher-precision $^{124}\rm{Sn}+^{124}\rm{Sn}$ result substantially strengthens this connection. Nevertheless, the quantitative inference remains conditional on the transport dynamics, the elementary $\rm{np}\rightarrow\rm{np}\gamma$ emission probability, and the assumed form and isospin dependence of the HMT. Further discussions of the momentum-space proton skin and SRCs as well as their connections to neutron skins are given in Section \ref{SEC_Pskin}.

\subsection{Fragments: parallel momentum distributions and isoscaling}

\begin{figure*}
\centering
\includegraphics[width=0.95\textwidth]{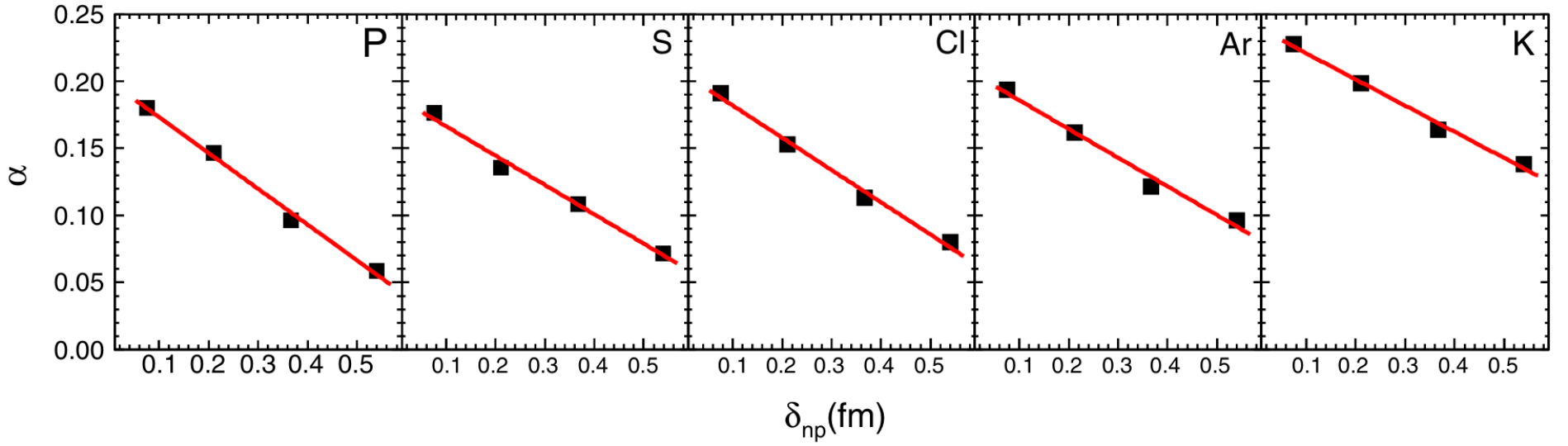}
\caption{(Color Online). Parameter $\alpha$ as a function of neutron skin thickness $R_{\rm{skin}}=\delta_{\rm{np}}$ for projectile-like
fragments with charge numbers $Z=15$-19 produced in
$^{50}$Ca+$^{12}$C collisions at 50\,MeV/nucleon. The approximately linear decrease of $\alpha$ with increasing $\delta_{\rm{np}}$ demonstrates the sensitivity of the isoscaling behavior of heavy projectile residues to the initial neutron distribution. Adapted from Ref.\,\cite{Dai2015PRC}.}
\label{fig:fragment-isoscaling}
\end{figure*}

Projectile-like fragments (PLFs) produced in peripheral and
semi-peripheral collisions can retain memory of the initial neutron
and proton distributions of the projectile, although their final
isotopic composition is modified by nucleon abrasion, transfer, and
subsequent evaporation\,\cite{Dai2015PRC,Ding2024NST}. Since these
processes preferentially sample the nuclear surface in peripheral
collisions, the neutron-skin thickness can influence the isotopic
composition and momentum distributions of PLFs. In particular,
changing the neutron skin modifies the number and spatial distribution
of neutrons in the surface region and therefore their probability of
being removed during the reaction. Fragment parallel-momentum
distributions provide another manifestation of this surface
sensitivity: recent LQMD calculations indicate that the width of the
low-momentum side of the PLF parallel-momentum distribution,
$\Gamma_{\rm{L}}$, is correlated with the neutron-skin thickness
\cite{Ma2024NST}.

An important observable characterizing the isotopic distributions of
fragments is the isoscaling behavior. For two similar reactions with
different isospin asymmetries, the ratio of fragment yields can be
approximately expressed as
\begin{equation}
R_{21}(N,Z)
=
\frac{Y_2(N,Z)}{Y_1(N,Z)}
=
C\exp(\alpha N+\beta Z),
\end{equation}
where $\alpha$ and $\beta$ are the isoscaling parameters and $C$ is
an overall normalization constant. Within a grand-canonical
description, $\alpha$ is related to the difference between the neutron chemical potentials of the two systems and thus provides a measure of their different neutron contents.
The sensitivity of isoscaling to the neutron skin was investigated
by Dai et al.\,\cite{Dai2015PRC} using the IQMD model followed by
GEMINI for peripheral $^{50}$Ca+$^{12}$C and
$^{48}$Ca+$^{12}$C collisions at 50\,MeV/nucleon. Different
neutron-skin thicknesses of $^{50}$Ca were generated by varying the
diffuseness of its neutron density distribution. The calculations
show that increasing the neutron-skin thickness suppresses the
production of neutron-rich PLFs. Consequently, the extracted
isoscaling parameter $\alpha$ decreases approximately linearly with
increasing $\delta_{\rm{np}}$, as shown in
FIG.\,\ref{fig:fragment-isoscaling}.

\begin{figure}[h!]
\centering
\includegraphics[width=0.42\textwidth]{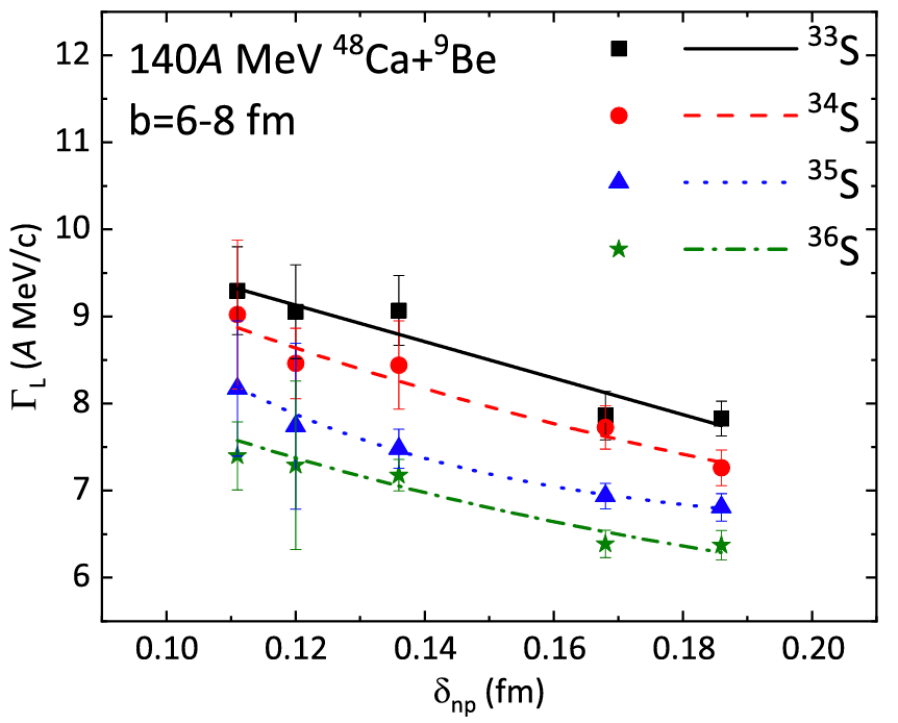}
\caption{(Color Online). Correlation between the width $\Gamma_{\rm{L}}$ of the
low-momentum side of the parallel momentum distribution
$p_{\parallel}$ and the neutron-skin thickness $\delta_{\rm{np}}$
of $^{48}$Ca for neutron-rich $^{33-36}$S fragments produced in
peripheral $^{48}$Ca+$^{9}$Be collisions at 140A\,MeV with
$b=6$-8\,fm. The lines represent exponential fits to the LQMD
calculations, illustrating the sensitivity of $\Gamma_{\rm{L}}$
to the neutron-skin thickness. Adapted from Ref.\,\cite{Ma2024NST}.}
\label{fig:fragment_parallel_momentum}
\end{figure}

The physical origin of this behavior can be understood from the
surface character of peripheral fragmentation. A larger
$\delta_{\rm{np}}$ redistributes more neutrons toward the nuclear
surface, where they are more loosely bound and can be more readily
abraded during the collision or emitted during subsequent
de-excitation. The surviving projectile residues consequently become
less neutron rich, resulting in smaller values of $\alpha$\,\cite{Dai2015PRC}. Heavy PLFs are particularly useful in this respect
because a substantial fraction of their nucleons remain spectators,
allowing the residues to preserve information on the initial neutron
and proton density distributions. The nearly linear
$\alpha$-$\delta_{\rm{np}}$ correlation therefore suggests that
isoscaling of projectile-like residues can provide a probe of the
neutron-skin thickness.

Complementary information may be obtained from the parallel momentum
distribution of projectile fragments. For $^{48}$Ca+$^{9}$Be
fragmentation at 140\,MeV/nucleon, Ref.\,\cite{Ma2024NST}
found that the asymmetric parallel-momentum distribution
$p_{\parallel}$ of PLFs can be characterized by different widths
$\Gamma_{\rm{L}}$ and $\Gamma_{\rm{R}}$ on its low- and
high-momentum sides. In particular, $\Gamma_{\rm{L}}$ for
neutron-rich fragments exhibits a sensitive correlation with
$\delta_{\rm{np}}$, shown in FIG.\,\ref{fig:fragment_parallel_momentum}, suggesting the parallel-momentum distribution
as an experimentally accessible complementary probe of the
neutron-rich surface. Thus, fragment isoscaling and momentum
distributions provide two related but distinct ways of encoding
neutron-skin information in peripheral projectile fragmentation
reactions.

\begin{figure}[h!]
\centering
\includegraphics[width=0.45\textwidth]{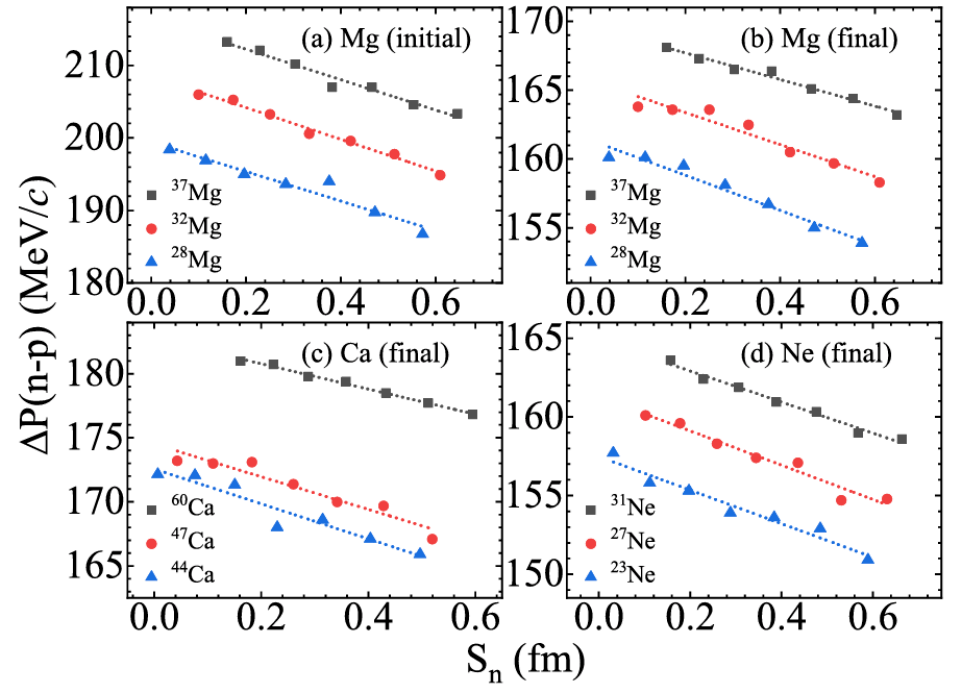}
\caption{(Color Online). Dependence of the neutron-proton momentum difference $P(\rm{n}-\rm{p})$ on the neutron-skin thickness $S_{\rm n}\equiv R_{\rm{skin}}$ for neutron-rich Mg, Ca, and Ne isotopes colliding with $^{12}$C at 50\,MeV/nucleon.
Panel (a) shows the correlation for the initialized Mg projectiles,
while panels (b)-(d) show the results for emitted neutrons and
protons in the final state. The dotted lines represent linear fits,
demonstrating the approximately linear decrease of $P(\rm{n}-\rm{p})$ with
increasing neutron-skin thickness. Adapted from
Ref.\,\cite{Ding2024PRC}.}
\label{fig:np-momentum-skin}
\end{figure}

\subsection{Pre-equilibrium emission and neutron-proton momentum differences}

Pre-equilibrium nucleons and light particles are emitted during the
early stage of a heavy-ion collision, before the reaction system
approaches statistical equilibrium. Since such early emissions can
preferentially sample the nuclear surface, they may retain information on the initial neutron and proton density distributions. In neutron-rich nuclei, weakly bound neutrons in the extended surface region can enhance neutron emission relative to proton emission. Early calculations by Ghosh et al.\,\cite{Ghosh1994PRC} investigated this effect explicitly and showed that the presence of a neutron-rich surface can modify pre-equilibrium nucleon-emission spectra. More generally, the different spatial distributions of neutrons and protons, together with the isospin-dependent mean field and the associated transport of isospin during the reaction, provide a connection between neutron-skin structure and the early emission
dynamics. Neutron-proton momentum observables provide complementary information on this isospin-dependent dynamics. In particular, neutron-proton differential flow was introduced to enhance the opposite responses of neutrons and protons to the isovector mean field while reducing
contributions common to the two species\,\cite{Li2000PRL}. Related
double differential-flow observables were subsequently proposed to
further reduce systematic uncertainties and enhance sensitivity to
the density dependence of the nuclear symmetry energy\,\cite{Yong2006PRC}. These studies establish the general importance of differences between neutron and proton momentum distributions as
probes of the isovector nuclear interaction.

A more direct connection between neutron-proton momentum differences
and the neutron-skin thickness was recently investigated by Ding
et al.\,\cite{Ding2024PRC}. Instead of the differential-flow
observable, they introduced the average neutron-proton relative
momentum
\begin{equation}
P(\rm{n}-\rm{p})
=
\frac{1}{N_{\rm n}N_{\rm p}}
\sum_{i=1}^{N_{\rm n}}
\sum_{j=1}^{N_{\rm p}}
\left|\vec{P}_{\rm{n}_i}-\vec{P}_{\rm{p}_j}\right|,
\end{equation}
where $N_{\rm n}$ and $N_{\rm p}$ denote the numbers of neutrons and protons, respectively. Within the IQMD calculations, $P(\rm{n}-\rm{p})$ exhibits a strong approximately linear anti-correlation with the neutron-skin thickness, both for the initialized projectile and for the emitted nucleons after the collision. As illustrated in FIG.\,\ref{fig:np-momentum-skin}, a thicker neutron skin systematically leads to a smaller neutron-proton momentum difference for the Mg, Ca, and Ne isotope chains considered in the calculations. This behavior reflects the complementary relation between the spatial and momentum distributions of neutrons and protons: As the neutron distribution becomes more extended relative to the proton distribution, the corresponding neutron-proton momentum difference is reduced.

\subsection{Coherent neutral-pion photoproduction}
\label{subsec:pion_photoproduction}

\begin{figure}[h!]
\centering
\includegraphics[width=0.48\textwidth]{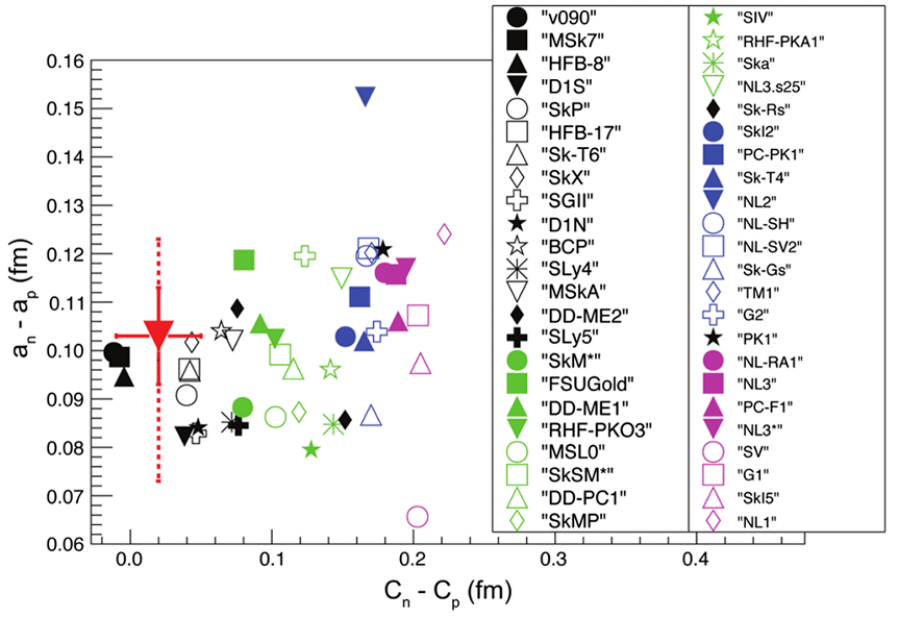}
\caption{(Color Online). Difference between the neutron and proton
surface diffuseness parameters, $a_{\rm n}-a_{\rm p}$, as a function of the corresponding difference between their half-density radii, $C_{\rm n}-C_{\rm p}$, for $^{208}\rm{Pb}$. The red inverted triangle denotes the result extracted from coherent $\pi^0$ photoproduction, with statistical and systematic uncertainties indicated by the solid and dashed error bars, respectively. The other symbols represent predictions from various nuclear structure models. The location of the experimental result suggests that the neutron distribution is more diffuse than the proton distribution, favoring a predominantly halo-type rather than skin-type surface profile. Adapted from
Ref.\,\cite{Tarbert2014PRL}.}
\label{fig:pion_photoproduction_skin}
\end{figure}

Coherent neutral-pion photoproduction has been proposed as an
alternative probe of neutron density distributions. In the coherent
reaction, the target nucleus remains in its ground state, and the
momentum-transfer dependence of the $\left(\gamma,\pi^0\right)$ cross section carries information about the nuclear form factor. By combining this information with the well-constrained proton distribution, Tarbert {et al.}\,\cite{Tarbert2014PRL} extracted
the neutron density of $^{208}\rm{Pb}$ from measurements performed
with the Crystal Ball detector at MAMI. Within
their reaction model and a two-parameter Fermi description (see Eq.\,(\ref{eq:2pf_density})), they
obtained a neutron half-density radius $C_{\rm n}\approx6.70\pm0.03\,\rm{fm}$, a diffuseness
$a_{\rm n}\approx0.55\pm0.01^{+0.02}_{-0.03}\,\rm{fm}$, and a neutron-skin thickness $R_{\rm{skin}}\approx0.15\pm0.03^{+0.01}_{-0.03}\,\rm{fm}$, where the first and second uncertainties are statistical and systematic,
respectively. FIG.\,\ref{fig:pion_photoproduction_skin} compares the extracted
differences in the neutron and proton half-density radii and
diffuseness parameters with predictions from a broad set of nuclear
structure models. The experimental result lies near
$C_{\rm n}-C_{\rm p}=0$ but gives a positive and appreciable
$a_{\rm n}-a_{\rm p}$. Within the adopted two-parameter Fermi
representation, this behavior favors a ``halo-type'' neutron
distribution, in which the neutron excess is accommodated primarily
through an enhanced surface diffuseness, rather than a ``skin-type'' distribution generated mainly by an outward displacement of the neutron half-density radius\,\cite{Tarbert2014PRL}. The comparison is therefore potentially sensitive not only to the rms neutron radius but also to how the neutron excess is distributed across the nuclear surface.

\begin{figure}[h!]
\centering
\includegraphics[width=0.42\textwidth]{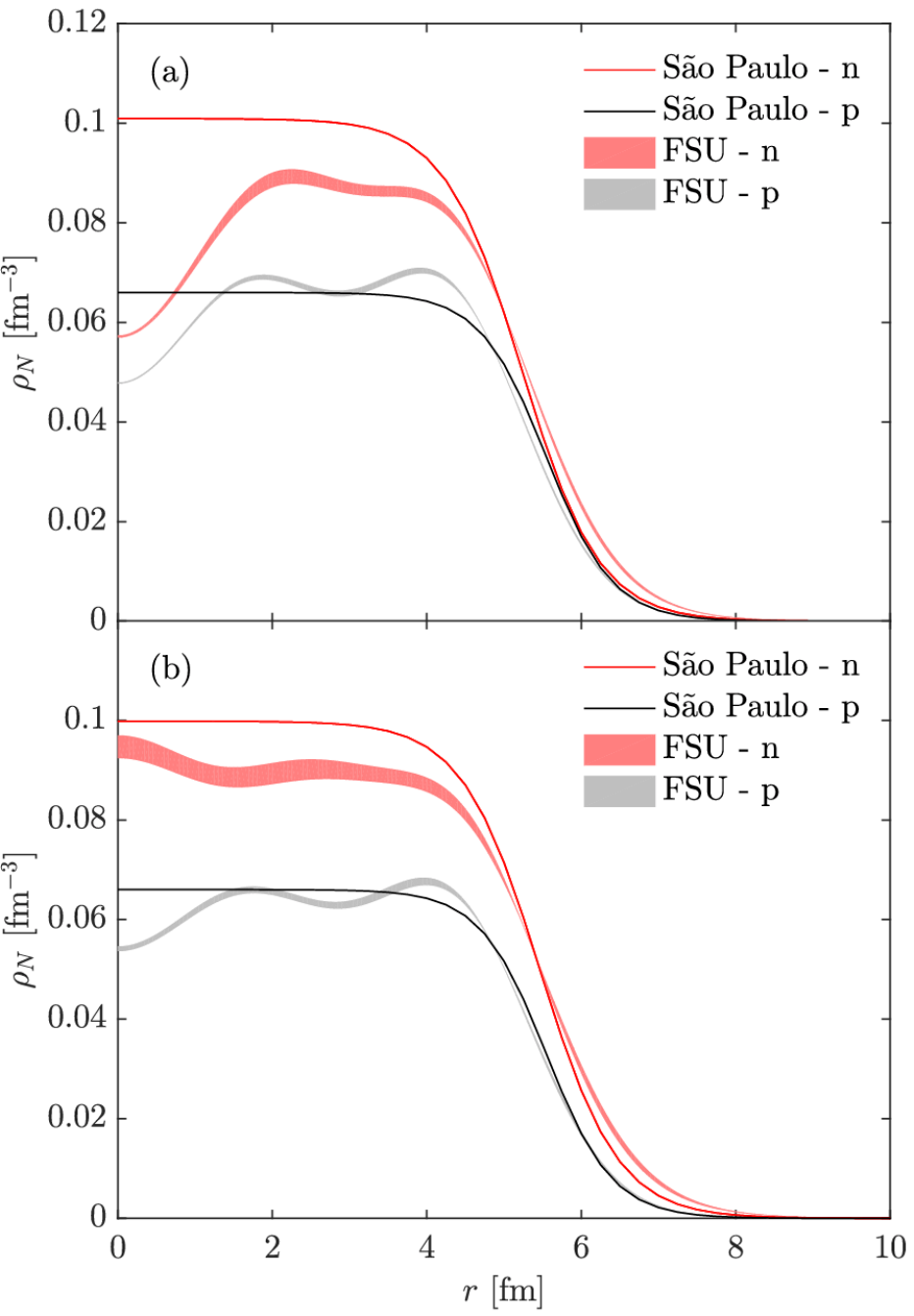}
\caption{(Color Online). Neutron (red) and proton (black or gray)
density distributions of (a) $^{116}\rm{Sn}$ and
(b) $^{124}\rm{Sn}$. The solid curves represent the phenomenological S\~ao Paulo parametrization, whereas the shaded bands show the range of predictions obtained with the FSU family of relativistic mean field models. The S\~ao Paulo densities exhibit a more rapid radial falloff and predict a negative neutron skin for $^{116}\rm{Sn}$ and a very small one for $^{124}\rm{Sn}$, whereas the FSU models yield more extended neutron distributions and appreciably larger neutron-skin thicknesses. The comparison illustrates the significant model dependence of the underlying density profiles used in calculations of coherent $\pi^0$ photoproduction. Adapted from
Ref.\,\cite{Colomer2022PRC}.}
\label{fig:colomer-sn-density}
\end{figure}

\begin{figure*}
\centering
\includegraphics[width=0.98\textwidth]{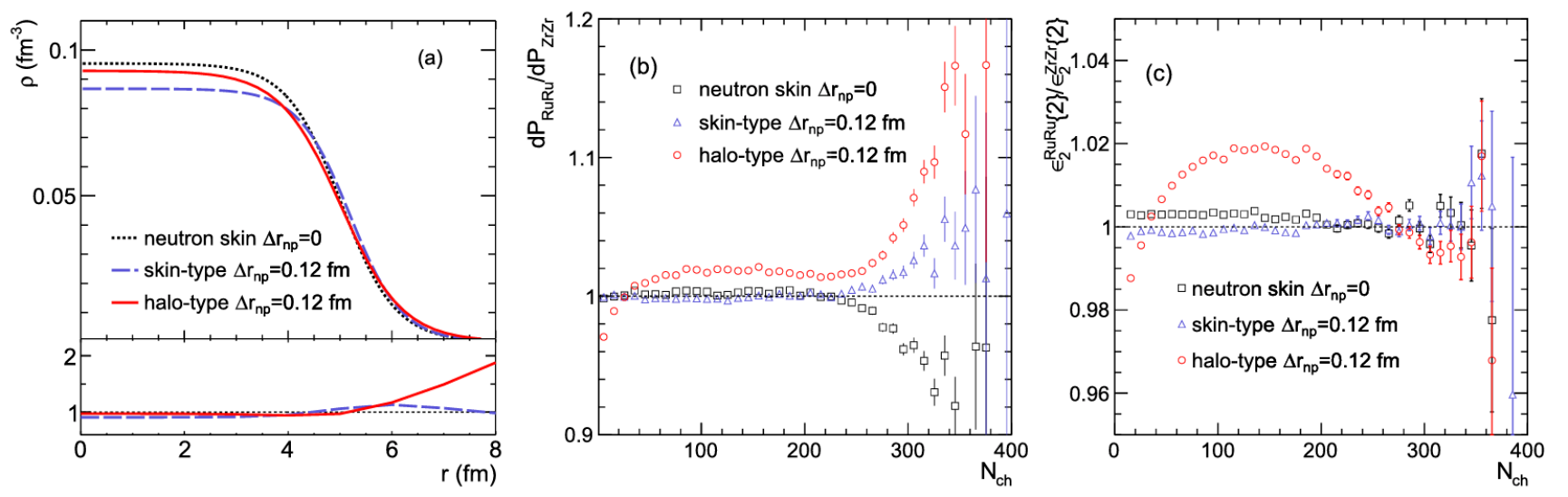}
\caption{(Color Online). Effects of the neutron density distribution on relativistic isobar collisions. (a) Neutron density distributions of $^{96}$Zr for different descriptions of the neutron skin. (b) Ratio of the charged-particle multiplicity distributions in $^{96}$Ru+$^{96}$Ru and $^{96}$Zr+$^{96}$Zr collisions. (c) Corresponding ratio of the initial eccentricity
$\varepsilon_2$. The comparison demonstrates that both the
multiplicity and the initial collision geometry are sensitive to
the detailed radial structure of the neutron distribution, in
particular to the distinction between skin-type and halo-type
neutron skins. Adapted from Ref.\,\cite{Xu2021PLB}.}
\label{fig:skin-eccentricity}
\end{figure*}

The claimed precision of this extraction should nevertheless be
interpreted cautiously. Pion propagation in nuclei involves
absorption, multiple scattering, charge exchange, Pauli blocking, and the in-medium modification of the $\Delta(1232)$ resonance\,
\cite{Ericson1988Book}. Miller showed that
final-state charge exchange can modify the calculated cross section
by several percent and substantially shift the extracted
$R_{\rm{skin}}$, thereby increasing its theoretical uncertainty\,\cite{Miller2019PRC}. 
A more systematic analysis by Colomer et al.\,\cite{Colomer2022PRC}, using both plane-wave and distorted-wave impulse approximations and several nuclear density models, further found that coherent $\pi^0$ photoproduction is predominantly sensitive to the isoscalar combination
$\rho_{\rm n}+\rho_{\rm p}$ and has only weak practical sensitivity to the isovector difference that determines $R_{\rm{skin}}$. As illustrated in FIG.\,\ref{fig:colomer-sn-density}, the phenomenological S\~ao Paulo parametrization and the FSU relativistic mean-field models predict visibly different neutron and proton density profiles for $^{116}\rm{Sn}$ and $^{124}\rm{Sn}$. In particular, the S\~ao Paulo densities yield a negative neutron skin for $^{116}\rm{Sn}$ and a very thin one for $^{124}\rm{Sn}$, whereas the FSU models predict more extended neutron distributions and significantly larger
neutron-skin thicknesses. Despite these substantial differences in the underlying density profiles, the calculated coherent
$\pi^0$-photoproduction cross sections remain remarkably similar.
The inclusion of pion final-state interactions reduces the sensitivity still further by partially washing out the differences between the density models. For the FSU densities, which span ranges of $R_{\rm{skin}}$ of approximately $0.07$\,fm for $^{116}\rm{Sn}$ and $0.10$\,fm for $^{124}\rm{Sn}$, the corresponding distorted-wave cross sections differ by less than about $1.5\%$ at the first diffraction maximum\,\cite{Colomer2022PRC}. Coherent pion
photoproduction therefore provides useful information on the overall nuclear density and surface profile, but a robust determination of the absolute neutron-skin thickness requires improved control of the reaction mechanism and, preferably, combined analyses with electron scattering or systematic measurements along an isotopic chain.

It is noteworthy that the value inferred from coherent pion
photoproduction, is smaller than but statistically compatible with the PREX-II result. Given the substantial and qualitatively different theoretical uncertainties affecting pion photoproduction, however, this comparison should not yet be regarded as evidence for a tension comparable to that between precise electroweak and nuclear-structure constraints.

\section{Collective Flow and Neutron Skin}
\label{sec:collective}

Collective flow is one of the most important probes of the dynamical
evolution of heavy-ion collisions, reflecting the conversion of the
initial spatial structure of the colliding nuclei into final-state
momentum distributions. Since the neutron skin modifies the spatial
distribution of neutrons relative to protons, it can influence both
the geometry of the participant region and the initial distribution
of isospin asymmetry. These effects can subsequently be carried into
collective observables through the collision dynamics, providing a
connection between the ground-state neutron distribution and the
measured particle emission.
The manifestation of neutron-skin effects in collective flow depends
strongly on the collision energy and the observable considered. At
intermediate energies, neutron-proton differences in directed and
differential flows are closely connected to isospin transport and the isovector mean field. At relativistic and ultra-relativistic energies, the neutron density profile instead enters primarily through the initial transverse geometry, which is converted by the subsequent collective expansion into anisotropic flow. In the following, we discuss these complementary aspects through elliptic and directed flows, relativistic isobar collisions, and the recent quantitative extraction of neutron-skin thickness from ultra-relativistic heavy-ion data.

\subsection{Elliptic flow $v_2$ and the early-stage asymmetry}

Collective flow observables characterize the anisotropic expansion of the fireball created in non central heavy ion collisions and retain information on the geometry of the system at the early stage of the reaction. In particular, the elliptic flow coefficient $v_2$ is closely related to the second-order eccentricity $\varepsilon_2$ of the initial overlap region,
approximately through $v_2 \approx \kappa_2 \varepsilon_2$, where $\kappa_2$ denotes the dynamical response of the expanding medium.
Consequently, modifications of the proton and neutron density distributions of the colliding nuclei can propagate through the initial geometry into the measured elliptic flow\,\cite{JiaZhang2023PRC}.

The neutron skin provides one such modification. A difference between the neutron and proton distributions changes the radial density profile and, therefore, the transverse distribution of matter participating in the collision. For a given impact parameter, variations of the neutron-skin thickness can modify both the effective size of the overlap region and its eccentricity. Although the resulting changes in $\varepsilon_2$ and $v_2$ are generally modest, ratios between closely related collision systems can substantially reduce common uncertainties associated with the subsequent dynamical evolution\,\cite{JiaZhang2023PRC}. This makes isobaric collisions particularly well suited for isolating such nuclear structure effects.

The $^{96}$Ru+$^{96}$Ru and $^{96}$Zr+$^{96}$Zr isobar systems provide an important example. The two nuclei have the same mass number but different proton and neutron distributions, and energy density functional calculations predict a more pronounced neutron skin in $^{96}$Zr. The sensitivity of relativistic isobar collisions to these differences was demonstrated in several studies\,\cite{Hammelmann2020PRC,Li2020PRL}. In particular, the
charged particle multiplicity difference between the two systems was shown to provide sensitivity to the underlying neutron density\,\cite{Li2020PRL}, while calculations of the initial eccentricity also demonstrated that nuclear deformation and neutron-skin effects need to be carefully distinguished\,\cite{Hammelmann2020PRC}. The latter point is
important because deformation can generate sizable modifications of the initial eccentricity even when the direct neutron-skin contribution to a given eccentricity observable is relatively small. 

A more direct connection between the neutron density profile and elliptic flow was investigated by Xu {et al.}\,\cite{Xu2021PLB}. They considered different neutron density distributions of $^{96}$Zr and calculated the resulting charged particle multiplicity and initial eccentricity in relativistic Ru+Ru and Zr+Zr collisions. As illustrated in FIG.\,\ref{fig:skin-eccentricity}, the Ru+Ru/Zr+Zr ratio of $\varepsilon_2$ is sensitive not only to the magnitude of the neutron-skin
thickness but also to the detailed radial form of the neutron distribution. In particular, a ``skin-type'' distribution, in which the neutron extension is mainly associated with an increased radius, and a ``halo-type'' distribution, characterized primarily by an increased surface diffuseness, can lead to different centrality dependences of the eccentricity ratio.
Since the final elliptic flow approximately follows the initial
$\varepsilon_2$, such differences can be transferred to the measured $v_2$ and thereby provide information beyond the neutron-skin thickness alone.

\begin{figure}[h!]
\centering
\includegraphics[width=0.45\textwidth]{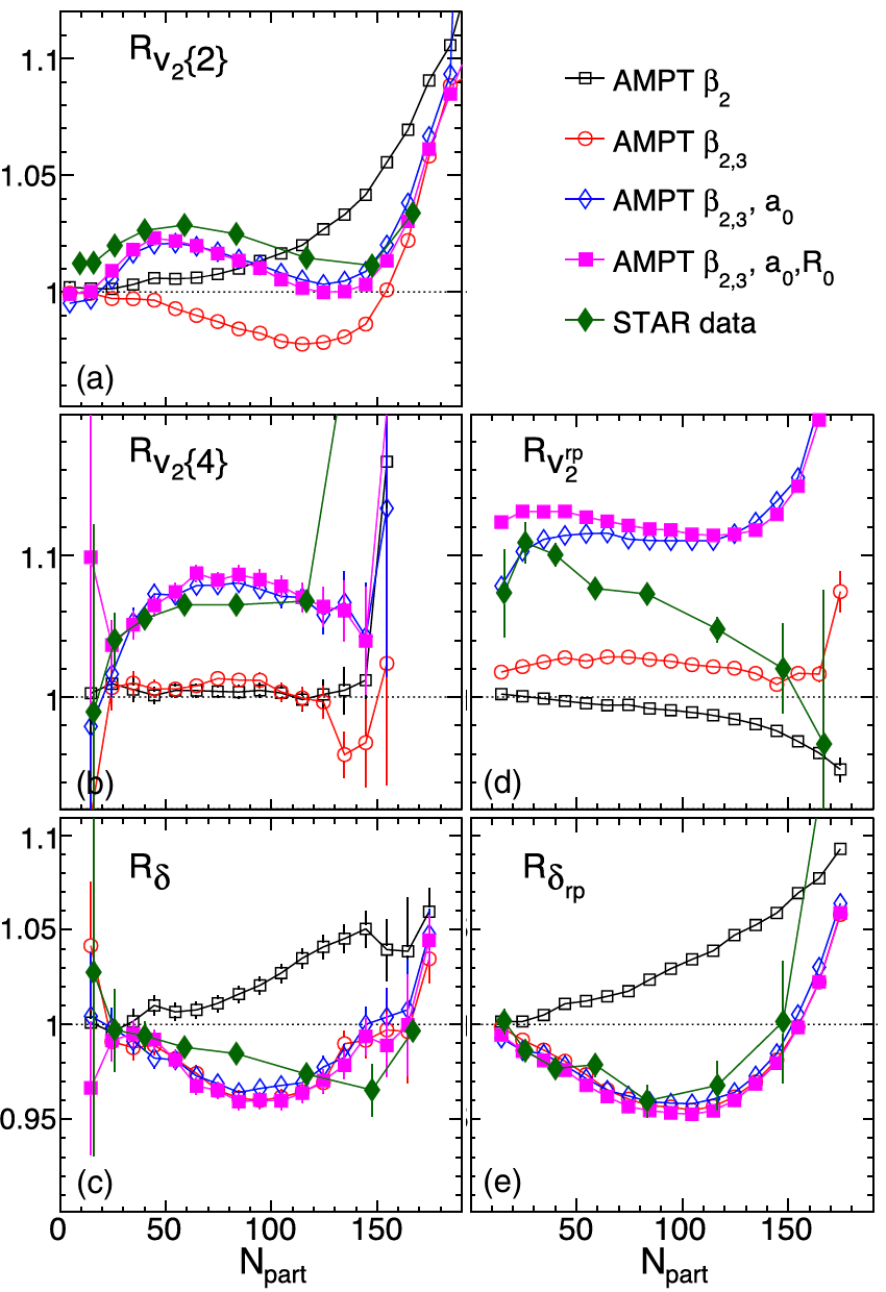}
\caption{(Color Online). Sensitivity of elliptic-flow observables to nuclear structure in $^{96}$Ru+$^{96}$Ru and $^{96}$Zr+$^{96}$Zr collisions. Shown are isobar ratios of elliptic-flow cumulants and flow-fluctuation observables as functions of the number of participating nucleons, with
the effects of nuclear deformation and radial density-profile parameters introduced successively. The radial surface profile, and hence the neutron-skin effect, predominantly modifies the reaction-plane component of elliptic flow, whereas nuclear deformation mainly affects elliptic-flow fluctuations. Adapted from Ref.\,\cite{Jia2023PRL}.}
\label{fig:skin-v2-isobar}
\end{figure}

An important issue in extracting neutron-skin information from $v_2$ is the simultaneous influence of nuclear deformation. Jia {et al.}\,\cite{Jia2023PRL} showed that these effects can be separated by considering different components and fluctuations of the elliptic flow. The elliptic flow can be viewed as containing a reaction-plane component associated with the average elliptic geometry and a fluctuating component generated by event-by-event variations of the initial configuration. Changes in the radial nuclear profile, including the surface diffuseness related to the neutron skin, predominantly affect the former, whereas quadrupole and octupole deformations mainly influence the flow fluctuations.
This separation is illustrated in FIG.\,\ref{fig:skin-v2-isobar}. In particular, the isobar ratio of the higher-order elliptic flow cumulant $v_2\{4\}$ is predominantly controlled by differences in the radial surface profile, while the fluctuation component obtained from $v_2\{2\}$ and $v_2\{4\}$ exhibits stronger sensitivity to the nuclear deformation parameters. The analysis therefore provides a strategy for disentangling neutron-skin and deformation effects rather than attributing the entire Ru/Zr difference in $v_2$ to a single nuclear-structure
parameter. Other collective observables can provide complementary constraints on the same initial geometry. Hydrodynamic calculations by Xu {et al.}\,\cite{Xu2023PRC}, for example, showed that the Ru+Ru/Zr+Zr ratio of the mean transverse momentum remains sensitive to the small differences in neutron skin and deformation while being relatively insensitive to the bulk evolution of the collision system. Moreover, its correlation with elliptic flow in central collisions can help constrain the deformation, which in turn reduces an important ambiguity in extracting neutron-skin
information.

These studies demonstrate that elliptic flow provides a connection
between the spatial structure of the colliding nuclei and final state momentum anisotropies. This broader concept has now been demonstrated experimentally by the STAR Collaboration through
collective-flow-assisted nuclear shape imaging in the reaction $^{238}\rm{U}+^{238}\rm{U}$ and $^{197}\rm{Au}+^{197}\rm{Au}$
collisions\,\cite{STAR2024Nature}. By combining elliptic-flow
fluctuations, mean-transverse-momentum fluctuations, and their
correlations, the STAR analysis extracted the quadrupole deformation and triaxiality of $^{238}\rm{U}$, providing direct evidence that ground-state nuclear geometry can survive as a measurable imprint in final state collective observables.
For neutron-skin studies, this result is important both as a
validation of the general imaging strategy and as a reminder that the neutron-skin contribution cannot be considered independently of other nuclear structure effects, particularly deformation. Ratios between isobaric systems, together with flow cumulants and complementary observables such as multiplicity and mean transverse momentum, therefore provide a promising way to isolate the subtle influence of the neutron density profile on the early-stage geometry of relativistic heavy-ion collisions.

\subsection{Directed flow $v_1$ and isospin transport}

This initial-state difference can be carried into the subsequent
sideward motion of neutrons and protons, characterized by $v_1^{\rm n}$ and $v_1^{\rm p}$. Their difference, $\Delta v_1=v_1^{\rm n}-v_1^{\rm p}$, provides a natural measure of the relative neutron-proton directed motion. Closely related observables can be constructed from the neutron and proton transverse momenta.
In particular, neutron-proton differential-flow observables additionally incorporate their relative abundances, thereby combining the relative sideward motion with isospin fractionation\,\cite{Li2000PRL,Li2002PRL,Yong2006PRC,Xie2015PRC}.
The interpretation of these observables necessarily involves the
isovector interaction. The neutron density profile, characterized in part by the neutron skin, determines the initial spatial distribution of the neutron excess, whereas the symmetry potential governs its subsequent dynamical evolution. In neutron-rich matter, the isovector mean field generates different forces on neutrons and protons, which can lead directly to a difference in their sideward motion and hence in $\Delta v_1$. At the same time, the symmetry potential modifies the neutron-proton composition of the emitted matter through isospin fractionation. Neutron-proton differential transverse flow combines these two effects, while partially suppressing contributions common to neutrons and protons. Such observables were therefore proposed as sensitive probes of the symmetry energy\,\cite{Li2000PRL}, particularly its behavior in the high-density region
reached in energetic heavy-ion collisions\,\cite{Li2002PRL}.

\begin{figure}[h!]
\centering
\includegraphics[width=0.48\textwidth]{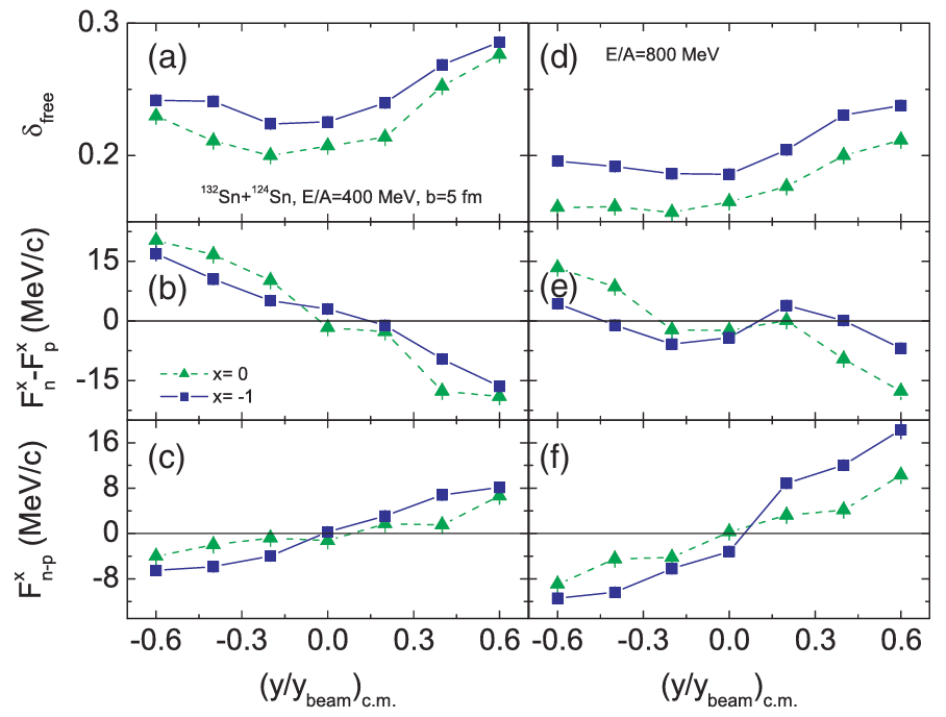}
\caption{(Color Online). Rapidity dependence of isospin-sensitive observables in $^{132}$Sn+$^{124}$Sn collisions at incident energies of 400 and 800\,MeV/nucleon. The upper panels show the isospin asymmetry of free nucleons, the middle panels the difference between the average neutron and proton transverse flows, and the lower panels the neutron-proton differential transverse flow. The comparison illustrates how isospin
fractionation and relative neutron-proton collective motion combine
to enhance the sensitivity to the density dependence of the symmetry energy characterized by the $x$ parameter. Adapted from Ref.\,\cite{Yong2006PRC}.}
\label{fig:np-differential-flow}
\end{figure}

The underlying mechanism is illustrated in
FIG.\,\ref{fig:np-differential-flow}. In the calculations of Yong
{et al.}\,\cite{Yong2006PRC}, changing the density dependence
of the symmetry energy (via the $x$ parameter) modifies both the isospin asymmetry of emitted free nucleons and the relative transverse motion of neutrons and protons. These two effects reinforce each other in the neutron-proton differential transverse flow, resulting in a stronger sensitivity than that obtained from the simple difference between the average neutron and proton transverse flows. The sensitivity also becomes more pronounced at the higher incident energy in these calculations, where higher densities are reached during the compression stage. A further
extension is the double neutron-proton differential transverse flow
between isotopic reaction systems, which retains sensitivity to the
symmetry energy while reducing Coulomb effects and other common
systematic uncertainties\,\cite{Yong2006PRC}.

From the perspective of neutron-skin physics, this mechanism is
particularly relevant because the neutron skin modifies the spatial
distribution of the initial isospin asymmetry. In semi-peripheral
collisions, where the surface regions of the nuclei play a relatively important role, variations of the neutron density profile can modify the local neutron-proton composition sampled by the participant matter. The subsequent isovector dynamics then
transports this initial spatial asymmetry and can convert it into
differences between neutron and proton emission and collective motion. Thus, the neutron skin and the symmetry potential enter the flow dynamics in complementary ways: the former modifies the initial
condition, while the latter controls the subsequent isospin transport.
The relation between the final flow and the symmetry energy is,
however, dynamical rather than associated with a single characteristic density. Particles emitted at different rapidities and transverse momenta can probe different stages of the collision, and the momentum dependence of the isovector mean field can further modify neutron and proton directed flows\,\cite{Xie2015PRC}. Consequently, the magnitude and even the detailed kinematic dependence of neutron-proton flow differences depend on the collision energy, density region, particle selection, and transport treatment. These effects need to be controlled when flow observables are used to isolate information associated with the initial neutron distribution.

\begin{figure}[h!]
\centering
\includegraphics[width=0.48\textwidth]{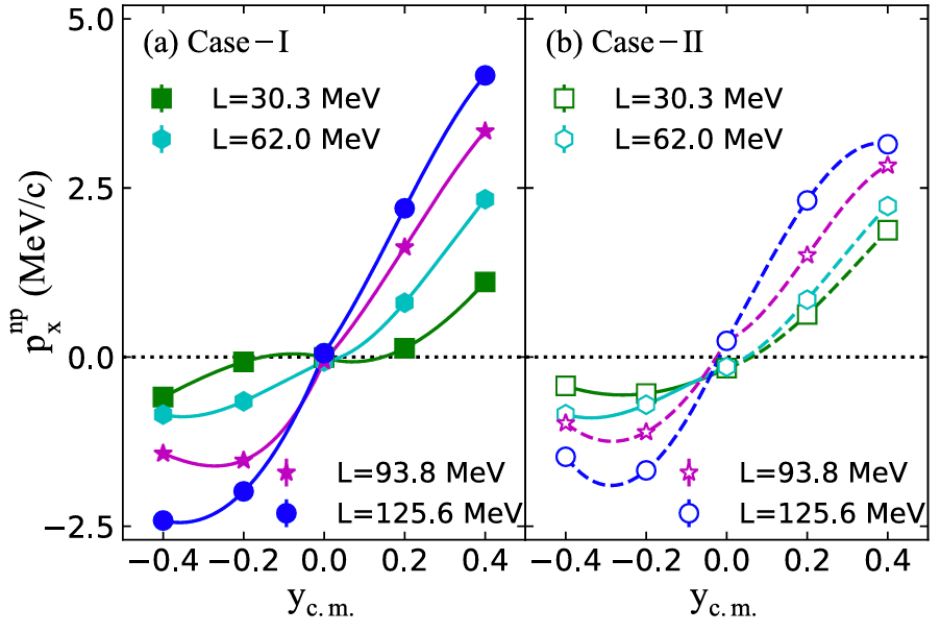}
\caption{(Color Online). Neutron-proton differential transverse flow
as a function of the center-of-mass rapidity in $^{132}$Sn+$^{124}$Sn collisions at 270\,MeV/nucleon for different
values of the symmetry energy slope parameter $L$ and two treatments
of the subsaturation symmetry energy. The comparison illustrates both the sensitivity of the differential flow to $L$ and the influence of the low-density symmetry potential on this sensitivity. Adapted from
Ref.\,\cite{Huang2022PRC}.}
\label{fig:np-flow-L}
\end{figure}

More recent transport calculations further demonstrate the sensitivity of neutron-proton differential flow to the symmetry energy. As shown in FIG.\,\ref{fig:np-flow-L}, calculations for
$^{132}$Sn+$^{124}$Sn collisions at 270\,MeV/nucleon exhibit an
appreciable dependence of the differential transverse flow on the
symmetry-energy slope parameter $L$\,\cite{Huang2022PRC}. The
differential observable shows a considerably stronger response than
the separate neutron and proton transverse flows, demonstrating the
advantage of isolating their relative isovector motion. The calculations also indicate that the subsaturation-density behavior of the symmetry energy can affect the final signal during the expansion stage. The observed flow therefore contains accumulated information on the isospin-dependent evolution over the collision history rather than exclusively on the highest-density stage.
This dynamical sensitivity also highlights an important issue in using collective flow to study neutron skins. A measured neutron-proton flow difference may contain contributions from both the initial neutron density profile and the subsequent symmetry energy driven evolution. These two ingredients are physically connected but play distinct roles: The neutron density profile determines the initial spatial distribution of the neutron excess, whereas the isovector mean field governs how this asymmetry is subsequently transported and converted into momentum-space observables. A consistent transport description of both effects is therefore essential for extracting neutron-skin information from isospin-sensitive flow measurements. Conversely, independent constraints on the neutron skin can reduce uncertainties in the initial isospin distribution and thereby improve the use of neutron-proton
flow observables as probes of the symmetry energy.

\subsection{Isobar collisions at RHIC: a case study}

The STAR Collaboration's isobar program at RHIC, involving
$^{96}$Ru+$^{96}$Ru and $^{96}$Zr+$^{96}$Zr collisions at
$\sqrt{s_{\rm{NN}}}=200$\,GeV, provides a unique opportunity to study
nuclear-structure effects, including the neutron skin, at relativistic energies. Since the two isobars have the same mass number but different proton and neutron numbers, many bulk properties of the two collision systems are similar, while differences associated with their nuclear density distributions can be enhanced in ratios of observables. Although these collisions occur at energies where the reaction dynamics is dominated by the formation and evolution of the quark-gluon plasma (QGP), initial-state differences between the two isobars can still affect final-state observables, including charged-particle multiplicity
ratios, elliptic and triangular flow coefficients, net-proton cumulants, and hyperon global polarization. In particular, collective flow provides a direct connection between the initial nuclear geometry and the momentum-space anisotropy of the produced particles. The interpretation of the isobar data requires a careful treatment of different nuclear-structure effects. In addition to their different neutron density distributions, $^{96}$Ru and $^{96}$Zr can have different intrinsic deformations. The quadrupole deformation $\beta_2$ primarily modifies the initial ellipticity and hence $v_2$, whereas the octupole deformation $\beta_3$ can generate additional triangularity and affect $v_3$. Transport calculations by Zhang and Jia\,\cite{Zhang2022PRL} showed that the observed ordering of the isobar flows in central collisions, $v_{2,\rm{Ru}}>v_{2,\rm{Zr}}$ and $v_{3,\rm{Ru}}<v_{3,\rm{Zr}}$, can be naturally understood in terms of a larger quadrupole deformation of $^{96}$Ru and stronger octupole correlations in $^{96}$Zr.

\begin{figure}[h!]
\centering
\includegraphics[width=0.4\textwidth]{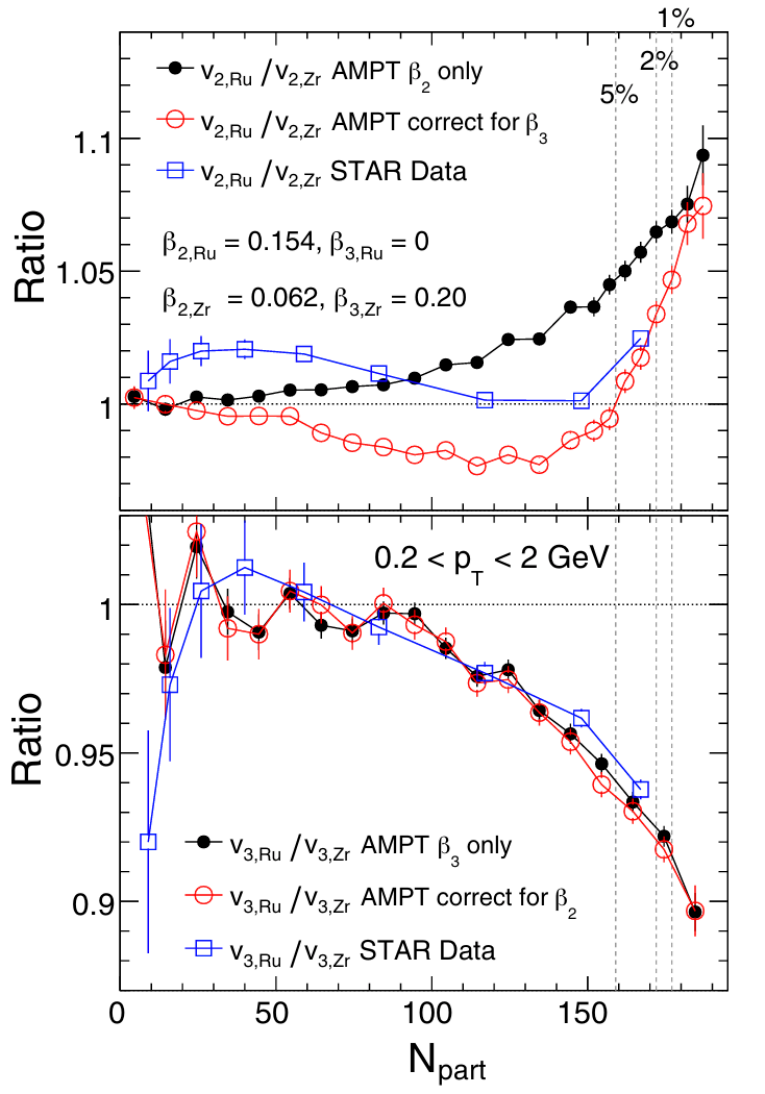}
\caption{(Color Online). Ratios of elliptic flow $v_2$ (upper panel)
and triangular flow $v_3$ (lower panel) between
$^{96}$Ru+$^{96}$Ru and $^{96}$Zr+$^{96}$Zr collisions at
$\sqrt{s_{\rm{NN}}}=200$\,GeV. The AMPT calculations incorporate the
different quadrupole and octupole deformations of the two isobars
and are compared with STAR data. The $v_2$ ratio is sensitive to
the interplay of different nuclear-structure effects, whereas the
$v_3$ ratio provides a particularly clear signature of the octupole
deformation of $^{96}$Zr. Adapted from Ref.\,\cite{Zhang2022PRL}.}
\label{fig:isobar-flow}
\end{figure}

As shown in FIG.\,\ref{fig:isobar-flow}, the different quadrupole and octupole deformations generate a characteristic centrality dependence of the flow ratios. The larger $\beta_2$ of $^{96}$Ru provides a positive contribution to the Ru/Zr ratio of $v_2$, whereas the sizable $\beta_3$ of $^{96}$Zr reduces the $v_2$ ratio over a broad centrality range. The latter effect becomes particularly important outside the most central collisions and produces a nonmonotonic centrality dependence of the $v_2$ ratio. In contrast, the calculated $v_3$ ratio is essentially insensitive to $\beta_2$ and is primarily controlled by the octupole deformation of $^{96}$Zr\,\cite{Zhang2022PRL}. The calculated $v_3$ ratio agrees well with the STAR data, supporting strong octupole correlations in $^{96}$Zr. This comparison demonstrates that collective flow at ultrarelativistic energies can retain information on the structure of the colliding nuclei despite the
subsequent QGP evolution.

Of particular relevance to neutron-skin physics is the remaining
difference between the calculated and measured $v_2$ ratios. The
deformation-based calculation of Ref.\,\cite{Zhang2022PRL} lies below the STAR data by up to about $2\%$ in the mid-central and peripheral regions. Zhang and Jia pointed out that this residual difference could be associated, at least partly, with the larger neutron skin of $^{96}$Zr, which modifies the radial density distribution and hence the initial eccentricity. This interpretation is consistent with calculations showing that the
Ru/Zr multiplicity and eccentricity ratios are sensitive to the
neutron density distributions and even to the detailed radial form
of the neutron skin\,\cite{Li2020PRL,Xu2021PLB}. The isobar comparison can therefore contain information not only on intrinsic nuclear deformation but also on the radial structure of the neutron
distribution.

The simultaneous sensitivity to deformation and neutron skin also
illustrates an important challenge in extracting neutron-density
information from relativistic isobar collisions. The two effects enter the initial geometry in different ways: nuclear deformation changes the intrinsic shape of the nucleus, whereas the neutron skin modifies its radial surface profile. Their contributions to a given observable, particularly $v_2$, can therefore coexist and need to be disentangled. A useful strategy is to combine observables with different sensitivities to the mean geometry and its fluctuations. In particular, Jia {et al.}\,\cite{Jia2023PRL} showed that differences in the radial density profile predominantly modify the intrinsic reaction-plane ellipticity, whereas nuclear deformation has a stronger influence on elliptic-flow fluctuations. Combining the mean elliptic response with its fluctuations therefore provides a possible means of separating radial neutron-skin effects from deformation effects.

The RHIC isobar program thus provides a particularly instructive
example of the connection between low-energy nuclear structure and
ultrarelativistic heavy-ion dynamics. The neutron density profile
enters as an initial-state property, modifying the transverse overlap geometry, while the subsequent QGP evolution converts these spatial differences into experimentally accessible momentum-space observables. Ratios between Ru+Ru and Zr+Zr collisions help suppress common dynamical uncertainties and enhance the sensitivity to the comparatively small structural differences between the two nuclei. At the same time, the coexistence of neutron-skin, radial-profile, and deformation effects demonstrates that neutron-skin information cannot generally be inferred from a single flow observable. A combined analysis of multiplicity, collective flow, flow fluctuations, and complementary observables is therefore important for constraining the neutron density distribution in relativistic isobar collisions. 

\subsection{Neutron-skin of $^{208}\rm{Pb}$ from ultra-relativistic collisions}\label{subs:Nk208-Ultra}

Beyond relative comparisons between isobaric systems,
ultra-relativistic heavy-ion collisions can also be used to
quantitatively constrain the neutron-skin thickness of an individual nucleus. Giacalone {et al.}\,\cite{Giacalone2023PRL} demonstrated this possibility by extracting the neutron skin of $^{208}\rm{Pb}$ from particle-production and collective-flow measurements in $^{208}\rm{Pb}+^{208}\rm{Pb}$ collisions at the LHC. At ultra-relativistic energies, particle production is mediated primarily by gluonic interactions and the initial energy deposition is therefore sensitive to the spatial distribution of all nucleons. The neutron density profile affects the transverse size, diffuseness, and eccentricity of the initially produced QGP, which are subsequently encoded in the multiplicity, transverse-momentum distributions, and
collective flow of the final-state particles.

\begin{figure}[h!]
\centering
\includegraphics[width=0.40\textwidth]{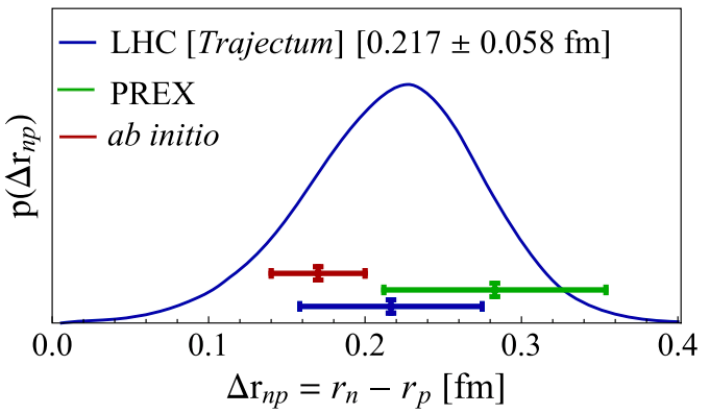}\\[0.5cm]
\includegraphics[width=0.45\textwidth]{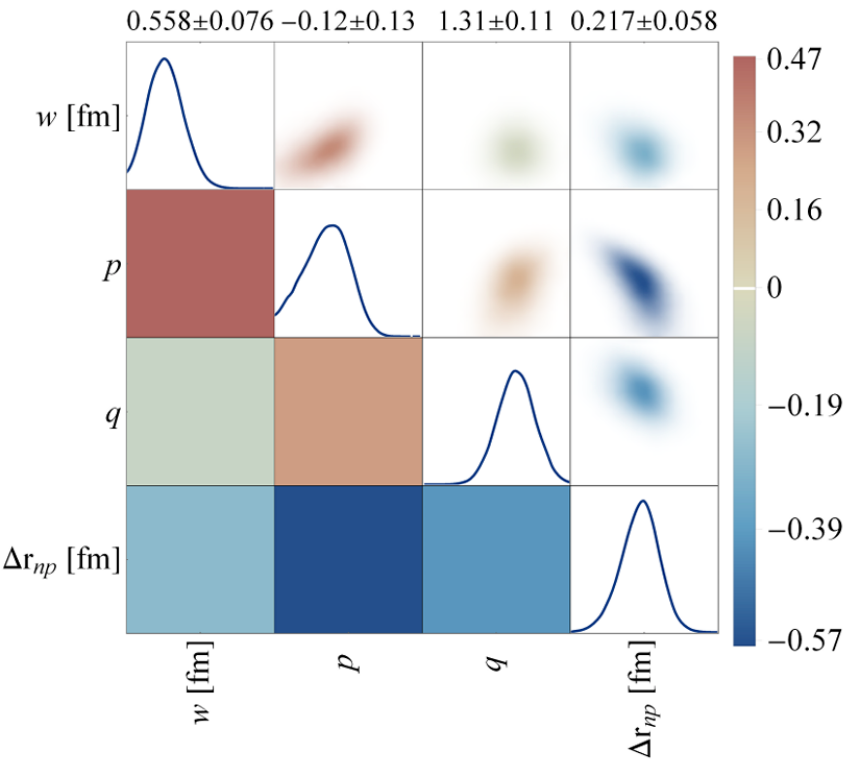}
\caption{(Color Online). Bayesian determination of the neutron-skin thickness of $^{208}\rm{Pb}$ from ultra-relativistic
$^{208}\rm{Pb}+^{208}\rm{Pb}$ collisions at the LHC.
Upper: Posterior probability distribution of
$\Delta r_{\rm{np}}$ inferred within the Trajectum framework (blue), compared with the PREX determination from parity-violating electron scattering (green) and an \textit{ab initio} nuclear-theory prediction (red). The horizontal intervals indicate the corresponding uncertainties. Lower: Marginalized posterior distributions and correlations among the effective nucleon width $w$, the TRENTo energy-deposition parameters $p$ and $q$, and $\Delta r_{\rm{np}}$. The diagonal entries show the one-dimensional posterior distributions, while the off-diagonal entries display the joint distributions and correlation coefficients. In particular, the strong anticorrelation between $p$ and $\Delta r_{\rm{np}}$ illustrates the importance of simultaneously constraining the neutron density profile and the initial energy-deposition mechanism. Adapted from Ref.\,\cite{Giacalone2023PRL}.}
\label{fig:Pb208-skin-LHC}
\end{figure}

In Ref.\,\cite{Giacalone2023PRL}, the point-neutron and point-proton densities of $^{208}\rm{Pb}$ were described by two-parameter Fermi distributions, see Eq.\,(\ref{eq:2pf_density}). The proton parameters were fixed at $C_{\rm p}\approx 6.680$\,fm and $a_{\rm p}\approx0.448$\,fm, corresponding to a point-proton rms radius of about $5.436$\,fm, while the neutron half-density
radius was set to $a_{\rm n}\approx6.690$\,fm. Motivated by experimental evidence that the neutron skin of $^{208}\rm{Pb}$ arises primarily from a more diffuse neutron distribution rather than a substantially larger half-density radius, the neutron skin was varied through the neutron diffuseness $a_{\rm n}$. Thus, the extracted $R_{\rm{skin}}$ should be understood within a predominantly ``halo-type'' variation of the neutron density profile. The neutron skin was promoted to a parameter in a global Bayesian analysis within the Trajectum hydrodynamic framework, together with parameters governing the initial energy deposition, nucleon width, and QGP transport properties. A larger neutron diffuseness produces a more extended nuclear profile, increasing the total hadronic cross section and generating a larger and more diffuse initial QGP. At a fixed centrality, this leads to a lower charged-particle multiplicity, weaker pressure gradients and hence a smaller mean transverse momentum, as well as a reduction of the initial eccentricity and the resulting elliptic flow. The analysis incorporated 653 data points from $^{208}\rm{Pb}+^{208}\rm{Pb}$ collisions and the total p$+^{208}\rm{Pb}$ cross section, yielding $
R_{\rm{skin}}(^{208}\rm{Pb})
\approx
0.217\pm0.058\,\rm{fm}$, or equivalently a point-neutron rms radius of about $5.653\pm0.058$\,fm. Importantly, this
constraint arises from the combined information carried by the total cross section, charged-particle multiplicity, mean transverse momentum, elliptic flow, and their centrality dependence, rather than from a direct mapping of any single observable onto $R_{\rm{skin}}$. Correlations with the parameters controlling the initial energy deposition, particularly their influence on the centrality dependence, are therefore incorporated explicitly in the reported uncertainty.

As shown in upper panel of FIG.\,\ref{fig:Pb208-skin-LHC}, the LHC determination is consistent with both the PREX measurement and the \textit{ab initio} calculation, while reaching a precision comparable to PREX\,\cite{PREX2021PRL,Hu2022NP}. This agreement is noteworthy because the underlying probes are very different: PVES accesses the neutron distribution through the weak interaction, whereas the heavy-ion approach exploits its influence on the initial collision geometry, energy deposition, and subsequent collective evolution. The Bayesian analysis also accounts for correlations between the neutron-skin thickness and parameters describing the initial state. As illustrated
in the lower panel of FIG.\,\ref{fig:Pb208-skin-LHC}, 
$R_{\rm{skin}}(^{208}\rm{Pb})\equiv\Delta r_{\rm{np}}$ is correlated with the effective nucleon width $w$ and the TRENTo
energy-deposition parameters $p$ and $q$. In particular, the strong anticorrelation between $\Delta r_{\rm{np}}$ and $p$ reflects the fact that both parameters influence the centrality dependence of particle-production-production and collective-flow observables. The simultaneous Bayesian determination of these quantities is therefore essential for separating nuclear-structure information from uncertainties in the initial energy deposition model.

Using the correlation between the neutron-skin thickness and the
symmetry energy slope parameter, the same analysis obtained
$L\approx79\pm39$\,MeV\,\cite{Giacalone2023PRL}, providing a
collider-based constraint on the isovector EOS\,\cite{Vinas2014EPJA}. This approach complements the relativistic isobar program discussed above. Isobar ratios provide relatively clean constraints on differences between the structures of two collision systems, whereas the Pb+Pb analysis demonstrates the possibility of extracting an absolute neutron-skin thickness from a single nuclear species through a global analysis of multiple observables. Together, these studies establish ultra-relativistic heavy-ion collisions as a
complementary probe of neutron distributions in finite nuclei, while also emphasizing the need to quantify correlations between nuclear density parameters and the modeling of the initial state.

\section{Neutron Skin, Proton Skin in Momentum Space and the Nucleon Short-Range Correlations}\label{SEC_Pskin}

Nuclear SRCs are compact two-nucleon
configurations generated by the repulsive core and/or tensor components of the nucleon-nucleon interaction. They are characterized by a large relative momentum, a comparatively small center-of-mass momentum, and hence an approximately back-to-back geometry. In the tensor-dominated momentum region, neutron-proton pairs are much more abundant than proton-proton pairs, revealing directly the strong spin-isospin dependence of the nuclear force\,\cite{Bethe71,Subedi08,Piasetzky06,Hen14,Hen17RMP}. This isovector structure has two complementary manifestations. In coordinate space, the competition between the symmetry energy, surface tension, and Coulomb interaction produces different neutron and proton density profiles and, in a neutron-rich nucleus, generally a neutron skin. In momentum space, neutron-proton SRCs redistribute neutron and proton occupations in opposite fractional proportions and generate a proton skin. The two skins are not unrelated phenomena, but distinct projections of isovector nuclear dynamics onto the full phase space of the nucleus.

\begin{figure}[h!]
\centering
\includegraphics[width=0.4\textwidth]{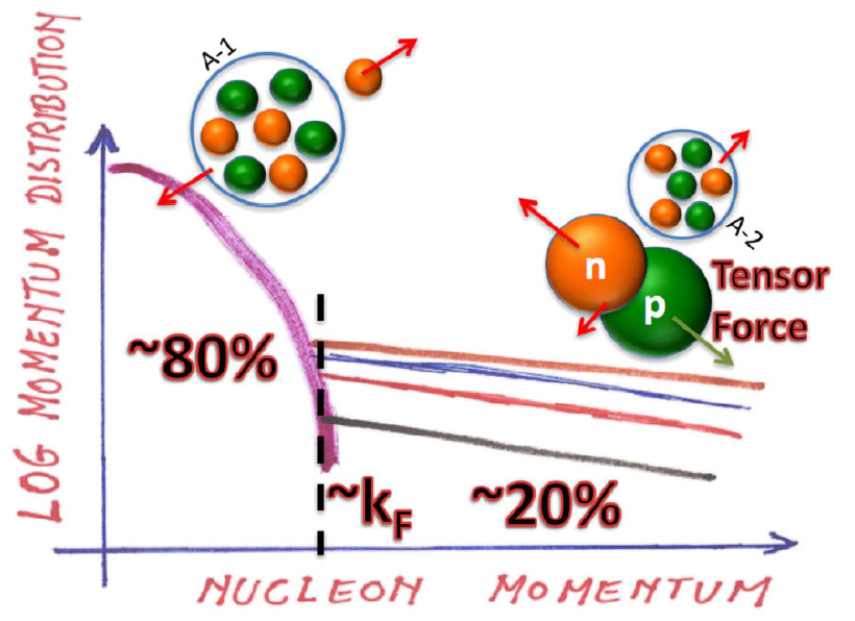}
\caption{(Color Online). Schematic single nucleon momentum distribution in a nucleus. SRCs deplete the Fermi sea and populate the HMT, which is dominated by nearly back-to-back neutron-proton pairs over the tensor force region. Adapted from Ref.\,\cite{Hen17RMP}.}
\label{fig:HMT-cartoon}
\end{figure}

The most direct one-body signature of SRCs is that they deplete states below the Fermi momentum and transfer a non-negligible fraction of nucleons to a high momentum tail (HMT), as established by nuclear many-body calculations and electron-scattering measurements\,\cite{Fantoni84,Benhar93,Pandharipande97,Subedi08,
Egiyan06,Shneor07,Piasetzky06,Korover14,Hen14,Fomin12,
Hen17RMP}. This basic structure is illustrated in FIG.\,\ref{fig:HMT-cartoon}\,\cite{Hen17RMP}. Over the momentum range dominated by two-nucleon SRCs, the tail is often represented approximately by $n_{\v{k}}=n(k)\sim k^{-4}$, with an upper cutoff of order 2-$3k_{\rm{F}}$. Thus, a simple schematic form is\,\cite{Cai2026EPJST,Cai15PRC,Cai2016PLB,Cai2016PRC,
Cai2022PRC,Cai2022AOP}
\begin{equation}
n_{\v{k}}^J=\left\{
\begin{array}{ll}
\Delta_J+\beta_J(k/k_{\rm{F}}^J)^2,
&0<k<k_{\rm{F}}^J,\\
&\\
C_J(k_{\rm{F}}^J/k)^4,
&k_{\rm{F}}^J<k<\phi_Jk_{\rm{F}}^J,
\end{array}\right.
\label{eq:skin-nk}
\end{equation}
where $J=\rm{n,p}$ and $C_J$ and $\phi_J$ determine the strength and
extent of the HMT. Related momentum distributions and their isospin
dependence have been investigated using generalized-contact, self-consistent Green's-function, Brueckner, variational, and other
many-body approaches\,\cite{Weiss15,CruzTorres18,Cosyn21}.
The fractional HMT population is
\begin{equation}
x_J^{\rm{HMT}}=3C_J\left(1-{\phi_J^{-1}}\right).
\label{eq:skin-hmt-fraction}
\end{equation}

The characteristic SRC configuration is a compact pair with an
inter-nucleon separation of order $1\,\rm{fm}$ or less, a large relative momentum, and a substantially smaller center-of-mass momentum. The two nucleons therefore emerge approximately back-to-back when the pair is resolved by a hard probe. This separation of momentum scales explains why the high-momentum parts of nuclear wave functions exhibit an approximately universal shape: the long-distance nuclear environment mainly controls how many correlated pairs are present, whereas their relative motion is governed primarily by the short-distance two-body interaction. The HMT is accompanied by a redistribution of spectral strength toward large removal energies; a high-momentum nucleon is thus not an  ndependent fast particle but one member of a correlated many-body configuration. This physics has consequences well beyond the momentum distribution itself. Within nuclear structure, SRCs modify occupation probabilities, kinetic energies, spectral functions, quasiparticle strengths, effective masses, pairing, and the isovector response. In nuclear reactions they affect energetic nucleon and cluster emission, subthreshold production, bremsstrahlung photons, and the initialization of transport simulations. At the quark level, the observed correlation between the SRC scaling factor and the EMC effect suggests that nucleons belonging to locally dense correlated configurations are especially sensitive to modification inside nuclei\,\cite{Weinstein11,Schmookler19}. In astrophysical matter,
the same redistribution of nucleon strength can influence composition, weak-interaction rates, neutrino transport, superfluidity, and cooling, although its broader consequences for the dense-matter EOS are outside the scope of the present section.
See Refs.\,\cite{Cai2026EPJST,Cai2026MPLA} for recent review.

The appearance of a high momentum power law tail is also part of a wider universality shared by strongly interacting Fermi systems. For systems with a short interaction range compared with the mean interparticle spacing, short-distance pair correlations generate a contact that fixes the leading large-$k$ behavior,  $n(k)\sim{C}/k^4$, and connects it to thermodynamic and response relations\,\cite{Tan08a,Tan08b,Tan08c}. Closely related behavior is observed in ultracold atomic Fermi gases, where the interaction and population imbalance can be controlled experimentally\,\cite{Giorgini08,Bloch08}. Nuclear forces contain additional spin, isospin, tensor, and finite-range structures, so this universality is not exact over all momenta. Nevertheless, it provides a deep reason why short-distance information can be encoded in a comparatively simple HMT and why SRC concepts connect nuclear physics, atomic many-body physics, and other strongly correlated quantum systems.

\begin{figure}[h!]
\centering
\includegraphics[width=0.38\textwidth]{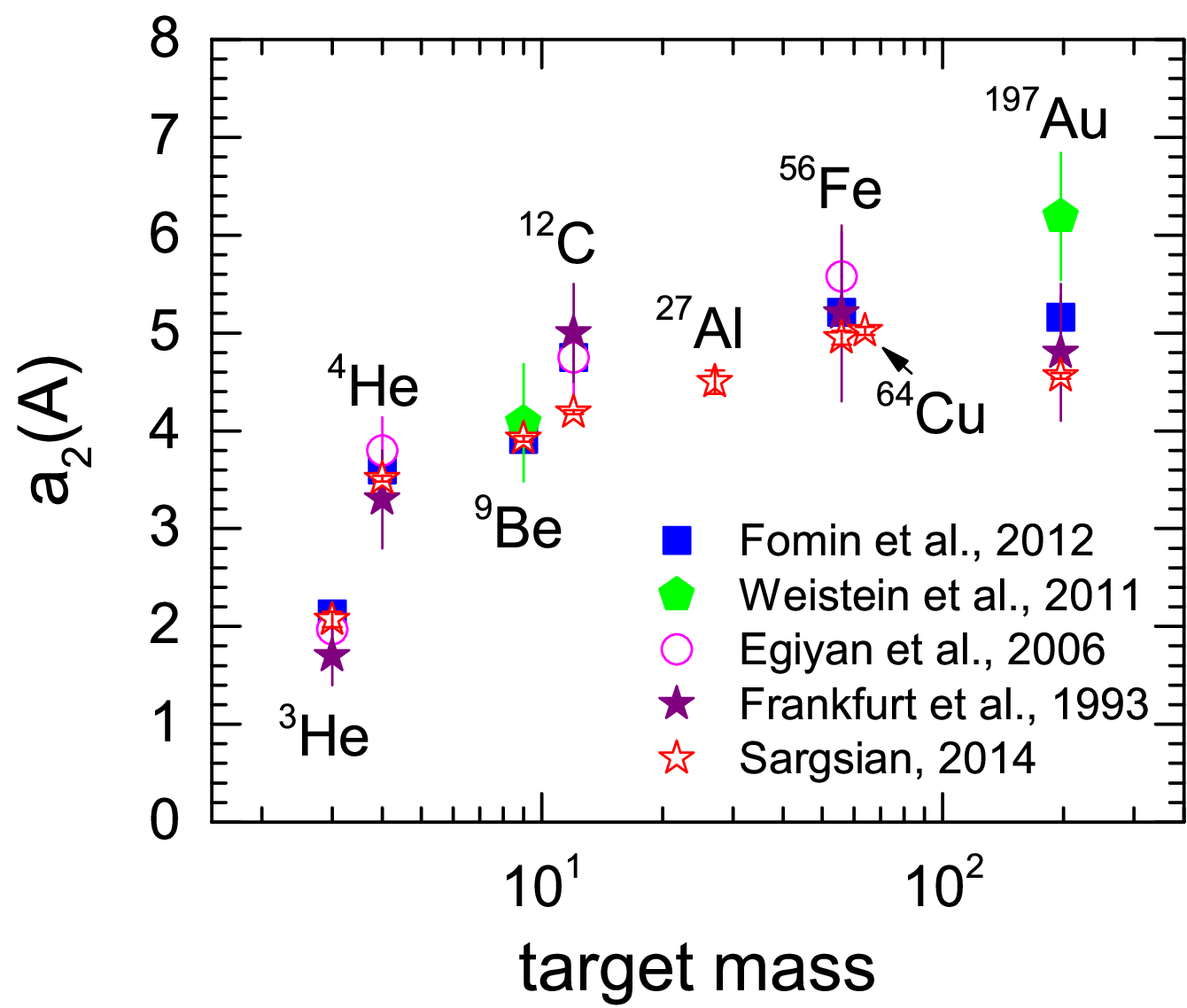}
\caption{(Color Online). Experimental SRC scaling factor $a_2(A)$ for representative nuclei as a function of the target mass. Adapted from Ref.\,\cite{Cai2026EPJST}.}
\label{fig:a2-scaling}
\end{figure}

Experimental evidence for SRCs and the associated HMT has been obtained from a broad range of reactions, most notably inclusive and exclusive electron-nucleus scattering at large momentum transfer, as well as proton-nucleus and nucleus-nucleus collisions\,\cite{Boffi96,Frois1987}.
In inclusive electron scattering, the approximately scaling region
$1.5\lesssim x_{\rm{Bj}}\lesssim1.9$ is characterized by a plateau in the per-nucleon cross-section ratio
\begin{equation}
a_2(A)=\frac{\sigma_A/A}{\sigma_{\rm{d}}/2},
\label{eq:a2-scaling}
\end{equation}
which measures, within reaction-model corrections, the relative abundance of two-nucleon SRC configurations in nucleus $A$ compared with the deuteron\,\cite{Frankfurt88,Frankfurt93,Weiss21}. As shown in FIG.\,\ref{fig:a2-scaling}, the measured $a_2(A)$ generally increases from light nuclei and tends toward saturation in heavy nuclei. Complementary coincidence measurements have further established the dominance of nearly back-to-back neutron-proton pairs and support a quasi-universal HMT extending from $k_{\rm{F}}$ to momenta of order $2k_{\rm{F}}$\,\cite{Hen14}. These results demonstrate that the HMT is not a small single-particle correction, but a generic many-body feature of nuclei.

Determining the microscopic origin, isospin dependence, and
nuclear-mass dependence of SRCs, together with their connection to the EMC effect and their broader implications for nuclear reactions and astrophysical matter, remains a major objective of current and future experimental programs\,\cite{Hauenstein02012021,Hen:2025rlk}.
Representative efforts include proposed SRC measurements at the
Electron-Ion  Collider (EIC) at  BNL\,\cite{Tu:2020ymk,Hauenstein:2021zql,
Boer:2011fh}, inverse-kinematics experiments by the R$^3$B Collaboration at GSI-FAIR\,\cite{SRC-r3b,r3b,myref}, and future programs at HIAF and the Electron-Ion Collider in China (EicC)\,\cite{Ye24,EicC-ref}. Complementary experiments at JLab, GSI, JINR, and IMP/Lanzhou continue to provide new information on correlated nucleon pairs in stable and neutron-rich nuclei\,\cite{Kahlbow:2023mtc,2023EPJA...59..188A,
2023EPJA...59..205F,JHXu25PRR}. These developments are extending SRC
studies toward increasingly asymmetric nuclei, where the isospin
dependence of the momentum distribution becomes especially important.

\begin{figure}[h!]
\centering
\includegraphics[width=0.4\textwidth]{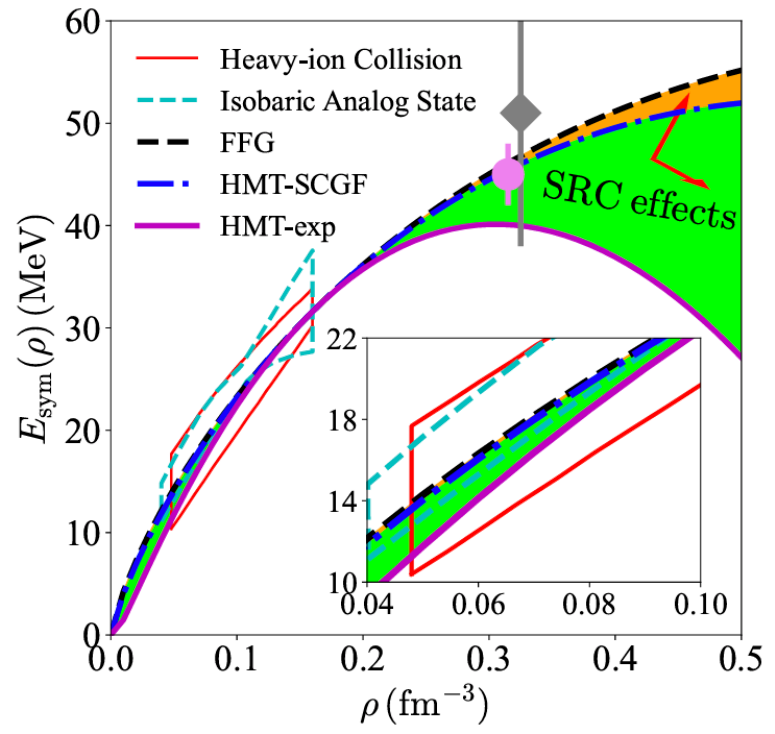}
\caption{(Color Online). Density dependence of the nuclear symmetry
energy in the FFG, HMT-SCGF, and HMT-exp models. Constraints from
heavy-ion collisions and analyses of isobaric analog
states are shown for comparison. Although all three
models are calibrated to reproduce empirical properties around
saturation density, the SRC-induced HMT modifies the balance between the kinetic and potential contributions to the symmetry energy leading to appreciable differences at both sub- and supra-saturation densities. The stronger and more isospin-dependent HMT adopted in the HMT-exp model produces the most pronounced softening at high density. At sub-saturation densities, these modifications are relevant to the isovector pressure governing neutron-skin formation and may therefore affect the $R_{\rm{skin}}$. Figure adapted from Ref.\,\cite{CaiLi22Gog}.}
\label{fig_ab_Esym}
\end{figure}

An important consequence of SRCs is their modification of the kinetic and potential components of the nuclear symmetry energy. Because neutron-proton correlations enhance the average kinetic energy of symmetric nuclear matter much more strongly than that of pure neutron matter, the SRC-induced HMT generally reduces the kinetic symmetry energy; the potential contribution must then be correspondingly readjusted to reproduce empirical constraints around $\rho_0$. As illustrated in FIG.\,\ref{fig_ab_Esym}, different assumptions about the magnitude and isospin dependence of the HMT can consequently produce markedly different density dependences of $E_{\rm{sym}}(\rho)$, particularly away from saturation density\,\cite{CaiLi22Gog}. Since neutron-skin formation is governed primarily by the isovector pressure
of neutron-rich matter at sub-saturation densities, these results
establish an important connection between SRC-driven momentum-space
correlations, the symmetry energy, and the predicted neutron-skin
thickness.

\begin{figure}[h!]
\centering
\includegraphics[width=0.48\textwidth]{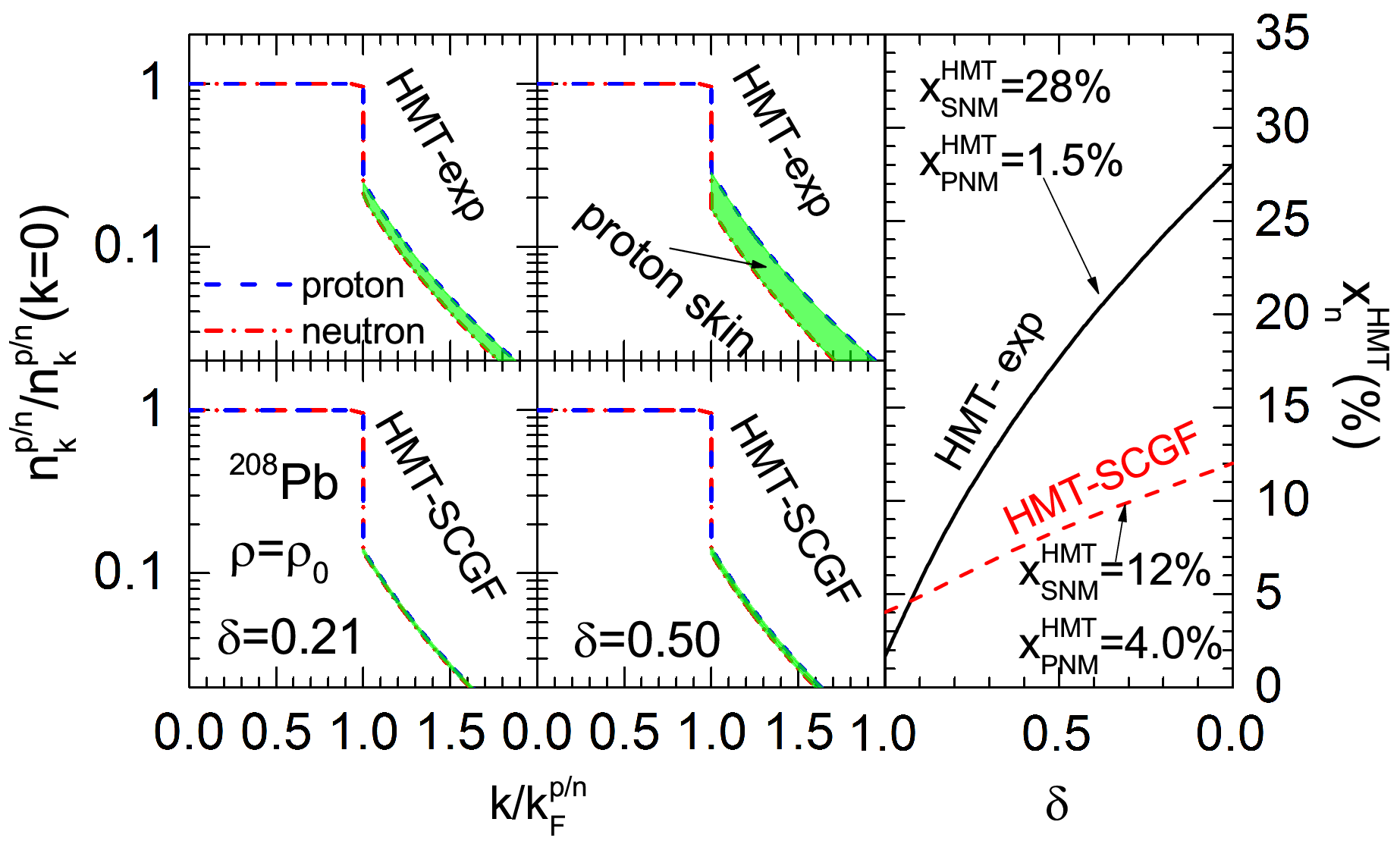}
\caption{(Color Online). Reduced neutron and proton momentum
distributions for isospin asymmetries $\delta\approx0.21$ and
$\delta\approx0.50$, together with the neutron fraction in the HMT for two representative SRC parameterizations. The broader normalized proton distribution defines the momentum-space proton skin. Adapted from Ref.\,\cite{Cai16b}.}
\label{fig:momentum-proton-skin}
\end{figure}

In the tensor dominated momentum region, neutron-proton pairs greatly outnumber proton-proton as well as neutron-neutron SRC pairs. Because each neutron-proton pair contributes one nucleon of each species, comparable numbers of neutrons and protons are promoted to high momentum. In a neutron-rich nucleus, however, protons are the minority component, so a larger fraction of protons resides in the HMT. Consequently, the normalized proton momentum distribution is broader and the average proton kinetic energy can exceed the neutron one\,\cite{Cai2026EPJST}.
This inversion defines a proton skin in momentum space; it does not imply that the absolute number of high-momentum protons exceeds that of high-momentum neutrons. As shown in FIG.\,\ref{fig:momentum-proton-skin}, the proton skin becomes more pronounced with increasing isospin asymmetry\,\cite{Hen15PRC,Cai15PRC,
Cai16b,Cai2026EPJST}. It therefore provides a momentum-space
manifestation of the isovector nuclear interaction: whereas the
coordinate-space neutron skin reflects the separation of the mean
neutron and proton density profiles, the momentum-space proton skin
reflects their unequal fractional occupation of low- and high-momentum states.

\begin{figure}[h!]
\centering
\includegraphics[width=0.43\textwidth]{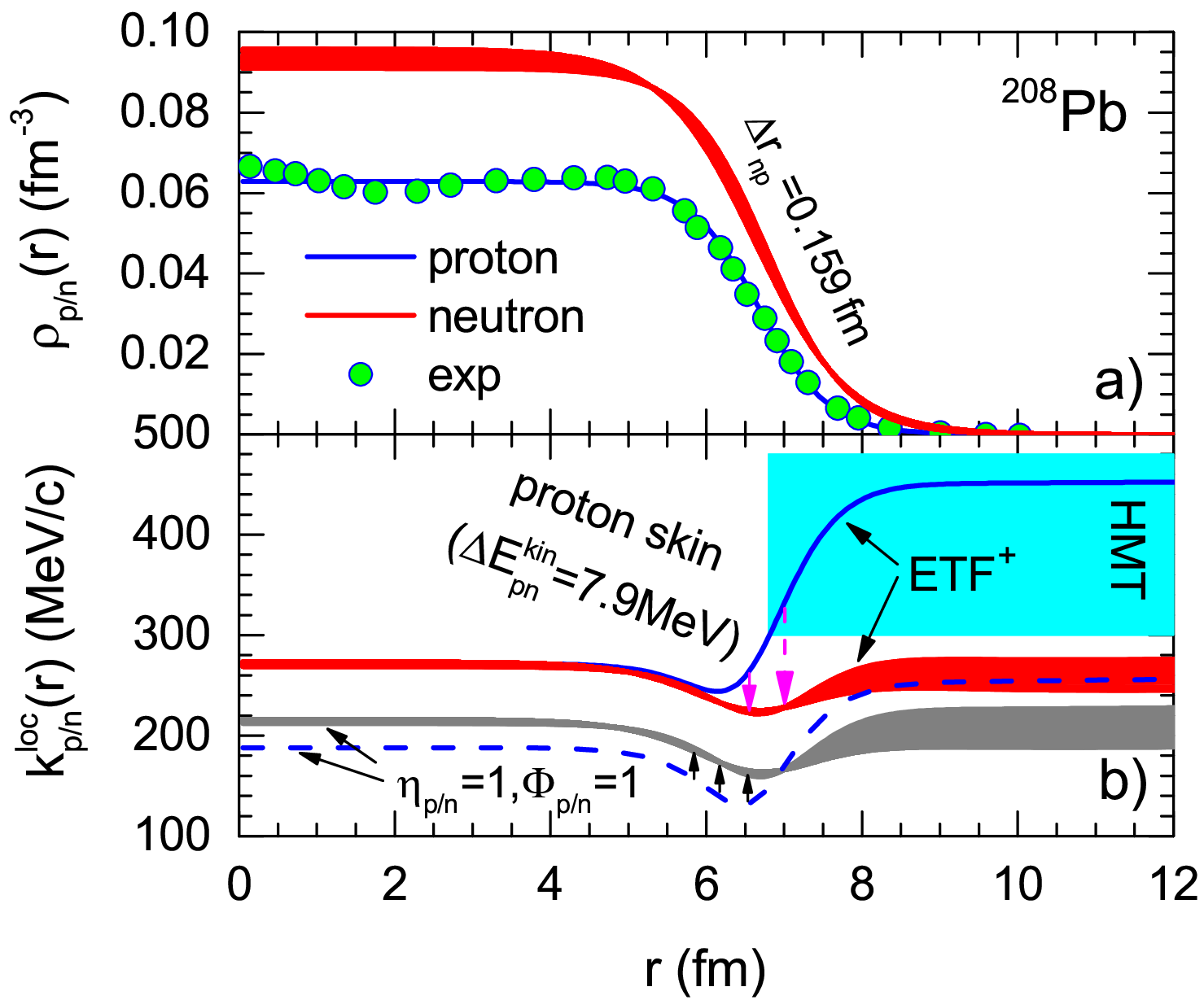}\\[0.25cm]
\hspace{-0.15cm}
\includegraphics[width=0.44\textwidth]{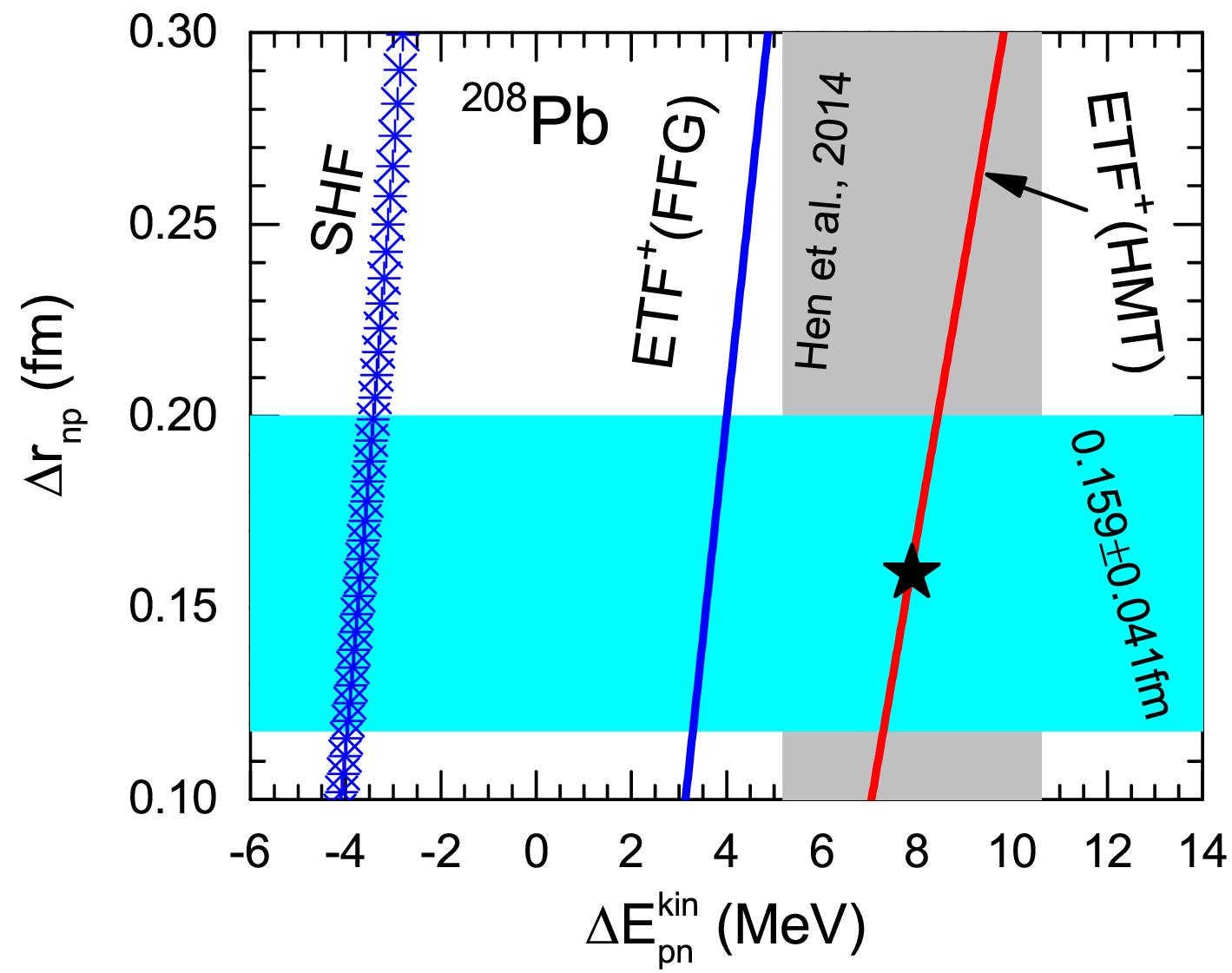}
\caption{(Color Online). Upper: neutron and proton density profiles and local momenta in $^{208}\rm{Pb}$. Lower: correlation between the
coordinate-space neutron skin $R_{\rm{skin}}\equiv\Delta r_{\rm{np}}$ and momentum-space proton skin $\Delta E_{\rm{pn}}^{\rm{kin}}$ in ETF$^+$ calculations with and without the HMT. Adapted from Ref.\,\cite{Cai16b}.}
\label{fig:skin-correlation}
\end{figure}

\begin{figure}[h!]
\centering
\includegraphics[width=0.48\textwidth]{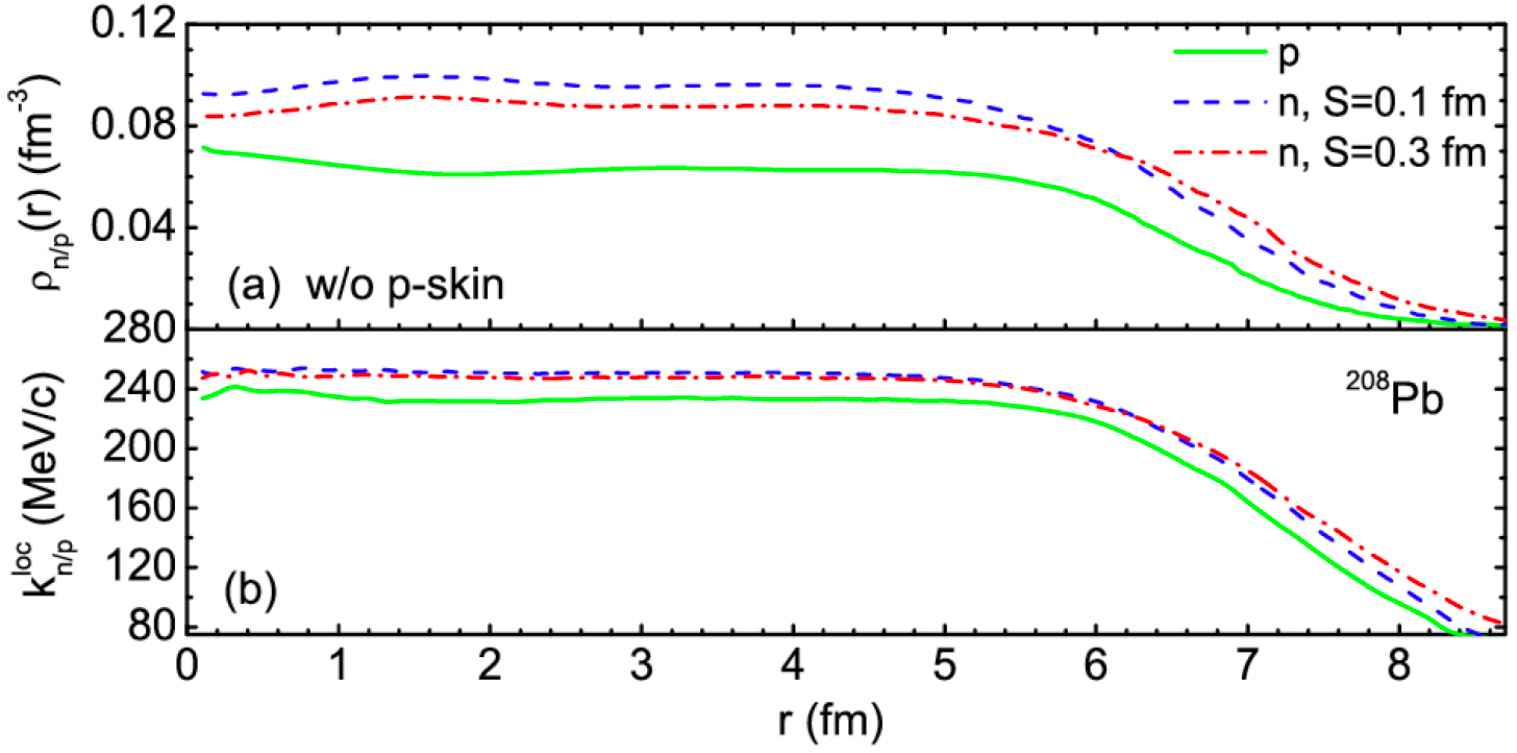}\\[0.25cm]
\includegraphics[width=0.48\textwidth]{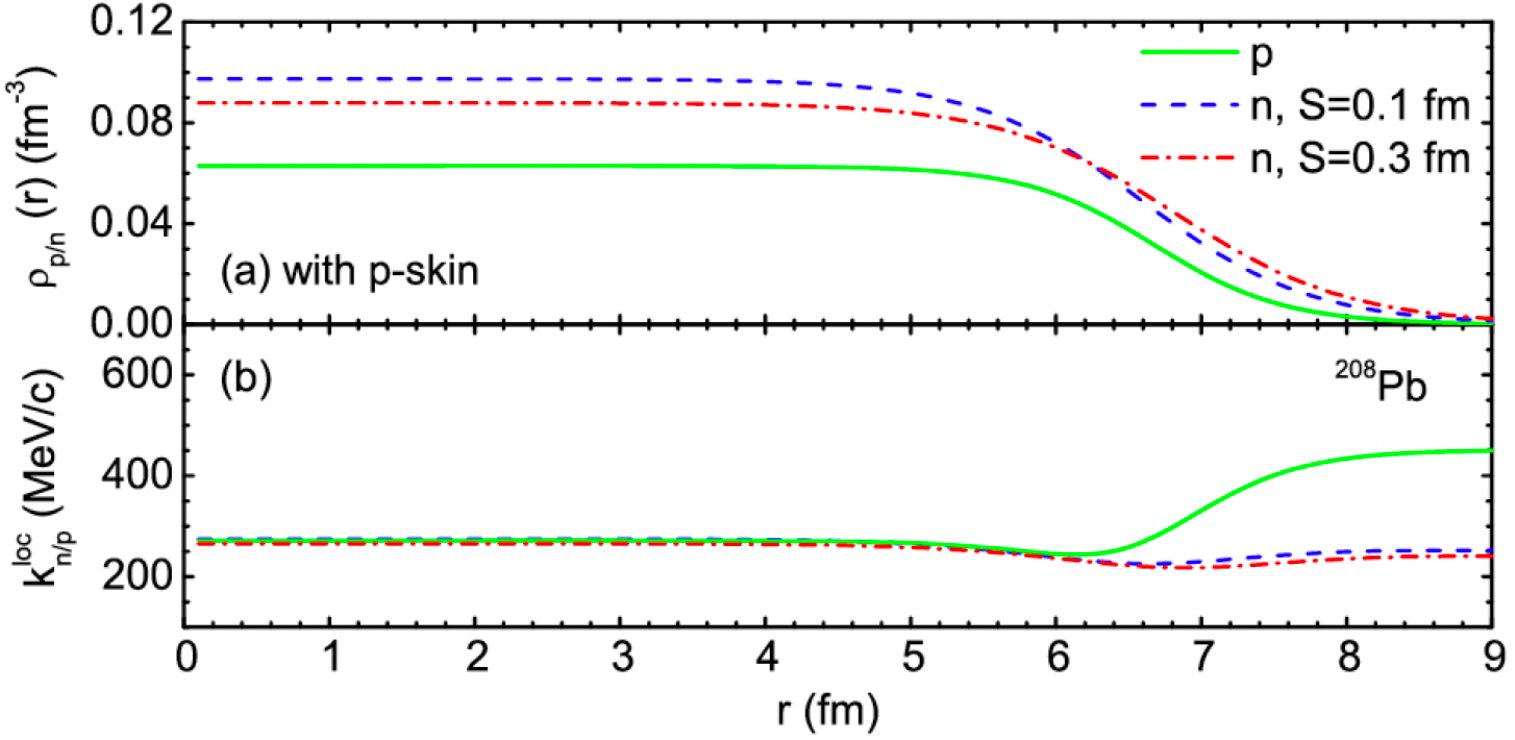}
\caption{(Color Online). 
Upper panels: input local-density profiles (a) and calculated
nucleon momentum distributions (b) for $^{208}\rm{Pb}$ obtained using the Skyrme functional, with neutron-skin thicknesses
$S=\langle\Delta r_{\rm{np}}\rangle=0.1$\,fm and $0.3$\,fm. Here, the nucleon kinetic-energy density of the colliding nuclei contains only the bulk contribution. Lower panels: corresponding results obtained after including the surface contribution to the nucleon kinetic-energy density. Figure taken from Ref.\,\cite{Guo2023PRC}.}
\label{fig_Guo23-nk}
\end{figure}

To connect momentum and coordinate spaces, it is useful to begin with the extended Thomas-Fermi kinetic energy density. To order $\hbar^2$ in the Wigner-Kirkwood expansion, it can be written as\,\cite{Cai16b,Brack1985PhysRep}
\begin{equation}
\varepsilon_J^{\rm{kin}}(r)=\frac{1}{2M_{\rm N}}
\left[\alpha_J^\infty\rho_J^{5/3}
+\frac{\eta_J}{36}\frac{(\nabla\rho_J)^2}{\rho_J}
+\frac{1}{3}\nabla^2\rho_J\right].
\label{eq:ETF-kinetic-density}
\end{equation}
The first term is the bulk contribution, the second is the
surface-sensitive Weizs\"acker term, and the last term is generally small for a smooth density profile. For the SRC-modified distribution in Eq.\,(\ref{eq:skin-nk}), the uniform-matter contribution becomes
\begin{equation}
\frac{2}{(2\pi)^3}\int_0^{\phi_Jk_{\rm{F}}^J}
\frac{k^2}{2M_{\rm N}}n_{\v{k}}^J\d\vec{k}
=\frac{1}{2M_{\rm N}}\frac{3}{5}(3\pi^2)^{2/3}
\rho_J^{5/3}\Phi_J,
\label{eq:SRC-kinetic-density}
\end{equation}
where $
\Phi_J=1+C_J(5\phi_J+{3}/{\phi_J}-8)>1$, charaterizing the SRC-HMT effect.
Thus $\alpha_J^\infty=(3/5)(3\pi^2)^{2/3}\Phi_J$. Since SRCs cause a
larger fractional depletion of protons in a neutron-rich system, the
enhancement factor is larger for protons. For $^{208}\rm{Pb}$, the
HMT-exp parametrization gives $\Phi_{\rm{p}}\approx2.09\pm0.50$ and
$\Phi_{\rm{n}}\approx1.60\pm0.33$\,\cite{Cai16b}. A convenient measure of the momentum-space proton skin is therefore\,\cite{Cai16b}
\begin{equation}
\Delta E_{\rm{pn}}^{\rm{kin}}
\equiv\langle E_{\rm{p}}^{\rm{kin}}\rangle
-\langle E_{\rm{n}}^{\rm{kin}}\rangle,~~
\langle E_J^{\rm{kin}}\rangle
=\frac{\int\varepsilon_J^{\rm{kin}}(r)\d\vec{r}}
{\int\rho_J(r)\d\vec{r}}
=\frac{\langle k_J^2\rangle}{2M_{\rm N}}.
\label{eq:proton-skin-definition}
\end{equation}
The upper part of FIG.\,\ref{fig:skin-correlation} shows the density and local momentum profiles obtained for $^{208}\rm{Pb}$, where $[k_J^{\rm{loc}}(r)]^2/2M_{\rm N}=\varepsilon_J^{\rm{kin}}(r)/\rho_J(r)$. The neutron density extends farther outward, whereas protons carry a larger local momentum near the surface once the HMT and surface term are included. The lower part exhibits an approximately linear correlation between the neutron skin
$R_{\rm{skin}}\equiv \Delta r_{\rm{np}}$ and $\Delta E_{\rm{pn}}^{\rm{kin}}$ within the ETF$^+$ model. Independent
Skyrme calculations likewise
demonstrate that the surface contribution is crucial for producing a
large proton momentum near the nuclear edge\,\cite{Guo2023PRC}, see FIG.\,\ref{fig_Guo23-nk}.

The physical origin of this complementarity can be clarified directly from phase space counting. Let $f(\v{r},\v{k})$ be normalized according to $
[{2}/{(2\pi)^3}]\int f(\vec{r},\vec{k})\d\vec{r}\d\vec{k}=N$.
For a uniform Fermi sphere,
$f=\Theta(R-r)\Theta(k_{\rm{F}}-k)$, this gives
$Rk_{\rm{F}}=(9\pi N/4)^{1/3}$ and
\begin{align}
\langle r^\ell\rangle\langle k^\ell\rangle
&=\left(\frac{3}{\ell+3}\right)^2R^\ell k_{\rm{F}}^\ell.
\label{eq:phase-space-general-moment}
\end{align}
If the momentum distribution contains the $k^{-4}$ tail of
Eq.\,(\ref{eq:skin-nk}), while the spatial distribution is kept uniform, the latter relation becomes, for $\ell\ne1$,
\begin{align}
\langle r^\ell\rangle\langle k^\ell\rangle
=&\left(\frac{3}{\ell+3}\right)^2R^\ell k_{\rm{F}}^\ell
\left[1+C_0\left(\frac{\ell+3}{\ell-1}\phi_0^{\ell-1}
+\frac{3}{\phi_0}-\frac{4\ell}{\ell-1}\right)\right].
\label{eq:HMT-general-moment}
\end{align}

Liouville's theorem states that Hamiltonian evolution preserves
phase space volume. For a degenerate system of spin half nucleons, the related semiclassical state-counting argument assigns two spin states to each phase-space cell of volume $h^3$. At fixed particle number, the effective coordinate- and momentum-space volumes are therefore linked: an increase of the characteristic spatial scale is accompanied, in this simple counting picture, by a reduction of the characteristic momentum scale. This inverse relation is the phase-space origin of the complementarity between spatial and momentum distributions. Strictly speaking, however, the cell-counting relation follows from quantum state counting and the Pauli principle, whereas Liouville's theorem ensures that the occupied phase-space volume is preserved during collisionless
Hamiltonian evolution. The uncertainty principle adds a genuinely quantum constraint, $\delta r\delta k\gtrsim\hbar/2$, preventing both distributions from becoming arbitrarily narrow at the same time. Consequently, $\langle r^2\rangle\langle k^2\rangle$ contains contributions from both the characteristic scales and their fluctuations\,\cite{Cai16b}. SRCs enlarge the momentum-space moments by transferring strength from below $k_{\rm{F}}$ to the HMT; their magnitude is controlled by the HMT strength $C_0$ and cutoff $\phi_0$. Coordinate-space diffuseness and momentum-space
broadening are therefore constrained by common phase-space considerations, although neither Liouville's theorem nor the uncertainty principle fixes their detailed functional forms.

For each nucleon species, this connection can be summarized by the nearly invariant combination $H_J\equiv\langle r_J^2\rangle\langle k_J^2\rangle$. Its residual dependence on the detailed density profile may be organized as an expansion in the small ratio of the surface-diffuseness scale to the bulk
length scale\,\cite{Cai2026EPJST},
\begin{equation}
H_J\approx\vartheta_0(N_J)
+\vartheta_1(N_J)\frac{\Lambda_{\rm{df}}^J}{\Lambda_{\rm{av}}^J}
+\vartheta_2(N_J)\left(\frac{\Lambda_{\rm{df}}^J}
{\Lambda_{\rm{av}}^J}\right)^2+\cdots.
\label{eq:HJ-expansion}
\end{equation}
Here $N_J$ is the number of nucleons of species $J$, while
$\Lambda_{\rm{av}}^J$ and $\Lambda_{\rm{df}}^J$ characterize its bulk and surface diffuseness scales, respectively. Since
$\Lambda_{\rm{df}}^J/\Lambda_{\rm{av}}^J$ is small, $H_J$ changes only moderately with the detailed surface profile. The coordinate- and momentum-space distributions are subject to common phase-space and quantum-mechanical constraints, but these constraints alone do not imply a universal relation between the neutron skin and the momentum-space proton skin. The latter is generated dynamically by the isospin dependence of the nuclear interaction, with SRCs playing a particularly important role through the predominance of neutron-proton pairs and the resulting larger fractional high-momentum occupation of minority protons in neutron-rich nuclei.

The isovector nuclear interaction supplies the dynamics that distinguishes neutrons from protons within this phase-space constraint. Its long- and intermediate-range mean-field components favor a more extended neutron density, whereas its short-range spin-isospin and tensor components preferentially correlate neutron-proton pairs and broaden the fractional proton momentum distribution. A neutron skin in coordinate space and a
proton skin in momentum space can therefore coexist without contradiction: the former records isovector separation in $r$ space, while the latter records isovector momentum fractionation in $k$ space. Their correlation is therefore not a universal consequence of phase-space conservation or quantum uncertainty, but reflects the isospin-dependent nuclear dynamics that determines the coordinate- and momentum-space distributions. Its precise slope and magnitude depend on the density profiles, surface terms, SRC strength, and isovector interaction. Measurements sensitive to both skins can consequently test the isovector nuclear force more stringently than either coordinate- or momentum-space information alone. The two effects should also be treated simultaneously when interpreting particle emission and hard-photon observables in heavy-ion collisions.

\section{Neutron Star and Neutron Skin in Multi-Messenger Era}
\label{sec:astrophysics}

Neutron-rich nuclei and NSs represent two remarkably different manifestations of strongly interacting matter with large isospin asymmetry. Their enormous difference in size and density notwithstanding, they are connected through the nuclear EOS, particularly its isovector
sector. Neutron-skin measurements therefore provide terrestrial information on neutron-rich matter around and below nuclear saturation density, while NS observations extend the exploration of the EOS to the much higher densities realized in stellar interiors. Establishing
the connection between these regimes has become an important theme in modern nuclear physics and nuclear astrophysics\,\cite{Li2022ApJ,Chatziioannou2025RMP,Drischler2021ARNPS,Lovato2022LRP,Li2025eXTP,
Sorensen2024PPNP,Li2019EPJA,Dexheimer2021JPG,
Lattimer2021ARNPS,Liu2024NSTAstro}.

The multi-messenger era has substantially broadened this connection. Measurements of neutron skins and other nuclear observables can now be combined with HIC experiments, precise pulsar masses, X-ray measurements
of NS radii, and GW constraints on tidal deformability. These probes are sensitive to different, but overlapping, density regions and thus provide complementary information on the EOS. In this section, we discuss how neutron skins are connected with NS radii and crustal properties through the density dependence of the symmetry energy, examine the information carried by tidal deformability, and finally
highlight how terrestrial nuclear experiments, HICs, and multi-messenger observations can be combined to constrain neutron-rich matter over a broad range of densities.

\subsection{From neutron skins to neutron star radii}

The connection between neutron skins in finite nuclei and NS radii originates from their common sensitivity to the pressure of neutron-rich matter. Although the two systems differ enormously in size and density, their properties are governed by the same underlying isovector nuclear interaction. In a neutron-rich nucleus, the pressure
associated with the symmetry energy tends to push excess neutrons toward the nuclear surface, thereby producing a neutron skin. In a NS, the pressure of neutron-rich matter provides the support against gravity and plays a central role in determining the stellar radius. This common dependence establishes an important physical link
between terrestrial measurements of neutron skins and astrophysical observations of NS radii\,\cite{Brown2000PRL,Horowitz2001PRL,Horowitz2001PRC,
Steiner2005PhysRep,Lattimer2014NPA,Wu2026AA,Yang2020ARNPS}.

A particularly transparent starting point was provided by
Brown\,\cite{Brown2000PRL}, who demonstrated a strong correlationbetween the neutron radius of $^{208}\rm{Pb}$ and the densitydependence of the neutron matter EOS. Around nuclear saturation density, this connection may be understood from the approximate relation Eq.\,(\ref{eq:symmetry_pressure}). A larger symmetry energy slope $L$ therefore generally corresponds to a larger pressure of neutron-rich matter around $\rho_0$. In a finite neutron-rich nucleus, this enhanced pressure favors a larger spatial separation between the neutron and proton distributions, leading to the well established correlation among $L$, $P_{\rm{PNM}}(\rho_0)$ and $R_{\rm{skin}}$.
This correlation has subsequently been explored using a wide variety of nuclear energy-density functionals and microscopic  calculations. The neutron-skin thickness can thus be regarded as an important terrestrial probe of the pressure of neutron-rich matter in the vicinity of nuclear density.

\begin{figure}[h!]
\centering
\includegraphics[width=0.48\textwidth]{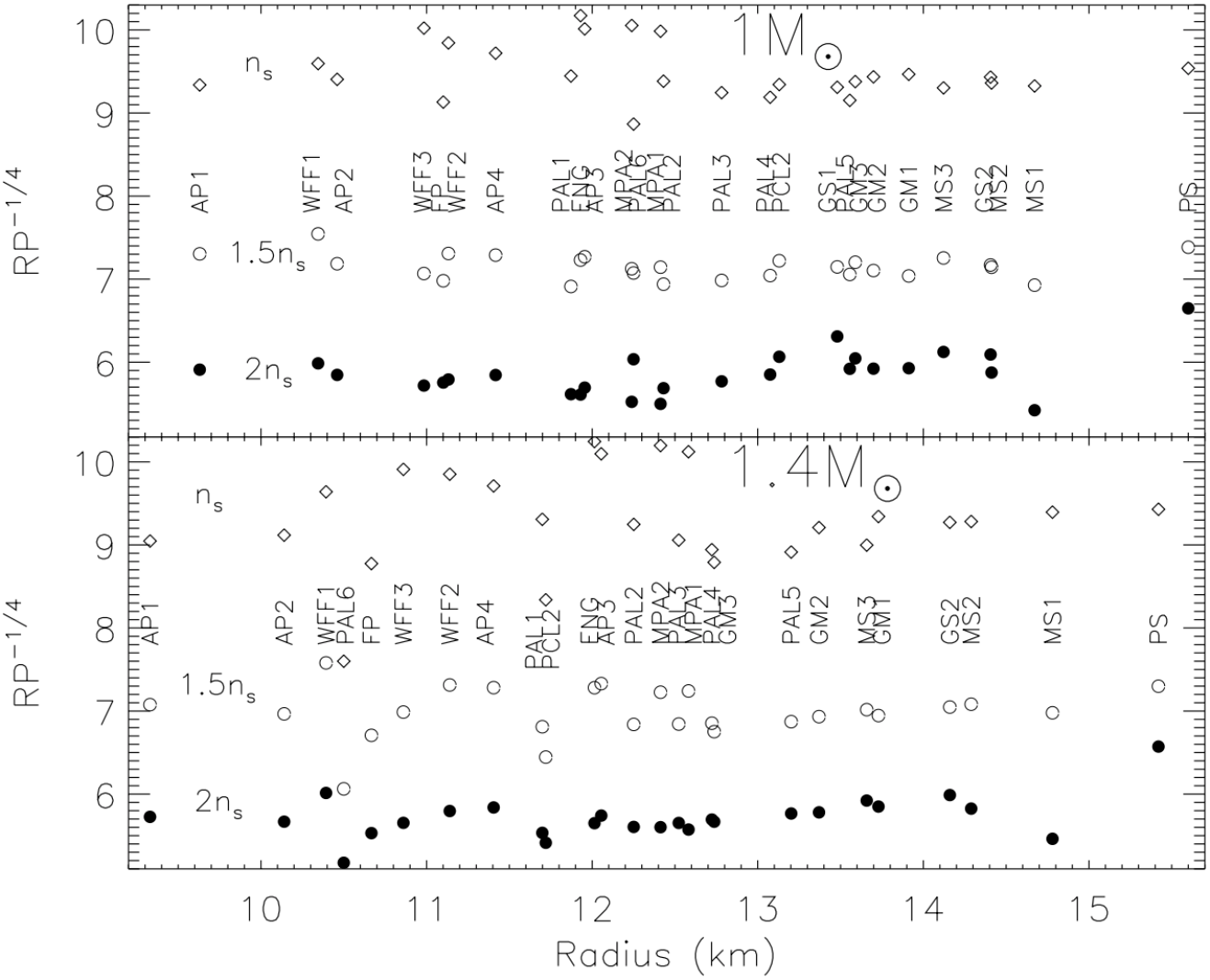}
\caption{Mass-radius relations for representative EOS 
of NS matter. The substantial variation of stellar radii
among different EOSs reflects their sensitivity to the
pressure of dense matter. Adapted from
Ref.\,\cite{Lattimer2001ApJ}.}
\label{fig:Lattimer2001_MR}
\end{figure}

The same physics has important consequences for NSs, which
may be viewed, in a broad sense, as gigantic neutron-rich nuclear systems bound by gravity. The density dependence of the symmetry energy that drives the formation of neutron skins also contributes substantially to the pressure of NS matter, particularly around nuclear saturation density. A stiffer symmetry energy, usually
associated with a larger $L$, generates a larger pressure in this density region and generally favors a larger stellar radius at a given mass\,\cite{Lattimer2001ApJ,Steiner2005PhysRep,Hebeler2010PRL,
Hebeler2014EPJA,Lattimer2014NPA}, namely increasing $L$ tends to shift the mass-radius (MR) relation toward larger radii. The effect is particularly evident for canonical-mass NSs, whereas the maximum mass is more strongly controlled by the EOS at higher densities.
The connection between pressure and stellar radius was quantified systematically by Lattimer and Prakash\,\cite{Lattimer2001ApJ}.
Considering a broad range of EOS, they found that the
radius of a NS of fixed mass is strongly correlated with
the pressure at densities around and moderately above saturation density. The correlation can be approximately expressed as $
R_M\approx C_MP^\delta(\rho)$ with $
\delta\approx0.23\sim0.26\approx1/4$, where $C_M$ depends only weakly on the stellar mass for ordinary NSs. The correlation becomes particularly tight when the pressure is evaluated at densities around $1$-$2\rho_0$. Physically, a larger pressure in this density regime provides stronger support against gravity and generally produces a larger stellar radius. The
broad range of MR relations shown in
FIG.\,\ref{fig:Lattimer2001_MR} illustrates the sensitivity of NS radii to the underlying dense matter EOS.

The pressure-radius correlation has subsequently become one of the basic ingredients in using NS observations to constrain dense matter\,\cite{Steiner2005PhysRep}. In particular, observational constraints on NS radii can be translated into constraints on the pressure at densities around one to two times nuclear saturation density. With the advent of gravitational wave events, the strong correlation between tidal deformability and stellar radius has further enabled tidal measurements to constrain the pressure in this density
regime through the conventional chain $
\Lambda
\leftrightarrow
R_{\rm{NS}}
\leftrightarrow
P(\rho\sim1\text{-}2\rho_0)$.
This interpretation has played an important role in extracting information on dense matter from NS observations\,\cite{Fattoyev2018PRL}. As will be discussed briefly in the following subsection, however, tidal deformability may also provide more direct information on the core EOS than suggested by this indirect chain alone.

\begin{figure}[h!]
\centering
\includegraphics[width=0.48\textwidth]{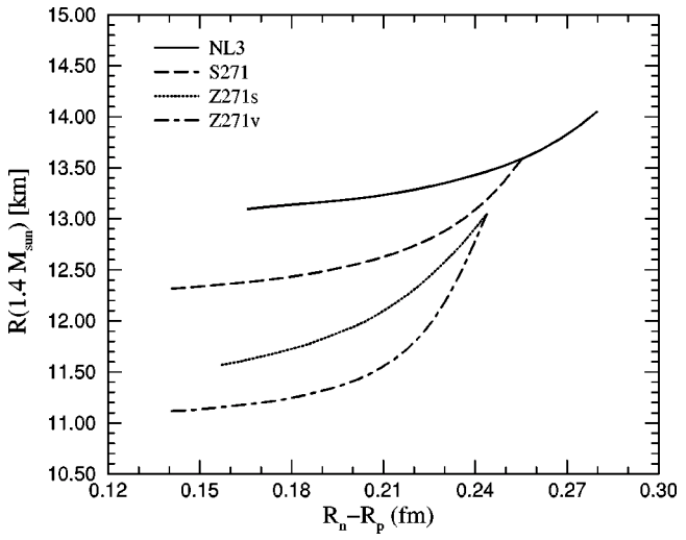}
\caption{Radius of a canonical NS as a function of the
neutron-skin thickness of $^{208}\rm{Pb}$ for several relativistic interactions. Within a given family of interactions. Adapted from
Ref.\,\cite{Horowitz2001PRC}.}
\label{fig:Horowitz2001_skin_radius}
\end{figure}

\begin{figure*}
\centering
\includegraphics[width=0.98\textwidth]{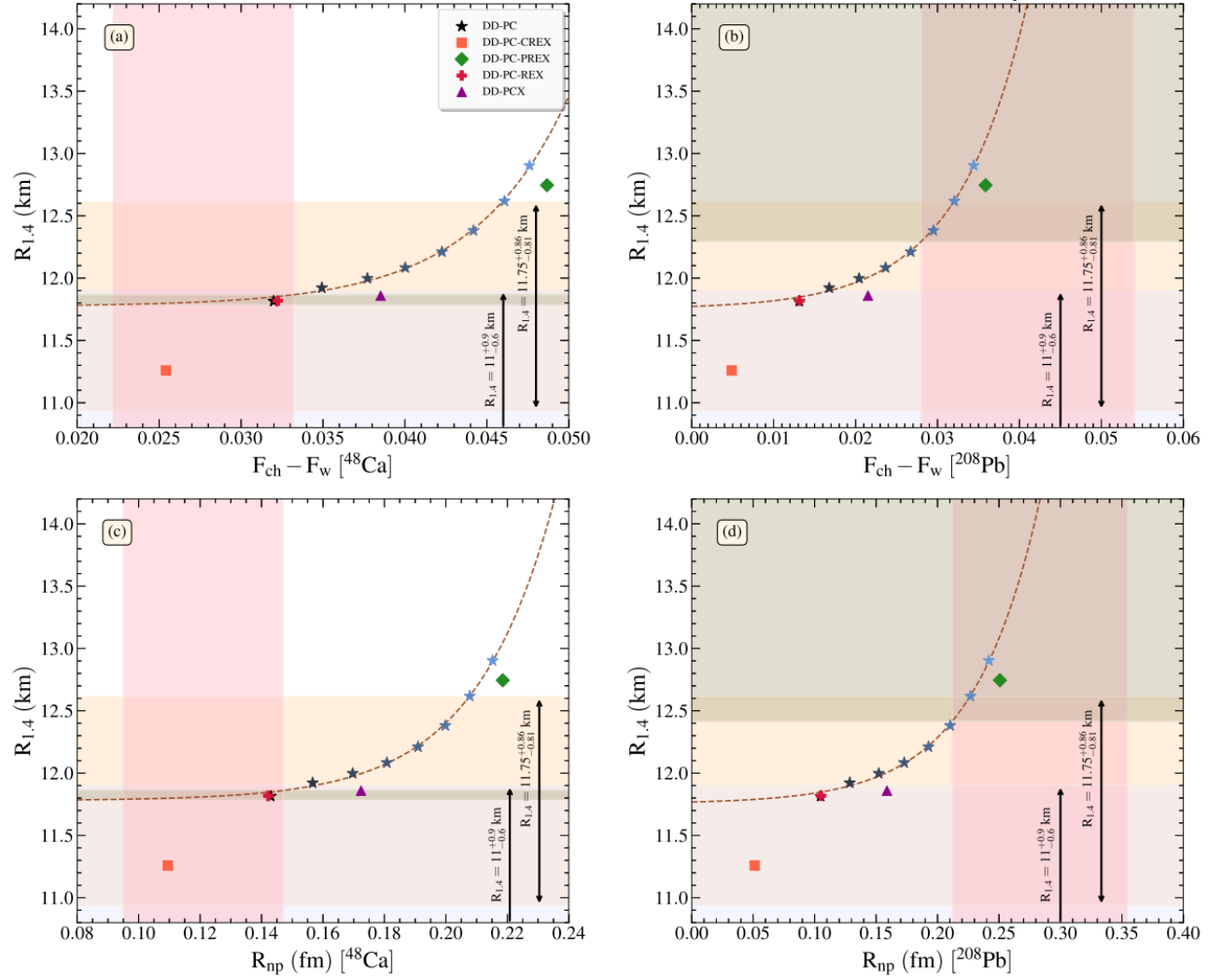}
\caption{(Color Online). Radius $R_{1.4}$ of a canonical NS as a function
of finite-nucleus observables for $^{48}\rm{Ca}$ and $^{208}\rm{Pb}$. The lower panels display the correlations between $R_{1.4}$ and the corresponding neutron-skin thickness. The vertical shaded regions represent constraints from CREX and PREX-II, while the horizontal
shaded regions indicate astrophysical constraints on the NS
radius. Adapted from Ref.\,\cite{Koliogiannis2025PLB}.}
\label{fig:Koliogiannis2025_skin_radius}
\end{figure*}

Combining the finite-nucleus and NS arguments leads to the
basic physical connection $
R_{\rm{skin}}
\leftrightarrow
L
\leftrightarrow
P(\rho\sim1\text{-}2\rho_0)
\leftrightarrow
R_{\rm{NS}}$.
A thicker neutron skin therefore generally favors a larger
NS radius, provided that the relatively large pressure
inferred around saturation density persists toward higher densities.
In this sense, the neutron skin of a heavy nucleus provides a terrestrial anchor for the low-density side of the same EOS that determines NS structure\,\cite{Horowitz2001PRC,Steiner2005PhysRep,Hebeler2014EPJA,
Fattoyev2018PRL}.
This connection was investigated explicitly by Horowitz and
Piekarewicz\,\cite{Horowitz2001PRC}. By systematically modifying the density dependence of the symmetry energy while retaining a successful description of finite nuclei, they studied the relation between the neutron-skin thickness of $^{208}\rm{Pb}$ and the radius of a
$1.4M_\odot$ NS. As shown in FIG.\,\ref{fig:Horowitz2001_skin_radius}, within a given family of interactions a larger neutron skin is associated with a larger $R_{1.4}$, providing a direct realization of the physical connection aforementioned.

\begin{figure}[h!]
\centering
\includegraphics[width=0.45\textwidth]{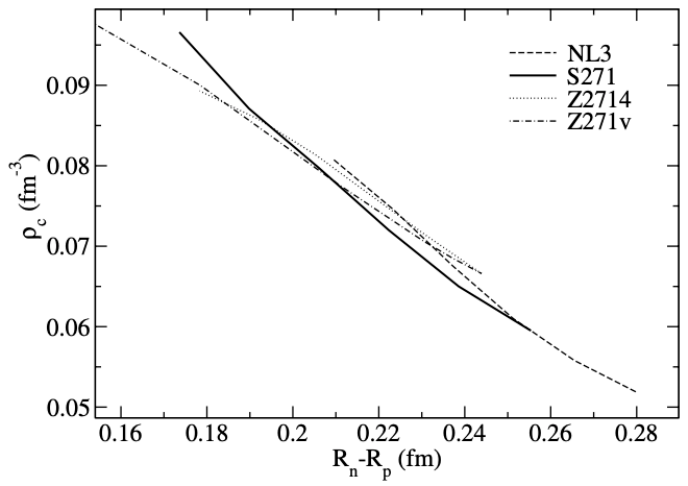}
\caption{(Color Online). Crust-core transition density $\rho_{\rm c}\equiv \rho_{\rm t}$ as a function of the neutron-skin thickness
$R_{\rm{skin}}$ of $^{208}\rm{Pb}$, obtained with several
relativistic effective interactions. A clear anti-correlation is found: a thicker neutron skin, associated with a more rapidly varying symmetry energy around sub-saturation density, generally corresponds to a lower transition density from nonuniform crustal matter to uniform neutron-rich matter. Adapted from
Ref.\,\cite{Horowitz2001PRL}.}
\label{fig:skin-crust-transition}
\end{figure}

\begin{figure}[h!]
\centering
\includegraphics[width=0.48\textwidth]{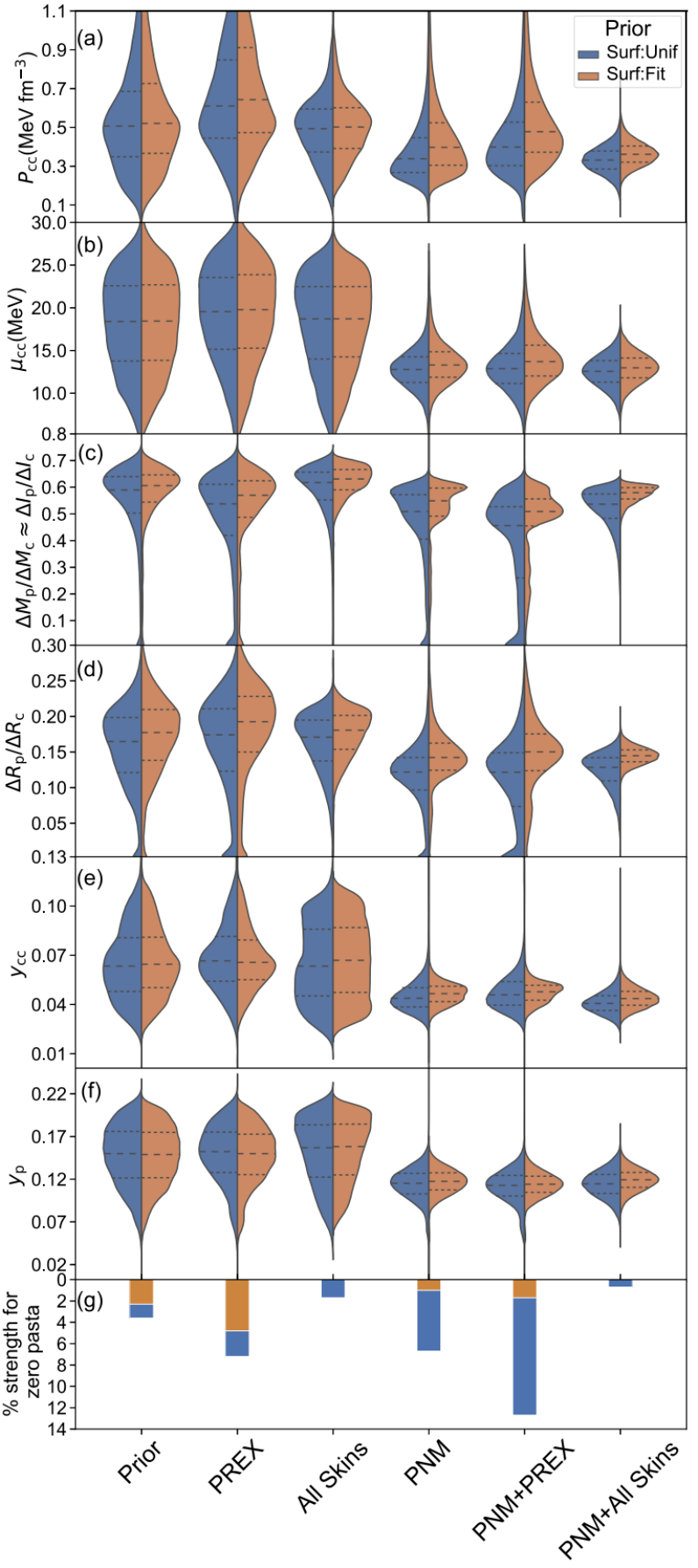}
\caption{(Color Online). Posterior distributions of selected NS crust properties. The comparison illustrates how neutron-skin measurements, including PREX, complement microscopic PNM information in constraining the structure and composition of the inner crust. Adapted from Ref.\,\cite{Newton2022PLB}.}
\label{fig:skin-crust}
\end{figure}

Importantly, the correlation between $R_{\rm{skin}}$ and $R_{1.4}$ does not imply a unique mapping between the two observables. The neutron skin of a heavy nucleus primarily constrains neutron-rich matter at subsaturation and near saturation densities, whereas the
radius of a canonical NS is particularly sensitive to the
pressure over a broader density interval extending to approximately $1$-$2\rho_0$.
Consequently, two EOSs may predict similar neutron skins
but develop different pressures at supra-saturation densities and hence produce different NS radii. Conversely, a relatively thick neutron skin accompanied by a small NS radius could indicate that an EOS that is relatively stiff around saturation density becomes softer as the density increases. Simultaneous information on
neutron skins and NS radii therefore probes not only the
pressure around $\rho_0$, but also its evolution toward higher densities.
The connection has acquired renewed significance with increasingly precise terrestrial and astrophysical measurements. PVES provides a clean electroweak determination of neutron distributions in finite nuclei, as demonstrated by PREX-II for
$^{208}\rm{Pb}$ and CREX for $^{48}\rm{Ca}$, while X-ray and GW observations provide complementary information on
NS radii. The apparently different implications of PREX-II
and CREX for the density dependence of the symmetry energy have motivated extensive recent investigations within energy density functional, microscopic, and Bayesian approaches\,\cite{Zhang2023PRC}.
A recent example was given in Ref.\,\cite{Koliogiannis2025PLB}, who connected the finite-nucleus
observables measured by CREX and PREX-II with NS properties
using a family of relativistic density dependent point coupling functionals. As shown in
FIG.\,\ref{fig:Koliogiannis2025_skin_radius}, the calculated $R_{1.4}$ increases systematically with the neutron-skin thicknesses of both $^{48}\rm{Ca}$ and $^{208}\rm{Pb}$. The terrestrial constraints from CREX and PREX-II can therefore be confronted directly
with astrophysical constraints on NS radii.

The overall physical picture is therefore clear, but should not be interpreted as a universal one-to-one correspondence. Neutron skins provide an important anchor on the isovector EOS around and below saturation density, whereas NS radii probe the
continuation of the pressure into the supra-saturation regime. Combining the two observables consequently provides information on how the EOS evolves
across these density regions. Heavy ion collisions offer an important complementary avenue in this respect, since they extend terrestrial access to neutron-rich matter toward supra-saturation densities and can therefore help bridge the density regimes probed by finite nuclei
and NSs. The additional information carried by tidal
deformability, including its relation to NS radii and to the core EOS, will be discussed in the following subsection.

Beyond the stellar radius, the neutron-skin thickness is also connected to several properties of the NS crust, since both probe the density dependence of the symmetry energy at sub-saturation densities\,\cite{Horowitz2001PRL,Xu2009PRC,Xu2009ApJ,Newton2013ApJS,
Newton2021PRC,Newton2022PLB,Xie2023NST,Grams2022PRC}.
A particularly important quantity is the crust-core transition density $\rho_{\rm t}$, at which nonuniform matter in the inner crust becomes unstable against uniform neutron-rich matter. The anti-correlation between $R_{\rm{skin}}$ and $\rho_{\rm t}$ was demonstrated by Horowitz and Piekarewicz\,\cite{Horowitz2001PRL} (FIG.\,\ref{fig:skin-crust-transition}) and subsequently investigated using different nuclear interactions and stability criteria\,\cite{Xu2009PRC,Xu2009ApJ}. The transition pressure $P_{\rm t}$ is also sensitive to the symmetry energy, although its correlation with $L$ or $R_{\rm{skin}}$ is generally less direct and more model dependent.
More recent statistical analyses have placed these connections on a quantitative footing. In particular, Bayesian studies combining neutron-skin measurements with constraints on pure neutron matter have shown that finite nucleus information can constrain not only the crust-core boundary but also the crust thickness and moment of inertia, as well as the amount and composition of nuclear pasta\,\cite{Newton2021PRC,Newton2022PLB,Newton2013ApJS,Xie2023NST,Grams2022PRC}.
As illustrated in FIG.\,\ref{fig:skin-crust}, Newton {et al.}\,\cite{Newton2022PLB} quantified how PREX and other neutron-skin data, when combined with chiral-EFT constraints on pure neutron matter, modify the posterior distributions of the crust-core transition pressure and chemical potential, the fractional mass and thickness of the pasta layers, and the proton fractions characterizing the pasta region. These results demonstrate that neutron-skin information can propagate well beyond the bulk radius of an NS into its detailed crustal structure. This connection is especially compelling because the densities characteristic of the inner crust overlap directly with the
sub-saturation regime probed by finite nuclei and terrestrial nuclear experiments.

\subsection{Tidal deformability and the dense matter EOS}

\begin{figure}[h!]
\centering
\includegraphics[width=0.48\textwidth]{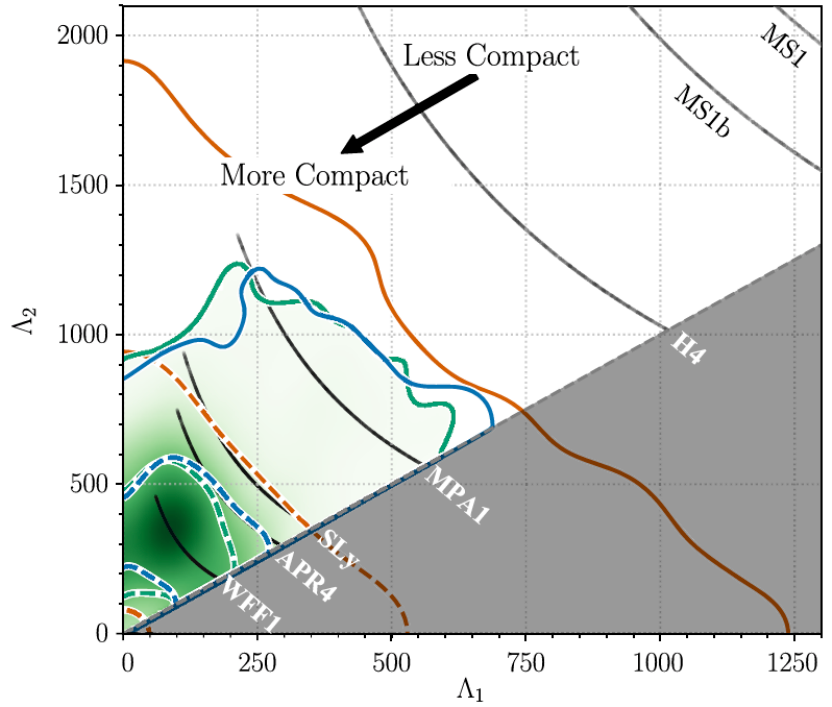}
\caption{(Color Online). Joint posterior distribution of the tidal deformabilities $\Lambda_1$ and $\Lambda_2$ of the two NSs in GW170817.  Results obtained by imposing a common EOS through an EOS-insensitive relation (green) and a parametrized EOS (blue) are compared with an analysis in which the two tidal deformabilities are treated independently (orange).  Representative EOS predictions are also shown, illustrating the sensitivity of the GW measurement to the stiffness of dense matter. Adapted from
Ref.\,\cite{Abbott2018PRL}.}
\label{fig:GW170817Lambda}
\end{figure}

\begin{figure}[h!]
\centering
\includegraphics[width=0.48\textwidth]{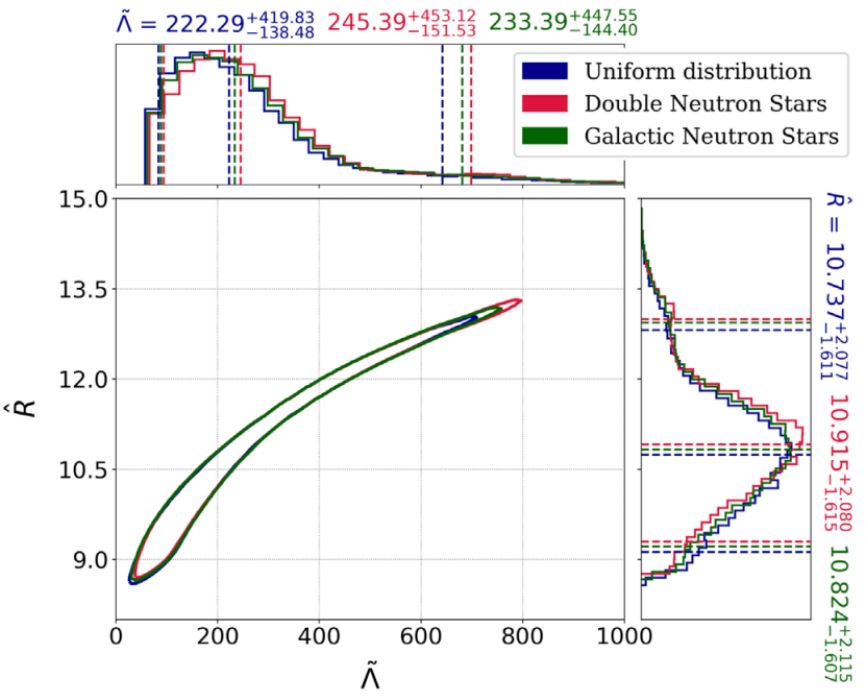}
\caption{(Color Online). Constraints on the binary tidal deformability and the common NS radius inferred from GW170817 for different assumptions about the component-mass distribution. The strong correlation illustrates how tidal information from GWs can be translated into constraints on NS radii. Adapted from
Ref.\,\cite{De2018PRL}.}
\label{fig:De2018_Lambda_R}
\end{figure}

The detection of GWs from binary NS mergers has
opened a new avenue for constraining the EOS of dense matter through the tidal response of NSs. During the inspiral, each star develops a quadrupole deformation in response to the gravitational field of its companion, leaving a characteristic imprint on the
GW signal\,\cite{Flanagan2008PRD,Hinderer2008ApJ,Hinderer2010PRD,
Abbott2017PRL}. The corresponding dimensionless tidal deformability is conventionally expressed as\,\cite{Hinderer2008ApJ}
\begin{equation}\label{eq:Lambda}
\Lambda=(2/3)k_2 \xi^{-5},
\end{equation}
where $k_2$ is the tidal Love number and $\xi\equiv M_{\rm{NS}}/R$ is the NS compactness. The strong dependence on compactness makes $\Lambda$ particularly
sensitive to the NS radius and hence to the underlying EOS.
GW170817 provided the first observational constraint on the tidal
deformability of NSs and demonstrated the potential of GWs for probing dense matter\,\cite{Abbott2017PRL,Abbott2018PRL}.  As illustrated in FIG.\,\ref{fig:GW170817Lambda}, the measured tidal deformabilities of the two binary components are strongly constrained once they are required to be described by a common EOS.  The resulting posterior distribution favors relatively small tidal deformabilities and disfavors several stiff EOSs predicting large values of $\Lambda$. According to Eq.\,(\ref{eq:Lambda}), tidal deformability is particularly sensitive to the stellar compactness and radius, allowing GW measurements to translate directly into constraints on NS radii and, ultimately, the dense matter EOS. De et al.\,\cite{De2018PRL}
showed that $\Lambda$ approximately scales with the sixth power of the radius at fixed mass, allowing the tidal information encoded in the GW signal to be translated into constraints on NS radii. As illustrated in FIG.\,\ref{fig:De2018_Lambda_R}, their analysis of GW170817 obtained a joint posterior distribution for the
binary tidal deformability and the characteristic radius, with the inferred radius remaining relatively insensitive to different assumptions about the component mass distribution.

The strong $\Lambda$-$R$ correlation has subsequently been explored using a variety of EOS constructions and inference schemes\,\cite{Annala2018PRL,Most2018PRL,Lim2018PRL,Raithel2018ApJL,
Abbott2018PRL,De2018PRL}. Although the quantitative results depend on the adopted EOS priors and assumptions concerning the high-density EOS, these analyses generally place the radius of a canonical NS in the range of about $11$-$13$\,km, with $R_{1.4}\approx 12$\,km representing
a useful characteristic scale. For example, De et
al.\,\cite{De2018PRL} inferred a common radius with a mean value of about $10.8$\,km and a relatively broad credible interval, whereas the LIGO-Virgo EOS analysis obtained a characteristic radius of about $11.9$\,km after imposing the observational lower limit on the maximum NS mass\,\cite{Abbott2018PRL}. Other analyses based on broad
EOS ensembles similarly favored canonical radii around $12$-$13$\,km\,\cite{Annala2018PRL,Most2018PRL}. These results established tidal deformability as one of the principal astrophysical observables for constraining NS radii and the EOS of dense matter.

The close relation between tidal deformability and NS radius also provides an important connection between GW
observations and terrestrial probes of neutron-rich matter. As discussed in the preceding subsection, neutron skins constrain the density dependence of the symmetry energy around and below saturation density, while NS radii are sensitive to the pressure over a broader density range extending above saturation. Since the $\Lambda$ is strongly correlated with the stellar radius, tidal
measurements provide an additional astrophysical constraint on this connection. Microscopic and phenomenological analyses have also demonstrated a strong correlation between $\Lambda_{1.4}$ and the
pressure of beta-equilibrated matter around twice saturation density\,\cite{Lim2018PRL,Abbott2018PRL}. In this sense, neutron-skin measurements and tidal observations probe different density regions
of the same underlying EOS and can therefore provide complementary information on its density dependence.

\begin{figure}[h!]
\centering
\includegraphics[width=0.38\textwidth]{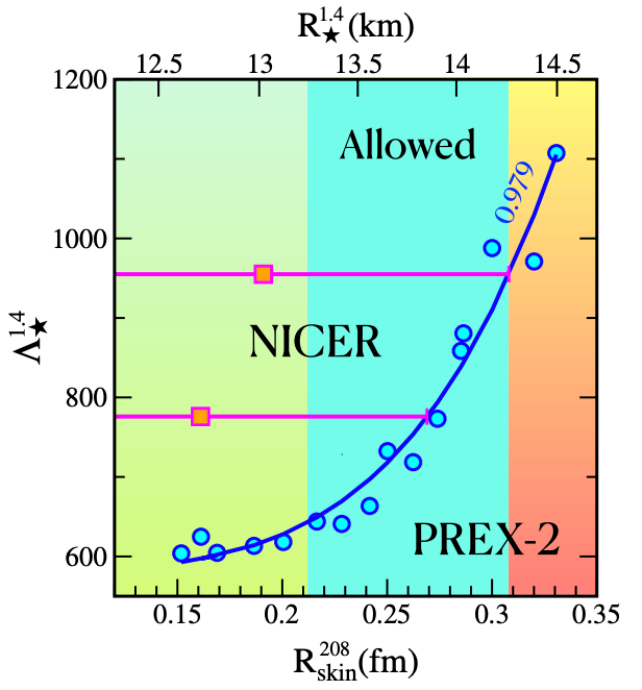}
\caption{(Color Online). Tidal deformability $\Lambda_{1.4}$ of a canonical NS as a function of its radius $R_{1.4}$ and the neutron-skin thickness of $^{208}\rm{Pb}$ for a representative set of relativistic energy-density functionals. The PREX-II and NICER
constraints illustrate the complementary connection between
terrestrial measurements of neutron-rich nuclei and astrophysical observations of NSs. Adapted from Ref.\,\cite{Reed2021PRL}.}
\label{fig:Reed2021_Lambda_Rskin}
\end{figure}

A particularly direct illustration of this connection was provided by Reed et al.\,\cite{Reed2021PRL}, see also  Ref.\,\cite{Fattoyev2018PRL}. As shown in
FIG.\,\ref{fig:Reed2021_Lambda_Rskin}, a representative set of relativistic energy density functionals exhibits strong correlations among the neutron-skin thickness of $^{208}\rm{Pb}$, the radius $R_{1.4}$, and the tidal deformability $\Lambda_{1.4}$ of a $1.4\,M_\odot$ NS. The figure also illustrates how the PREX-II constraint on $R_{\rm{skin}}$ can be confronted with the NICER
constraint on the stellar radius, thereby connecting terrestrial information on the symmetry energy with astrophysical measurements of NS structure. The initial analysis of GW170817 constrained $\Lambda_{1.4}\lesssim800$ for slowly spinning
NSs\,\cite{Abbott2017PRL}, while subsequent analyses based on a common EOS for the two stars yielded a tighter upper bound of approximately $\Lambda_{1.4}\lesssim580$ at the 90\% credible level\,\cite{Abbott2018PRL,De2018PRL}. Within the class of models considered in Ref.\,\cite{Reed2021PRL}, the combined PREX-II and NICER
constraints favor relatively large values of $R_{1.4}$ and
$\Lambda_{1.4}$, making comparison with the tighter GW170817 constraint particularly interesting. These correlations should nevertheless be interpreted with some care. Neutron skins primarily probe the EOS around and below saturation density, whereas NS radii and tidal deformabilities depend on its extension to higher densities. Indeed, the strong correlation between $R_{\rm{skin}}$ and $R_{1.4}$ obtained within a restricted family of energy-density functionals may become weaker when more general high-density EOS behavior is allowed
\,\cite{Fattoyev2018PRL,Lim2018PRL,Reed2021PRL}. Consequently, a large neutron skin accompanied by a relatively small NS radius or tidal deformability could signal significant softening
of the EOS above saturation density. If established by increasingly precise terrestrial and astrophysical measurements, such behavior could point to a qualitative change in the high-density EOS, potentially associated with the emergence of new degrees of freedom, a phase transition or crossover\,\cite{Alford2013PRD,
Benic2015AA,AlvarezCastillo2017PRC,Masuda2013ApJ,
Masuda2013PTEP,Baym2019ApJ,McLerran2019PRL,
Annala2020NatPhys,Annala2023NatCommun,Bauswein2019PRL,
Most2019PRL,Lin2011PRC,XieLi2021PRC,ZhangLi2023PRC,
Li2024PRD,GrundlerLi2026PRD,Tang2021PRD-aa,
Han2023SciBull,Tang2025PRD,Liu2016PRD,ChuChen2017PRD,
CaoChen2023,CaoChen2026,ZhangLi2025EPJA,GrundlerLi2025PRD,Li2026ApJ,LiGrundler2026PLB}. Such a pattern, however, would not by itself constitute evidence for a phase transition, since a sufficiently flexible purely nucleonic EOS, for example through different higher-order symmetry energy coefficients, could produce similar behavior.

\begin{figure}[h!]
\centering
\includegraphics[width=0.4\textwidth]{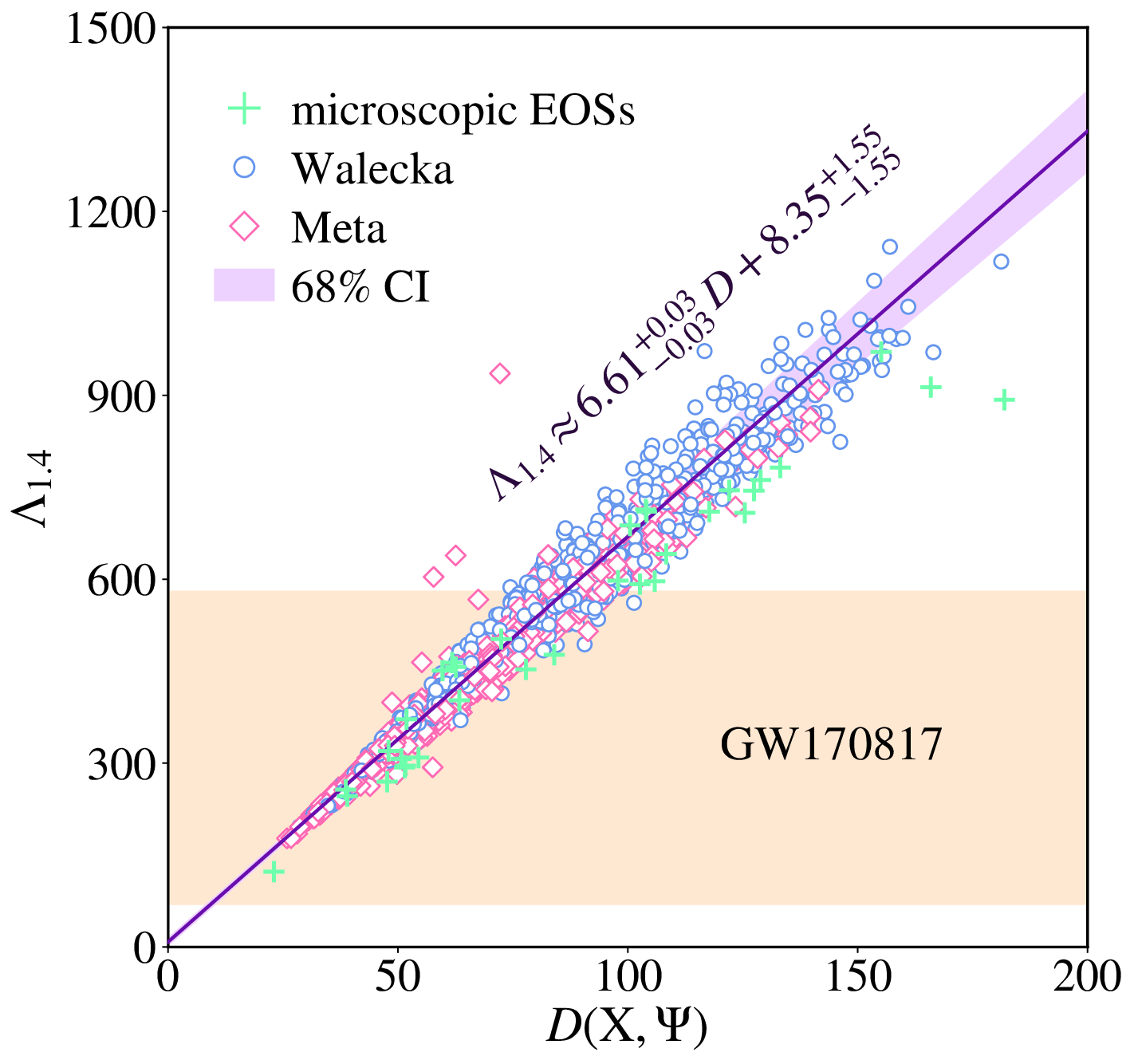}\\[0.25cm]
\includegraphics[width=0.4\textwidth]{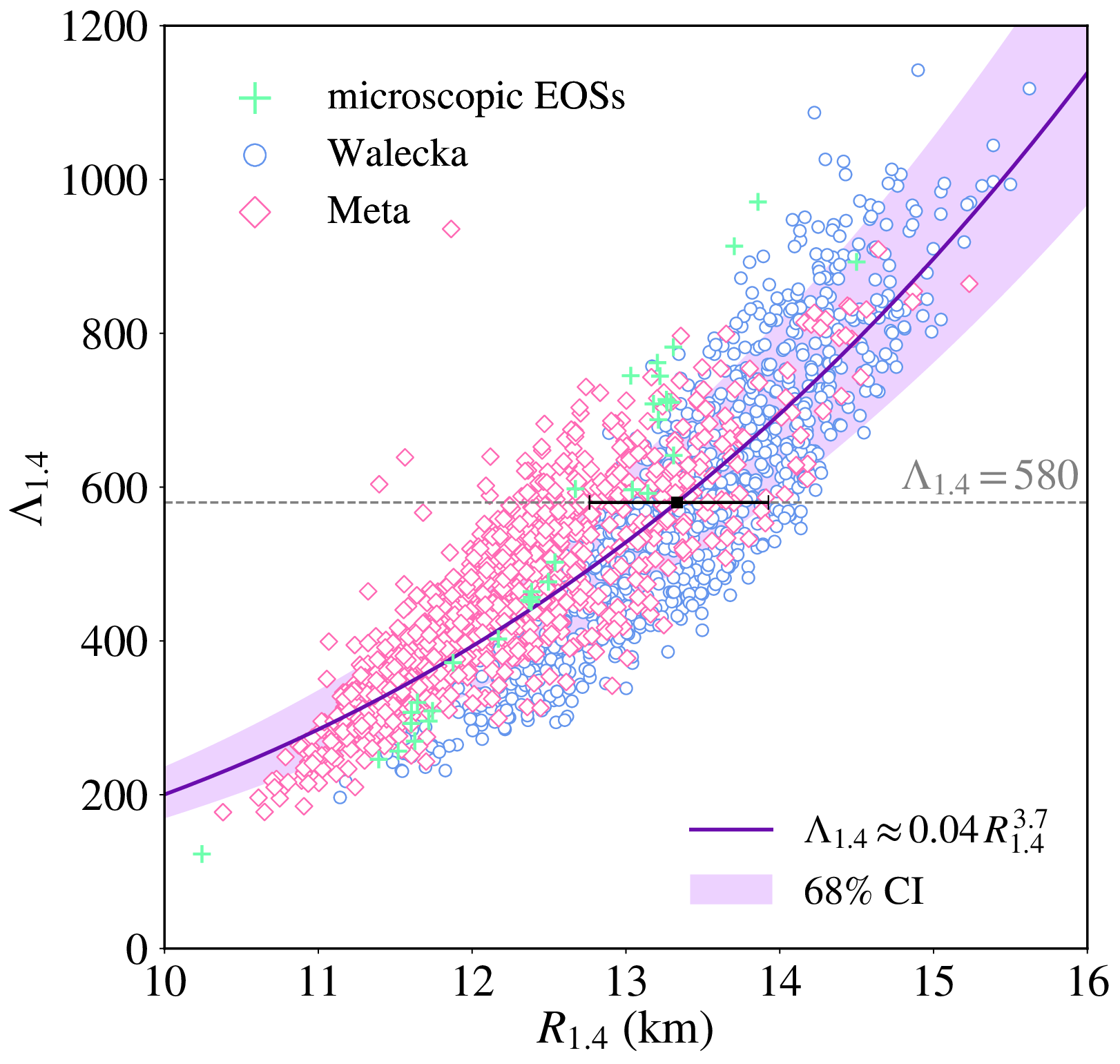}
\caption{(Color Online). Correlations of $\Lambda_{1.4}$ with the intrinsic tidal-response function $D(\rm{X},\Psi)$ (upper) and with $R_{1.4}$
(lower), where $\rm{X}=P_{\rm c}/\varepsilon_{\rm c}$ is the central EOS parameter.
The conventional $\Lambda_{1.4}$-$R_{1.4}$ relation exhibits larger scatter than the $\Lambda$-$D$ scaling, indicating a more direct connection between tidal deformability and the core EOS. Adapted from Ref.\,\cite{Shi2026xxx}.}
\label{fig:Lambda_core_scaling}
\end{figure}

Traditionally, the EOS information carried by $\Lambda_{1.4}$ has been interpreted mainly through the strong $\Lambda$-$R$ correlation
and the connection of the NS radius with the pressure around $1$-$2\rho_0$\,\cite{Lattimer2001ApJ,Lim2018PRL,Abbott2018PRL,
Raithel2018ApJL}. The robustness and model dependence of this connection have been investigated using different EOS constructions. Moreover, strong correlations of
$\Lambda_{1.4}$ with the pressure and nuclear matter parameters at supra-saturation densities have been found in a number of studies\,\cite{Tews2018PRC,Zhao2018PRD,Malik2018PRC,Kim2018PRC,Carson2019PRD,Malik2019PRC,
Tong2020PRC}. The tidal response, however, contains information beyond this indirect $\Lambda$-$R$-pressure connection. 
Recently, Shi et al.\,\cite{Shi2026xxx} identified a previously unrecognized, approximately EOS-insensitive scaling between $\Lambda$ and the intrinsic tidal response function $D(\rm{X},\Psi)$, in the framework of IPAD-TOV\,\cite{Cai2023ApJ,Cai2023PRD,Cai2024PRD,
Cai2024Frontiers,CaiLi2025EPJA,Cai2025PRD,
Cai2026Trace,Cai2025Phase,Cai2026Bound},
where $\rm{X}=P_{\rm c}/\varepsilon_{\rm c}$ is the central EOS parameter and $\Psi$
characterizes the stability of the stellar configuration. This scaling enables an observed $\Lambda$ to directly constrain $\rm{X}$, providing access to the EOS in the NS core without relying solely on an intermediate determination of the stellar radius.
This new connection is illustrated in
FIG.\,\ref{fig:Lambda_core_scaling}. The upper panel shows the tight $\Lambda_{1.4}$-$D(\rm{X},\Psi)$ scaling obtained from a large EOS ensemble. For comparison, the lower panel shows the conventional $\Lambda_{1.4}$-$R_{1.4}$ correlation obtained
from the same EOS ensemble. Although the latter remains strong and underlies the radius constraints discussed above, it exhibits appreciably larger scatter than the $\Lambda$-$D$ scaling\,\cite{Shi2026xxx}. Thus, the familiar $\Lambda$-$R$ correlation,
as also illustrated in FIGs.\,\ref{fig:De2018_Lambda_R} and
\ref{fig:Reed2021_Lambda_Rskin}, does not exhaust the EOS information encoded in the tidal response: $\Lambda$ also carries a more direct imprint of the core EOS.
Tidal deformability should therefore not be regarded merely as an indirect probe of the NS radius. Besides constraining the EOS around $1$-$2\rho_0$ through the conventional $\Lambda$-$R$ relation, it provides direct information on the central EOS parameter $\rm{X}$. Combined with the corresponding mass scaling, this connection can further constrain the central pressure and energy density\,\cite{Shi2026xxx}. Tidal observations thus extend the EOS information obtained from neutron skins and heavy-ion collisions toward the high-density NS core.
Recently, the $D$ function\,\cite{Shi2026xxx} has also been adopted to study the post-merger dynamics of binary NS mergers, revealing correlations between the high-density behavior of the sound speed and the post-merger GW frequency\,\cite{Stroud2026}.

\subsection{Multi-messenger constraints and heavy-ion collisions}

Constraining the EOS over the broad density range relevant to NSs requires complementary information from terrestrial experiments and astrophysical observations. HICs provide a unique terrestrial laboratory for creating compressed nuclear matter under controlled conditions. Depending on the beam energy and observable, densities from around saturation density to several times $\rho_0$ can be
explored. Collective-flow measurements constrain the pressure of compressed nuclear matter, while isospin-sensitive flows and pion production provide important information on the symmetry energy at supra-saturation densities\,\cite{Danielewicz2002Science,LeFevre2016NPA,Russotto2016PRC,
Estee2021PRL,Sorensen2024PPNP}. In particular, present HIC constraints are most effective around $1$-$2\rho_0$, while selected observables can retain sensitivity toward $2$-$3\rho_0$ and beyond.

\begin{figure}[h!]
\centering
\includegraphics[width=0.46\textwidth]{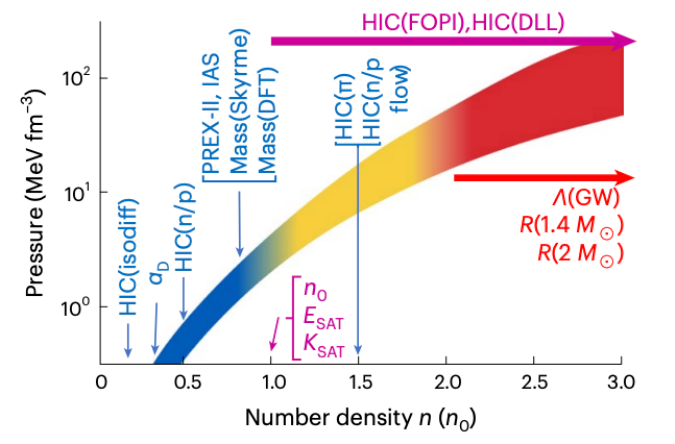}
\caption{(Color Online). Density regions probed by representative nuclear experiments and astrophysical observations used to constrain the nuclear EOS. The different observables provide complementary information over a broad range of densities extending from finite
nuclei to NS interiors. Adapted from Ref.\,\cite{Tsang2024NatAstron}.}
\label{fig:Tsang2024_density}
\end{figure}

\begin{figure*}
\centering
\includegraphics[width=0.9\textwidth]{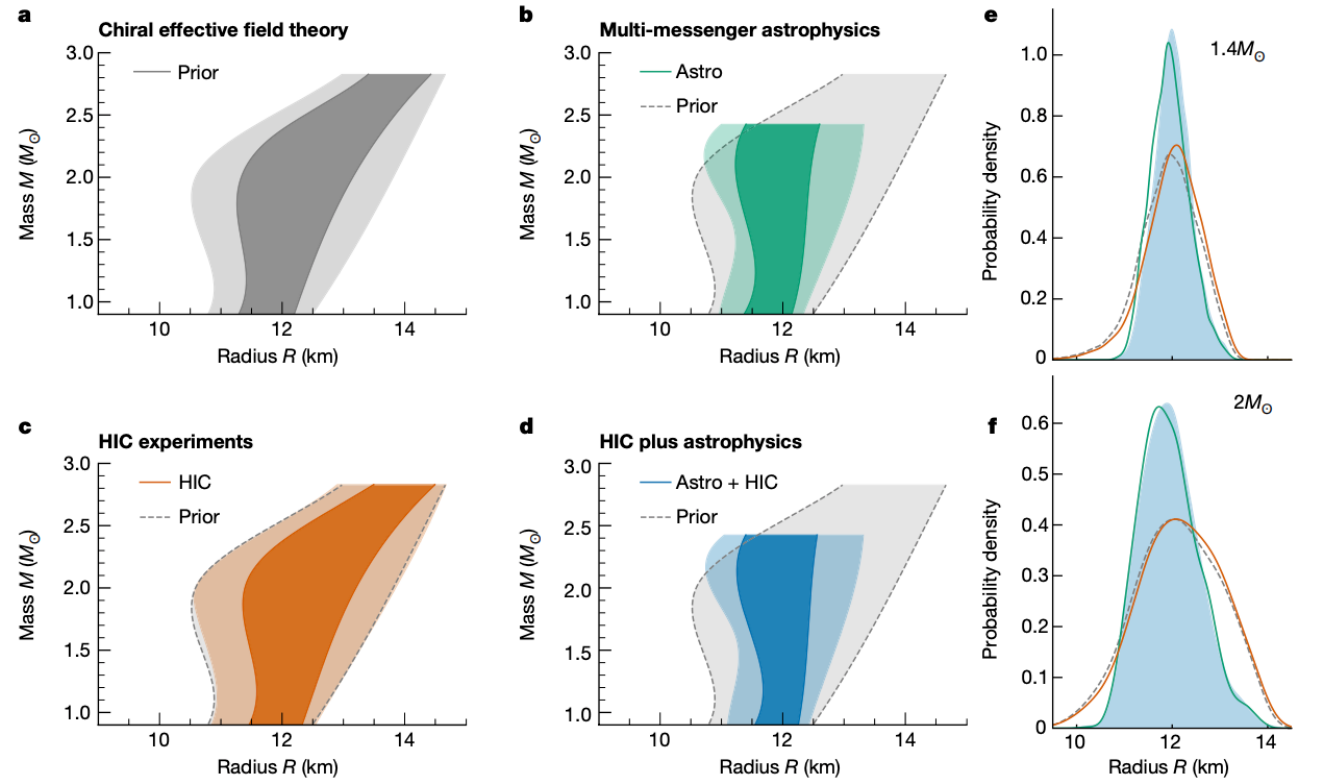}
\caption{(Color Online). Constraints on the NS MR relation from multi-messenger observations and HIC experiments, shown separately and in combination. The comparison illustrates the complementary impact of terrestrial HIC information and astrophysical observations on the dense-matter EOS. Adapted from Ref.\,\cite{Huth2022Nature}.}
\label{fig:Huth2022_MR}
\end{figure*}

\begin{figure*}
\centering
\includegraphics[width=0.95\textwidth]{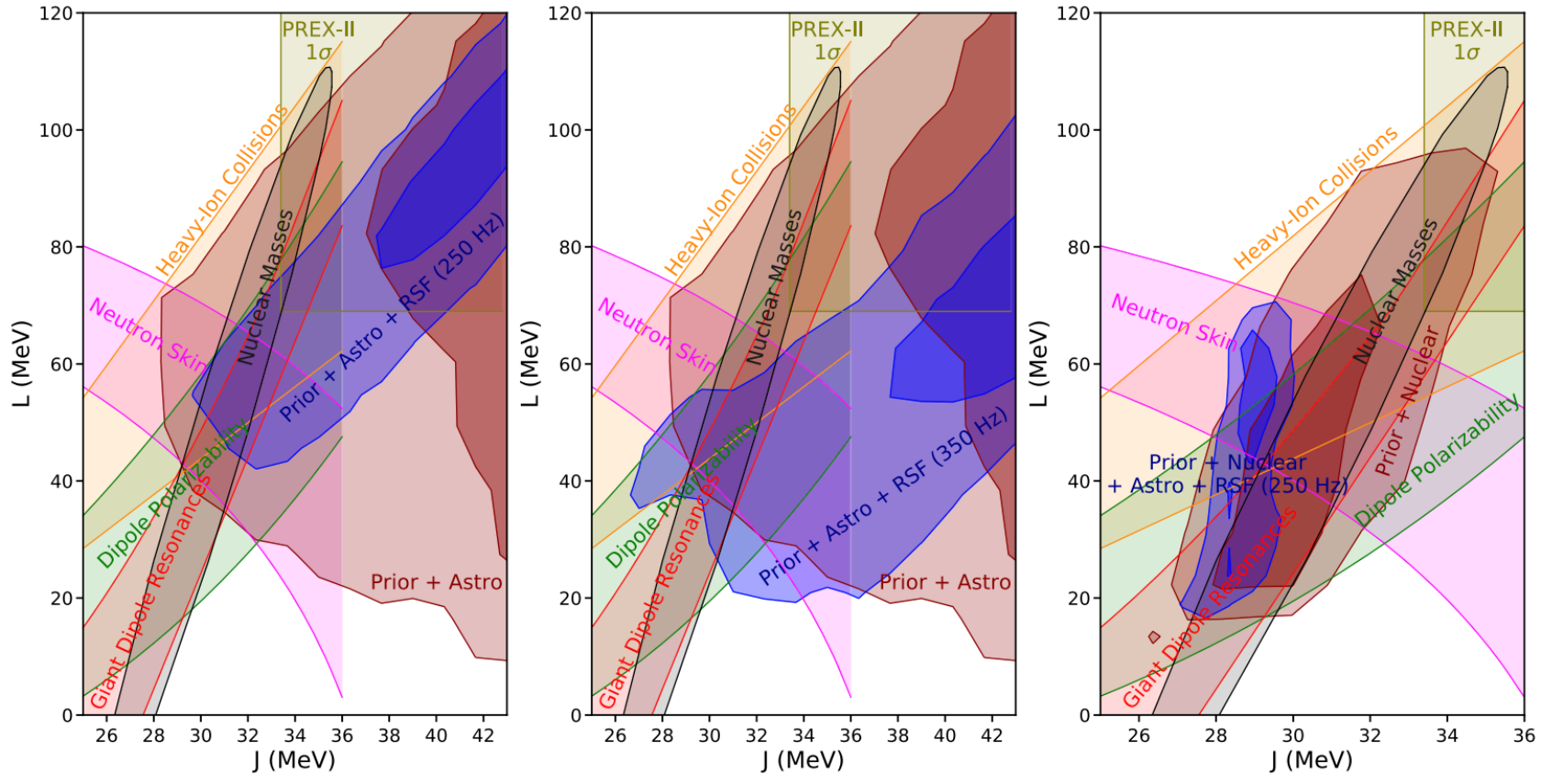}
\caption{(Color Online). Constraints on the symmetry energy $S_{\rm v}\equiv J\equiv E_{\rm{sym}}(\rho_0)$ and its slope parameter $L$ from nuclear experiments and astrophysical observations. Constraints from neutron skins, giant resonances, dipole polarizability, HICs, and PREX-II are compared with astrophysical constraints and the projected sensitivity of multimessenger resonant shattering flares (RSFs). The latter illustrate the potential of crust-sensitive multimessenger observations to provide constraints on $L$ complementary to terrestrial nuclear experiments. Adapted from Ref.\,\cite{Neill2023PRL}.}
\label{fig:SvL_summary}
\end{figure*}

The density regions explored by terrestrial and astrophysical observables are illustrated in FIG.\,\ref{fig:Tsang2024_density}.
The compilation of Tsang et al.\,\cite{Tsang2024NatAstron} shows that nuclear experiments provide a sequence of constraints extending from sub-saturation to supra-saturation densities. Nuclear masses, isobaric analog states, neutron skins, and related structure observables
primarily constrain the EOS around and below $\rho_0$, whereas HIC observables, including collective flow and particle production, extend the experimentally accessible region to higher densities. Astrophysical observations, on the other hand, provide increasingly important constraints at densities characteristic of NS interiors. The overlap among these density windows makes a joint analysis of
terrestrial and astrophysical information particularly valuable.
The astrophysical side of this program has developed rapidly in the multi-messenger era. GW observations of binary NS mergers constrain tidal deformabilities, while X-ray pulse-profile modeling with NICER provides independent mass-radius information. The first NICER measurements of PSR J0030+0451\,\cite{Riley19,Miller19} were followed
by measurements of the massive PSR J0740+6620\,\cite{Fonseca21,Riley21,Miller21,Salmi22,Salmi24,Ditt24} and PSR J0437-4715\,\cite{Choud24}. Together with precise pulsar masses and GW observations, these measurements provide complementary constraints on the EOS over the density range encountered in NSs. Recent NICER observations of additional pulsars are further enlarging the mass range over which radius information is available.
More recently, NICER measurements have been extended to
PSR J0437-4715\,\cite{Choud24}, the massive PSR J1614-2230\,\cite{Mauviard26}, and PSR J0614-3329\,\cite{Mauviard25,Miller26}.
Together, these measurements increasingly sample NSs over a broad range of masses and provide complementary constraints on the EOS.

HICs, on the other hand, complement these observations in an essential way. Astrophysical measurements probe cold, $\beta$-equilibrated matter in gravitational equilibrium, whereas HICs create compressed nuclear matter dynamically
in the laboratory. The two approaches therefore involve different physical conditions and systematic uncertainties. Importantly, HICs provide direct terrestrial information in the intermediate-density
region where microscopic calculations become increasingly uncertain, thereby helping to bridge conventional nuclear constraints around $\rho_0$ and astrophysical information from the deeper NS interior.
This complementarity was demonstrated quantitatively by Huth et al.\,\cite{Huth2022Nature}, who combined microscopic nuclear theory, HIC experiments, and multi-messenger NS observations within a common Bayesian framework. Their HIC constraints include information from
collective-flow measurements of symmetric nuclear matter and isospin-sensitive observables. The HIC data have their largest impact at intermediate densities and favor somewhat larger pressures around $1$-$2\rho_0$, while astrophysical observations provide stronger
constraints at higher densities. The remarkable consistency between the two approaches provides an important cross-check of the EOS over the density range relevant to NSs\,\cite{Huth2022Nature}.
The corresponding impact on NS structure is shown in
FIG.\,\ref{fig:Huth2022_MR}. HIC data alone already restrict the allowed MR region relative to the EOS prior, particularly through their constraint on the EOS at intermediate densities. Combining HIC information with multi-messenger observations further reduces the allowed region. The effect is especially informative for lower-mass NSs, whose structure is more strongly connected to the
EOS at densities accessible to present HIC experiments, while the properties of more massive NSs increasingly probe higher-density matter.

The combined picture therefore connects several complementary regimes: finite nuclei and neutron skins constrain the symmetry energy around and below saturation density, HICs extend the experimental sensitivity
toward supra-saturation densities, and NS observations probe the EOS over a broader range extending into the high-density core\,\cite{LynchTsang2022PLB,Tsang2024NatAstron,Huth2022Nature}.
Rather than providing redundant information, these probes constrain different density regions and different aspects of the EOS, making their combination particularly powerful.
This complementarity can also be illustrated in terms of the symmetry energy parameters around saturation density. In particular, the symmetry energy $S_{\rm v}\equiv E_{\rm{sym}}(\rho_0)\equiv J$ and its slope $L$ provide a common plane for comparing constraints from nuclear structure, neutron skins, HICs, microscopic theory, and astrophysical
observations\,\cite{Tsang2019PLB,Essick2021PRC,Yue2022PRR,Lattimer2023}. As shown in FIG.\,\ref{fig:SvL_summary}, the constraints obtained from neutron skins, giant resonances, dipole polarizability, and HICs occupy complementary regions of the $S_{\rm v}$-$L$ plane. Astrophysical
observations provide additional information, although their mapping onto $S_{\rm v}$ and $L$ generally involves assumptions about the EOS away from saturation density. Future multimessenger observables sensitive to the NS crust may provide a more direct connection: the projected
constraints from resonant shattering flares (RSFs), inferred through coincident GW and gamma-ray observations, can reach a sensitivity to $L$ comparable to that of several terrestrial probes\,\cite{Neill2023PRL}. The overall consistency and complementarity of these different constraints illustrate the importance of combining
nuclear experiments, HICs, microscopic theory, and multimessenger observations in determining the density dependence of the symmetry energy.

\section{Challenges and Potential Future Directions}
\label{sec:challenges}

The preceding discussions demonstrate that neutron skins connect nuclear structure, reaction dynamics, and the EOS of neutron-rich matter. Converting this broad sensitivity into a precise determination of the NS thickness nevertheless remains challenging. Nuclear deformation, surface diffuseness, shell structure, clustering, and many-body correlations can modify the same observables used to infer neutron skins, while reaction-based extractions introduce additional dependence on transport dynamics, initialization, in-medium interactions, and non-equilibrium effects. Consequently, no single observable or simple correlation with one EOS parameter can provide a universally model-independent determination.

Future progress will require coordinated analyses of complementary observables across isotopic chains, beam energies, and experimental facilities. Radioactive-beam and collider measurements, dipole responses, electroweak probes such as coherent elastic neutrino-nucleus scattering, and astrophysical observations can constrain different aspects of neutron distributions and isovector interactions. Combining these data with ab initio calculations, improved many-body descriptions, and Bayesian uncertainty quantification will be essential for disentangling nuclear-structure and reaction-model effects, identifying the relevant density dependence of the symmetry energy, and establishing a quantitatively controlled connection between finite nuclei and neutron-rich matter.

\subsection{Disentangling neutron skin and deformation effects}

A major challenge in extracting neutron-skin information from HICs is to disentangle it from other aspects of nuclear structure, particularly nuclear deformation and the detailed surface density profile. Both neutron skin and deformation modify the initial nuclear geometry and can therefore leave correlated signatures in particle production and collective flow. This issue is particularly important for the $^{96}\rm{Ru}+{}^{96}\rm{Ru}$ and $^{96}\rm{Zr}+{}^{96}\rm{Zr}$ isobar collisions, where differences in neutron distributions coexist with differences in nuclear  deformation\,\cite{Hammelmann2020PRC,Xu2021PLB}. Recent studies have shown that these effects can, in principle, be separated by exploiting their different signatures: the neutron-skin difference between the Ru and Zr isobars predominantly affects the reaction-plane component of elliptic flow, whereas deformation differences are reflected more strongly in flow fluctuations\,\cite{Jia2023PRL}.

A related ambiguity concerns the distinction between neutron-skin and halo-type density distributions. A larger $R_{\rm{skin}}$ may originate primarily from an increase of the neutron half-density radius, from an enhanced surface diffuseness leading to a more extended neutron tail, or from a combination of both. As illustrated in FIG.\,\ref{fig:skin-eccentricity}, skin-type and halo-type neutron distributions can produce appreciable differences in the initial eccentricity even for a similar $R_{\rm{skin}}$\,\cite{Xu2021PLB}. Since many HIC observables preferentially probe the low-density nuclear surface, fixed assumptions about deformation or surface diffuseness may introduce systematic model dependence in the extracted $R_{\rm{skin}}$. Combining observables with complementary sensitivities across different centralities will therefore be important for simultaneously constraining neutron skin, deformation, and halo-like surface structure\,\cite{Yang2023Universe,Wang2026PRC,Xi2025NST,Li2022PRC,Ma2024Entropy,Wei2024NST,Abdulhamid2024PRR,Wang2026CPL}.

\begin{figure}[h!]
\centering
\includegraphics[width=0.4\textwidth]{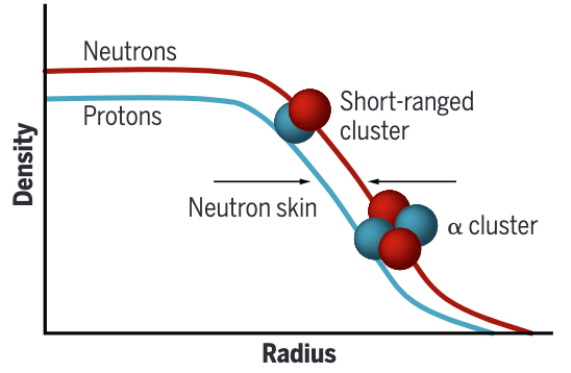}
\caption{(Color Online). Schematic illustration of the interplay among the neutron skin, SRCs, and $\alpha$-cluster formation in the nuclear surface. Surface clustering and correlations can modify the neutron density distribution and thus affect the interpretation of
$R_{\rm{skin}}$ and its correlation with the symmetry energy\,\cite{Tanaka2021Science,Hen2021Science}.}
\label{fig:skin-clusteraa}
\end{figure}

Beyond the distinction between skin- and halo-type density profiles, many-body correlations and clustering introduce an additional degree of complexity in the neutron-rich nuclear surface\,\cite{He2014PRL,Shi2021NST,He2021PRC,Wang2023PRC,Cao2023PRC,Wang2026PRL,Zhou2026PRL,Li2020PRCCluster,Xu2018NSTCluster,
Zhang2024EPJACluster,RWang2026PRC,Sun2024NatCommun,RWang2023PRC}. As schematically illustrated in FIG.\,\ref{fig:skin-clusteraa}, the dilute surface region
may support correlations ranging from short-range correlated nucleon pairs to light nuclear clusters. The quasi-free $\alpha$-knockout measurements of Tanaka {et al.}\,\cite{Tanaka2021Science} provided direct evidence for
$\alpha$-cluster formation at the surface of neutron-rich Sn isotopes and revealed a close interplay between surface clustering and the neutron   skin, see also Ref.\,\cite{Hen2021Science}. The accompanying theoretical analysis indicates that including $\alpha$ clustering can reduce the predicted $R_{\rm{skin}}$, while short-range correlated
nucleon pairs may produce a related effect\,\cite{Miller2019PLB}.
These results suggest that the neutron-rich surface should not always be viewed simply as a smooth mean-field density distribution: correlations and cluster formation may modify both its spatial structure and the inferred relation between $R_{\rm{skin}}$ and the symmetry energy. Quantifying these effects therefore represents an important direction for future neutron-skin studies.

\begin{figure}[h!]
\centering
\includegraphics[width=0.35\textwidth]{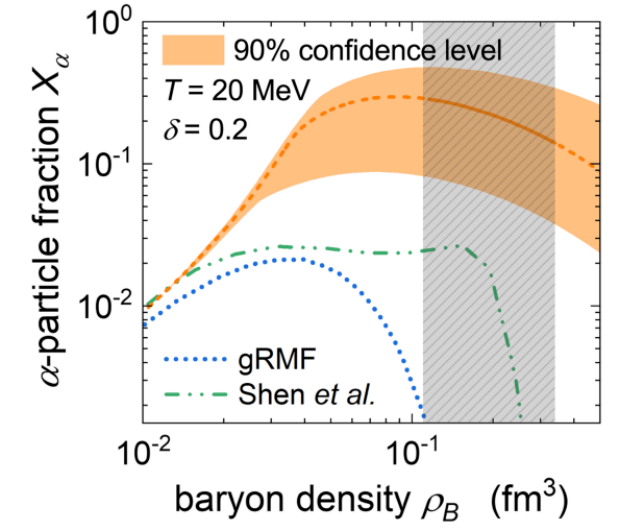}
\caption{(Color Online). Fractions of light clusters in hot and dense nuclear matter as functions of baryon density and temperature, including d, t, $^3{\rm{He}}$, and $\alpha$ particles. The extracted cluster fractions demonstrate that light-cluster correlations, particularly $\alpha$ clustering, can remain appreciable at densities around and above nuclear saturation density. Such clustering introduces additional
degrees of freedom into the EOS and may be relevant to the hot
neutron-rich matter encountered in supernovae and NS mergers. Adapated from Ref.\,\cite{Wang2026PRL}.}
\label{fig:dense-clustering}
\end{figure}

More broadly, clustering may not be restricted to the dilute nuclear surface. Recent analyses of intermediate energy heavy-ion collisions suggest that light clusters can survive in hot and compressed nuclear matter at densities extending beyond $\rho_0$\,\cite{Wang2026PRL}.
Using a kinetic approach that dynamically incorporates cluster formation, dissociation, and the Mott effect, Wang {et al.}\,\cite{Wang2026PRL} extracted the abundances of light clusters from FOPI data on central Au+Au collisions.
As shown in FIG.\,\ref{fig:dense-clustering}, appreciable fractions of d, t, $^3{\rm{He}}$, and particularly $\alpha$ particles can persist in hot nuclear matter, with the inferred $\alpha$ fraction reaching about 0.2 at densities around $1$-$2\rho_0$. This finding indicates that cluster degrees of freedom may remain relevant over a considerably broader density range than usually assumed in a purely nucleonic mean-field description. The possible persistence of clustering at such densities also provides an interesting connection between neutron-skin physics and astrophysical matter. Neutron skins constrain the isospin-dependent EOS primarily through the properties of neutron-rich matter around and below saturation density, whereas cluster formation can modify the composition
and thermodynamics in this same density regime. A consistent treatment of cluster correlations may therefore be important when extrapolating constraints inferred from neutron skins toward the EOS of astrophysical matter, particularly under the hot conditions encountered in supernovae and NS mergers\,\cite{Wang2026PRL}.

\subsection{Dipole response as a probe of neutron skins}

\begin{figure}[h!]
\centering
\includegraphics[width=0.48\textwidth]{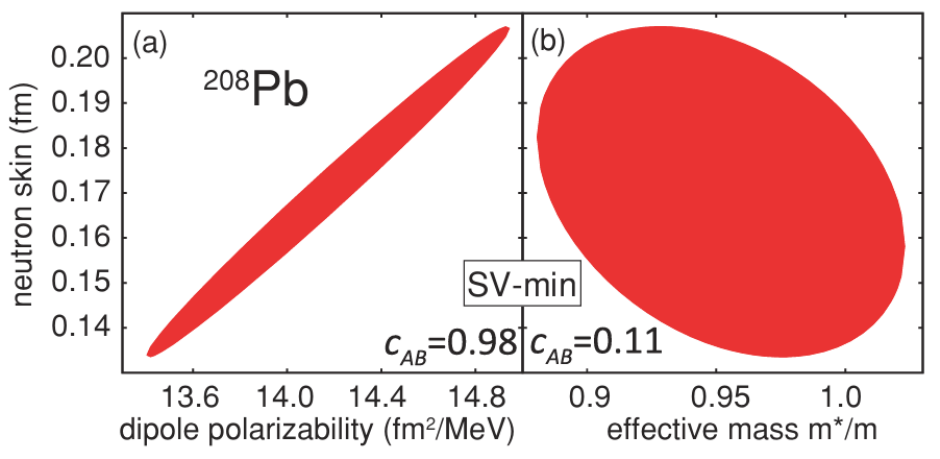}\\[0.5cm]
\includegraphics[width=0.4\textwidth]{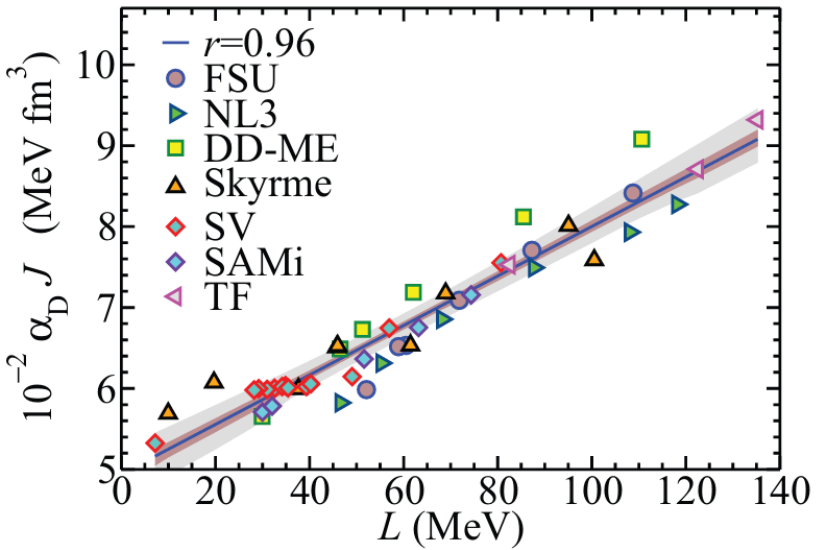}
\caption{(Color Online). Correlations of the electric dipole polarizability $\alpha_{\rm D}$ with the neutron skin and the symmetry energy. Upper: Covariance correlation between $\alpha_{\rm D}$ and the neutron-skin thickness of $^{208}\rm{Pb}$ and the effective mass, demonstrating their strong correlation within the SV-min energy density
functional. Lower: Correlation between $\alpha_{\rm D}J$ and the symmetry energy slope parameter $L$ for a representative set of nuclear energy-density functionals. Adapted from
Refs.\,\cite{Reinhard2010PRC,RocaMaza2013PRC}.}
\label{fig:alphaD-skin-symmetry}
\end{figure}

Electric dipole response provides an important complementary probe of neutron skins and the isovector properties of nuclei. A particularly useful observable is the electric dipole polarizability $\alpha_{\rm D}$, which is sensitive to the inverse energy-weighted $E1$ strength and has been shown to correlate with $R_{\rm{skin}}$ and the density dependence of the symmetry energy. Covariance analyses of nuclear energy-density functionals already demonstrated that neutron radii are strongly connected with isovector observables and the neutron-matter equation of state, while also emphasizing the importance of identifying observables that provide genuinely independent information on the neutron skin\,\cite{Reinhard2010PRC}, see the upper panel of FIG.\,\ref{fig:alphaD-skin-symmetry}. The benchmark measurement in $^{208}\rm{Pb}$ subsequently demonstrated that $\alpha_{\rm D}$ can provide a quantitative constraint on the neutron skin\,\cite{Tamii2011PRL,Tamii2014EPJA}, while theoretical studies further clarified its correlations with the symmetry energy\,\cite{RocaMaza2013PRC,Zhang2014PRC}, see the lower panel of FIG.\,\ref{fig:alphaD-skin-symmetry}. In particular, Zhang and Chen showed that $\alpha_{\rm D}$ in $^{208}\rm{Pb}$ is most strongly correlated with the symmetry energy, or nearly equivalently the pure-neutron-matter equation of state, at a characteristic subsaturation density around $\rho_0/3$\,\cite{Zhang2015PRC}. This finding provides a more direct physical interpretation of the density region probed by $\alpha_{\rm D}$ and helps establish the connection between dipole response, low-density isovector properties, and neutron skins. Measurements in $^{48}\rm{Ca}$, combined with microscopic calculations, have further extended this approach toward lighter nuclei\,\cite{Birkhan2017PRL,Zhang2018PLB}.

Despite this progress, translating the dipole response into $R_{\rm{skin}}$ remains subject to theoretical uncertainties. The sensitivity of $\alpha_{\rm D}$ to the symmetry energy at subsaturation densities implies that its connection to $R_{\rm{skin}}$ is mediated by the underlying isovector interaction rather than being a simple one-to-one correspondence. Moreover, low-energy $E1$ strength associated with the pygmy dipole resonance can depend on shell structure, pairing, collectivity, continuum effects, and the nuclear interaction. More generally, dipole excitations encode other aspects of nuclear structure: giant dipole resonances have, for example, been proposed as fingerprints of $\alpha$-cluster configurations\,\cite{He2014PRL}, while recent precision measurements of the deuteron electric polarizability demonstrate the increasing experimental capability to characterize dipole response\,\cite{Hao2026PRL}, although these studies are not directly related to neutron skins. Future progress will benefit from systematic dipole-response measurements across nuclei, improved microscopic calculations, and their combination with complementary neutron-radius measurements to isolate and constrain the neutron-skin information.

\subsection{Model dependence of transport calculations}

\begin{figure}[h!]
\centering
\includegraphics[width=0.46\textwidth]{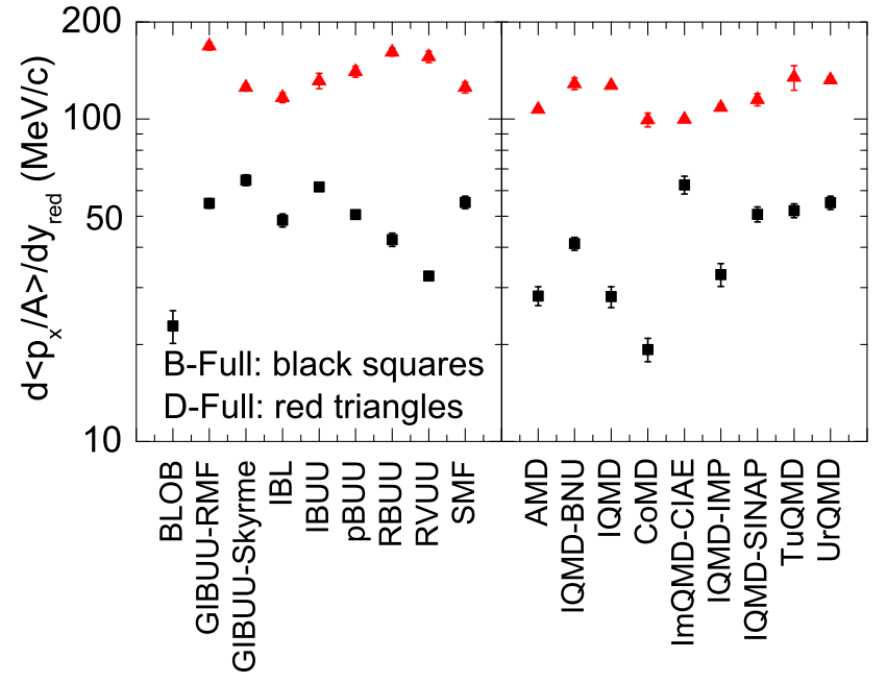}
\caption{(Color Online). Slope parameters of the transverse flow predicted by nine BUU-type and nine QMD-type transport models for Au+Au collisions at 100$A$ and 400$A$\,MeV under controlled physical inputs. The spread among the calculations illustrates the systematic model dependence of transport predictions. Adapted from
Ref.\,\cite{Xu2016PRC}.}
\label{fig:transport-comparison}
\end{figure}

A major challenge for extracting neutron-skin information from HICs is the model dependence of the reaction dynamics. This issue has been particularly well documented at intermediate energies, where controlled comparisons among different transport approaches have revealed
appreciable differences even when common physical inputs and collision conditions are employed\,\cite{Xu2016PRC,Ono2019PRC,
Colonna2021PRC,Xu2024PRC}. As illustrated in
FIG.\,\ref{fig:transport-comparison}, the predicted transverse flow can exhibit a substantial spread among both BUU- and QMD-type models\,\cite{Xu2016PRC}. Such differences originate from the numerical and physical implementation of the transport dynamics, including initialization, mean-field propagation, nucleon-nucleon collisions and Pauli blocking, as well as cluster and pion production. They should therefore be regarded as a source of systematic theoretical uncertainty when experimental observables are used to infer nuclear properties.

\begin{figure}[h!]
\centering
\includegraphics[width=0.4\textwidth]{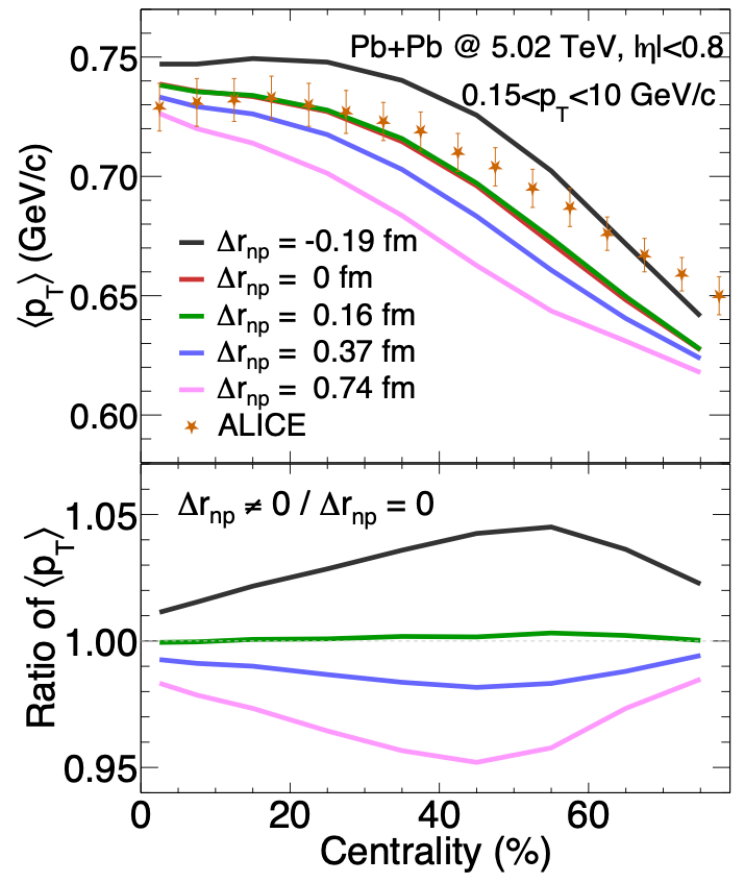}
\caption{(Color Online). Centrality dependence of the mean transverse momentum $\langle p_{\rm T}\rangle$ in $^{208}\rm{Pb}+^{208}\rm{Pb}$ collisions at
$\sqrt{s_{\rm{NN}}}=5.02$ TeV for different neutron-skin
configurations. The lower panel shows the ratios relative to the $R_{\rm{skin}}=0$ configuration. Adapted from
Ref.\,\cite{Zhao2026PRC}.}
\label{fig:skin-pt-model}
\end{figure}

The importance of this issue becomes particularly evident for neutron-skin studies, because the signal associated with $R_{\rm{skin}}$ can itself be relatively small. At ultra-relativistic energies, the relevant dynamics is different: the neutron skin modifies the initial collision geometry, and this information must survive the subsequent partonic and hadronic evolution before appearing in
final-state observables. Recent AMPT calculations of
$^{208}\rm{Pb}+^{208}\rm{Pb}$ collisions at $\sqrt{s_{\rm{NN}}}=5.02$\,TeV demonstrate such a sensitivity\,\cite{Zhao2026PRC}. As shown in FIG.\,\ref{fig:skin-pt-model}, a larger $R_{\rm{skin}}$ systematically reduces $\langle p_{\rm T}\rangle$, but for realistic moderate neutron skins the effect is only at the
few-percent level. This illustrates a central difficulty: a genuine neutron-skin signal may be comparable to uncertainties associated with the dynamical modeling itself.

Reducing model dependence is therefore not merely a technical improvement but a prerequisite for a quantitative determination of $R_{\rm{skin}}$ from HICs. Systematic benchmarking across independent transport frameworks, together with observables or ratios in which
dynamical uncertainties partially cancel, will be essential for establishing whether the predicted neutron-skin response is robust against the underlying reaction model.

\subsection{High-energy collider probes}

\begin{figure}[h!]
\centering
\includegraphics[width=0.48\textwidth]{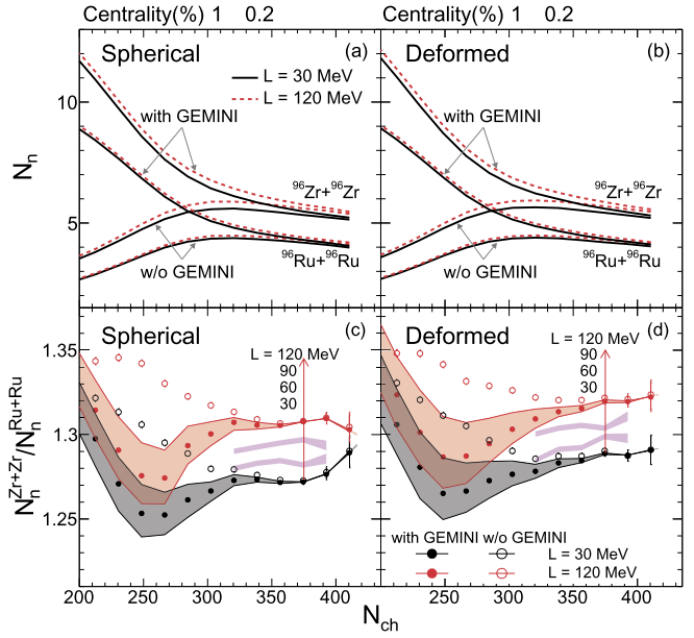}
\caption{(Color Online). Free spectator-neutron multiplicities in $^{96}\rm{Zr}+^{96}\rm{Zr}$ and $^{96}\rm{Ru}+^{96}\rm{Ru}$ collisions (upper panels) and their ratios (lower panels), calculated with spherical and deformed nuclear density distributions for different values of the symmetry energy slope parameter $L$.
In ultracentral collisions, the ratio exhibits enhanced sensitivity to the neutron-skin difference while reducing uncertainties associated with cluster deexcitation and nuclear deformation. Adapted from Ref.\,\cite{Liu2022PLB}.}
\label{fig:spectator-neutron-skin}
\end{figure}

Ultra-relativistic heavy-ion collisions provide a qualitatively different avenue for probing neutron skins. At intermediate energies, neutron-skin effects are transmitted mainly through isospin-dependent
reaction dynamics, whereas at RHIC and LHC energies the neutron distribution can be encoded in the initial geometry and density profile and subsequently reflected in particle production and collective evolution. Beyond these bulk observables, several complementary strategies are emerging. Free spectator neutrons in ultracentral isobar collisions provide a particularly direct probe of
the neutron-rich nuclear surface because they largely avoid the complicated dynamics of the mid-rapidity QGP\,\cite{Liu2022PLB}. As illustrated in
FIG.\,\ref{fig:spectator-neutron-skin}, the ratio of free spectator neutrons in $^{96}\rm{Zr}+^{96}\rm{Zr}$ to
$^{96}\rm{Ru}+^{96}\rm{Ru}$ collisions retains appreciable sensitivity to the symmetry energy slope $L$, and hence to the neutron-skin difference, while substantially reducing uncertainties associated with cluster deexcitation and nuclear deformation\,\cite{Liu2022PLB}. Other proposed directions include grazing collisions that enhance sensitivity to the
nuclear surface\,\cite{Xu2022PRC}, hard probes whose yields provide a different weighting of the nuclear overlap geometry \,\cite{VanderSchee2024PLB}, and spectator observables designed to access possible deformation of the neutron skin\,\cite{Liu2023PLB}.

More broadly, high-energy collisions may connect neutron skin physics to aspects of nuclear structure that are difficult to access at intermediate energies. Measurements involving different isotopes, such as $^{40}\rm{Ca}$ and $^{48}\rm{Ca}$, could extend collider imaging across isotopic chains and provide direct connections with
low-energy neutron-radius  measurements\,\cite{Vitsos2026EPJC}, while
conserved-charge observables in asymmetric collision systems offer another possible handle on the neutron distribution\,\cite{Pihan2025}. At still higher energies, an especially interesting possibility is the connection to small-$x$ nuclear structure. In the color-glass-condensate (CGC) picture\,\cite{McLerran1994PRD,McLerran1994PRD2,
JalilianMarian1997PRD,Kovchegov1999PRD,
Kovner1995PRD,Krasnitz1999NPB,
Lappi2003PRC,Lappi2006NPA,Gelis2010ARNPS}, the transverse nuclear density enters the local saturation scale $Q_{\rm s}$, so that the neutron-rich
surface can leave an imprint on the spatial distribution and fluctuations of the small-$x$ gluon fields. Coherent and incoherent $J/\psi$ photoproduction in ultra-peripheral collisions has recently been proposed as a means of exploiting this connection\,\cite{Li2026CPL}. Thus, high-energy neutron-skin studies may evolve
from imaging the spatial distribution of nucleons toward probing its connection with the gluonic structure of nuclei. Combining such observables with intermediate-energy measurements, which probe the same neutron distribution through very different reaction dynamics,
offers an important route toward crossvalidating extracted
$R_{\rm{skin}}$ and reducing model dependence.

\subsection{Opportunities with radioactive beams}

\begin{figure}[h!]
\centering
\includegraphics[width=0.48\textwidth]{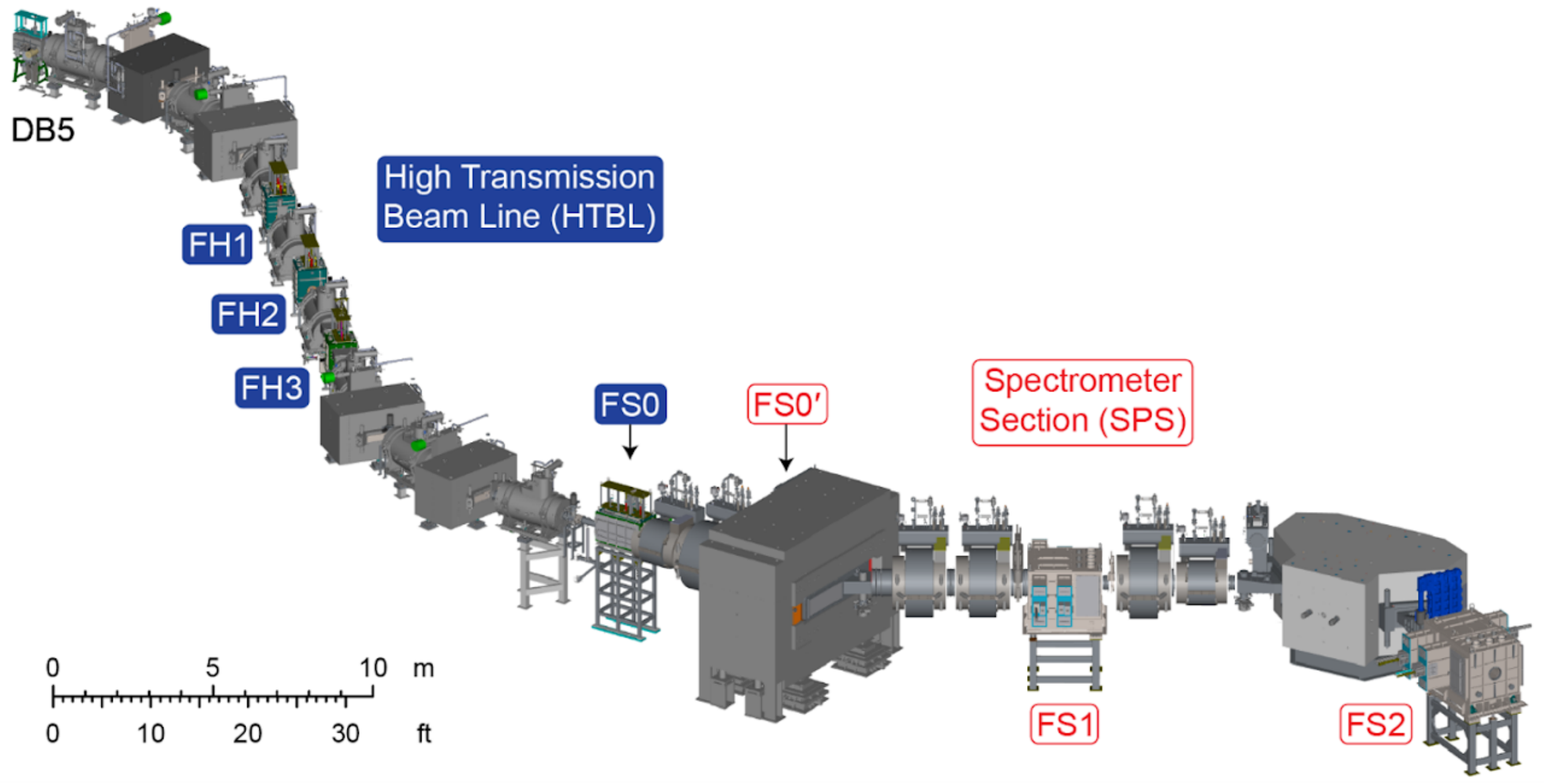}
\caption{(Color Online). Layout of the High Rigidity Spectrometer (HRS) at the Facility for Rare Isotope Beams (FRIB), including the High Transmission Beam Line (HTBL) and the Spectrometer Section (SPS). HRS will substantially extend the experimental reach of FRIB toward the most neutron-rich isotopes, providing new opportunities for reaction studies of nuclei with large neutron excess and pronounced
neutron skins. Adapted from Ref.\,\cite{FRIB2022}.}
\label{fig:frib}
\end{figure}

The continuing development of radioactive-beam facilities opens a particularly important opportunity for neutron-skin studies. Facilities such as FRIB in the United States, RIBF in Japan, FAIR in Germany, RAON in Korea, and HIRFL-CSR and other emerging facilities in China
will provide access to increasingly neutron-rich nuclei far from stability\,\cite{FRIB2022,myref,Ye24,NSAC2023}. In particular, the High Rigidity Spectrometer (HRS) at FRIB, shown in FIG.\,\ref{fig:frib}, will substantially extend the experimental capabilities for reaction studies
with fast, neutron-rich rare-isotope beams\,\cite{Noji2023NIMA}.
Rather than determining $R_{\rm{skin}}$ only for a few benchmark nuclei, radioactive beams make it possible to follow the evolution of neutron and proton distributions systematically along isotopic chains. Such measurements can reveal how neutron-rich surfaces develop with
increasing isospin asymmetry and provide a broader test of the connection between $R_{\rm{skin}}$ and the density dependence of the symmetry energy. Interaction and fragmentation cross sections are particularly promising in this respect. Recent measurements with R$^3$B at GSI/FAIR have demonstrated the precision attainable for nuclear interaction cross sections and are being developed explicitly toward neutron-skin  determination\,\cite{Ponnath2025NPA}. These
measurements also emphasize an important challenge: reaction-model uncertainties must be sufficiently controlled before precise neutron radii can be reliably extracted.

\begin{figure*}
\centering
\includegraphics[width=0.95\textwidth]{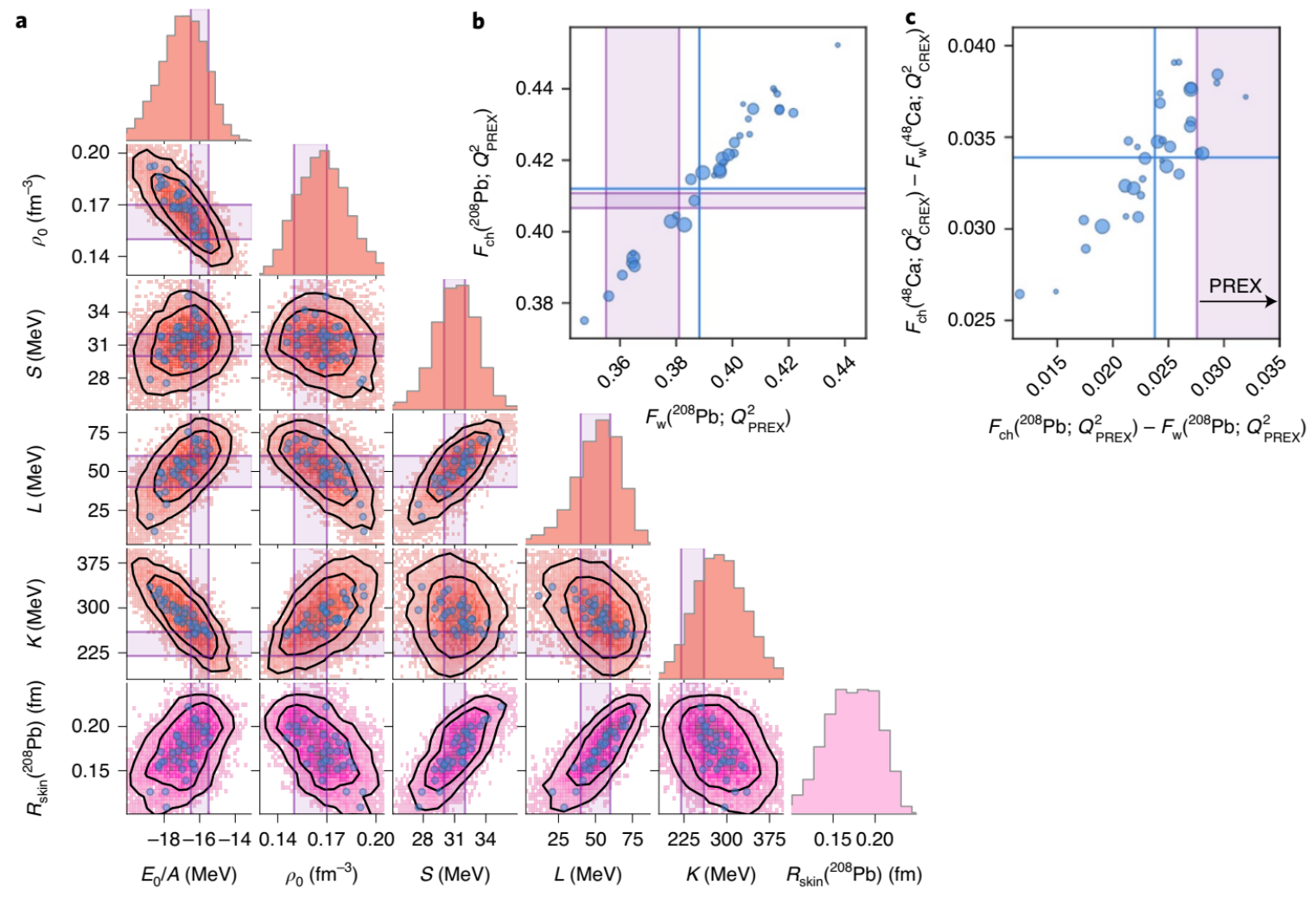}
\caption{(Color Online). Posterior predictive distributions for the neutron-skin thickness
$R_{\rm{skin}}(^{208}\rm{Pb})$, nuclear-matter properties at saturation density, and the charge and weak form factors relevant to PREX and CREX. Panel (a) demonstrates how the predicted neutron skin connects finite-nucleus
structure with the symmetry energy $S$ and its slope $L$, while panels (b) and (c) confront the calculated form factors with the electroweak measurements. The contours denote the $68\%$ and $90\%$ credible regions. Reproduced from Ref.\,\cite{Hu2022NP}.}
\label{fig:Hu2022_abinitio}
\end{figure*}

Radioactive beams will also greatly enlarge the range of reaction observables available for studying neutron-rich surfaces. Measurements of neutron and proton emission, light clusters, pion and photon production, and projectile-like fragments can be extended toward systems with much larger neutron excess, where neutron-skin effects
are expected to become more pronounced. At the same time,
inverse-kinematics experiments at facilities such as R$^3$B provide access to the microscopic structure of unstable neutron-rich nuclei \,\cite{r3b,Kahlbow:2023mtc,SRC-r3b}. In particular, measurements of short-range correlated nucleon pairs along isotopic chains may help clarify how short-distance nuclear correlations evolve together with the long-range neutron distribution, providing a complementary
connection between microscopic nuclear dynamics and neutron-skin formation. The major opportunity offered by radioactive beams is therefore not simply access to nuclei with larger $R_{\rm{skin}}$, but the possibility of combining several complementary observables for the same sequence of isotopes. Such a systematic program could
help disentangle neutron-skin effects from deformation, surface diffuseness, clustering, and reaction dynamics, and ultimately map the evolution of neutron-rich surfaces toward the limits of nuclear stability.

\subsection{Ab initio approaches and Bayesian inference}

A central challenge for future neutron-skin studies is to move from identifying observables that are qualitatively sensitive to the isovector interaction toward  quantitatively extracting the density dependence of the
symmetry energy with controlled uncertainties. Modern \textit{ab initio} calculations based on two- and three-nucleon interactions derived from chiral effective field theory (EFT) can now predict neutron distributions in nuclei ranging from $^{48}\rm{Ca}$ to $^{208}\rm{Pb}$\,\cite{Hagen2016NatPhys,Birkhan2017PRL,Hu2022NP,Novario2023PRL,Xu2024NSTabinitio,Yang2025CPL}.
The neutron skin is particularly valuable in this framework because it connects an experimentally accessible finite-nucleus observable with the isovector components of microscopic nuclear forces and the pressure of neutron-rich matter around and below saturation density. As illustrated in FIG.\,\ref{fig:Hu2022_abinitio}, combining many-body calculations with history
matching, Bayesian calibration, and fast emulators produces posterior predictive distributions for $R_{\rm{skin}}(^{208}\rm{Pb})$ and its correlations with $E_{\rm{sym}}(\rho_0)$ and $L$, rather than a single
theoretical prediction. The uncertainties entering such calculations include the truncation and regularization of chiral EFT, uncertainties in the fitted low-energy constants, many-body and model-space approximations, and emulator errors\,\cite{Furnstahl2015PRC,Melendez2019PRC,Drischler2020PRL,
Drischler2020PRC,LiXie2025PRC,Wang2025CPC}. Their correlations across nuclei, observables, and densities must be retained, since treating theoretical errors as independent may produce
artificially narrow constraints on the neutron skin and symmetry energy.

\begin{figure}[h!]
\centering
\includegraphics[width=0.42\textwidth]{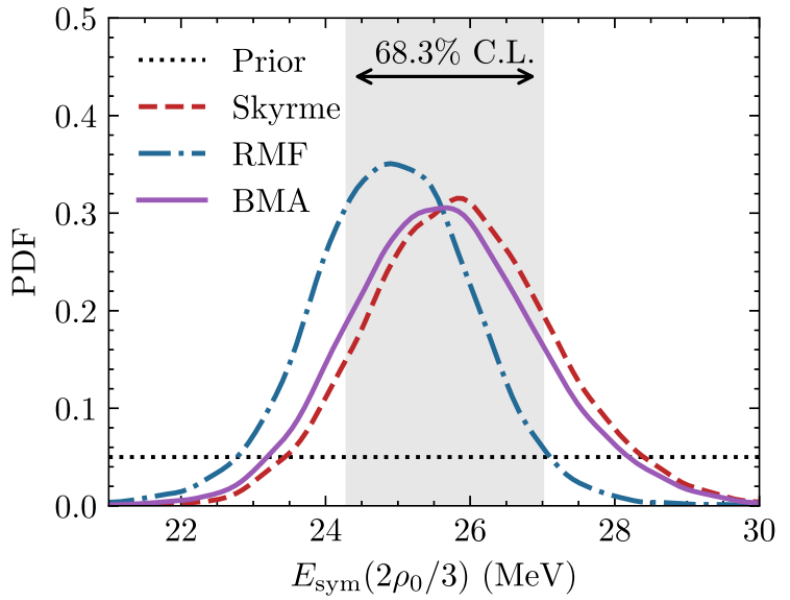}
\caption{(Color Online). Posterior probability distributions of
$E_{\rm{sym}}(2\rho_0/3)$ inferred using the nonrelativistic Skyrme EDF, the relativistic mean field model, and Bayesian model averaging (BMA). The prior distribution is also shown, while the shaded region denotes the $68.3\%$ credible interval of the BMA posterior. The comparison illustrates how BMA combines information from different model classes and incorporates inter-model uncertainty into the inference of the symmetry energy. Adapted from Ref.\,\cite{Qiu2024PLB}.}
\label{fig:BMA-Qiu}
\end{figure}

Bayesian inference provides a natural framework for combining neutron-skin information from PREX-II and CREX with dipole responses, hadronic probes, and neutron-skin sensitive observables in heavy-ion collisions\,\cite{Morfouace2019PLB,Xu2020PRC,Newton2021PRC,Essick2021PRL}. In particular, the analyses of Zhang and collaborators illustrate several levels of uncertainty quantification relevant to neutron-skin physics. Bayesian fits of nuclear resonances and neutron-skin data quantify correlations among isoscalar and isovector nuclear-matter parameters\,\cite{Xu2021PRC}, while the joint analysis of the CREX and PREX-II weak form factor measurements demonstrates how different experimental precisions and parameter correlations shape the posterior distributions of $R_{\rm{skin}}(^{48}\rm{Ca})$, $R_{\rm{skin}}(^{208}\rm{Pb})$, and $E_{\rm{sym}}(\rho)$\,\cite{Zhang2023PRC}. 
Bayesian model averaging (BMA) across non-relativistic and relativistic energy density functionals further provides a way to incorporate inter-model uncertainty rather than reporting the result of a single model\,\cite{Qiu2024PLB,Neufcourt2018PRC,Neufcourt2019PRL,
Connell2021JPG,Kejzlar2023SciRep,Saito2024PRC}. As an example,
FIG.\,\ref{fig:BMA-Qiu} shows the posterior distributions of
$E_{\rm{sym}}(2\rho_0/3)$ obtained by Qiu {et al.}\,\cite{Qiu2024PLB} from non-relativistic Skyrme and relativistic mean-field models, together with their Bayesian model-averaged result\,\cite{Qiu2024PLB}. Although the two model classes favor somewhat different posterior distributions, BMA combines them according to their statistical support from the data and yields $E_{\rm{sym}}(2\rho_0/3)\approx25.6^{+1.4}_{-1.3}$\,MeV at the $68.3\%$
credible level. This provides a representative example of how
inter-model uncertainty can be incorporated explicitly into the
extraction of isovector properties.
These studies emphasize that uncertainty in an extracted neutron skin is not determined by the experimental error alone, but also by parameter degeneracies, prior choices, correlations among observables, and model discrepancy. Microscopic neutron-matter calculations and NS observations can supply complementary constraints, but their different density sensitivities should be preserved instead of assuming a universal mapping from $R_{\rm{skin}}$ to NS radii or tidal deformabilities.

Machine learning methods, including Gaussian-process and reduced-basis emulators, can make the large ensembles of nuclear-structure and reaction calculations required for Bayesian analyses computationally feasible\,\cite{Boehnlein2022RMP,He2023SCPMA,He2023NST,Zhou2024PPNP,
Ma2023CPL,Konig2020PLB,Duguet2024RMP,You2025NST,Guo2025CPL,Zhang2025CPLML,RWang2020PRR,Wang2022PLBML,WangLi2023FrontPhys}. Deep neural networks may also learn nonlinear mappings between measured observables and EOS parameters, as already explored for the reconstruction of the NS EOS\,\cite{Fujimoto2018PRD,Soma2022JCAP}. For neutron-skin studies, their most promising role is to emulate expensive many-body or transport calculations and to enable joint inference from several complementary probes. Nevertheless, their results remain conditional on the training data and physical priors; quantifying epistemic uncertainty, detecting extrapolation beyond the training domain, and imposing physical consistency are therefore essential. Future analyses should combine these tools with model comparison, model averaging, and posterior predictive checks so that the inferred precision of $R_{\rm{skin}}$, $L$, and $E_{\rm{sym}}(\rho)$ is not dominated by a particular interaction, transport model, or EOS parameterization.

\begin{figure}[h!]
\centering
\includegraphics[width=0.47\textwidth]{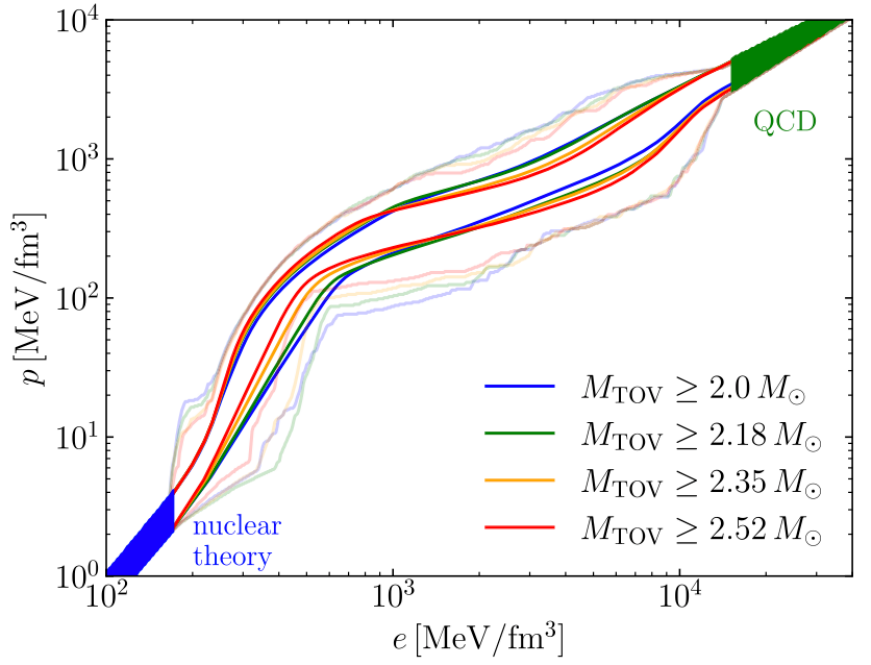}
\caption{(Color Online). Probability distributions of viable EOSs in the pressure-energy density plane for different lower bounds imposed
on the maximum mass of a nonrotating NS. The blue and green shaded
regions indicate the low-density nuclear-theory and asymptotically
high-density perturbative-QCD constraints, respectively. The colored curves delimit the 95\% credible regions obtained for different maximum-mass requirements, while the lighter curves mark the
boundaries of the corresponding excluded regions. Increasing the
minimum allowed maximum mass eliminates softer EOSs and narrows the
available pressure range, particularly at the intermediate densities relevant to NS cores. Adapted from 
Ref.\,\cite{Ecker2023MNRAS}.}
\label{fig:dense_matter_EOS}
\end{figure}

This requirement becomes particularly important when bridging the density regimes probed by finite nuclei and NSs. Nuclear-structure observables and ab initio calculations based on chiral effective field theory provide a controlled anchor mainly at sub-saturation and near-saturation densities, including correlated truncation and many-body  uncertainties\,\cite{Hebeler2013ApJ,Drischler2020PRL,Drischler2020PRC}. By contrast, the interiors of massive NSs may reach several times the saturation density, where the convergence of the effective-field-theory expansion becomes uncertain and additional degrees of freedom or phase transitions may emerge. This change of regime is illustrated in FIG.\,\ref{fig:dense_matter_EOS}: the EOS ensemble is tightly anchored by nuclear theory at low energy densities but broadens substantially toward the densities relevant to NS cores, while increasingly stringent lower bounds on the maximum NS mass progressively eliminate soft high-density EOSs and narrow the allowed pressure of energy-density region\,\cite{Ecker2023MNRAS}. A robust joint analysis should therefore propagate the low-density theoretical covariance into a sufficiently flexible, causal, and thermodynamically consistent high-density EOS, which can then be constrained by HIC experiments and multi-messenger observations\,\cite{Essick2021PRL,Huth2022Nature}. The aim is not to impose a unique extrapolation from $R_{\rm{skin}}$ or $L$ to NS properties, but to identify the density intervals informed by each observable and those in which the inference remains dominated by priors or model assumptions. Such a density-resolved framework would provide a more controlled connection between neutron-rich nuclei and dense matter while making the remaining gap in empirical information explicit.

\subsection{Isovector spin-orbit interaction as a clue to solving the PREX-CREX puzzle}

The PREX-II and CREX measurements provide relatively model independent constraints on the differences between the charge and weak form factors of $^{208}\rm{Pb}$ and $^{48}\rm{Ca}$, respectively. However, their simultaneous
interpretation poses a challenge to conventional nuclear energy density functionals. PREX-II favors a relatively thick neutron skin in $^{208}\rm{Pb}$, whereas CREX indicates a much thinner neutron skin in $^{48}\rm{Ca}$\,\cite{PREX2021PRL,CREX}. Since most conventional functionals predict a strong correlation between the neutron skins of these two nuclei, they generally cannot reproduce both measurements within their one standard
deviation uncertainties. Bayesian analyses further show that the two measurements favor different regions of the symmetry energy parameter space, suggesting that neutron skins may depend not only on the bulk symmetry energy but also on shell structure and other components of the nuclear interaction\,\cite{Zhang2023PRC}.

\begin{figure}[h!]
\centering
\includegraphics[width=0.45\textwidth]{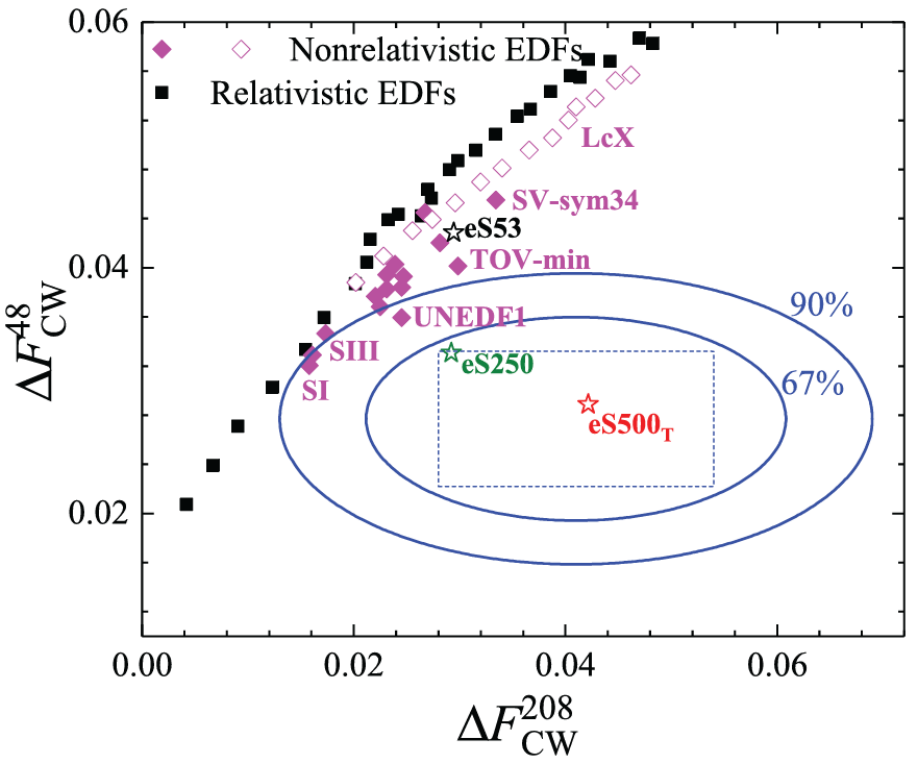}
\caption{(Color Online). Differences between the charge and weak form factors of $^{208}\rm{Pb}$ and $^{48}\rm{Ca}$ predicted by different energy density
functionals, compared with the joint PREX-II and CREX probability regions. The extended Skyrme functionals containing a strong isovector spin-orbit interaction primarily modify the $^{48}\rm{Ca}$ observable while producing a much smaller change in $^{208}\rm{Pb}$. This different sensitivity provides a possible mechanism for reconciling the neutron skins inferred from PREX-II
and CREX. Reproduced from Ref.\,\cite{Yue2026SciBull}.}
\label{fig:Yue2026_IVSO}
\end{figure}

\begin{figure}[h!]
\centering
\includegraphics[width=0.48\textwidth]{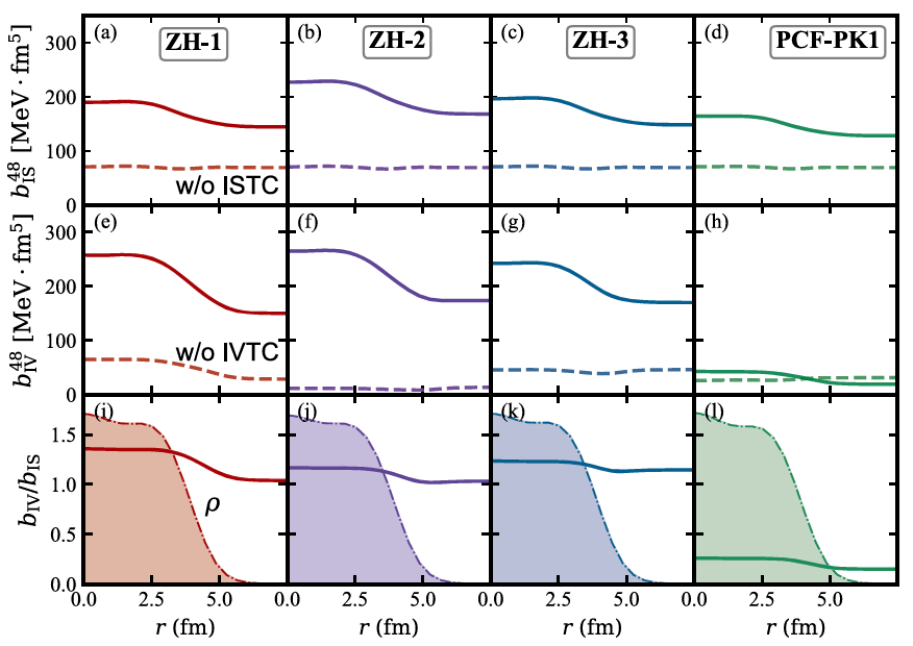}
\caption{(Color Online). Radial dependence of the isoscalar
spin-orbit strength $b_{\rm{IS}}$ (top), isovector spin-orbit
strength $b_{\rm{IV}}$ (middle), and their ratio
$b_{\rm{IV}}/b_{\rm{IS}}$ (bottom) in $^{48}\rm{Ca}$ for the
ZH-1, ZH-2, ZH-3, and PCF-PK1 covariant EDFs. Dashed curves show
results with the corresponding tensor coupling removed. The enhanced isovector tensor coupling strongly increases $b_{\rm{IV}}$, driving $b_{\rm{IV}}/b_{\rm{IS}}$ above unity over much of the nucleus and providing a covariant mechanism for the strong isovector spin-orbit interaction suggested by the CREX and PREX-II data. Adapted
from Ref.\,\cite{Qiu2026PLB}.}
\label{fig:IVSO-covariant}
\end{figure}

A possible resolution has recently been proposed in terms of a substantially enhanced isovector spin-orbit interaction\,\cite{Yue2026SciBull}. As shown in
FIG.\,\ref{fig:Yue2026_IVSO} (see also FIG.\,\ref{fig:prex_crex}), the difference between the charge and weak form factors of $^{48}\rm{Ca}$ is particularly sensitive to the isovector spin-orbit strength, mainly because its eight excess neutrons occupy the spin-orbit unsaturated $1f_{7/2}$ orbital. By contrast, the corresponding observable in $^{208}\rm{Pb}$ is much less affected because of cancellations
among occupied spin-orbit partner orbitals. Strengthening the isovector spin-orbit interaction can therefore reduce
$R_{\rm{skin}}(^{48}\rm{Ca})$ while leaving
$R_{\rm{skin}}(^{208}\rm{Pb})$ nearly unchanged, moving the theoretical predictions toward the joint PREX-II and CREX probability region. Extended Skyrme calculations indicate that an isovector spin-orbit interaction about
four times stronger than that in conventional parametrizations can reproduce both measurements while retaining a reasonable description of nuclear bulk
and shell properties\,\cite{Yue2026SciBull}. A covariant realization has also been proposed in which an enhanced isovector tensor coupling generates the required isovector spin-orbit interaction\,\cite{Qiu2026PLB}.
The microscopic origin of this enhancement is illustrated in FIG.\,\ref{fig:IVSO-covariant}.  A non-relativistic reduction of the covariant functional shows that the tensor coupling produces only a moderate modification of the isoscalar spin-orbit strength $b_{\rm{IS}}$, but strongly enhances its isovector
counterpart $b_{\rm{IV}}$.  In the resulting ZH functionals,
$b_{\rm{IV}}/b_{\rm{IS}}$ exceeds unity over much of the nuclear
interior and surface, compared with about $1/3$ in the original
PCF-PK1 functional\,\cite{Qiu2026PLB}.  Such a large isovector
spin-orbit strength is not excluded by present constraints, since the isovector tensor sector of covariant EDFs remains relatively poorly determined.  Nevertheless, its magnitude is substantially larger than in conventional functionals and should therefore be regarded as an important prediction to be tested, rather than an already established feature of the nuclear interaction.  Further constraints from spin-orbit sensitive observables and neutron-rich nuclei\,\cite{Yue2026SciBull,Yue2026Particles,Zhao2025PRR}, together with global statistical calibrations including tensor couplings, will
be important for determining whether such a strong isovector
interaction is compatible with nuclear data more generally.

The universality and microscopic origin of this mechanism nevertheless require systematic investigation. Kunjipurayil, Piekarewicz, and Salinas confirmed
within a relativistic framework that enhancing the isovector spin-orbit potential can reduce the neutron skin of $^{48}\rm{Ca}$ without substantially changing that of $^{208}\rm{Pb}$. However, they cautioned that a sufficiently large phenomenological enhancement may alter the established ordering of spin-orbit partners and disturb conventional shell closures\,\cite{Kunjipurayil2025PRC}. Future calculations should therefore refit the central, tensor, and spin-orbit sectors simultaneously within Skyrme, Gogny, and covariant functionals and test them against charge and weak form factors, neutron skins, electric dipole polarizabilities, binding energies, charge radii, single particle spectra, spin-orbit splittings, and shell gaps across a wide range of nuclei. Precise treatment of the electroweak reaction
mechanism is equally important. 

Finally, the precision interpretation of PREX and CREX also requires a consistent treatment of electroweak radiative corrections. Roca-Maza and Jakubassa-Amundsen calculated an enhancement of the parity violating asymmetry by approximately $5\%$ from the QED corrections included in their analysis, which would increase the inferred neutron skin thicknesses\,\cite{RocaMaza2025PRL}. Reed and Horowitz subsequently argued that the finite momentum transfer axial vector electron vertex correction to $Z$ exchange was omitted and largely cancels the corresponding vector vertex contribution\,\cite{Reed2026Comment}. In their estimate, the dominant remaining contribution is vacuum polarization, which reduces the asymmetry by approximately $0.7\%$ and produces slightly smaller extracted neutron skins. The resulting changes are less than one half of the quoted systematic uncertainties and less than one quarter of the statistical uncertainties. These corrections therefore appear too small to resolve the PREX-CREX puzzle at the present experimental precision. Nevertheless, the ongoing discussion demonstrates that complete and internally consistent radiative corrections will be necessary for future high precision extractions of neutron skins.

\subsection{Beyond the symmetry energy slope parameter}

Although the neutron skin thickness is often interpreted through its correlation with the symmetry energy slope parameter $L$, it generally depends on the density dependence of the symmetry energy beyond the linear term. Because the nuclear surface samples a finite range of subsaturation densities, the curvature coefficient $K_{\rm{sym}}$ can make a non-negligible contribution to $R_{\rm{skin}}$. This contribution may be partially hidden by correlations between $L$ and $K_{\rm{sym}}$ inherent in restricted families of conventional energy density functionals\,\cite{Raduta2018PRC}. Within a selected set of covariant energy density functionals, Reed et al.\,\cite{Reed2024PRC} found that the CREX weak skin form factor of $^{48}\rm{Ca}$ correlates strongly with the combination $K_{\rm{sym}}-6L$, suggesting that PREX and CREX may constrain complementary combinations of symmetry-energy parameters. The relevance of $K_{\rm{sym}}$ becomes more evident when neutron skins are considered together with NS observables. The radius and tidal deformability of a canonical NS probe the pressure over a broader density interval extending above saturation density and therefore depend on correlated combinations of $L$, $K_{\rm{sym}}$, and higher order EOS parameters\,\cite{Zhang2019JPG,Li2020PRC}. Different combinations of these parameters may produce similar values of $\Lambda_{1.4}$, demonstrating that tidal deformability alone cannot uniquely determine the density dependence of the symmetry energy\,\cite{Krastev2019JPG,Li2021Universe}.

\begin{figure}[h!]
\centering
\includegraphics[width=0.45\textwidth]{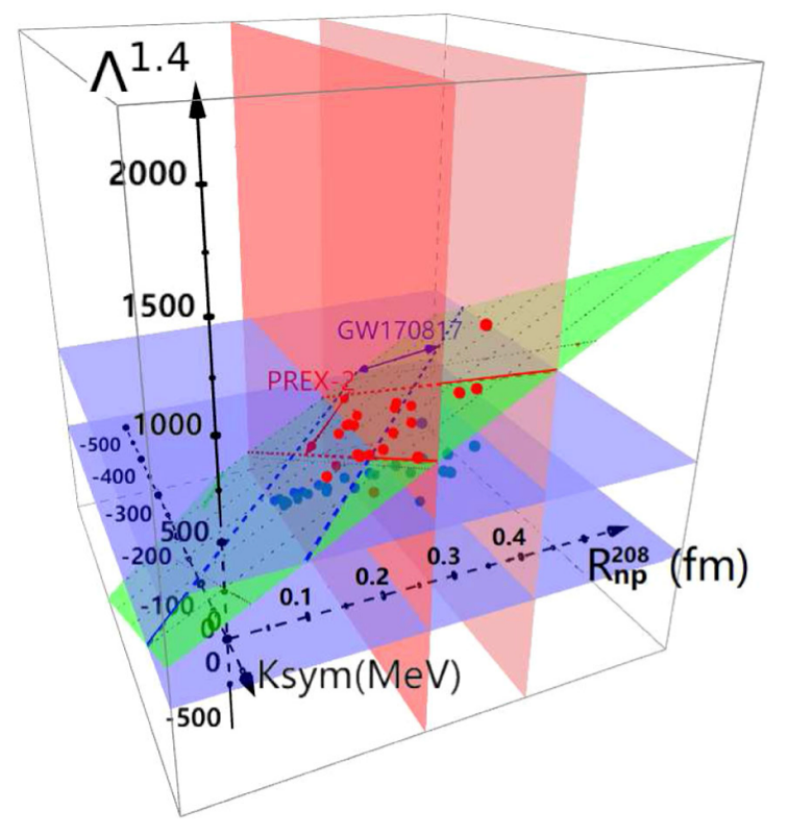}
\caption{(Color Online). Tidal deformability $\Lambda_{1.4}$ as a function of the neutron skin thickness $R_{\rm{skin}}^{208}$ of $^{208}\rm{Pb}$ and the symmetry-energy curvature coefficient $K_{\rm{sym}}$. Results from relativistic and nonrelativistic energy density functionals are represented by red and blue symbols, respectively. The green surface denotes a bivariate regression to the complete set of models considered, while the red and blue planes represent the constraints from PREX-II and GW170817, respectively. Adapted from Ref.\,\cite{Guan2025PRL}.}
\label{fig:Guan2025_Ksym}
\end{figure}

This complementarity is illustrated in FIG.\,\ref{fig:Guan2025_Ksym}. Guan and Niu\,\cite{Guan2025PRL} showed that the direct correlation between $R_{\rm{skin}}^{208}$ and $\Lambda_{1.4}$ differs between the relativistic and nonrelativistic energy density functionals considered in their analysis. By introducing $K_{\rm{sym}}$ as an additional variable, they obtained the approximate regression plane
$\Lambda_{1.4}\approx2.168K_{\rm{sym}}+1978R_{\rm{skin}}^{208}+291.1$,
with a coefficient of determination $\rm{R}^2\approx0.83$. Within this ensemble of energy density functionals, the combined PREX-II and GW170817 constraints yield
$-441\lesssim K_{\rm{sym}}/\rm{MeV}\lesssim-63$.
The result indicates that freedom in $K_{\rm{sym}}$ may account for part of the apparent tension between a thick neutron skin and a moderate tidal deformability. However, the fitted relation is specific to the models and assumptions employed and should not be regarded as universal. Besides $K_{\rm{sym}}$, the predicted neutron skin depends on the surface symmetry energy, surface stiffness, and the difference between neutron and proton surface widths\,\cite{Warda2009PRC}. Shell structure and the isovector spin-orbit interaction may be especially important for $^{48}\rm{Ca}$\,\cite{Yue2026SciBull}. Pairing, nuclear deformation, clustering, many body correlations, and the functional form of the isovector interaction may provide additional nucleus dependent contributions. Systematic measurements across isotopic chains, combined with dipole polarizabilities, heavy-ion collisions, and NS observations, are therefore needed to separate these effects and determine whether the inferred sensitivity to $K_{\rm{sym}}$ persists beyond specific theoretical frameworks.

\subsection{From short-range correlations and spectral functions to long-range neutron skins}

\begin{figure}[h!]
\centering
\includegraphics[width=0.48\textwidth]{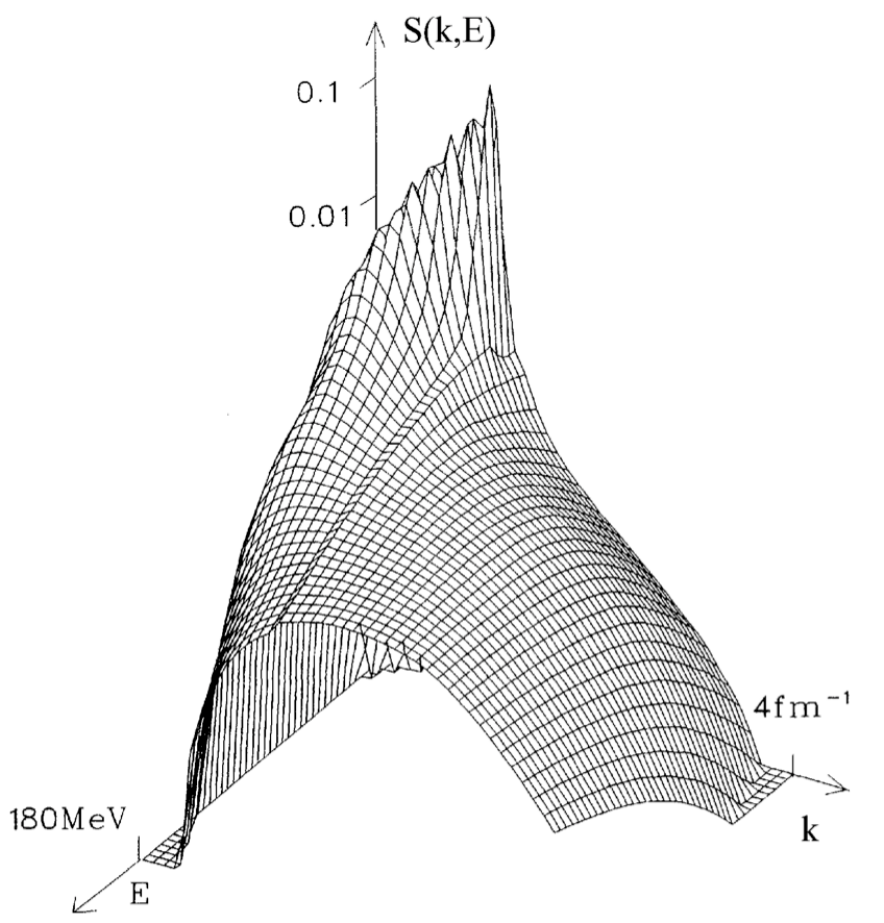}
\caption{Energy dependence of the hole spectral function of symmetric nuclear matter at equilibrium density for several representative nucleon momenta. Many-body correlations broaden and fragment the
single-particle strength relative to the free Fermi-gas result.
Figure taken from Ref.\,\cite{Benhar2008RMP}.}
\label{fig:hole-spectral-function}
\end{figure}

A microscopic connection between SRCs and long-range neutron skins requires a consistent description of single particle dynamics and bulk thermodynamics. In the Green function formalism, the nucleon momentum distribution is obtained from the hole spectral function according to\,\cite{Lehmann1954NC,Migdal1957JETP,Galitskii1958JETP,Abrikosov1963Book,Dickhoff2025Book}, namely $
n_{J}(k)=\int_{-\infty}^{\mu_{J}}
S_{J}^{\rm{h}}(\mathbf{k},E)\d E$, where $
J=\rm{n},\rm{p}$ and $\mu_{J}$ denotes the corresponding chemical potential. The spectral function contains both the quasiparticle contribution and the fragmented background generated by many-body correlations, while its energy-weighted integral connects single-particle dynamics to the total ground-state energy through the Galitskii-Koltun sum rule\,\cite{Koltun1972PRL}. Nuclear Green function and, the dispersive optical-model studies, for example, provide practical frameworks for constraining spectral strength, nucleon self-energies, occupation probabilities, and coordinate-space densities using nuclear structure and reaction data\,\cite{Polls1995PPNP,Soma2006PRC,Mahzoon2017PRL,Atkinson2020PRC,Atkinson2024FrontPhys}. Determining these quantities and their uncertainties is essential for understanding how SRC-induced depletion below the Fermi surface and the associated high-momentum tail modify the kinetic and potential contributions to the symmetry energy\,\cite{Hen17RMP}. Although the Hugenholtz--Van Hove (HVH) theorem remains an exact thermodynamic relation\,\cite{Hugenholtz1958Physica}, its familiar mean-field decomposition\,\cite{Xu2010PRC,Xu2011NPA,Chen2012PRC,
Li2015PLB,Cai2019PRC,Cai2018PRC},
\begin{equation}
E_{\rm{sym}}(\rho)
=
\frac{k_{\rm{F}}^2}{6M_0^{\ast}}
+\frac{1}{2}U_{\rm{sym}}(\rho,k_{\rm{F}}),
\end{equation}
becomes more involved when correlations fragment the single-particle strength\,\cite{Xu2010PRC}, here $M_0^{\ast}$ is the nucleon Dirac mass in symmetric matter\cite{Cai2012PRC,Cai2012PLB,Cai2015PRC,Cai2017NST}, and $U_{\rm{sym}}$ is the isospin-related potential. A generalized formulation in terms of consistently dressed propagators, self-energies, and spectral functions will therefore be necessary to preserve thermodynamic consistency beyond the mean-field approximation\,\cite{Baym1961PR,Baym1962PR}.
An example of single-particle strength is
illustrated in FIG.\,\ref{fig:hole-spectral-function}; in addition to the concentrated quasiparticle strength, many-body correlations generate a broad background extending toward large momenta and removal energies\,\cite{Benhar1989NPA,Benhar2008RMP}.

A related challenge is to determine momentum- and coordinate-space distributions consistently from an underlying phase-space description. Although SRCs characterize short-distance and high-momentum dynamics, they can influence long-range one-body observables through the correlated many-body wave function. Using a schematic two-component model and a chiral-dynamics estimate, Miller et al.\,\cite{Miller2019PLB} showed that tensor-dominated $\rm{np}$ correlations can generate non-negligible corrections to nuclear charge radii, with their magnitude and sign depending on the single-particle orbital structure. Their analysis also emphasizes that, when nuclear interactions are softened through unitary transformations, long-range operators such as the mean-square-radius operator must be transformed consistently. Wigner-function calculations further indicate that SRC-generated high-momentum components originate predominantly from the nuclear interior and can induce a small but nonzero modification of the neutron skin of $^{48}\rm{Ca}$\,\cite{Cosyn21}, while calculations beyond the independent-particle approximation show that both short- and long-range correlations can modify neutron skins\,\cite{Co2022PRC}. The coexistence of a proton skin in momentum space and a neutron skin in coordinate space provides another manifestation of this momentum-coordinate connection and may have observable consequences in nuclear reactions\,\cite{Cai16b,Guo2023PRC}. Nevertheless, SRCs are not expected to be the primary origin of neutron skins, which are governed mainly by the isovector mean field, surface symmetry energy, shell structure, and possibly clustering effects. Instead, correlations constitute an additional many-body correction and a source of theoretical uncertainty. Their effects should therefore be quantified in future interpretations of PREX and CREX and in assessments of whether proposed resolutions of their apparent tension remain robust beyond conventional mean-field descriptions\,\cite{PREX2021PRL,CREX,Reinhard2022PRL,Hu2022NP,Zhang2023PRC,Yang2023PRC}.

The symmetry energy provides the essential bridge between these
SRCs and the long-range neutron skin. As illustrated in FIG.\,\ref{fig_ab_Esym}, SRC-induced high-momentum
components redistribute the kinetic and potential contributions to $E_{\rm{sym}}(\rho)$ and can significantly alter its density
dependence away from saturation\,\cite{CaiLi22Gog}. The resulting
change in the isovector pressure at sub-saturation densities may, in turn, modify the formation of the neutron skin. Thus,
$R_{\rm{skin}}$ is not a direct image of SRC pairs, but a long-range structural observable that may retain their imprint through the correlated many-body dynamics and the symmetry energy\,\cite{Cai2026EPJST}.
A complementary connection has recently been explored in
ultrarelativistic HICs. Pei Li {et al.}\,\cite{LiP2025SRC}
demonstrated that incorporating nucleon-nucleon SRCs into the initial nuclear configurations can appreciably modify higher-order fluctuations of the transverse size, which are subsequently reflected in final state mean-$p_{\rm T}$ fluctuations. In particular, the SRC effects exhibit a
systematic scaling with nuclear size and density, suggesting that
relativistic HICs may provide a novel probe of two-body correlations beyond the conventional sensitivity to one-body nuclear structure such as neutron skins and deformation.

\subsection{Coherent elastic neutrino-nucleus scattering, dark matter, neutrino transport in neutron stars, and neutron skins}

\begin{figure}[h!]
\centering
\includegraphics[width=0.48\textwidth]{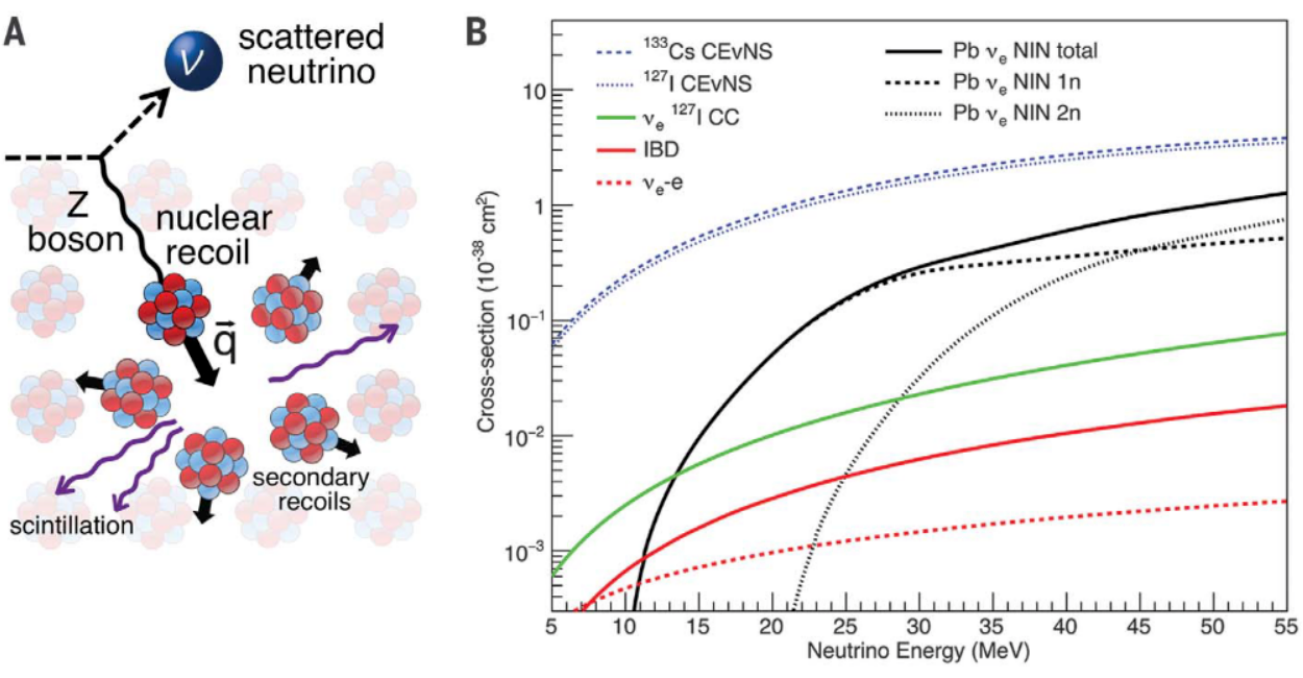}
\caption{(Color Online). Schematic illustration and characteristic cross sections of coherent elastic neutrino-nucleus scattering. (a) For $qR<1$, a neutrino probes the entire nucleus coherently through neutral-current exchange, producing a low-energy nuclear recoil and an interaction probability enhanced approximately as $N^2$. (b) Total CE$\nu$NS cross sections for $^{133}\rm{Cs}$ and $^{127}\rm{I}$ compared with neutrino-electron scattering, the charged-current interaction on $^{127}\rm{I}$, inverse beta decay, and neutrino-induced neutron production on $^{208}\rm{Pb}$. Reproduced from Ref.\,\cite{Akimov2017Science}.}
\label{fig:Akimov2017_CEvNS}
\end{figure}

Coherent elastic neutrino-nucleus scattering (CE$\nu$NS) provides a promising electroweak approach to determining neutron distributions. At sufficiently low momentum transfer, $qR<1$, the wavelength associated with the exchanged neutral current is larger than the nuclear size, and the scattering amplitudes from individual nucleons add coherently. As illustrated in FIG.\,\ref{fig:Akimov2017_CEvNS}, this produces a cross section dominated by the nuclear weak charge and approximately proportional to $N^2$, while the experimentally observable signal is a low-energy nuclear recoil\,\cite{Freedman1974PRD,Akimov2017Science}. In the neutrino-energy range relevant to the Spallation Neutron Source, the CE$\nu$NS cross section is substantially larger than those of several other low-energy neutrino reactions, making this process particularly attractive for precision studies with compact, low-threshold detectors.

\begin{figure}[h!]
\centering
\includegraphics[width=0.42\textwidth]{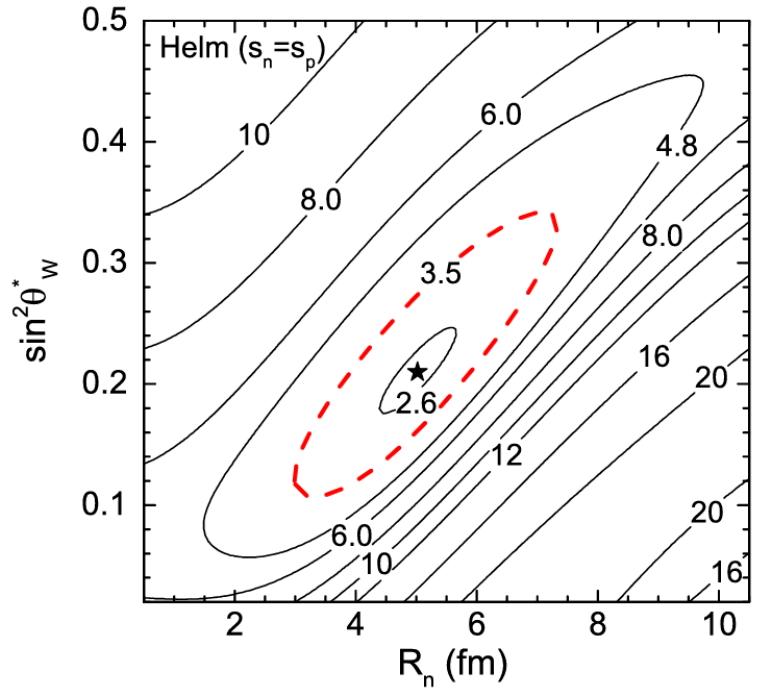}
\caption{(Color Online). The $\chi^2$ contours in the $R_{\rm{n}}$-$\sin^2\theta_{\rm W}^{*}$ plane obtained from a two-dimensional fit to the COHERENT data using the Helm form factor with $s_{\rm{n}}=s_{\rm{p}}$. The star denotes the best-fit values $R_{\rm{n}}\approx5.02$\,fm and
$\sin^2\theta_{\rm W}^{*}\approx0.21$, while the dashed contour corresponds to
$\chi^2=\chi_{\min}^2+1$. Adapted from
Ref.\,\cite{Huang2019PRD}.}
\label{fig:CsI-Rn-weak-angle}
\end{figure}

The recoil-energy spectrum depends on the weak form factor and is therefore sensitive to the neutron radius and, when combined with the accurately known proton radius, the neutron-skin thickness. Existing COHERENT data have already been used to constrain the average neutron distribution of CsI\,\cite{Cadeddu2018PRL}, while theoretical studies have examined the extraction of neutron form factors and quantified the associated nuclear-structure uncertainties\,\cite{Ciuffoli2018PRD,Payne2019PRC,Coloma2020JHEP}. In particular, joint analyses of the CsI neutron skin and the low-energy weak mixing angle demonstrate the need to account for correlations between nuclear and electroweak parameters\,\cite{Huang2019PRD}.
As illustrated in FIG.\,\ref{fig:CsI-Rn-weak-angle}, the resulting
two-dimensional fit exhibits a strong positive correlation between
$R_{\rm{n}}$ and $\sin^2\theta_{\rm W}^{*}$. Fixing either quantity can therefore significantly affect the inferred constraint on the other.
Future measurements with improved recoil thresholds, controlled neutrino fluxes, and multiple targets spanning different masses and isospin asymmetries could help separate the neutron radius from surface diffuseness and test the predicted evolution of neutron skins across isotopic chains. The broad energy spectrum of supernova neutrinos may provide an additional opportunity to probe the momentum dependence of the weak form factor and improve the sensitivity to neutron skins in large low-threshold detectors\,\cite{Huang2022PRD}.

Neutrino transport in neutron-rich matter offers a complementary but more indirect probe of the same isovector physics. Neutrino opacities and mean free paths depend on nucleon effective masses, chemical potentials, density and spin-density response functions, many-body correlations, and the composition of matter\,\cite{Reddy1998PRD,Burrows1998PRC,Burrows2006NPA}. In particular, the isovector mean-field difference $U_{\rm{n}}-U_{\rm{p}}$, which is closely connected to the nuclear symmetry energy, modifies the kinematics and rates of charged-current neutrino reactions in hot neutron-rich matter\,\cite{Roberts2012PRC}. The same isovector interaction can influence neutron skins in finite nuclei as well as neutrino diffusion, deleptonization, and emission from proto-NSs, although no unique mapping between these observables should be assumed. Future progress will require microscopic response calculations constrained simultaneously by nuclear-structure, CE$\nu$NS, and nucleon-scattering data, together with consistent uncertainty propagation in neutrino-transport simulations\,\cite{Hutauruk2022PRC,Lin2023PRC}. Combining laboratory CE$\nu$NS measurements with a future high-statistics Galactic supernova-neutrino signal could therefore test whether a common isovector interaction can consistently describe neutron distributions in finite nuclei and weak-interaction processes in dense neutron-rich matter.

\begin{figure}[h!]
\centering
\includegraphics[width=0.48\textwidth]{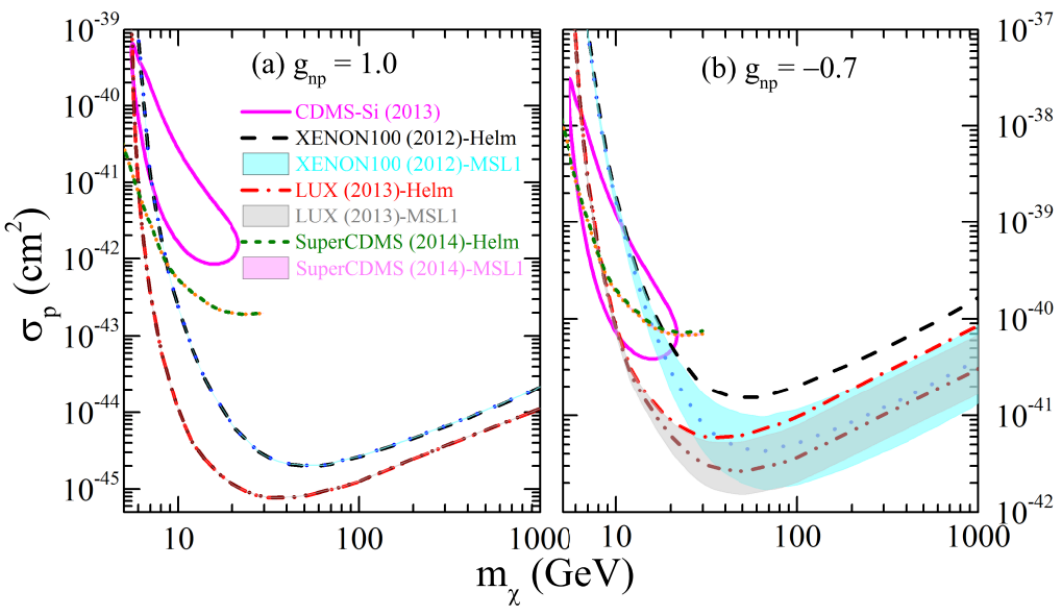}
\caption{(Color Online). Constraints on the spin-independent
DM-proton cross section as a function of the DM mass for
isospin-invariant DM with $f_{\rm{n}}/f_{\rm{p}}=1$ (left) and
isospin-violating DM with $f_{\rm{n}}/f_{\rm{p}}=-0.7$ (right).
Results obtained with the empirical Helm form factor are compared
with SHF calculations with different symmetry energy slopes
$L(\rho_{\rm{c}})$ and, consequently, different neutron skins and
neutron form factors.  The neutron-skin dependence is weak for
$f_{\rm{n}}/f_{\rm{p}}=1$, but is strongly amplified by the
destructive interference between proton and neutron amplitudes for $f_{\rm{n}}/f_{\rm{p}}=-0.7$. Adapted from
Ref.\,\cite{Zheng2014JCAP}.}
\label{fig:DMskin}
\end{figure}

The same sensitivity to neutron distributions also establishes an
interesting connection to direct searches for dark matter (DM).
As discussed above for CE$\nu$NS, scattering from a nucleus at finite momentum transfer depends on nuclear response functions that encode its underlying proton and neutron structure.  Large-scale shell-model calculations have provided weak form factors and nuclear responses for a broad range of experimentally relevant targets, including Cs, I,
Ar, F, Na, Ge, and Xe\,\cite{Hoferichter2020PRD}.  Beyond the
leading coherent response, vector, axial-vector, scalar, tensor, and dipole interactions involve different combinations of nuclear
structure factors, with additional corrections arising from
chiral-EFT two-body currents\,\cite{Hoferichter2020PRD}.
The same framework also enters DM-nucleus scattering, where the
relative weighting of proton and neutron responses depends on the
underlying DM interaction.  Consequently, uncertainties in the
neutron density and neutron skin can propagate into the interpretation of direct-detection experiments, providing another connection between neutron-skin physics and searches for physics beyond the Standard Model. A particularly transparent example is spin-independent isospin-violating dark matter\,\cite{Zheng2014JCAP}.
The coherent scattering amplitude involves the combination
$ZF_{\rm{p}}(q)+(f_{\rm{n}}N/f_{\rm{p}})F_{\rm{n}}(q)$, making the difference between proton and neutron form factors relevant whenever their couplings are weighted differently.  As illustrated in FIG.\,\ref{fig:DMskin}, Zheng {et al.}\,\cite{Zheng2014JCAP} varied the symmetry-energy slope $L(\rho_{\rm{c}})$ in Skyrme-Hartree-Fock (SHF) calculations, thereby changing the neutron skin and neutron form factor of neutron-rich Xe isotopes.  For the isospin-invariant case
$f_{\rm{n}}/f_{\rm{p}}=1$, the resulting form-factor dependence is small.  In contrast, for $f_{\rm{n}}/f_{\rm{p}}=-0.7$, destructive interference between the proton and neutron amplitudes amplifies the otherwise modest difference between $F_{\rm{p}}(q)$ and $F_{\rm{n}}(q)$, producing a pronounced neutron-skin dependence of the inferred DM-proton cross section.  The sensitivity of Xe-based detectors can change by a factor of a few and, for sufficiently heavy DM, by more than an order of magnitude\,\cite{Zheng2014JCAP}. Microscopic calculations of DM nuclear responses, including shell-model
and chiral-EFT treatments\,\cite{Baudis2013PRD,Vietze2015PRD,
Hoferichter2016PRD,Gazda2017PRD,Hu2022PRLDM,Hu2026FrontPhys},
therefore provide an important avenue for propagating increasingly precise constraints on neutron radii and density profiles into direct-detection analyses and for identifying the interaction scenarios in which neutron-skin uncertainties become significant.

\section{Conclusions}
\label{sec:conclusion}

The neutron skin is a fundamental manifestation of isospin asymmetry in finite nuclei and an important probe of the isovector nuclear interaction. Its thickness is closely related to the density dependence of the symmetry energy, particularly around and below saturation density, while also depending on surface properties, shell structure, deformation, and many-body correlations. Heavy-ion collisions provide a distinctive approach to studying neutron skins because the initial neutron and proton distributions are dynamically transformed into measurable particle-emission and collective-flow observables. They therefore complement static nuclear-structure measurements and offer access to the interplay among the neutron skin, reaction dynamics, and the symmetry energy over a broad range of densities and collision energies.

The central focus of this review has been the effects of neutron skins on particle emission in heavy-ion collisions. At intermediate energies, neutron-to-proton and $\rm{t}/^3\rm{He}$ yield ratios, light-cluster production, fragment momentum distributions, isoscaling, and pre-equilibrium emission retain information about the neutron-rich nuclear surface and the isospin composition of the participant region. The $R_{\rm{np}}$ and $\rm{t}/^3\rm{He}$ ratios are especially promising because their correlations with neutron-skin thickness have been demonstrated for several neutron-rich reaction systems. Fragment observables provide complementary sensitivity to the projectile surface, while neutron-proton momentum differences and rapidity-dependent yields can help identify when and where the initial skin information is released during the reaction.
Pion and hard-photon production extend these studies to different stages and density regimes of the collision. The $\pi^-/\pi^+$ ratio is primarily employed as a probe of the high-density symmetry energy, but its interpretation may also depend on the initial neutron skin, particularly in peripheral reactions where surface nucleons contribute more strongly. Hard photons from $\rm{np}$ bremsstrahlung exhibit an anticorrelation with the neutron-skin thickness in model calculations and provide an electromagnetic probe less affected by final-state strong interactions. Their production also illustrates the possible interplay between neutron skins in coordinate space and SRC-induced proton skins in momentum space. A consistent treatment of nuclear density profiles, momentum distributions, and reaction dynamics is therefore required before these observables can be used for precision extraction.

Collective flows offer a complementary manifestation of neutron-skin effects. At intermediate energies, directed and elliptic flows are influenced by the neutron-rich surface, isospin transport, and the isovector mean field. At relativistic energies, the neutron skin modifies the initial geometry and eccentricity fluctuations, which are subsequently reflected in multiplicities and anisotropic flows. Isobar collisions involving Ru and Zr demonstrate both the potential and the complexity of this approach, since neutron-skin and nuclear-deformation effects can contribute simultaneously to the same observables. Ultra-relativistic Pb+Pb collisions provide another possible route to constraining the neutron distribution of $^{208}\rm{Pb}$, although the extraction depends on the modeling of the initial state and the hydrodynamic response. These developments extend neutron-skin studies continuously from intermediate-energy reactions to high-energy collider experiments.

The principal challenge is to distinguish neutron-skin signatures from other nuclear-structure and reaction-model effects. Nuclear deformation, surface diffuseness, halo and clustering phenomena, shell structure, and SRCs may generate correlated changes in particle emission and collective flow. Transport-model uncertainties arise additionally from the initialization of nuclei, isovector mean fields, in-medium nucleon-nucleon scattering, Pauli blocking, cluster formation, resonance dynamics, and particle-production mechanisms. Because different observables probe different stages and density regions of a collision, their neutron-skin sensitivities cannot be interpreted through a single universal correlation with the symmetry-energy slope parameter $L$. Coordinated model comparisons and simultaneous analyses of several observables will be essential for separating the initial-state neutron skin from the subsequent reaction dynamics.

Future radioactive-beam facilities will enable systematic measurements along longer isotopic chains and with larger neutron excesses, while improved intermediate-energy experiments and high-energy colliders will provide higher statistics and a wider range of collision systems and centralities. The resulting heavy-ion constraints should be combined with parity-violating electron scattering, dipole responses, nucleon scattering, and coherent elastic neutrino-nucleus scattering. Ab initio calculations, microscopic spectral-function and phase-space descriptions, and Bayesian inference will be important for incorporating these complementary data and propagating their uncertainties consistently. Connections to NS radii, tidal deformabilities, and neutrino transport can then place the heavy-ion results within the broader density evolution of neutron-rich matter, without assuming a simple one-to-one correspondence between finite nuclei and NSs. In this way, particle emission and collective dynamics in heavy-ion collisions can become an increasingly quantitative bridge between neutron distributions at the nuclear surface and the EOS of neutron-rich matter.

\section*{Acknowledgments}

The author gratefully acknowledges the members of the STAR Collaboration, the low-energy nuclear physics community, and the astrophysics community for stimulating discussions that have shaped this review. 
This work was supported in part by the National Natural Science Foundation of China under contract No. 12547102, and the STCSM under Grant No. 23590780100 and 23JC1400200.


\begin{thebibliography}{99}


\bibitem{MaYG_arxiv}
Y.-G. Ma,
\textit{Frontier Questions and Emerging Directions in Nuclear Science and Technology},
Nucl. Sci. Tech. \textbf{37}, in press (2026);
\href{https://arxiv.org/abs/2608.26207}
{arXiv:2608.26207}.

\bibitem{Steiner2005PhysRep}
A.W. Steiner, M. Prakash, J.M. Lattimer, and P.J. Ellis,
\textit{Isospin Asymmetry in Nuclei and Neutron Stars},
\href{https://doi.org/10.1016/j.physrep.2005.02.004}
{Phys. Rep. \textbf{411}, 325 (2005).}

\bibitem{Yang2020ARNPS}
J. Yang and J. Piekarewicz,
\textit{Covariant Density Functional Theory in Nuclear Physics and
Astrophysics},
\href{https://doi.org/10.1146/annurev-nucl-101918-023608}
{Annu. Rev. Nucl. Part. Sci. \textbf{70}, 1--27 (2020).}

\bibitem{Tsang2024NatAstron}
C.Y. Tsang et al.,
\textit{Determination of the equation of state from nuclear experiments and neutron star observations},
\href{https://doi.org/10.1038/s41550-023-02161-z}
{Nat. Astron. \textbf{8}, 328 (2024).}

\bibitem{Drischler2021ARNPS}
C. Drischler, J.W. Holt, and C. Wellenhofer,
\textit{Chiral Effective Field Theory and the High-Density Nuclear
Equation of State},
\href{https://doi.org/10.1146/annurev-nucl-102419-041903}
{Annu. Rev. Nucl. Part. Sci. \textbf{71}, 403 (2021).}

\bibitem{Lovato2022LRP}
A. Lovato et al.,
\textit{Long Range Plan: Dense Matter Theory for Heavy-Ion
Collisions and Neutron Stars},
\href{https://arxiv.org/abs/2211.02224}
{arXiv:2211.02224 [nucl-th] (2022).}


\bibitem{Fukushima2011RPP}
K. Fukushima and T. Hatsuda,
\textit{The Phase Diagram of Dense QCD},
\href{https://doi.org/10.1088/0034-4885/74/1/014001}
{Rep. Prog. Phys. \textbf{74}, 014001 (2011).}

\bibitem{Ghiglieri2020PhysRep}
J. Ghiglieri, A. Kurkela, M. Strickland, and A. Vuorinen,
\textit{Perturbative Thermal QCD: Formalism and Applications},
\href{https://doi.org/10.1016/j.physrep.2020.07.004}
{Phys. Rep. \textbf{880}, 1 (2020).}


\bibitem{Haque2025PPNP}
N. Haque and M.G. Mustafa,
\textit{Hard Thermal Loop: Theory and Applications},
\href{https://doi.org/10.1016/j.ppnp.2024.104136}
{Prog. Part. Nucl. Phys. \textbf{140}, 104136 (2025).}

\bibitem{Laine2016Book}
M. Laine and A. Vuorinen,
\textit{Basics of Thermal Field Theory: A Tutorial on Perturbative
Computations},
\href{https://doi.org/10.1007/978-3-319-31933-9}
{Lecture Notes in Physics \textbf{925}, Springer, Cham (2016).}

\bibitem{Romatschke2019Book}
P. Romatschke and U. Romatschke,
\textit{Relativistic Fluid Dynamics In and Out of Equilibrium:
And Applications to Relativistic Nuclear Collisions},
\href{https://doi.org/10.1017/9781108651998}
{Cambridge University Press, Cambridge (2019).}


\bibitem{Li2008PhysRep}
B.-A. Li, L.-W. Chen, and C.~M. Ko,
\textit{Recent progress and new challenges in isospin physics with
heavy-ion reactions},
\href{https://doi.org/10.1016/j.physrep.2008.04.005}
{Phys. Rep. \textbf{464}, 113 (2008)}.

\bibitem{Sorensen2024PPNP}
A. Sorensen et al.,
\textit{Dense nuclear matter equation of state from heavy-ion collisions},
\href{https://doi.org/10.1016/j.ppnp.2023.104080}
{Prog. Part. Nucl. Phys. \textbf{134}, 104080 (2024).}

\bibitem{BraunMunzinger2016PhysRep}
P. Braun-Munzinger, V. Koch, T. Sch\"afer, and J. Stachel,
\textit{Properties of Hot and Dense Matter from Relativistic
Heavy-Ion Collisions},
\href{https://doi.org/10.1016/j.physrep.2015.12.003}
{Phys. Rep. \textbf{621}, 76 (2016).}

\bibitem{Shuryak2017RMP}
E. Shuryak,
\textit{Strongly Coupled Quark-Gluon Plasma in Heavy-Ion
Collisions},
\href{https://doi.org/10.1103/RevModPhys.89.035001}
{Rev. Mod. Phys. \textbf{89}, 035001 (2017).}

\bibitem{Busza2018ARNPS}
W. Busza, K. Rajagopal, and W. van der Schee,
\textit{Heavy Ion Collisions: The Big Picture and the Big
Questions},
\href{https://doi.org/10.1146/annurev-nucl-101917-020852}
{Annu. Rev. Nucl. Part. Sci. \textbf{68}, 339 (2018).}

\bibitem{Heinz2013ARNPS}
U. Heinz and R. Snellings,
\textit{Collective Flow and Viscosity in Relativistic Heavy-Ion
Collisions},
\href{https://doi.org/10.1146/annurev-nucl-102212-170540}
{Annu. Rev. Nucl. Part. Sci. \textbf{63}, 123 (2013).}

\bibitem{Gale2013IJMPA}
C. Gale, S. Jeon, and B. Schenke,
\textit{Hydrodynamic Modeling of Heavy-Ion Collisions},
\href{https://doi.org/10.1142/S0217751X13400113}
{Int. J. Mod. Phys. A \textbf{28}, 1340011 (2013).}

\bibitem{Florkowski2018RPP}
W. Florkowski, M.P. Heller, and M. Spali\'nski,
\textit{New Theories of Relativistic Hydrodynamics in the LHC Era},
\href{https://doi.org/10.1088/1361-6633/aaa091}
{Rep. Prog. Phys. \textbf{81}, 046001 (2018).}

\bibitem{Elfner2023JPG}
H. Elfner and B. M\"uller,
\textit{The Exploration of Hot and Dense Nuclear Matter:
Introduction to Relativistic Heavy-Ion Physics},
\href{https://doi.org/10.1088/1361-6471/ace824}
{J. Phys. G \textbf{50}, 103001 (2023).}

\bibitem{Chen2018PhysRep}
J. H. Chen, D. Keane, Y. G. Ma, A. H. Tang, and Z. B. Xu,
\textit{Antinuclei in Heavy-Ion Collisions},
\href{https://doi.org/10.1016/j.physrep.2018.07.002}
{Phys. Rep. \textbf{760}, 1 (2018).}


\bibitem{Ma2018PPNP}
C. W. Ma and Y. G. Ma,
\textit{Shannon Information Entropy in Heavy-Ion Collisions},
\href{https://doi.org/10.1016/j.ppnp.2018.01.002}
{Prog. Part. Nucl. Phys. \textbf{99}, 120 (2018).}

\bibitem{Deng2024PPNP}
X.G. Deng, D.Q. Fang, and Y.G. Ma,
\textit{Shear Viscosity of Nucleonic Matter},
\href{https://doi.org/10.1016/j.ppnp.2023.104095}
{Prog. Part. Nucl. Phys. \textbf{136}, 104095 (2024).}

\bibitem{Chen2024NST}
J. H. Chen et al.,
\textit{Properties of the QCD Matter: Review of Selected Results
from the Relativistic Heavy-Ion Collider Beam Energy Scan
(RHIC BES) Program},
\href{https://doi.org/10.1007/s41365-024-01591-2}
{Nucl. Sci. Tech. \textbf{35}, 214 (2024).}

\bibitem{Shou2024NST}
Q. Y. Shou, Y. G. Ma, S. Zhang, et al.,
\textit{Properties of QCD Matter: A Review of Selected Results
from ALICE Experiment},
\href{https://doi.org/10.1007/s41365-024-01583-2}
{Nucl. Sci. Tech. \textbf{35}, 219 (2024)}.

\bibitem{Zhao2024NSTEM}
J. Zhao, J. H. Chen, X. G. Huang, et al.,
\textit{Electromagnetic Fields in Ultra-Peripheral
Relativistic Heavy-Ion Collisions},
\href{https://doi.org/10.1007/s41365-024-01374-9}
{Nucl. Sci. Tech. \textbf{35}, 20 (2024)}.

\bibitem{Xu2025CPL}
J. Xu, Z. Qin, R. Zou, D. Si, S. Xiao, B. Tian,
Y. Wang, and Z. Xiao,
\textit{Imaging Freeze-Out Sources and Extracting Strong Interaction
Parameters in Relativistic Heavy-Ion Collisions},
\href{https://doi.org/10.1088/0256-307X/42/3/031401}
{Chin. Phys. Lett. \textbf{42}, 031401 (2025)}.

\bibitem{Ma2025NST}
Y. G. Ma,
\textit{Multi-proton emission at the limits of nuclear stability: challenges for extreme open quantum systems},
\href{https://doi.org/10.1007/s41365-025-01831-z }
{Nucl. Sci. Tech. \textbf{36}, 236 (2025)}.


\bibitem{Oertel2017RMP}
M. Oertel, M. Hempel, T. Kl\"ahn, and S. Typel,
\textit{Equations of State for Supernovae and Compact Stars},
\href{https://doi.org/10.1103/RevModPhys.89.015007}
{Rev. Mod. Phys. \textbf{89}, 015007 (2017).}

\bibitem{Vidana2018PRSA}
I. Vida\~na,
\textit{Hyperons: The Strange Ingredients of the Nuclear Equation
of State},
\href{https://doi.org/10.1098/rspa.2018.0145}
{Proc. R. Soc. A \textbf{474}, 20180145 (2018).}

\bibitem{Furusawa2023PPNP}
K. Furusawa and H. Nagakura,
\textit{Nuclei in Core-Collapse Supernovae Engine},
\href{https://doi.org/10.1016/j.ppnp.2022.104018}
{Prog. Part. Nucl. Phys. \textbf{129}, 104018 (2023).}

\bibitem{Mezzacappa2020LRCA}
A. Mezzacappa, E. Endeve, O.E.B. Messer, and S.W. Bruenn,
\textit{Physical, Numerical, and Computational Challenges of Modeling
Neutrino Transport in Core-Collapse Supernovae},
\href{https://doi.org/10.1007/s41115-020-00010-8}
{Living Rev. Comput. Astrophys. \textbf{6}, 4 (2020).}

\bibitem{Fischer2024PPNP}
T. Fischer, G. Guo, K. Langanke, G. Martínez-Pinedo, Y.-Z. Qian, and M.-R. Wu,
\textit{Neutrinos and nucleosynthesis of elements},
\href{https://doi.org/10.1016/j.ppnp.2024.104107}
{Prog. Part. Nucl. Phys. \textbf{137}, 104107 (2024).}

\bibitem{Li2021Universe}
B.A. Li, B.J. Cai, W.J. Xie, and N.B. Zhang,
\textit{Progress in Constraining Nuclear Symmetry Energy Using
Neutron Star Observables Since GW170817},
\href{https://doi.org/10.3390/universe7060182}
{Universe \textbf{7}, 182 (2021).}

\bibitem{Alford2008RMP}
M.G. Alford, A. Schmitt, K. Rajagopal, and T. Sch\"afer,
\textit{Color Superconductivity in Dense Quark Matter},
\href{https://doi.org/10.1103/RevModPhys.80.1455}
{Rev. Mod. Phys. \textbf{80}, 1455 (2008).}

\bibitem{Watts2016RMP}
A.L. Watts et al.,
\textit{Colloquium: Measuring the Neutron Star Equation of State
Using X-Ray Timing},
\href{https://doi.org/10.1103/RevModPhys.88.021001}
{Rev. Mod. Phys. \textbf{88}, 021001 (2016).}

\bibitem{Burgio2021PPNP}
G.F. Burgio, H.J. Schulze, I. Vida\~na, and J.B. Wei,
\textit{Neutron Stars and the Nuclear Equation of State},
\href{https://doi.org/10.1016/j.ppnp.2021.103879}
{Prog. Part. Nucl. Phys. \textbf{120}, 103879 (2021).}

\bibitem{Baym2018RPP}
G. Baym, T. Hatsuda, T. Kojo, P.D. Powell, Y. Song, and
T. Takatsuka,
\textit{From Hadrons to Quarks in Neutron Stars: A Review},
\href{https://doi.org/10.1088/1361-6633/aaae14}
{Rep. Prog. Phys. \textbf{81}, 056902 (2018).}

\bibitem{Baiotti2019PPNP}
L. Baiotti,
\textit{Gravitational Waves from Neutron Star Mergers and Their
Relation to the Nuclear Equation of State},
\href{https://doi.org/10.1016/j.ppnp.2019.103714}
{Prog. Part. Nucl. Phys. \textbf{109}, 103714 (2019).}

\bibitem{Orsaria2019JPG}
M.G. Orsaria, G. Malfatti, M. Mariani, I.F. Ranea-Sandoval,
F. Garc\'ia, W.M. Spinella, G.A. Contrera, G. Lugones, and
F. Weber,
\textit{Phase Transitions in Neutron Stars and Their Links to
Gravitational Waves},
\href{https://doi.org/10.1088/1361-6471/ab1d81}
{J. Phys. G \textbf{46}, 073002 (2019).}

\bibitem{Li2019EPJA}
B.A. Li, P.G. Krastev, D.H. Wen, and N.B. Zhang,
\textit{Towards Understanding Astrophysical Effects of Nuclear
Symmetry Energy},
\href{https://doi.org/10.1140/epja/i2019-12780-8}
{Eur. Phys. J. A \textbf{55}, 117 (2019).}

\bibitem{Dexheimer2021JPG}
V. Dexheimer, J. Noronha, J. Noronha-Hostler, N. Yunes, and
C. Ratti,
\textit{Future Physics Perspectives on the Equation of State from
Heavy-Ion Collisions to Neutron Stars},
\href{https://doi.org/10.1088/1361-6471/abe104}
{J. Phys. G \textbf{48}, 073001 (2021).}

\bibitem{Lattimer2021ARNPS}
J.M. Lattimer,
\textit{Neutron Stars and the Nuclear Matter Equation of State},
\href{https://doi.org/10.1146/annurev-nucl-102419-124827}
{Annu. Rev. Nucl. Part. Sci. \textbf{71}, 433 (2021).}


\bibitem{Chen2025SCPMA}
J.-H. Chen, Z.-T. Liang, Y.-G. Ma, X.-L. Sheng, and Q. Wang,
\textit{Vector Meson's Spin Alignments in High Energy Reactions},
\href{https://doi.org/10.1007/s11433-024-2495-1}
{Sci. China Phys. Mech. Astron. \textbf{68}, 211001 (2025)}.

\bibitem{Shen2025Research}
D. Shen, J. Chen, X.-G. Huang, Y.-G. Ma, A. Tang, and G. Wang,
\textit{A Review of Intense Electromagnetic Fields in Heavy-Ion
Collisions: Theoretical Predictions and Experimental Results},
\href{https://doi.org/10.34133/research.0726}
{Research \textbf{8}, 0726 (2025)}.

\bibitem{He2023NST}
W.-B. He, Y.-G. Ma, L.-G. Pang, H.-C. Song, and K. Zhou,
\textit{High-Energy Nuclear Physics Meets Machine Learning},
\href{https://doi.org/10.1007/s41365-023-01233-z}
{Nucl. Sci. Tech. \textbf{34}, 88 (2023).}

\bibitem{Ma2023CPL}
Y.-G. Ma, L.-G. Pang, R. Wang, and K. Zhou,
\textit{Phase Transition Study Meets Machine Learning},
\href{https://doi.org/10.1088/0256-307X/40/12/122101}
{Chin. Phys. Lett. \textbf{40}, 122101 (2023).}


\bibitem{He2026SCPMA}
W. B. He, Q. F. Li, Y. G. Ma, Z. M. Niu, J. C.  Pei, Y. X.  Zhang, Phase Transition Study Meets Machine Learning,
\href{}
{Science China Phys. Mech. \& Astro., (2026).}

\bibitem{Wang2025NST}
Y. J. Wang, S. Xiao, M. T.  Wan et al., 
\textit{The enhancement of neutron-rich particle emission from out-of-fission-plane in Fermi energy heavy ion reactions},
\href{https://doi.org/10.1007/s41365-025-01742-z}
{Nucl. Sci. Tech.  \textbf{36}, 155 (2025).}

\bibitem{Fang2026NST}
L. M. Fang, Y. G. Ma, S. Zhang, 
\textit{Correlations among proton collective flows in Au plus Au collisions at E=1.23A GeV within the isospin-dependent quantum molecular dynamics model},
\href{https://doi.org/10.1007/s41365-026-02046-6}
{Nucl. Sci. Tech.  \textbf{37}, 193 (2026).}



\bibitem{Brown2000PRL}
B.A. Brown,
\textit{Neutron Radii in Nuclei and the Neutron Equation of State},
\href{https://journals.aps.org/prl/abstract/10.1103/PhysRevLett.85.5296}
{Phys. Rev. Lett. \textbf{85}, 5296 (2000).}

\bibitem{Typel2001PRC}
S. Typel and B.A. Brown,
\textit{Neutron Radii and the Neutron Equation of State in Relativistic Models},
\href{https://journals.aps.org/prc/abstract/10.1103/PhysRevC.64.027302}
{Phys. Rev. C \textbf{64}, 027302 (2001).}

\bibitem{Centelles2009PRL}
M. Centelles, X. Roca-Maza, X. Vi\~nas, and M. Warda,
\textit{Nuclear Symmetry Energy Probed by Neutron Skin Thickness of Nuclei},
\href{https://journals.aps.org/prl/abstract/10.1103/PhysRevLett.102.122502}
{Phys. Rev. Lett. \textbf{102}, 122502 (2009).}


\bibitem{Horowitz2001PRL}
C.J. Horowitz and J. Piekarewicz,
\textit{Neutron Star Structure and the Neutron Radius of $^{208}\rm{Pb}$},
\href{https://journals.aps.org/prl/abstract/10.1103/PhysRevLett.86.5647}
{Phys. Rev. Lett. \textbf{86}, 5647 (2001).}


\bibitem{Horowitz2014JPG}
C.J. Horowitz, E.F. Brown, Y. Kim, W.G. Lynch, R. Michaels,
A. Ono, J. Piekarewicz, M.B. Tsang, and H.H. Wolter,
\textit{A Way Forward in the Study of the Symmetry Energy:
Experiment, Theory, and Observation},
\href{https://doi.org/10.1088/0954-3899/41/9/093001}
{J. Phys. G \textbf{41}, 093001 (2014).}

\bibitem{Thiel2019JPG}
M. Thiel, C. Sfienti, J. Piekarewicz, C.J. Horowitz, and
M. Vanderhaeghen,
\textit{Neutron Skins of Atomic Nuclei: Per Aspera Ad Astra},
\href{https://doi.org/10.1088/1361-6471/ab2c6d}
{J. Phys. G \textbf{46}, 093003 (2019).}

\bibitem{BohrMottelson1998Book}
A. Bohr and B.R. Mottelson,
\textit{Nuclear Structure, Vol. I: Single-Particle Motion},
\href{https://doi.org/10.1142/3530}
{World Scientific, Singapore (1998).}

\bibitem{RingSchuck1980Book}
P. Ring and P. Schuck,
\textit{The Nuclear Many-Body Problem},
\href{https://books.google.com/books?id=PTynSM-nMA8C}
{Springer (1980).}

\bibitem{DeVries1987ADNDT}
H. De Vries, C.W. De Jager, and C. De Vries,
\textit{Nuclear Charge-Density-Distribution Parameters from Elastic
Electron Scattering},
\href{https://doi.org/10.1016/0092-640X(87)90013-1}
{At. Data Nucl. Data Tables \textbf{36}, 495 (1987).}

\bibitem{Warda2009PRC}
M. Warda, X. Vi\~nas, X. Roca-Maza, and M. Centelles,
\textit{Neutron Skin Thickness in the Droplet Model with
Surface Width Dependence: Indications of Softness of the
Nuclear Symmetry Energy},
\href{https://doi.org/10.1103/PhysRevC.80.024316}
{Phys. Rev. C \textbf{80}, 024316 (2009).}



\bibitem{Sotani2022PTEP}
H. Sotani, N. Nishimura, and T. Naito,
\textit{New Constraints on the Neutron-Star Mass and Radius Relation
from Terrestrial Nuclear Experiments},
\href{https://doi.org/10.1093/ptep/ptac055}
{Prog. Theor. Exp. Phys. \textbf{2022}, 041D01 (2022)}.




\bibitem{Zhang2023PRC}
Z. Zhang and L.W. Chen,
\textit{Bayesian Inference of the Symmetry Energy and the Neutron Skin in
$^{48}\rm{Ca}$ and $^{208}\rm{Pb}$ from CREX and PREX-2},
\href{https://journals.aps.org/prc/abstract/10.1103/PhysRevC.108.024317}
{Phys. Rev. C \textbf{108}, 024317 (2023).}




\bibitem{RocaMaza2011PRL}
X. Roca-Maza, M. Centelles, X. Vi\~nas, and M. Warda,
\textit{Neutron Skin of $^{208}\rm{Pb}$, Nuclear Symmetry Energy,
and the Parity Radius Experiment},
\href{https://doi.org/10.1103/PhysRevLett.106.252501}
{Phys. Rev. Lett. \textbf{106}, 252501 (2011)}.


\bibitem{Bertsch1988PhysRep}
G.~F. Bertsch and S. Das Gupta,
\textit{A guide to microscopic models for intermediate energy
heavy-ion collisions},
\href{https://doi.org/10.1016/0370-1573(88)90170-6}
{Phys. Rep. \textbf{160}, 189 (1988)}.

\bibitem{Aichelin1991PhysRep}
J. Aichelin,
\textit{``Quantum'' molecular dynamics---A dynamical microscopic
$n$-body approach to investigate fragment formation and the nuclear
equation of state in heavy-ion collisions},
\href{https://doi.org/10.1016/0370-1573(91)90094-3}
{Phys. Rep. \textbf{202}, 233 (1991)}.

\bibitem{Botermans1990PhysRep}
W. Botermans and R. Malfliet,
\textit{Quantum transport theory of nuclear matter},
\href{https://doi.org/10.1016/0370-1573(90)90174-Z}
{Phys. Rep. \textbf{198}, 115 (1990)}.

\bibitem{Danielewicz1984AnnPhys}
P. Danielewicz,
\textit{Quantum theory of nonequilibrium processes, I},
\href{https://doi.org/10.1016/0003-4916(84)90092-7}
{Ann. Phys. \textbf{152}, 239 (1984)}.

\bibitem{Bonasera1994PhysRep}
M. Bonasera, F. Gulminelli, and J. Molitoris,
\textit{The Boltzmann equation at the borderline: A decade of
Monte Carlo simulations of a quantum kinetic equation},
\href{https://doi.org/10.1016/0370-1573(94)90108-2}
{Phys. Rep. \textbf{243}, 1 (1994)}.

\bibitem{Ko1996JPhysG}
C.~M. Ko and G.~Q. Li,
\textit{Medium effects in high energy heavy-ion collisions},
\href{https://doi.org/10.1088/0954-3899/22/12/002}
{J. Phys. G \textbf{22}, 1673 (1996)}.


\bibitem{Bass1998PPNP}
S.~A. Bass \textit{et al.},
\textit{Microscopic models for ultrarelativistic heavy-ion
collisions},
\href{https://doi.org/10.1016/S0146-6410(98)00058-1}
{Prog. Part. Nucl. Phys. \textbf{41}, 255 (1998)}.

\bibitem{Cassing1999PhysRep}
W. Cassing and E.~L. Bratkovskaya,
\textit{Hadronic and electromagnetic probes of hot and dense
nuclear matter},
\href{https://doi.org/10.1016/S0370-1573(98)00028-3}
{Phys. Rep. \textbf{308}, 65 (1999)}.

\bibitem{Li1998IJMPE}
B.-A. Li, C.~M. Ko, and W. Bauer,
\textit{Isospin physics in heavy-ion collisions at intermediate
energies},
\href{https://doi.org/10.1142/S0218301398000087}
{Int. J. Mod. Phys. E \textbf{7}, 147 (1998)}.


\bibitem{Buss2012PhysRep}
O. Buss, T. Gaitanos, K. Gallmeister, H. van Hees,
M. Kaskulov, O. Lalakulich, A.~B. Larionov, T. Leitner,
J. Weil, and U. Mosel,
\textit{Transport-theoretical description of nuclear reactions},
\href{https://doi.org/10.1016/j.physrep.2011.12.001}
{Phys. Rep. \textbf{512}, 1 (2012)}.

\bibitem{Xiao2009PRL}
Z. Xiao, B.-A. Li, L.-W. Chen, G.-C. Yong, and M. Zhang,
\textit{Circumstantial Evidence for a Soft Nuclear Symmetry Energy
at Suprasaturation Densities},
\href{https://journals.aps.org/prl/abstract/10.1103/PhysRevLett.102.062502}
{Phys. Rev. Lett. \textbf{102}, 062502 (2009).}


\bibitem{Tsang2009PRL}
M.B. Tsang et al.,
\textit{Constraints on the Density Dependence of the Symmetry Energy},
\href{https://journals.aps.org/prl/abstract/10.1103/PhysRevLett.102.122701}
{Phys. Rev. Lett. \textbf{102}, 122701 (2009).}


\bibitem{Li2016NST}
X.-F. Li, D.-Q. Fang, and Y.-G. Ma,
\textit{Determination of the Neutron Skin Thickness from Interaction
Cross Section and Charge-Changing Cross Section for B, C, N, O, and
F Isotopes},
\href{https://link.springer.com/article/10.1007/s41365-016-0064-z}
{Nucl. Sci. Tech. \textbf{27}, 71 (2016).}


\bibitem{Ma2013PRC}
C.W. Ma, H.L. Wei, and Y.G. Ma,
\textit{Neutron-Skin Effects in Isobaric Yield Ratios for Mirror
Nuclei in a Statistical Abrasion-Ablation Model},
\href{https://journals.aps.org/prc/abstract/10.1103/PhysRevC.88.044612}
{Phys. Rev. C \textbf{88}, 044612 (2013).}

\bibitem{Fang2010PRC}
D.Q. Fang, Y.G. Ma, X.Z. Cai, W.D. Tian, and H.W. Wang,
\textit{Neutron Removal Cross Section as a Measure of Neutron Skin},
\href{https://journals.aps.org/prc/abstract/10.1103/PhysRevC.81.047603}
{Phys. Rev. C \textbf{81}, 047603 (2010).}

\bibitem{Schwinger1961JMP}
J. Schwinger,
\textit{Brownian Motion of a Quantum Oscillator},
\href{https://doi.org/10.1063/1.1703727}
{J. Math. Phys. \textbf{2}, 407 (1961).}

\bibitem{Keldysh1965JETP}
L.V. Keldysh,
\textit{Diagram Technique for Nonequilibrium Processes},
\href{https://www.jetp.ras.ru/cgi-bin/e/index/e/20/4/p1018?a=list}
{Sov. Phys. JETP \textbf{20}, 1018 (1965).}

\bibitem{Chou1985PhysRep}
K.-C. Chou, Z.-B. Su, B.-L. Hao, and L. Yu,
\textit{Equilibrium and Nonequilibrium Formalisms Made Unified},
\href{https://doi.org/10.1016/0370-1573(85)90136-X}
{Phys. Rep. \textbf{118}, 1 (1985).}

\bibitem{KadanoffBaym1962Book}
L.P. Kadanoff and G. Baym,
\textit{Quantum Statistical Mechanics: Green's Function Methods in
Equilibrium and Nonequilibrium Problems}
\href{https://www.routledge.com/Quantum-Statistical-Mechanics-Greens-Function-Methods-in-Equilibrium/Kadanoff-Baym/p/book/9780201410464}
{(W.A. Benjamin, New York, 1962; CRC Press, Boca Raton, 2018)}.

\bibitem{Ivanov1999NPA}
Yu.B. Ivanov, J. Knoll, and D.N. Voskresensky,
\textit{Self-Consistent Approximations to Non-Equilibrium Many-Body
Theory},
\href{https://arxiv.org/abs/hep-ph/9807351}
{Nucl. Phys. A \textbf{657}, 413 (1999).}

\bibitem{Rios2011AnnPhys}
A. Rios, B. Barker, M. Buchler, and P. Danielewicz,
\textit{Towards a Nonequilibrium Green's Function Description of
Nuclear Reactions: One-Dimensional Mean-Field Dynamics},
\href{https://www.sciencedirect.com/science/article/pii/S000349161000223X}
{Ann. Phys. \textbf{326}, 1274 (2011).}


\bibitem{Yang2023Universe}
J. Yang, X. Chen, Y. Cui, Z. Li, and Y. Zhang,
\textit{Probing the Neutron Skin of Unstable Nuclei with Heavy-Ion
Collisions},
\href{https://www.mdpi.com/2218-1997/9/5/206}
{Universe \textbf{9}, 206 (2023).}

\bibitem{Colonna2020PPNP}
M. Colonna,
\textit{Collision Dynamics at Medium and Relativistic Energies},
\href{https://doi.org/10.1016/j.ppnp.2020.103775}
{Prog. Part. Nucl. Phys. \textbf{113}, 103775 (2020).}

\bibitem{Wolter2022PPNP}
H. Wolter, M. Colonna, D. Cozma, P. Danielewicz, C.M. Ko,
A. Ono, M.B. Tsang, R. Wang, Y.-X. Zhang, J. Xu, et al.,
\textit{Transport Model Comparison Studies of Intermediate-Energy
Heavy-Ion Collisions},
\href{https://doi.org/10.1016/j.ppnp.2022.103962}
{Prog. Part. Nucl. Phys. \textbf{125}, 103962 (2022).}

\bibitem{Giacalone2023PRL}
G. Giacalone, G. Nijs, and W. van der Schee,
\textit{Determination of the Neutron Skin of $^{208}$Pb from
Ultrarelativistic Nuclear Collisions},
\href{https://doi.org/10.1103/PhysRevLett.131.202302}
{Phys. Rev. Lett. \textbf{131}, 202302 (2023).}

\bibitem{CREX} D. Adhikari et al., \textit{Precision Determination of the Neutral Weak Form Factor of\; $^{48}$Ca}, \href{https://journals.aps.org/prl/abstract/10.1103/PhysRevLett.129.042501}{Phys. Rev. Lett. \textbf{129}, 042501 (2022).}


\bibitem{PREX2021PRL}
D. Adhikari et al. (PREX Collaboration),
\textit{Accurate Determination of the Neutron Skin Thickness of
$^{208}$Pb through Parity-Violation in Electron Scattering},
\href{https://doi.org/10.1103/PhysRevLett.126.172502}
{Phys. Rev. Lett. \textbf{126}, 172502 (2021).}



\bibitem{Chatziioannou2025RMP}
K. Chatziioannou, H.T. Cromartie, S. Gandolfi, I. Tews, D. Radice,
A.W. Steiner, and A.L. Watts,
\textit{Neutron Stars and the Dense Matter Equation of State},
\href{https://doi.org/10.1103/ymsq-cfcw}
{Rev. Mod. Phys. \textbf{97}, 045007 (2025).}

\bibitem{Li2025eXTP}
A. Li, A.L. Watts, G. Zhang, S. Guillot, Y. Xu, A. Santangelo,
S. Zane, H. Feng, S.-N. Zhang, M. Ge, et al.,
\textit{Dense Matter in Neutron Stars with eXTP},
\href{https://doi.org/10.1007/s11433-025-2761-4}
{Sci. China Phys. Mech. Astron. \textbf{68}, 119503 (2025)}.


\bibitem{Essick2021PRC}
R. Essick, P. Landry, A. Schwenk, and I. Tews,
\textit{Detailed examination of astrophysical constraints on the symmetry
energy and the neutron skin of $^{208}$Pb with minimal modeling assumptions},
\href{https://doi.org/10.1103/PhysRevC.104.065804}
{Phys. Rev. C \textbf{104}, 065804 (2021).}

\bibitem{Lattimer2023}
J.M. Lattimer,
\textit{Constraints on Nuclear Symmetry Energy Parameters},
\href{https://doi.org/10.3390/particles6010003}
{Particles \textbf{6}, 30 (2023).}

\bibitem{Mammei2024ARNPS}
J.M. Mammei, C.J. Horowitz, J. Piekarewicz, B.T. Reed, and
C. Sfienti,
\textit{Neutron Skins: Weak Elastic Scattering and Neutron Stars},
\href{https://doi.org/10.1146/annurev-nucl-102122-024207}
{Annu. Rev. Nucl. Part. Sci. \textbf{74}, 321 (2024).}

\bibitem{Sammarruca2024Symmetry}
F. Sammarruca,
\textit{The Neutron Skin of $^{48}\rm{Ca}$ and $^{208}\rm{Pb}$:
A Critical Analysis},
\href{https://doi.org/10.3390/sym16010034}
{Symmetry \textbf{16}, 34 (2024).}

\bibitem{LimHolt22}
Y. Lim and J. W. Holt,
\textit{Neutron Star Radii, Deformabilities, and Moments of Inertia
from Experimental and Ab Initio Theory Constraints of the
$^{208}$Pb Neutron Skin Thickness},
\href{https://doi.org/10.3390/galaxies10050099}
{Galaxies \textbf{10}, 99 (2022)}.

\bibitem{Atkinson2024FrontPhys}
M.C. Atkinson and W.H. Dickhoff,
\textit{Neutron Skins: A Perspective from Dispersive Optical Models},
\href{https://doi.org/10.3389/fphy.2024.1487314}
{Front. Phys. \textbf{12}, 1487314 (2024).}

\bibitem{Tanaka2024FrontPhys}
M. Tanaka, W. Horiuchi, and M. Fukuda,
\textit{Unveiling Radii and Neutron Skins of Unstable Atomic Nuclei
via Nuclear Collisions},
\href{https://doi.org/10.3389/fphy.2024.1488428}
{Front. Phys. \textbf{12}, 1488428 (2024).}

\bibitem{Ding2024NST}
M.-Q. Ding, D.-Q. Fang, and Y.-G. Ma,
\textit{Neutron Skin and Its Effects in Heavy-Ion Collisions},
\href{https://doi.org/10.1007/s41365-024-01584-1}
{Nucl. Sci. Tech. \textbf{35}, 211 (2024).}



\bibitem{Miyagi2025FrontPhys}
T. Miyagi,
\textit{Nuclear Radii from First Principles},
\href{https://doi.org/10.3389/fphy.2025.1581854}
{Front. Phys. \textbf{13}, 1581854 (2025).}

\bibitem{NeumannCosel2025FrontPhys}
P. von Neumann-Cosel and A. Tamii,
\textit{Electric Dipole Polarizability Constraints on Neutron Skin
and Symmetry Energy},
\href{https://doi.org/10.3389/fphy.2025.1629987}
{Front. Phys. \textbf{13}, 1629987 (2025).}

\bibitem{Jia2025RPP}
J. Jia,
\textit{Imaging Nuclei by Smashing Them at High Energies:
How Are Their Shapes Revealed after Destruction?},
\href{https://doi.org/10.1088/1361-6633/ae0654}
{Rep. Prog. Phys. \textbf{88}, 092301 (2025).}

\bibitem{Horowitz2001PRC-a}
C.J. Horowitz, S.J. Pollock, P.A. Souder, and R. Michaels,
\textit{Parity Violating Measurements of Neutron Densities},
\href{https://journals.aps.org/prc/abstract/10.1103/PhysRevC.63.025501}
{Phys. Rev. C \textbf{63}, 025501 (2001).}

\bibitem{Clark2003PRC}
B.~C. Clark, L.~J. Kerr, and S. Hama,
\textit{Neutron densities from a global analysis of medium-energy
proton--nucleus elastic scattering},
\href{https://doi.org/10.1103/PhysRevC.67.054605}
{Phys. Rev. C \textbf{67}, 054605 (2003)}.

\bibitem{Terashima2008PRC}
S. Terashima, H. Sakaguchi, H. Takeda, T. Ishikawa, M. Itoh,
T. Kawabata, T. Murakami, M. Uchida, Y. Yasuda, M. Yosoi,
J. Zenihiro, H.~P. Yoshida, T. Noro, T. Ishida, S. Asaji,
and T. Yonemura,
\textit{Proton elastic scattering from tin isotopes at 295 MeV
and systematic change of neutron density distributions},
\href{https://doi.org/10.1103/PhysRevC.77.024317}
{Phys. Rev. C \textbf{77}, 024317 (2008)}.

\bibitem{Zenihiro2010PRC}
J. Zenihiro, H. Sakaguchi, T. Murakami, M. Yosoi, Y. Yasuda,
S. Terashima, Y. Iwao, H. Takeda, M. Itoh, H.~P. Yoshida,
and M. Uchida,
\textit{Neutron density distributions of
$^{204,206,208}$Pb deduced via proton elastic scattering at
$E_p=295$ MeV},
\href{https://doi.org/10.1103/PhysRevC.82.044611}
{Phys. Rev. C \textbf{82}, 044611 (2010)}.

\bibitem{KanadaEnyo2021arXiv}
Y. Kanada-En'yo,
\textit{Isotopic analysis of 295 MeV proton scattering off
$^{204,206,208}$Pb for improvement of neutron densities and radii},
\href{https://doi.org/10.48550/arXiv.2106.00151}
{arXiv:2106.00151 [nucl-th] (2021)}.

\bibitem{Tarbert2014PRL}
C.~M. Tarbert et al.,
\textit{Neutron Skin of $^{208}$Pb from Coherent Pion Photoproduction},
\href{https://doi.org/10.1103/PhysRevLett.112.242502}
{Phys. Rev. Lett. \textbf{112}, 242502 (2014)}.

\bibitem{Miller2019PRC}
G.~A. Miller,
\textit{Coherent-Nuclear Pion Photoproduction and Neutron Radii},
\href{https://doi.org/10.1103/PhysRevC.100.044608}
{Phys. Rev. C \textbf{100}, 044608 (2019)}.

\bibitem{Colomer2022PRC}
F. Colomer, P. Capel, M. Ferretti, J. Piekarewicz, C. Sfienti,
M. Thiel, V. Tsaran, and M. Vanderhaeghen,
\textit{Theoretical Analysis of the Extraction of Neutron Skin
Thickness from Coherent $\pi^0$ Photoproduction off Nuclei},
\href{https://doi.org/10.1103/PhysRevC.106.044318}
{Phys. Rev. C \textbf{106}, 044318 (2022)}.

\bibitem{Ericson1988Book}
T.~E.~O. Ericson and W. Weise,
\textit{Pions and Nuclei},
\href{https://books.google.com/books?vid=ISBN9780198520085}
{International Series of Monographs on Physics, Vol.~74
(Clarendon Press, Oxford, 1988)}.

\bibitem{XuWang2024SCPMA}
H. Xu and F. Wang,
\textit{Determine the neutron skin thickness and nuclear symmetry
energy at relativistic heavy ion collisions},
\href{https://doi.org/10.1360/SSPMA-2024-0023}
{Sci. Sin. Phys. Mech. Astron. \textbf{54} (2024)}.


\bibitem{Xu2016PRC}
J. Xu et al.,
\textit{Understanding Transport Simulations of Heavy-Ion Collisions at
100$A$ and 400$A$ MeV: Comparison of Heavy-Ion Transport Codes under
Controlled Conditions},
\href{https://journals.aps.org/prc/abstract/10.1103/PhysRevC.93.044609}
{Phys. Rev. C \textbf{93}, 044609 (2016).}

\bibitem{Jia2023PRL}
J. Jia, G. Giacalone, and C. Zhang,
\textit{Separating the Impact of Nuclear Skin and Nuclear Deformation
in High-Energy Isobar Collisions},
\href{https://journals.aps.org/prl/abstract/10.1103/PhysRevLett.131.022301}
{Phys. Rev. Lett. \textbf{131}, 022301 (2023).}

\bibitem{Li2020PRL}
H. Li, H.J. Xu, Y. Zhou, X.B. Wang, J. Zhao, L.W. Chen, and F. Wang,
\textit{Probing the Neutron Skin with Ultrarelativistic Isobaric Collisions},
\href{https://journals.aps.org/prl/abstract/10.1103/PhysRevLett.125.222301}
{Phys. Rev. Lett. \textbf{125}, 222301 (2020).}

\bibitem{Lattimer2014NPA}
J.M. Lattimer,
\textit{Symmetry Energy in Nuclei and Neutron Stars},
\href{https://doi.org/10.1016/j.nuclphysa.2014.04.008}
{Nucl. Phys. A \textbf{928}, 276 (2014).}




\bibitem{Lattimer2001ApJ}
J.M. Lattimer and M. Prakash,
\textit{Neutron Star Structure and the Equation of State},
\href{https://doi.org/10.1086/319702}
{Astrophys. J. \textbf{550}, 426 (2001).}


\bibitem{Huth2022Nature}
S. Huth et al.,
\textit{Constraining neutron-star matter with microscopic and macroscopic collisions},
\href{https://doi.org/10.1038/s41586-022-04750-w}
{Nature \textbf{606}, 276 (2022).}


\bibitem{Tsang2019PLB}
M.B. Tsang, W.G. Lynch, P. Danielewicz, and C.Y. Tsang,
\textit{Symmetry energy constraints from GW170817 and laboratory experiments},
\href{https://doi.org/10.1016/j.physletb.2019.06.059}
{Phys. Lett. B \textbf{795}, 533 (2019).}


\bibitem{Riley19}
T.E. Riley et al.,
\textit{A NICER View of PSR J0030+0451: Millisecond Pulsar Parameter Estimation},
\href{https://doi.org/10.3847/2041-8213/ab481c}
{Astrophys. J. Lett. \textbf{887}, L21 (2019).}

\bibitem{Miller19}
M.C. Miller et al.,
\textit{PSR J0030+0451 Mass and Radius from NICER Data and Implications for the Properties of Neutron Star Matter},
\href{https://doi.org/10.3847/2041-8213/ab50c5}
{Astrophys. J. Lett. \textbf{887}, L24 (2019).}

\bibitem{Fonseca21}
E. Fonseca et al.,
\textit{Refined Mass and Geometric Measurements of the High-mass PSR J0740+6620},
\href{https://doi.org/10.3847/2041-8213/ac03b8}
{Astrophys. J. Lett. \textbf{915}, L12 (2021).}


\bibitem{Riley21}
T.E. Riley et al.,
\textit{A NICER View of the Massive Pulsar PSR J0740+6620 Informed by Radio Timing and XMM-Newton Spectroscopy},
\href{https://doi.org/10.3847/2041-8213/ac0a81}
{Astrophys. J. Lett. \textbf{918}, L27 (2021).}

\bibitem{Miller21}
M.C. Miller et al.,
\textit{The Radius of PSR J0740+6620 from NICER and XMM-Newton Data},
\href{https://doi.org/10.3847/2041-8213/ac089b}
{Astrophys. J. Lett. \textbf{918}, L28 (2021).}

\bibitem{Salmi22}
T. Salmi et al.,
\textit{The Radius of PSR J0740+6620 from NICER with NICER Background Estimates},
\href{https://doi.org/10.3847/1538-4357/ac983d}
{Astrophys. J. \textbf{941}, 150 (2022).}

\bibitem{Salmi24}
T. Salmi et al.,
\textit{The Radius of the High Mass Pulsar PSR J0740+6620 With 3.6 Years of NICER Data},
\href{https://arxiv.org/abs/2406.14466}
{Astrophys. J. (2024).}

\bibitem{Ditt24}
A.J. Dittmann et al.,
\textit{A More Precise Measurement of the Radius of PSR J0740+6620 Using Updated NICER Data},
\href{https://arxiv.org/abs/2406.14467}
{Astrophys. J. (2024).}

\bibitem{Choud24}
D. Choudhury et al.,
\textit{A NICER View of the Nearest and Brightest Millisecond Pulsar: PSR J0437-4715},
\href{https://doi.org/10.3847/2041-8213/ad5a6f}
{Astrophys. J. Lett. \textbf{971}, L20 (2024).}

\bibitem{Mauviard26}
L. Mauviard et al.,
\textit{A NICER view of PSR J1614-2230: a massive and compact millisecond pulsar},
\href{https://arxiv.org/abs/2609.00172}
{arXiv:2609.00172 [astro-ph.HE] (2026).}


\bibitem{Mauviard25}
L. Mauviard et al.,
\textit{A NICER View of the 1.4 $M_{\odot}$ Edge-on Pulsar PSR J0614-3329},
\href{https://doi.org/10.3847/1538-4357/ae145d}
{Astrophys. J. \textbf{995}, 60 (2025).}


\bibitem{Miller26}
M.C. Miller et al.,
\textit{The Radius of the Neutron Star PSR J0614-3329 from NICER Data},
\href{https://arxiv.org/abs/2609.00965}
{arXiv:2609.00965 [astro-ph.HE] (2026).}


\bibitem{Abbott2017PRL}
B.P. Abbott et al. (LIGO Scientific Collaboration and Virgo Collaboration),
\textit{GW170817: Observation of Gravitational Waves from a Binary Neutron Star Inspiral},
\href{https://journals.aps.org/prl/abstract/10.1103/PhysRevLett.119.161101}
{Phys. Rev. Lett. \textbf{119}, 161101 (2017).}

\bibitem{Abbott2018PRL}
B.P. Abbott et al. (LIGO Scientific Collaboration and Virgo Collaboration),
\textit{GW170817: Measurements of Neutron Star Radii and Equation of State},
\href{https://journals.aps.org/prl/abstract/10.1103/PhysRevLett.121.161101}
{Phys. Rev. Lett. \textbf{121}, 161101 (2018).}


\bibitem{Tang2021PRD}
S.-P. Tang, J.-L. Jiang, M.-Z. Han, Y.-Z. Fan, and D.-M. Wei,
\textit{Constraints on the Phase Transition and Nuclear Symmetry
Parameters from PSR J0740+6620 and Multimessenger Data of Other
Neutron Stars},
\href{https://doi.org/10.1103/PhysRevD.104.063032}
{Phys. Rev. D \textbf{104}, 063032 (2021)}.

\bibitem{Kumar2024LRR}
R. Kumar, V. Dexheimer, J. Jahan, et al.,
\textit{Theoretical and experimental constraints for the equation of
state of dense and hot matter},
\href{https://doi.org/10.1007/s41114-024-00049-6}
{Living Rev. Relativ. \textbf{27}, 3 (2024)}.

\bibitem{Burgio2024FASS}
G. F. Burgio, H. C. Das, and I. Vida\~na,
\textit{The nuclear symmetry energy and the neutron skin thickness
in nuclei},
\href{https://doi.org/10.3389/fspas.2024.1505560}
{Front. Astron. Space Sci. \textbf{11}, 1505560 (2024)}.

\bibitem{Miyatsu2025FrontPhys}
T. Miyatsu, M.-K. Cheoun, K. Kim, and K. Saito,
\textit{Novel features of asymmetric nuclear matter from terrestrial
experiments and astrophysical observations of neutron stars},
\href{https://doi.org/10.3389/fphy.2024.1531475}
{Front. Phys. \textbf{12}, 1531475 (2025)}.


\bibitem{Sedrakian2023PPNP}
A. Sedrakian, J. J. Li, and F. Weber,
\textit{Heavy baryons in compact stars},
\href{https://doi.org/10.1016/j.ppnp.2023.104041}
{Prog. Part. Nucl. Phys. \textbf{131}, 104041 (2023)}.

\bibitem{Sammarruca2025FASS}
F. Sammarruca and T. Ajagbonna,
\textit{General features of the stellar matter equation of state from
microscopic theory, new maximum-mass constraints, and causality},
\href{https://doi.org/10.3389/fspas.2025.1554123}
{Front. Astron. Space Sci. \textbf{12}, 1554123 (2025)}.

\bibitem{Li2026PRL}
B.-A. Li,
\textit{Trace Anomaly of Cold Dense Matter Constrained by Collective Flow},
\href{https://doi.org/10.1103/b2t4-km2r}
{Phys. Rev. Lett. \textbf{136}, 242301 (2026)}.

\bibitem{Xie2026SCPMA}
W.-J. Xie,
\textit{Study of the nuclear matter equation of state and neutron star
properties based on latest NICER radius observations and chiral
effective field theory results},
\href{https://doi.org/10.1360/SSPMA-2025-0387}
{Sci. Sin. Phys. Mech. Astron. (2026)}.

\bibitem{Tong2025FASS}
H. Tong, S. Wang, and J. Meng,
\textit{Relativistic ab initio calculations for static and rotating
neutron stars},
\href{https://doi.org/10.3389/fspas.2025.1666331}
{Front. Astron. Space Sci. \textbf{12}, 1666331 (2025)}.

\bibitem{Yunes2022NRP}
N. Yunes, M. C. Miller, and K. Yagi,
\textit{Gravitational-wave and X-ray probes of the neutron star
equation of state},
\href{https://doi.org/10.1038/s42254-022-00420-y}
{Nat. Rev. Phys. \textbf{4}, 237--246 (2022)}.



\bibitem{Cui2025NST}
Y. Cui, Y. Tian, C.-J. Xia, Y.-X. Zhang, and Z.-X. Li,
\textit{Constraint on the symmetry energy at high densities from
neutron star observations using relativistic mean-field models},
\href{https://doi.org/10.1007/s41365-025-01734-z}
{Nucl. Sci. Tech. \textbf{36}, 141 (2025)}.

\bibitem{Xu2026SCPMA}
J. Xu, J. Zhou, M.-Y. Qiu, and Z. Zhang,
\textit{Constraints on the nuclear matter equation of state from
neutron-star observables and nuclear structure data},
\href{https://doi.org/10.1360/SSPMA-2025-0160}
{Sci. Sin. Phys. Mech. Astron. \textbf{56}, 272006 (2026)}.



\bibitem{CaiLi2025EPJA}
B.-J. Cai and B.-A. Li,
\textit{Novel Scalings of Neutron Star Properties from Analyzing
Dimensionless Tolman--Oppenheimer--Volkoff Equations},
\href{https://doi.org/10.1140/epja/s10050-025-01507-7}
{Eur. Phys. J. A \textbf{61}, 55 (2025)}.

\bibitem{Zhang2013PLB}
Z. Zhang and L.W. Chen,
\textit{Constraining the Symmetry Energy at Subsaturation Densities
Using Isotope Binding Energy Difference and Neutron Skin Thickness},
\href{https://www.sciencedirect.com/science/article/pii/S037026931300628X}
{Phys. Lett. B \textbf{726}, 234 (2013).}


\bibitem{Mondal2016PRC}
C. Mondal, B.K. Agrawal, M. Centelles, G. Col\`o, X. Roca-Maza,
N. Paar, X. Vi\~nas, S.K. Singh, and S.K. Patra,
\textit{Model Dependence of the Neutron-Skin Thickness on the Symmetry Energy},
\href{https://journals.aps.org/prc/abstract/10.1103/PhysRevC.93.064303}
{Phys. Rev. C \textbf{93}, 064303 (2016).}

\bibitem{Mondal2022PRC}
C. Mondal,
\textit{Density Dependence of Symmetry Energy and Neutron Skin
Thickness Revisited Using Relativistic Mean Field Models with
Nonlinear Couplings},
\href{https://journals.aps.org/prc/abstract/10.1103/PhysRevC.105.034305}
{Phys. Rev. C \textbf{105}, 034305 (2022).}



\bibitem{Ma_2023} Y. G. Ma, S. Zhang, \textit{Influence of Nuclear Structure in Relativistic Heavy-Ion Collisions},  \href{https://doi.org/10.1007/978-981-19-6345-2_5}{A Chapter in Handbook of Nuclear Physics, Springer, Singapore.  (2023)}.

\bibitem{Sun2010PLB}X.Y. Sun, D.Q. Fang, Y.G. Ma, X.Z. Cai, J.G. Chen, W. Guo, W.D. Tian, and H.W. Wang, \textit{Neutron/proton Ratio of Nucleon Emissions as a Probe of Neutron Skin}, \href{https://www.sciencedirect.com/science/article/pii/S037026930901377X}{Phys. Lett. \textbf{B682}, 396 (2010).}

\bibitem{Dai2014PRC}Z.T. Dai et al., \textit{Triton/\,$^3\rm{He}$ Ratio as an Observable for Neutron-skin Thickness},\href{https://journals.aps.org/prc/abstract/10.1103/PhysRevC.89.014613}{Phys. Rev. C \textbf{89}, 014613 (2014).}

\bibitem{Ding2024PRC}M.Q. Ding, D.Q. Fang, and Y.G. Ma, \textit{Effects of Neutron-skin Thickness on Light-particle Production}, \href{https://journals.aps.org/prc/abstract/10.1103/PhysRevC.109.024616}{Phys. Rev. C \textbf{109}, 024616 (2024).}


\bibitem{Jia2024NST}
J. Jia, G. Giacalone, B. Bally, J.D. Brandenburg, U. Heinz,
S. Huang, D. Lee, Y.-J. Lee, C. Loizides, W. Li, M. Luzum,
G. Nijs, J. Noronha-Hostler, M. Ploskon, W. van der Schee,
B. Schenke, C. Shen, V. Som\`a, A. Timmins, Z. Xu, and Y. Zhou,
\textit{Imaging the Initial Condition of Heavy-Ion Collisions and
Nuclear Structure across the Nuclide Chart},
\href{https://doi.org/10.1007/s41365-024-01589-w}
{Nucl. Sci. Tech. \textbf{35}, 220 (2024).}

\bibitem{GiacaloneNST} G. Giacalone, \textit{Beyond axial symmetry: high-energy collisions unveil the ground-state shape of 238U}, \href{https://doi.org/10.1007/s41365-024-01582-3}{Nucl. Sci. Tech. {\bf 35}, 218 (2024)}.

\bibitem{SchenkeeNST} B. Schenkee, \textit{Violent collisions can reveal hexadecapole deformation of nuclei}, \href{https://doi.org/10.1007/s41365-024-01509-y}{Nucl. Sci. Tech. {\bf 35}, 115 (2024)}.


\bibitem{Ma_2017}S. Zhang, Y. G. Ma, J. H. Chen, W. B. He, and C. Zhong,  \textit{Nuclear cluster structure effect on elliptic and triangular flows in heavy-ion collisions},  \href{https://doi.org/10.1103/PhysRevC.95.064904}{Phys. Rev.  C {\bf 95}, 064904 (2017)}.


\bibitem{Li2002PRL}
B.A. Li, \textit{Probing the High Density Behavior of the Nuclear Symmetry
Energy with High Energy Heavy-Ion Collisions},
\href{https://doi.org/10.1103/PhysRevLett.88.192701}
{Phys. Rev. Lett. \textbf{88}, 192701 (2002).}


\bibitem{Dai2015PRC}
Z.T. Dai, D.Q. Fang, Y.G. Ma, X.G. Cao, G.Q. Zhang, and W.Q. Shen,
\textit{Effect of Neutron Skin Thickness on Projectile Fragmentation},
\href{https://journals.aps.org/prc/abstract/10.1103/PhysRevC.91.034618}
{Phys. Rev. C \textbf{91}, 034618 (2015).}


\bibitem{Wei2014PRC}
G.-F. Wei, B.-A. Li, J. Xu, and L.-W. Chen,
\textit{Influence of neutron-skin thickness on $\pi^{-}/\pi^{+}$
ratio in Pb+Pb collisions},
\href{https://doi.org/10.1103/PhysRevC.90.014610}
{Phys. Rev. C \textbf{90}, 014610 (2014).}

\bibitem{Wei2015PRC}
G.F. Wei, \textit{Probing the Neutron-skin Thickness by Photon Production from Reactions Induced by Intermediate-energy Protons},
\href{https://journals.aps.org/prc/abstract/10.1103/PhysRevC.92.014614}
{Phys. Rev. C \textbf{92}, 014614 (2015).}

\bibitem{Wang2022PRC}
S.S. Wang, Y.G. Ma, D.Q. Fang, and X.G. Cao,
\textit{Effects of Neutron-skin Thickness on Direct Hard Photon Emission from Reactions Induced by the Neutron-rich Projectile $^{50}$Ca},
\href{https://journals.aps.org/prc/abstract/10.1103/PhysRevC.105.034616}
{Phys. Rev. C \textbf{105}, 034616 (2022).}

\bibitem{Guo2023PRC}
W.M. Guo, B.A. Li, and G.C. Yong,
\textit{Interplay of Effects of Neutron Skins in Coordinate Space and Proton Skins in Momentum Space on Emission of Hard Photons in Heavy-ion Collisions near the Fermi Energy},
\href{https://journals.aps.org/prc/abstract/10.1103/PhysRevC.108.034617}
{Phys. Rev. C \textbf{108}, 034617 (2023).}


\bibitem{Ma2024NST}
C.W. Ma, Y.J. Duan, Y.F. Guo, C.Y. Qiao, Y.T. Wang, J. Pu, K.X. Cheng, and H.L. Wei,
\textit{A Possible Probe to Neutron-skin Thickness by Fragment Parallel Momentum Distribution in Projectile Fragmentation Reactions},
\href{https://doi.org/10.1007/s41365-024-01455-9}
{Nucl. Sci. Tech. \textbf{35}, 99 (2024).}


\bibitem{Ono2019PRC}
A. Ono et al.,
\textit{Comparison of Heavy-Ion Transport Simulations: Collision Integral
with Pions and $\Delta$ Resonances in a Box},
\href{https://journals.aps.org/prc/abstract/10.1103/PhysRevC.100.044617}
{Phys. Rev. C \textbf{100}, 044617 (2019).}


\bibitem{Yong2006PRC}
G.C. Yong, B.A. Li, and L.W. Chen,
\textit{Double Neutron-Proton Differential Transverse Flow as a Probe
for the High Density Behavior of the Nuclear Symmetry Energy},
\href{https://journals.aps.org/prc/abstract/10.1103/PhysRevC.74.064617}
{Phys. Rev. C \textbf{74}, 064617 (2006).}



\bibitem{Li2003PRC}
B.-A. Li,
\textit{Isospin dependence of the $\pi^-/\pi^+$ ratio and density
dependence of the nuclear symmetry energy},
\href{https://doi.org/10.1103/PhysRevC.67.017601}
{Phys. Rev. C \textbf{67}, 017601 (2003).}



\bibitem{Xiao2014EPJA}
Z.-G. Xiao, G.-C. Yong, L.-W. Chen, B.-A. Li, M. Zhang,
G.-Q. Xiao, and N. Xu,
\textit{Probing nuclear symmetry energy at high densities using pion,
kaon, eta and photon productions in heavy-ion collisions},
\href{https://doi.org/10.1140/epja/i2014-14037-6}
{Eur. Phys. J. A \textbf{50}, 37 (2014).}

\bibitem{Li2015PRC}
B.-A. Li,
\textit{Symmetry potential of the $\Delta(1232)$ resonance and its
effects on the $\pi^-/\pi^+$ ratio in heavy-ion collisions},
\href{https://doi.org/10.1103/PhysRevC.92.034603}
{Phys. Rev. C \textbf{92}, 034603 (2015).}


\bibitem{Cai16b}
B.J. Cai, B.A. Li, and L.W. Chen,
\textit{Proton Skins in Momentum Space and Neutron Skins in Coordinate Space in Heavy Nuclei},
\href{https://doi.org/10.1103/PhysRevC.94.061302}
{Phys. Rev. C \textbf{94}, 061302(R) (2016).}

\bibitem{JHXu25PRR}
J.-H. Xu et al.,
\textit{Precise Measurement of Short-Range Correlations in Nuclei from Bremsstrahlung Gamma-Ray Emission in Low-Energy Heavy-Ion Collisions},
\href{https://doi.org/10.1103/jw1p-36pb}
{Phys. Rev. Res. \textbf{7}, 043174 (2025).}

\bibitem{JHXu2026PRC}
J.-H. Xu et al.,
\textit{Experimental study of bremsstrahlung $\gamma$-ray emission
and short-range correlations in $^{124}$Sn+$^{124}$Sn collisions at
25 MeV/nucleon},
\href{https://doi.org/10.1103/dhz2-nl56}
{Phys. Rev. C \textbf{113}, 044613 (2026).}

\bibitem{YHQin24PLB-a}
Y. Qin et al.,
\textit{Probing high-momentum component in nucleon momentum
distribution by neutron-proton bremsstrahlung $\gamma$ rays in
heavy-ion reactions},
\href{https://doi.org/10.1016/j.physletb.2024.138514}
{Phys. Lett. B \textbf{850}, 138514 (2024).}

\bibitem{JHXu24PLB}
J.-H. Xu et al.,
\textit{Reconstruction of bremsstrahlung $\gamma$-ray spectrum in
heavy-ion reactions with the Richardson--Lucy algorithm},
\href{https://doi.org/10.1016/j.physletb.2024.139009}
{Phys. Lett. B \textbf{857}, 139009 (2024).}

\bibitem{Ghosh1994PRC}
S. Ghosh, M. Nandy, P.K. Sarkar, and N. Chakravarty,
\textit{Neutron Skin Effect in Preequilibrium Nucleon Emissions},
\href{https://journals.aps.org/prc/abstract/10.1103/PhysRevC.49.1059}
{Phys. Rev. C \textbf{49}, 1059 (1994).}

\bibitem{Li2000PRL}
B.A. Li,
\textit{Neutron-Proton Differential Flow as a Probe of Isospin-Dependence of the Nuclear Equation of State},
\href{https://journals.aps.org/prl/abstract/10.1103/PhysRevLett.85.4221}
{Phys. Rev. Lett. \textbf{85}, 4221 (2000).}


\bibitem{JiaZhang2023PRC}
J. Jia and C. Zhang,
\textit{Scaling Approach to Nuclear Structure in High-Energy
Heavy-Ion Collisions},
\href{https://doi.org/10.1103/PhysRevC.107.L021901}
{Phys. Rev. C \textbf{107}, L021901 (2023)}.


\bibitem{Hammelmann2020PRC}
J. Hammelmann, A. Soto-Ontoso, M. Alvioli, H. Elfner, and M. Strikman,
\textit{Influence of the Neutron-skin Effect on Nuclear Isobar Collisions
at Energies Available at the BNL Relativistic Heavy Ion Collider},
\href{https://journals.aps.org/prc/abstract/10.1103/PhysRevC.101.061901}
{Phys. Rev. C \textbf{101}, 061901(R) (2020).}


\bibitem{Xu2021PLB}
H.J. Xu, H. Li, X.B. Wang, C. Shen, and F. Wang,
\textit{Determine the Neutron Skin Type by Relativistic Isobaric Collisions},
\href{https://doi.org/10.1016/j.physletb.2021.136453}
{Phys. Lett. B \textbf{819}, 136453 (2021).}

\bibitem{Xu2023PRC}
H.J. Xu, W. Zhao, H. Li, Y. Zhou, L.W. Chen, and F. Wang,
\textit{Probing Nuclear Structure with Mean Transverse Momentum
in Relativistic Isobar Collisions},
\href{https://journals.aps.org/prc/abstract/10.1103/PhysRevC.108.L011902}
{Phys. Rev. C \textbf{108}, L011902 (2023).}



\bibitem{STAR2024Nature}
M.~I. Abdulhamid et al. (STAR Collaboration),
\textit{Imaging Shapes of Atomic Nuclei in High-Energy Nuclear
Collisions},
\href{https://doi.org/10.1038/s41586-024-08097-2}
{Nature \textbf{635}, 67 (2024)}.



\bibitem{Xie2015PRC}
W.J. Xie, Z.Q. Feng, J. Su, and F.S. Zhang,
\textit{Probing the Momentum-dependent Symmetry Potential via Nuclear Collective Flows},
\href{https://doi.org/10.1103/PhysRevC.91.054609}
{Phys. Rev. C \textbf{91}, 054609 (2015).}


\bibitem{Huang2022PRC}
X. Huang, G.F. Wei, Q.J. Zhi, Y.C. Yang, and Z.W. Long,
\textit{Neutron-proton Differential Transverse Flow in
$^{132}$Sn+$^{124}$Sn Collisions at 270 MeV/nucleon},
\href{https://doi.org/10.1103/PhysRevC.106.014604}
{Phys. Rev. C \textbf{106}, 014604 (2022).}

\bibitem{Zhang2022PRL}
C. Zhang and J. Jia,
\textit{Evidence of Quadrupole and Octupole Deformations in
$^{96}$Zr+$^{96}$Zr and $^{96}$Ru+$^{96}$Ru Collisions at
Ultrarelativistic Energies},
\href{https://doi.org/10.1103/PhysRevLett.128.022301}
{Phys. Rev. Lett. \textbf{128}, 022301 (2022).}



\bibitem{Hu2022NP}
B. Hu, W. Jiang, T. Miyagi, Z. Sun, A. Ekstr\"om, C. Forss\'en,
G. Hagen, J.D. Holt, T. Papenbrock, S.R. Stroberg, and I. Vernon,
\textit{Ab Initio Predictions Link the Neutron Skin of $^{208}$Pb
to Nuclear Forces},
\href{https://doi.org/10.1038/s41567-022-01715-8}
{Nat. Phys. \textbf{18}, 1196 (2022).}

\bibitem{Vinas2014EPJA}
X. Vi\~nas, M. Centelles, X. Roca-Maza, and M. Warda,
\textit{Density Dependence of the Symmetry Energy from Neutron
Skin Thickness in Finite Nuclei},
\href{https://doi.org/10.1140/epja/i2014-14027-8}
{Eur. Phys. J. A \textbf{50}, 27 (2014).}

\bibitem{Bethe71}
H.A. Bethe,
\textit{Theory of Nuclear Matter},
\href{https://doi.org/10.1146/annurev.ns.21.120171.000521}
{Annu. Rev. Nucl. Part. Sci. \textbf{21}, 93--244 (1971).}

\bibitem{Subedi08}
R. Subedi et al.,
\textit{Probing Cold Dense Nuclear Matter},
\href{https://doi.org/10.1126/science.1156675}
{Science \textbf{320}, 1476--1478 (2008).}

\bibitem{Piasetzky06}
E. Piasetzky, M. Sargsian, L. Frankfurt, M. Strikman, and J.W. Watson,
\textit{Evidence for Strong Dominance of Proton--Neutron Correlations in Nuclei},
\href{https://doi.org/10.1103/PhysRevLett.97.162504}
{Phys. Rev. Lett. \textbf{97}, 162504 (2006).}

\bibitem{Hen14}
O. Hen et al.,
\textit{Momentum Sharing in Imbalanced Fermi Systems},
\href{https://doi.org/10.1126/science.1256785}
{Science \textbf{346}, 614--617 (2014).}

\bibitem{Hen17RMP}
O. Hen, G.A. Miller, E. Piasetzky, and L.B. Weinstein,
\textit{Nucleon--Nucleon Correlations, Short-Lived Excitations, and the Quarks Within},
\href{https://doi.org/10.1103/RevModPhys.89.045002}
{Rev. Mod. Phys. \textbf{89}, 045002 (2017).}

\bibitem{Fantoni84}
S. Fantoni and V.R. Pandharipande,
\textit{Momentum Distribution of Nucleons in Nuclear Matter},
\href{https://doi.org/10.1016/0375-9474(84)90226-4}
{Nucl. Phys. A \textbf{427}, 473--492 (1984).}

\bibitem{Benhar93}
O. Benhar, V.R. Pandharipande, and S.C. Pieper,
\textit{Electron-Scattering Studies of Correlations in Nuclei},
\href{https://doi.org/10.1103/RevModPhys.65.817}
{Rev. Mod. Phys. \textbf{65}, 817--828 (1993).}

\bibitem{Pandharipande97}
V.R. Pandharipande, I. Sick, and P.K.A. deWitt Huberts,
\textit{Independent-Particle Motion and Correlations in Fermion Systems},
\href{https://doi.org/10.1103/RevModPhys.69.981}
{Rev. Mod. Phys. \textbf{69}, 981--991 (1997).}

\bibitem{Egiyan06}
K.S. Egiyan et al.,
\textit{Measurement of Two- and Three-Nucleon Short-Range Correlation Probabilities in Nuclei},
\href{https://doi.org/10.1103/PhysRevLett.96.082501}
{Phys. Rev. Lett. \textbf{96}, 082501 (2006).}

\bibitem{Shneor07}
R. Shneor et al.,
\textit{Investigation of Proton--Proton Short-Range Correlations via the $^{12}\rm{C}(e,e^{\prime}pp)$ Reaction},
\href{https://doi.org/10.1103/PhysRevLett.99.072501}
{Phys. Rev. Lett. \textbf{99}, 072501 (2007).}

\bibitem{Korover14}
I. Korover et al.,
\textit{Probing the Repulsive Core of the Nucleon--Nucleon Interaction via the $^4\rm{He}(e,e^{\prime}pN)$ Triple-Coincidence Reaction},
\href{https://doi.org/10.1103/PhysRevLett.113.022501}
{Phys. Rev. Lett. \textbf{113}, 022501 (2014).}

\bibitem{Fomin12}
N. Fomin et al.,
\textit{New Measurements of High-Momentum Nucleons and Short-Range Structures in Nuclei},
\href{https://doi.org/10.1103/PhysRevLett.108.092502}
{Phys. Rev. Lett. \textbf{108}, 092502 (2012).}

\bibitem{Cai2026EPJST}
B.J. Cai,
\textit{Nucleon Short-Range Correlations and High-Momentum Dynamics: Implications on the Equation of State of Dense Matter},
\href{https://doi.org/10.1140/epjs/s11734-026-02227-9}
{Eur. Phys. J. Spec. Top. (2026).}


\bibitem{Cai15PRC}
B.J. Cai and B.A. Li,
\textit{Isospin Quartic Term in the Kinetic Energy of Neutron-Rich Nucleonic Matter},
\href{https://doi.org/10.1103/PhysRevC.92.011601}
{Phys. Rev. C \textbf{92}, 011601(R) (2015).}


\bibitem{Cai2016PRC}
B.-J. Cai and B.-A. Li,
\textit{Symmetry energy of cold nucleonic matter within a relativistic
mean field model encapsulating effects of high-momentum nucleons
induced by short-range correlations},
\href{https://doi.org/10.1103/PhysRevC.93.014619}
{Phys. Rev. C \textbf{93}, 014619 (2016)}.

\bibitem{Cai2016PLB}
B.-J. Cai and B.-A. Li,
\textit{Nucleon effective E-mass in neutron-rich matter from the
Migdal--Luttinger jump},
\href{https://doi.org/10.1016/j.physletb.2016.03.059}
{Phys. Lett. B \textbf{757}, 79 (2016)}.

\bibitem{Cai2022PRC}
B.-J. Cai and B.-A. Li,
\textit{Investigating effects of relativistic kinematics,
dimensionality, interactions, and short-range correlations on the ratio
of quartic over quadratic nuclear symmetry energies},
\href{https://doi.org/10.1103/PhysRevC.105.064607}
{Phys. Rev. C \textbf{105}, 064607 (2022)}.

\bibitem{Cai2022AOP}
B.-J. Cai and B.-A. Li,
\textit{Equation of state of neutron-rich matter in $d$ dimensions},
\href{https://doi.org/10.1016/j.aop.2022.169062}
{Ann. Phys. \textbf{444}, 169062 (2022)}.


\bibitem{Weiss15}
R. Weiss, B. Bazak, and N. Barnea,
\textit{Generalized Nuclear Contacts and Momentum Distributions},
\href{https://doi.org/10.1103/PhysRevC.92.054311}
{Phys. Rev. C \textbf{92}, 054311 (2015).}

\bibitem{CruzTorres18}
R. Cruz-Torres, A. Schmidt, G.A. Miller, L.B. Weinstein, N. Barnea, R. Weiss, E. Piasetzky, and O. Hen,
\textit{Short-Range Correlations and the Isospin Dependence of Nuclear Correlation Functions},
\href{https://doi.org/10.1016/j.physletb.2018.07.069}
{Phys. Lett. B \textbf{785}, 304--308 (2018).}

\bibitem{Cosyn21}
W. Cosyn and J. Ryckebusch,
\textit{Phase-Space Distributions of Nuclear Short-Range Correlations},
\href{https://doi.org/10.1016/j.physletb.2021.136526}
{Phys. Lett. B \textbf{820}, 136526 (2021).}

\bibitem{Weinstein11}
L.B. Weinstein, E. Piasetzky, D.W. Higinbotham, J. Gomez, O. Hen, and R. Shneor,
\textit{Short-Range Correlations and the EMC Effect},
\href{https://doi.org/10.1103/PhysRevLett.106.052301}
{Phys. Rev. Lett. \textbf{106}, 052301 (2011).}

\bibitem{Schmookler19}
B. Schmookler et al.,
\textit{Modified Structure of Protons and Neutrons in Correlated Pairs},
\href{https://doi.org/10.1038/s41586-019-0925-9}
{Nature \textbf{566}, 354--358 (2019).}




\bibitem{Cai2026MPLA}
B.-J. Cai, B.-A. Li, and Y.-G. Ma,
\textit{Neutron star equation of state with nucleon short-range
correlations: A concise review and open issues},
\href{https://doi.org/10.1142/S0217732326300053}
{Mod. Phys. Lett. A, 2630005 (2026).}


\bibitem{Tan08a}
S. Tan,
\textit{Energetics of a Strongly Correlated Fermi Gas},
\href{https://doi.org/10.1016/j.aop.2008.03.004}
{Ann. Phys. \textbf{323}, 2952--2970 (2008).}

\bibitem{Tan08b}
S. Tan,
\textit{Large Momentum Part of a Strongly Correlated Fermi Gas},
\href{https://doi.org/10.1016/j.aop.2008.03.005}
{Ann. Phys. \textbf{323}, 2971--2986 (2008).}

\bibitem{Tan08c}
S. Tan,
\textit{Generalized Virial Theorem and Pressure Relation for a Strongly Correlated Fermi Gas},
\href{https://doi.org/10.1016/j.aop.2008.03.003}
{Ann. Phys. \textbf{323}, 2987--2990 (2008).}

\bibitem{Giorgini08}
S. Giorgini, L.P. Pitaevskii, and S. Stringari,
\textit{Theory of Ultracold Atomic Fermi Gases},
\href{https://doi.org/10.1103/RevModPhys.80.1215}
{Rev. Mod. Phys. \textbf{80}, 1215--1274 (2008).}

\bibitem{Bloch08}
I. Bloch, J. Dalibard, and W. Zwerger,
\textit{Many-Body Physics with Ultracold Gases},
\href{https://doi.org/10.1103/RevModPhys.80.885}
{Rev. Mod. Phys. \textbf{80}, 885--964 (2008).}

\bibitem{Boffi96}
S. Boffi, C. Giusti, F.D. Pacati, and M. Radici,
\textit{Electromagnetic Response of Atomic Nuclei},
\href{https://global.oup.com/academic/product/electromagnetic-response-of-atomic-nuclei-9780198517740}
{Clarendon Press, Oxford (1996).}

\bibitem{Frois1987}
B. Frois and C.N. Papanicolas,
\textit{Electron Scattering and Nuclear Structure},
\href{https://doi.org/10.1146/annurev.nucl.37.1.133}
{Annu. Rev. Nucl. Part. Sci. \textbf{37}, 133--176 (1987).}


\bibitem{Frankfurt88}
L.L. Frankfurt and M.I. Strikman,
\textit{Hard Nuclear Processes and Microscopic Nuclear Structure},
\href{https://doi.org/10.1016/0370-1573(88)90179-2}
{Phys. Rep. \textbf{160}, 235--427 (1988).}

\bibitem{Frankfurt93}
L.L. Frankfurt, M.I. Strikman, D.B. Day, and M.M. Sargsyan,
\textit{Evidence for Short-Range Correlations from High-$Q^2$ $(e,e^{\prime})$ Reactions},
\href{https://doi.org/10.1103/PhysRevC.48.2451}
{Phys. Rev. C \textbf{48}, 2451--2461 (1993).}

\bibitem{Weiss21}
R. Weiss, A.W. Denniston, J.R. Pybus, O. Hen, E. Piasetzky, A. Schmidt, L.B. Weinstein, and N. Barnea,
\textit{Extracting the Number of Short-Range Correlated Nucleon Pairs from Inclusive Electron Scattering Data},
\href{https://doi.org/10.1103/PhysRevC.103.L031301}
{Phys. Rev. C \textbf{103}, L031301 (2021).}

\bibitem{Hauenstein02012021}
F. Hauenstein, J. Kahlbow, and O. Hen,
\textit{From Quarks to Nuclei: Short-Range Correlations Studies across the Globe},
\href{https://doi.org/10.1080/10619127.2020.1832819}
{Nucl. Phys. News \textbf{31}, 19--24 (2021).}

\bibitem{Hen:2025rlk}
O. Hen, D.W. Higinbotham, E. Piasetzky, and A. Schmidt,
\textit{Topical Issue on Short-Range Correlations and the EMC Effect},
\href{https://doi.org/10.1140/epja/s10050-025-01578-6}
{Eur. Phys. J. A \textbf{61}, 109 (2025).}

\bibitem{Tu:2020ymk}
Z. Tu, A. Jentsch, M. Baker, L. Zheng, J.H. Lee, R. Venugopalan, O. Hen, D. Higinbotham, E.C. Aschenauer, and T. Ullrich,
\textit{Probing Short-Range Correlations in the Deuteron via Incoherent Diffractive J/$\psi$ Production with Spectator Tagging at the EIC},
\href{https://doi.org/10.1016/j.physletb.2020.135877}
{Phys. Lett. B \textbf{811}, 135877 (2020).}

\bibitem{Hauenstein:2021zql}
F. Hauenstein et al.,
\textit{Measuring Recoiling Nucleons from the Nucleus with the Future Electron--Ion Collider},
\href{https://doi.org/10.1103/PhysRevC.105.034001}
{Phys. Rev. C \textbf{105}, 034001 (2022).}

\bibitem{Boer:2011fh}
D. Boer et al.,
\textit{Gluons and the Quark Sea at High Energies: Distributions, Polarization, Tomography},
\href{https://arxiv.org/abs/1108.1713}
{arXiv:1108.1713 [nucl-th] (2011).}

\bibitem{SRC-r3b}
R$^3$B Collaboration,
\textit{R$^3$B SRC Proposal},
\href{https://indico.ph.tum.de/event/7050/contributions/6385/attachments/4485/5708/NuPECC_SRC.pdf}
{Proposal document (2025).}

\bibitem{r3b}
H. Qi, J. Kahlbow, and the R$^3$B S522 Collaboration Team,
\textit{Short-Range Correlation Studies in Inverse Kinematics to Access Unstable Nuclei},
\href{https://ui.adsabs.harvard.edu/abs/2023APS..HAWE11003Q}
{APS Meeting Abstracts, E11.003 (2023).}

\bibitem{myref}
Facility for Antiproton and Ion Research in Europe,
\textit{FAIR Website},
\href{https://fair-center.eu/}
{https://fair-center.eu/ (2025).}

\bibitem{Ye24}
Z. Ye, H. Zhang, Y. Zhang, and H. Zhao,
\textit{New Chinese Facilities for Short-Range Correlation Physics},
\href{https://doi.org/10.1140/epja/s10050-024-01343-1}
{Eur. Phys. J. A \textbf{60}, 126 (2024).}

\bibitem{EicC-ref}
D.P. Anderle et al.,
\textit{Electron--Ion Collider in China},
\href{https://doi.org/10.1007/s11467-021-1062-0}
{Front. Phys. \textbf{16}, 64701 (2021).}

\bibitem{Kahlbow:2023mtc}
J. Kahlbow,
\textit{Study of SRCs in Neutron-Rich Nuclei with Inverse Kinematics Measurements},
\href{https://doi.org/10.1140/epja/s10050-023-01028-1}
{Eur. Phys. J. A \textbf{59}, 128 (2023).}

\bibitem{2023EPJA...59..188A}
J. Arrington, R. Cruz-Torres, T.J. Hague, L. Kurbany, S. Li, D. Meekins, and N. Santiesteban,
\textit{The Jefferson Lab Tritium Program of Nucleon and Nuclear Structure Measurements},
\href{https://doi.org/10.1140/epja/s10050-023-01085-6}
{Eur. Phys. J. A \textbf{59}, 188 (2023).}

\bibitem{2023EPJA...59..205F}
N. Fomin, J. Arrington, and S. Li,
\textit{Searching for Three-Nucleon Short-Range Correlations},
\href{https://doi.org/10.1140/epja/s10050-023-01112-6}
{Eur. Phys. J. A \textbf{59}, 205 (2023).}


\bibitem{CaiLi22Gog}
B.-J. Cai and B.-A. Li,
\textit{Nuclear equation of state and single-nucleon potential from
Gogny-like energy density functionals encapsulating effects of
nucleon-nucleon short-range correlations},
\href{https://arxiv.org/abs/2210.10924}
{arXiv:2210.10924 [nucl-th] (2022)}.

\bibitem{Hen15PRC}
O. Hen, B.A. Li, W.J. Guo, L.B. Weinstein, and E. Piasetzky,
\textit{Symmetry Energy of Nucleonic Matter with Tensor Correlations},
\href{https://doi.org/10.1103/PhysRevC.91.025803}
{Phys. Rev. C \textbf{91}, 025803 (2015).}

\bibitem{Brack1985PhysRep}
M. Brack, C. Guet, and H.-B. H{\aa}kansson,
\textit{Selfconsistent semiclassical description of average nuclear
properties---a link between microscopic and macroscopic models},
\href{https://doi.org/10.1016/0370-1573(86)90078-5}
{Phys. Rep. \textbf{123}, 275--364 (1985).}


\bibitem{Li2022ApJ}
F. Li, B.-J. Cai, Y. Zhou, W.-Z. Jiang, and L.-W. Chen,
\textit{Effects of Isoscalar- and Isovector-Scalar Meson Mixing on
Neutron Star Structure},
\href{https://iopscience.iop.org/article/10.3847/1538-4357/ac5e2a}
{Astrophys. J. \textbf{929}, 183 (2022).}

\bibitem{Liu2024NSTAstro}
W.-P. Liu, B. Guo, Z. An, et al.,
\textit{Recent Progress in Nuclear Astrophysics Research and Its
Astrophysical Implications at the China Institute of Atomic Energy},
\href{https://doi.org/10.1007/s41365-024-01590-3}
{Nucl. Sci. Tech. \textbf{35}, 217 (2024)}.


\bibitem{Horowitz2001PRC}
C.J. Horowitz and J. Piekarewicz,
\textit{Neutron Radii of $^{208}\rm{Pb}$ and Neutron Stars},
\href{https://journals.aps.org/prc/abstract/10.1103/PhysRevC.64.062802}
{Phys. Rev. C \textbf{64}, 062802(R) (2001).}



\bibitem{Wu2026AA}
X. Wu, S. Bao, M. Ju, J. Hu, and H. Shen,
\textit{Symmetry Energy Effect on Rotating Neutron Stars},
\href{https://www.aanda.org/articles/aa/full_html/2026/02/aa52766-24/aa52766-24.html}
{Astron. Astrophys. \textbf{706}, A132 (2026).}



\bibitem{Hebeler2014EPJA}
K. Hebeler and A. Schwenk,
\textit{Symmetry Energy, Neutron Skin, and Neutron Star Radius from Chiral Effective Field Theory Interactions},
\href{https://doi.org/10.1140/epja/i2014-14011-4}
{Eur. Phys. J. A \textbf{50}, 11 (2014).}



\bibitem{Hebeler2010PRL}
K. Hebeler, J.M. Lattimer, C.J. Pethick, and A. Schwenk,
\textit{Constraints on Neutron Star Radii Based on Chiral Effective Field Theory Interactions},
\href{https://journals.aps.org/prl/abstract/10.1103/PhysRevLett.105.161102}
{Phys. Rev. Lett. \textbf{105}, 161102 (2010).}

\bibitem{Fattoyev2018PRL}
F.J. Fattoyev, J. Piekarewicz, and C.J. Horowitz,
\textit{Neutron Skins and Neutron Stars in the Multimessenger Era},
\href{https://journals.aps.org/prl/abstract/10.1103/PhysRevLett.120.172702}
{Phys. Rev. Lett. \textbf{120}, 172702 (2018).}


\bibitem{Koliogiannis2025PLB}
P.S. Koliogiannis, E. Y\"uksel, and N. Paar,
\textit{Constraining Neutron Star Properties through Parity-Violating
Electron Scattering Experiments and Relativistic Point Coupling Interactions},
\href{https://doi.org/10.1016/j.physletb.2025.139362}
{Phys. Lett. B \textbf{862}, 139362 (2025).}


\bibitem{Xu2009PRC}
J. Xu, L.W. Chen, B.A. Li, and H.R. Ma,
\textit{Locating the Inner Edge of the Neutron Star Crust Using
Terrestrial Nuclear Laboratory Data},
\href{https://journals.aps.org/prc/abstract/10.1103/PhysRevC.79.035802}
{Phys. Rev. C \textbf{79}, 035802 (2009).}

\bibitem{Xu2009ApJ}
J. Xu, L.W. Chen, B.A. Li, and H.R. Ma,
\textit{Nuclear Constraints on Properties of Neutron Star Crusts},
\href{https://doi.org/10.1088/0004-637X/697/2/1549}
{Astrophys. J. \textbf{697}, 1549 (2009).}


\bibitem{Newton2022PLB}
W.G. Newton, R. Preston, L. Balliet, and M. Ross,
\textit{From Neutron Skins and Neutron Matter to the Neutron Star Crust},
\href{https://doi.org/10.1016/j.physletb.2022.137481}
{Phys. Lett. B \textbf{834}, 137481 (2022).}

\bibitem{Newton2021PRC}
W.G. Newton and G. Crocombe,
\textit{Nuclear Symmetry Energy from Neutron Skins and Pure Neutron Matter in
a Bayesian Framework},
\href{https://doi.org/10.1103/PhysRevC.103.064323}
{Phys. Rev. C \textbf{103}, 064323 (2021).}

\bibitem{Newton2013ApJS}
W.~G. Newton, M. Gearheart, J. Hooker, and B.-A. Li,
\textit{The Nuclear Symmetry Energy, the Inner Crust, and Global
Neutron Star Modeling},
\href{https://doi.org/10.1088/0067-0049/204/1/9}
{Astrophys. J. Suppl. Ser. \textbf{204}, 9 (2013)}.

\bibitem{Grams2022PRC}
G. Grams, R. Somasundaram, J. Margueron, and S. Reddy,
\textit{Properties of the Neutron Star Crust: Quantifying and
Correlating Uncertainties with Improved Nuclear Physics},
\href{https://doi.org/10.1103/PhysRevC.105.035806}
{Phys. Rev. C \textbf{105}, 035806 (2022)}.

\bibitem{Xie2023NST}
W.-J. Xie, Z.-W. Ma, and J.-H. Guo,
\textit{Bayesian Inference of the Crust-Core Transition Density via
the Neutron-Star Radius and Neutron-Skin Thickness Data},
\href{https://doi.org/10.1007/s41365-023-01239-7}
{Nucl. Sci. Tech. \textbf{34}, 91 (2023)}.

\bibitem{Flanagan2008PRD}
E.E. Flanagan and T. Hinderer,
\textit{Constraining neutron-star tidal Love numbers with gravitational-wave detectors},
\href{https://journals.aps.org/prd/abstract/10.1103/PhysRevD.77.021502}
{Phys. Rev. D \textbf{77}, 021502(R) (2008).}

\bibitem{Hinderer2008ApJ}
T. Hinderer,
\textit{Tidal Love Numbers of Neutron Stars},
\href{https://doi.org/10.1086/533487}
{Astrophys. J. \textbf{677}, 1216--1220 (2008).}

\bibitem{Hinderer2010PRD}
T. Hinderer, B.D. Lackey, R.N. Lang, and J.S. Read,
\textit{Tidal deformability of neutron stars with realistic equations of state and their gravitational wave signatures in binary inspiral},
\href{https://journals.aps.org/prd/abstract/10.1103/PhysRevD.81.123016}
{Phys. Rev. D \textbf{81}, 123016 (2010).}


\bibitem{De2018PRL}
S. De, D. Finstad, J.M. Lattimer, D.A. Brown, E. Berger, and C.M. Biwer,
\textit{Tidal Deformabilities and Radii of Neutron Stars from the Observation of GW170817},
\href{https://journals.aps.org/prl/abstract/10.1103/PhysRevLett.121.091102}
{Phys. Rev. Lett. \textbf{121}, 091102 (2018).}

\bibitem{Raithel2018ApJL}
C.A. Raithel, F. \"Ozel, and D. Psaltis,
\textit{Tidal Deformability from GW170817 as a Direct Probe of the Neutron Star Radius},
\href{https://doi.org/10.3847/2041-8213/aabcbf}
{Astrophys. J. Lett. \textbf{857}, L23 (2018).}

\bibitem{Most2018PRL}
E.R. Most, L.R. Weih, L. Rezzolla, and J. Schaffner-Bielich,
\textit{New Constraints on Radii and Tidal Deformabilities of Neutron Stars from GW170817},
\href{https://journals.aps.org/prl/abstract/10.1103/PhysRevLett.120.261103}
{Phys. Rev. Lett. \textbf{120}, 261103 (2018).}

\bibitem{Annala2018PRL}
E. Annala, T. Gorda, A. Kurkela, and A. Vuorinen,
\textit{Gravitational-Wave Constraints on the Neutron-Star-Matter Equation of State},
\href{https://journals.aps.org/prl/abstract/10.1103/PhysRevLett.120.172703}
{Phys. Rev. Lett. \textbf{120}, 172703 (2018).}

\bibitem{Lim2018PRL}
Y. Lim and J.W. Holt,
\textit{Neutron Star Tidal Deformabilities Constrained by Nuclear Theory and Experiment},
\href{https://journals.aps.org/prl/abstract/10.1103/PhysRevLett.121.062701}
{Phys. Rev. Lett. \textbf{121}, 062701 (2018).}

\bibitem{Reed2021PRL}
B.T. Reed, F.J. Fattoyev, C.J. Horowitz, and J. Piekarewicz,
\textit{Implications of PREX-2 on the Equation of State of Neutron-Rich Matter},
\href{https://journals.aps.org/prl/abstract/10.1103/PhysRevLett.126.172503}
{Phys. Rev. Lett. \textbf{126}, 172503 (2021).}


\bibitem{Alford2013PRD}
M. G. Alford, S. Han, and M. Prakash,
\textit{Generic conditions for stable hybrid stars},
\href{https://doi.org/10.1103/PhysRevD.88.083013}
{Phys. Rev. D \textbf{88}, 083013 (2013)}.


\bibitem{Benic2015AA}
S. Beni\'c, D. Blaschke, D. E. Alvarez-Castillo, T. Fischer,
and S. Typel,
\textit{A new quark-hadron hybrid equation of state for astrophysics.
I. High-mass twin compact stars},
\href{https://doi.org/10.1051/0004-6361/201425318}
{Astron. Astrophys. \textbf{577}, A40 (2015)}.

\bibitem{AlvarezCastillo2017PRC}
D. E. Alvarez-Castillo and D. B. Blaschke,
\textit{High-mass twin stars with a multipolytrope equation of state},
\href{https://doi.org/10.1103/PhysRevC.96.045809}
{Phys. Rev. C \textbf{96}, 045809 (2017)}.

\bibitem{Masuda2013ApJ}
K. Masuda, T. Hatsuda, and T. Takatsuka,
\textit{Hadron-Quark Crossover and Massive Hybrid Stars with Strangeness},
\href{https://doi.org/10.1088/0004-637X/764/1/12}
{Astrophys. J. \textbf{764}, 12 (2013)}.

\bibitem{Masuda2013PTEP}
K. Masuda, T. Hatsuda, and T. Takatsuka,
\textit{Hadron--quark crossover and massive hybrid stars},
\href{https://doi.org/10.1093/ptep/ptt045}
{Prog. Theor. Exp. Phys. \textbf{2013}, 073D01 (2013)}.


\bibitem{Baym2019ApJ}
G. Baym, S. Furusawa, T. Hatsuda, T. Kojo, and H. Togashi,
\textit{New Neutron Star Equation of State with Quark--Hadron Crossover},
\href{https://doi.org/10.3847/1538-4357/ab441e}
{Astrophys. J. \textbf{885}, 42 (2019)}.

\bibitem{McLerran2019PRL}
L. McLerran and S. Reddy,
\textit{Quarkyonic Matter and Neutron Stars},
\href{https://doi.org/10.1103/PhysRevLett.122.122701}
{Phys. Rev. Lett. \textbf{122}, 122701 (2019)}.

\bibitem{Annala2020NatPhys}
E. Annala, T. Gorda, A. Kurkela, J. N\"attil\"a, and A. Vuorinen,
\textit{Evidence for quark-matter cores in massive neutron stars},
\href{https://doi.org/10.1038/s41567-020-0914-9}
{Nat. Phys. \textbf{16}, 907--910 (2020)}.

\bibitem{Annala2023NatCommun}
E. Annala, T. Gorda, J. Hirvonen, O. Komoltsev, A. Kurkela,
J. N\"attil\"a, and A. Vuorinen,
\textit{Strongly interacting matter exhibits deconfined behavior
in massive neutron stars},
\href{https://doi.org/10.1038/s41467-023-44051-y}
{Nat. Commun. \textbf{14}, 8451 (2023)}.

\bibitem{Bauswein2019PRL}
A. Bauswein, N.-U. F. Bastian, D. B. Blaschke,
K. Chatziioannou, J. A. Clark, T. Fischer, and M. Oertel,
\textit{Identifying a First-Order Phase Transition in Neutron-Star
Mergers through Gravitational Waves},
\href{https://doi.org/10.1103/PhysRevLett.122.061102}
{Phys. Rev. Lett. \textbf{122}, 061102 (2019)}.

\bibitem{Most2019PRL}
E. R. Most, L. J. Papenfort, V. Dexheimer, M. Hanauske,
S. Schramm, H. St\"ocker, and L. Rezzolla,
\textit{Signatures of Quark-Hadron Phase Transitions in
General-Relativistic Neutron-Star Mergers},
\href{https://doi.org/10.1103/PhysRevLett.122.061101}
{Phys. Rev. Lett. \textbf{122}, 061101 (2019)}.

\bibitem{Lin2011PRC}
W. Lin, B.-A. Li, J. Xu, C. M. Ko, and D. H. Wen,
\textit{Energy release from hadron-quark phase transition in neutron
stars and the axial $w$ mode of gravitational waves},
\href{https://doi.org/10.1103/PhysRevC.83.045802}
{Phys. Rev. C \textbf{83}, 045802 (2011)}.

\bibitem{XieLi2021PRC}
W.-J. Xie and B.-A. Li,
\textit{Bayesian inference of the dense-matter equation of state
encapsulating a first-order hadron-quark phase transition from
observables of canonical neutron stars},
\href{https://doi.org/10.1103/PhysRevC.103.035802}
{Phys. Rev. C \textbf{103}, 035802 (2021)}.

\bibitem{ZhangLi2023PRC}
N.-B. Zhang and B.-A. Li,
\textit{Properties of first-order hadron-quark phase transition
from inverting neutron star observables},
\href{https://doi.org/10.1103/PhysRevC.108.025803}
{Phys. Rev. C \textbf{108}, 025803 (2023)}.

\bibitem{Li2024PRD}
B.-A. Li, X. Grundler, W.-J. Xie, and N.-B. Zhang,
\textit{Bayesian inference of fine features of the nuclear equation
of state from future neutron star radius measurements to 0.1 km
accuracy},
\href{https://doi.org/10.1103/PhysRevD.110.103040}
{Phys. Rev. D \textbf{110}, 103040 (2024)}.

\bibitem{GrundlerLi2026PRD}
X. Grundler and B.-A. Li,
\textit{Bayesian constraints on the neutron star equation of state
with a smooth hadron-quark crossover},
\href{https://doi.org/10.1103/pn2b-vls3}
{Phys. Rev. D \textbf{113}, 103012 (2026)}.

\bibitem{Tang2021PRD-aa}
S.-P. Tang, J.-L. Jiang, W.-H. Gao, Y.-Z. Fan, and D.-M. Wei,
\textit{Constraint on phase transition with the multimessenger data
of neutron stars},
\href{https://doi.org/10.1103/PhysRevD.103.063026}
{Phys. Rev. D \textbf{103}, 063026 (2021)}.

\bibitem{Han2023SciBull}
M.-Z. Han, Y.-J. Huang, S.-P. Tang, and Y.-Z. Fan,
\textit{Plausible presence of new state in neutron stars with masses
above $0.98M_{\rm TOV}$},
\href{https://doi.org/10.1016/j.scib.2023.04.007}
{Sci. Bull. \textbf{68}, 913--919 (2023)}.

\bibitem{Tang2025PRD}
S.-P. Tang, Y.-J. Huang, and Y.-Z. Fan,
\textit{Phase transition and nuclear symmetry energy from neutron
star observations: Constraints in light of PSR J0614-3329},
\href{https://doi.org/10.1103/bmsk-8n85}
{Phys. Rev. D \textbf{112}, 083009 (2025)}.

\bibitem{Liu2016PRD}
H. Liu, J. Xu, L.-W. Chen, and K.-J. Sun,
\textit{Isospin properties of quark matter from a 3-flavor NJL model},
\href{https://doi.org/10.1103/PhysRevD.94.065032}
{Phys. Rev. D \textbf{94}, 065032 (2016)}.

\bibitem{ChuChen2017PRD}
P.-C. Chu and L.-W. Chen,
\textit{Isovector properties of quark matter and quark stars in an
isospin-dependent confining model},
\href{https://doi.org/10.1103/PhysRevD.96.083019}
{Phys. Rev. D \textbf{96}, 083019 (2017)}.

\bibitem{ZhangLi2025EPJA}
N.-B. Zhang and B.-A. Li,
\textit{Impact of the nuclear equation of state on the formation
of twin stars},
\href{https://doi.org/10.1140/epja/s10050-025-01497-6}
{Eur. Phys. J. A \textbf{61}, 31 (2025)}.


\bibitem{CaoChen2023}
Z. Cao and L.-W. Chen,
\textit{Neutron Star vs Quark Star in the Multimessenger Era},
\href{https://arxiv.org/abs/2308.16783}
{arXiv:2308.16783 (2023)}.

\bibitem{CaoChen2026}
Z. Cao and L.-W. Chen,
\textit{On the Possibility of a Strong First-Order Phase Transition
in Neutron Stars},
\href{https://arxiv.org/abs/2606.06378}
{arXiv:2606.06378 (2026)}.
\bibitem{GrundlerLi2025PRD}
X. Grundler and B.-A. Li,
\textit{Bayesian quantification of observability and equation of state
of twin stars},
\href{https://doi.org/10.1103/hsd4-j54y}
{Phys. Rev. D \textbf{112}, 103012 (2025)}.

\bibitem{Li2026ApJ}
B.-A. Li, X. Grundler, W.-J. Xie, and N.-B. Zhang,
\textit{Bayesian Inference of Hybrid Star Properties from Future
High-precision Measurements of Their Radii},
\href{https://doi.org/10.3847/1538-4357/ae38c0}
{Astrophys. J. \textbf{998}, 262 (2026)}.

\bibitem{LiGrundler2026PLB}
B.-A. Li and X. Grundler,
\textit{Quantifying the information gain from future high-precision
radius measurements for identifying twin neutron stars},
\href{https://doi.org/10.1016/j.physletb.2026.140863}
{Phys. Lett. B \textbf{880}, 140863 (2026)}.

\bibitem{Tews2018PRC}
I. Tews, J. Margueron, and S. Reddy,
\textit{Critical examination of constraints on the equation of state of dense matter obtained from GW170817},
\href{https://journals.aps.org/prc/abstract/10.1103/PhysRevC.98.045804}
{Phys. Rev. C \textbf{98}, 045804 (2018).}

\bibitem{Zhao2018PRD}
T. Zhao and J.M. Lattimer,
\textit{Tidal deformabilities and neutron star mergers},
\href{https://journals.aps.org/prd/abstract/10.1103/PhysRevD.98.063020}
{Phys. Rev. D \textbf{98}, 063020 (2018).}

\bibitem{Malik2018PRC}
T. Malik, B.K. Agrawal, J.N. De, S.K. Samaddar, C. Provid\^encia,
C. Mondal, and T.K. Jha,
\textit{GW170817: Constraining the nuclear matter equation of state
from the neutron star tidal deformability},
\href{https://journals.aps.org/prc/abstract/10.1103/PhysRevC.98.035804}
{Phys. Rev. C \textbf{98}, 035804 (2018).}

\bibitem{Kim2018PRC}
Y.M. Kim, Y. Lim, K. Kwak, C.H. Hyun, and C.H. Lee,
\textit{Tidal deformability of neutron stars with realistic nuclear
energy density functionals},
\href{https://journals.aps.org/prc/abstract/10.1103/PhysRevC.98.065805}
{Phys. Rev. C \textbf{98}, 065805 (2018).}

\bibitem{Carson2019PRD}
Z. Carson, A.W. Steiner, and K. Yagi,
\textit{Constraining nuclear matter parameters with GW170817},
\href{https://journals.aps.org/prd/abstract/10.1103/PhysRevD.99.043010}
{Phys. Rev. D \textbf{99}, 043010 (2019).}

\bibitem{Malik2019PRC}
T. Malik, B.K. Agrawal, J.N. De, S.K. Samaddar, C. Provid\^encia,
C. Mondal, and T.K. Jha,
\textit{Tides in merging neutron stars: Consistency of the GW170817
event with experimental data on finite nuclei},
\href{https://journals.aps.org/prc/abstract/10.1103/PhysRevC.99.052801}
{Phys. Rev. C \textbf{99}, 052801(R) (2019).}

\bibitem{Tong2020PRC}
H. Tong, P.W. Zhao, and J. Meng,
\textit{Symmetry energy at supra-saturation densities via the
gravitational waves from GW170817},
\href{https://journals.aps.org/prc/abstract/10.1103/PhysRevC.101.035802}
{Phys. Rev. C \textbf{101}, 035802 (2020).}

\bibitem{Shi2026xxx}
J.H. Shi, B.J. Cai, B.A. Li, and Y.G. Ma,
\textit{A New Scaling of Neutron Star Tidal Deformability for
Directly Probing the Core Equation of State},
\href{https://arxiv.org/abs/2606.21402}
{arXiv:2606.21402 [astro-ph.HE] (2026).}

\bibitem{Cai2023ApJ}
B.-J. Cai, B.-A. Li, and Z. Zhang,
\textit{Core States of Neutron Stars from Anatomizing Their Scaled
Structure Equations},
\href{https://doi.org/10.3847/1538-4357/acdef0}
{Astrophys. J. \textbf{952}, 147 (2023)}.

\bibitem{Cai2023PRD}
B.-J. Cai, B.-A. Li, and Z. Zhang,
\textit{Central Speed of Sound, the Trace Anomaly, and Observables
of Neutron Stars from a Perturbative Analysis of Scaled
Tolman-Oppenheimer-Volkoff Equations},
\href{https://doi.org/10.1103/PhysRevD.108.103041}
{Phys. Rev. D \textbf{108}, 103041 (2023)}.

\bibitem{Cai2024PRD}
B.-J. Cai and B.-A. Li,
\textit{Strong Gravity Extruding Peaks in Speed of Sound Profiles
of Massive Neutron Stars},
\href{https://doi.org/10.1103/PhysRevD.109.083015}
{Phys. Rev. D \textbf{109}, 083015 (2024)}.

\bibitem{Cai2024Frontiers}
B.-J. Cai and B.-A. Li,
\textit{New Insights into Supradense Matter from Dissecting
Scaled Stellar Structure Equations},
\href{https://doi.org/10.3389/fspas.2024.1502888}
{Front. Astron. Space Sci. \textbf{11}, 1502888 (2024)}.

\bibitem{Cai2025PRD}
B.-J. Cai and B.-A. Li,
\textit{Unraveling Trace Anomaly of Supradense Matter via
Neutron Star Compactness Scaling},
\href{https://doi.org/10.1103/3p2p-p3d4}
{Phys. Rev. D \textbf{112}, 023023 (2025)}.

\bibitem{Cai2026Trace}
B.-J. Cai, B.-A. Li, and Y.-G. Ma,
\textit{Is the Trace Anomaly at Its Minimum Value at Neutron Star
Centers?},
\href{https://doi.org/10.1103/nbmw-k5fs}
{Phys. Rev. D \textbf{113}, 023002 (2026)}.

\bibitem{Cai2025Phase}
B.-J. Cai, B.-A. Li, and Y.-G. Ma,
\textit{Revisiting the Possibility of a Sharp Phase Transition
in Cold Neutron Stars},
\href{https://arxiv.org/abs/2511.08380}
{arXiv:2511.08380 [astro-ph.HE]}.

\bibitem{Cai2026Bound}
B.-J. Cai, B.-A. Li, and Y.-G. Ma,
\textit{Effective Upper Bound on the Pressure-to-Energy Density
Ratio in Neutron Stars},
\href{https://doi.org/10.1103/1c3x-5w3k}
{Phys. Rev. D \textbf{114}, 043062 (2026)}.


\bibitem{Stroud2026}
J. Stroud, D. Radice, and S. Reddy,
\textit{High-Density Sound Speed and Post-Merger Dynamics},
\href{https://arxiv.org/abs/2607.15588}
{arXiv:2607.15588 [astro-ph.HE] (2026).}

\bibitem{Danielewicz2002Science}
P. Danielewicz, R. Lacey, and W.G. Lynch,
\textit{Determination of the Equation of State of Dense Matter},
\href{https://doi.org/10.1126/science.1078070}
{Science \textbf{298}, 1592 (2002).}

\bibitem{LeFevre2016NPA}
A. Le F\`evre, Y. Leifels, W. Reisdorf, J. Aichelin, and C. Hartnack,
\textit{Constraining the nuclear matter equation of state around twice saturation density},
\href{https://doi.org/10.1016/j.nuclphysa.2015.09.015}
{Nucl. Phys. A \textbf{945}, 112 (2016).}



\bibitem{Estee2021PRL}
J. Estee et al. (S$\pi$RIT Collaboration),
\textit{Probing the Symmetry Energy with the Spectral Pion Ratio},
\href{https://journals.aps.org/prl/abstract/10.1103/PhysRevLett.126.162701}
{Phys. Rev. Lett. \textbf{126}, 162701 (2021).}


\bibitem{Russotto2016PRC}
P. Russotto et al.,
\textit{Results of the ASY-EOS Experiment at GSI: The Symmetry Energy at Suprasaturation Density},
\href{https://doi.org/10.1103/PhysRevC.94.034608}
{Phys. Rev. C \textbf{94}, 034608 (2016).}

\bibitem{LynchTsang2022PLB}
W.G. Lynch and M.B. Tsang,
\textit{Decoding the density dependence of the nuclear symmetry energy},
\href{https://doi.org/10.1016/j.physletb.2022.137098}
{Phys. Lett. B \textbf{830}, 137098 (2022).}



\bibitem{Yue2022PRR}
T.G. Yue, L.W. Chen, Z. Zhang, and Y. Zhou,
\textit{Constraints on the symmetry energy from PREX-II in the multimessenger era},
\href{https://doi.org/10.1103/PhysRevResearch.4.L022054}
{Phys. Rev. Research \textbf{4}, L022054 (2022).}



\bibitem{Neill2023PRL}
D. Neill, R. Preston, W.G. Newton, and D. Tsang,
\textit{Constraining the Nuclear Symmetry Energy with
Multimessenger Resonant Shattering Flares},
\href{https://journals.aps.org/prl/abstract/10.1103/PhysRevLett.130.112701}
{Phys. Rev. Lett. \textbf{130}, 112701 (2023).}

\bibitem{Wang2026PRC}
Z. Wang, J. Chen, J. Jia, Y.-G. Ma, and C. Zhang,
\textit{Disentangling Nuclear Structure through Multiparticle
Azimuthal Correlations in High-Energy Isobar Collisions},
\href{https://journals.aps.org/prc/abstract/10.1103/47jc-xzxm}
{Phys. Rev. C \textbf{113}, 034905 (2026).}

\bibitem{Xi2025NST}
B.-S. Xi, J.-H. Chen, L. Ma, Y.-G. Ma, and T.-T. Wang,
\textit{Study of the Momentum Correlation of Nucleons in
$^{96}\rm{Ru}$+$^{96}\rm{Ru}$ and $^{96}\rm{Zr}$+$^{96}\rm{Zr}$
Collisions at $\sqrt{s_{\rm{NN}}}=7.7$ and 200 GeV from a Multiphase
Transport Model},
\href{https://link.springer.com/article/10.1007/s41365-025-01826-w}
{Nucl. Sci. Tech. \textbf{36}, 228 (2025).}



\bibitem{Li2022PRC}
F. Li, Y.-G. Ma, S. Zhang, G.-L. Ma, and Q.-Y. Shou,
\textit{Impact of Nuclear Structure on the Background in the Chiral
Magnetic Effect in $^{96}\rm{Ru}$+$^{96}\rm{Ru}$ and
$^{96}\rm{Zr}$+$^{96}\rm{Zr}$ Collisions at
$\sqrt{s_{\rm{NN}}}=7.7$--200 GeV from a Multiphase Transport Model},
\href{https://journals.aps.org/prc/abstract/10.1103/PhysRevC.106.014906}
{Phys. Rev. C \textbf{106}, 014906 (2022).}


\bibitem{Ma2024Entropy}
W.-H. Ma and Y.-G. Ma,
\textit{Exploring the Diversity of Nuclear Density through Information
Entropy},
\href{https://www.mdpi.com/1099-4300/26/9/763}
{Entropy \textbf{26}, 763 (2024).}


\bibitem{Abdulhamid2024PRR}
M.I. Abdulhamid et al. (STAR Collaboration),
\textit{Upper Limit on the Chiral Magnetic Effect in Isobar Collisions
at the Relativistic Heavy-Ion Collider},
\href{https://journals.aps.org/prresearch/abstract/10.1103/PhysRevResearch.6.L032005}
{Phys. Rev. Research \textbf{6}, L032005 (2024).}

\bibitem{Wei2024NST}
K. Wei, Y.-L. Ye, and Z.-H. Yang,
\textit{Clustering in Nuclei: Progress and Perspectives},
\href{https://doi.org/10.1007/s41365-024-01588-x}
{Nucl. Sci. Tech. \textbf{35}, 216 (2024)}.

\bibitem{Wang2026CPL}
Q. Wang, L.-G. Pang, and X.-N. Wang,
\textit{Addressing the Ultra-Central Puzzle with Initial-State
Nuclear Structures},
\href{https://doi.org/10.1088/0256-307X/43/2/020101}
{Chin. Phys. Lett. \textbf{43}, 020101 (2026)}.

\bibitem{Shi2021NST}
C.Z. Shi and Y.G. Ma,
\textit{$\alpha$-Clustering Effect on Flows of Direct Photons in
Heavy-Ion Collisions},
\href{https://doi.org/10.1007/s41365-021-00897-9}
{Nucl. Sci. Tech. \textbf{32}, 66 (2021)}.

\bibitem{He2021PRC}
J. He, W.B. He, Y.G. Ma, and S. Zhang,
\textit{Machine-Learning-Based Identification for Initial Clustering
Structure in Relativistic Heavy-Ion Collisions},
\href{https://doi.org/10.1103/PhysRevC.104.044902}
{Phys. Rev. C \textbf{104}, 044902 (2021)}.

\bibitem{Wang2023PRC}
R. Wang, Y.G. Ma, L.W. Chen, C.M. Ko, K.J. Sun, and Z. Zhang,
\textit{Kinetic Approach of Light-Nuclei Production in
Intermediate-Energy Heavy-Ion Collisions},
\href{https://doi.org/10.1103/PhysRevC.108.L031601}
{Phys. Rev. C \textbf{108}, L031601 (2023)}.

\bibitem{Cao2023PRC}
R.X. Cao, S. Zhang, and Y.G. Ma,
\textit{Effects of the $\alpha$-Cluster Structure and the Intrinsic
Momentum Component of Nuclei on the Longitudinal Asymmetry in
Relativistic Heavy-Ion Collisions},
\href{https://doi.org/10.1103/PhysRevC.108.064906}
{Phys. Rev. C \textbf{108}, 064906 (2023)}.

\bibitem{Wang2026PRL}
R. Wang, Z. Zhang, Y.G. Ma, L.W. Chen, C.M. Ko, and K.J. Sun,
\textit{Clustered Nature of Hot and Dense Nuclear Matter:
Signatures from Heavy-Ion Collisions},
\href{https://doi.org/10.1103/ffxr-mmlg}
{Phys. Rev. Lett. \textbf{136}, 172301 (2026)}.

\bibitem{Zhou2026PRL}
P. Li, B. Zhou, and G.L. Ma,
\textit{Identifying $\alpha$-Cluster Configurations in $^{20}\rm{Ne}$
via Ultracentral Ne+Ne Collisions},
\href{https://doi.org/10.1103/tffz-8q1m}
{Phys. Rev. Lett. \textbf{136}, 082302 (2026)}.


\bibitem{He2014PRL}
W.B. He, Y.G. Ma, X.G. Cao, X.Z. Cai, and G.Q. Zhang,
\textit{Giant Dipole Resonance as a Fingerprint of $\alpha$ Clustering
Configurations in $^{12}\rm{C}$ and $^{16}\rm{O}$},
\href{https://journals.aps.org/prl/abstract/10.1103/PhysRevLett.113.032506}
{Phys. Rev. Lett. \textbf{113}, 032506 (2014).}

\bibitem{Li2020PRCCluster}
Y.-A. Li, S. Zhang, and Y.-G. Ma,
\textit{Signatures of $\alpha$-clustering in $^{16}\rm{O}$
by using a multiphase transport model},
\href{https://doi.org/10.1103/PhysRevC.102.054907}
{Phys. Rev. C \textbf{102}, 054907 (2020)}.

\bibitem{Zhang2024EPJACluster}
Y.-X. Zhang, S. Zhang, and Y.-G. Ma,
\textit{Signatures of an $\alpha$ + core structure in
$^{44}\rm{Ti}+^{44}\rm{Ti}$ collisions at
$\sqrt{s_{\rm{NN}}}=5.02$ TeV by a multiphase transport model},
\href{https://doi.org/10.1140/epja/s10050-024-01290-x}
{Eur. Phys. J. A \textbf{60} (2024)}.

\bibitem{Xu2018NSTCluster}
Z.-W. Xu, S. Zhang, Y.-G. Ma, J.-H. Chen, and C. Zhong,
\textit{Influence of $\alpha$-clustering nuclear structure
on the rotating collision system},
\href{https://doi.org/10.1007/s41365-018-0523-9}
{Nucl. Sci. Tech. \textbf{29}, 186 (2018)}.

\bibitem{RWang2026PRC}
R. Wang, Z. Zhang, S. Burrello, M. Colonna, and E.G. Lanza,
\textit{Phase-space excluded-volume approach for light clusters
in a nuclear medium},
\href{https://doi.org/10.1103/4py6-jyk8}
{Phys. Rev. C \textbf{113}, 034624 (2026).}

\bibitem{Sun2024NatCommun}
K.-J. Sun, R. Wang, C.M. Ko, Y.-G. Ma, and C. Shen,
\textit{Unveiling the dynamics of little-bang nucleosynthesis},
\href{https://doi.org/10.1038/s41467-024-45474-x}
{Nat. Commun. \textbf{15}, 1074 (2024).}

\bibitem{RWang2023PRC}
R. Wang, Y.-G. Ma, L.-W. Chen, C.M. Ko, K.-J. Sun, and Z. Zhang,
\textit{Kinetic approach of light-nuclei production in
intermediate-energy heavy-ion collisions},
\href{https://doi.org/10.1103/PhysRevC.108.L031601}
{Phys. Rev. C \textbf{108}, L031601 (2023).}

\bibitem{Tanaka2021Science}
J. Tanaka et al.,
\textit{Formation of $\alpha$ Clusters in Dilute Neutron-Rich Matter},
\href{https://www.science.org/doi/10.1126/science.abe4688}
{Science \textbf{371}, 260--264 (2021).}

\bibitem{Hen2021Science}
O. Hen,
\textit{From Nuclear Clusters to Neutron Stars},
\href{https://www.science.org/doi/10.1126/science.abf2427}
{Science \textbf{371}, 232 (2021).}


\bibitem{Miller2019PLB}
G.A. Miller, A. Beck, S. May-Tal Beck, L.B. Weinstein,
E. Piasetzky, and O. Hen,
\textit{Can Long-Range Nuclear Properties Be Influenced by Short Range
Interactions? A Chiral Dynamics Estimate},
\href{https://www.sciencedirect.com/science/article/pii/S0370269319303193}
{Phys. Lett. B \textbf{793}, 360--364 (2019).}

\bibitem{Reinhard2010PRC}
P.G. Reinhard and W. Nazarewicz,
\textit{Information Content of a New Observable: The Case of the Nuclear Neutron Skin},
\href{https://journals.aps.org/prc/abstract/10.1103/PhysRevC.81.051303}
{Phys. Rev. C \textbf{81}, 051303(R) (2010).}

\bibitem{Tamii2011PRL}
A. Tamii et al.,
\textit{Complete Electric Dipole Response and the Neutron Skin in
$^{208}\rm{Pb}$},
\href{https://journals.aps.org/prl/abstract/10.1103/PhysRevLett.107.062502}
{Phys. Rev. Lett. \textbf{107}, 062502 (2011).}

\bibitem{Tamii2014EPJA}
A. Tamii, P. von Neumann-Cosel, and I. Poltoratska,
\textit{Electric Dipole Response of $^{208}\rm{Pb}$ from Proton
Inelastic Scattering: Constraints on Neutron Skin Thickness and
Symmetry Energy},
\href{https://doi.org/10.1140/epja/i2014-14028-7}
{Eur. Phys. J. A \textbf{50}, 28 (2014)}.

\bibitem{RocaMaza2013PRC}
X. Roca-Maza, M. Brenna, G. Col\`o, M. Centelles, X. Vi\~nas,
B.K. Agrawal, N. Paar, D. Vretenar, and J. Piekarewicz,
\textit{Electric Dipole Polarizability in $^{208}\rm{Pb}$:
Insights from the Droplet Model},
\href{https://journals.aps.org/prc/abstract/10.1103/PhysRevC.88.024316}
{Phys. Rev. C \textbf{88}, 024316 (2013).}

\bibitem{Zhang2014PRC}
Z. Zhang and L.W. Chen,
\textit{Constraining the Density Slope of Nuclear Symmetry Energy at
Subsaturation Densities Using Electric Dipole Polarizability in
$^{208}\rm{Pb}$},
\href{https://journals.aps.org/prc/abstract/10.1103/PhysRevC.90.064317}
{Phys. Rev. C \textbf{90}, 064317 (2014).}

\bibitem{Zhang2015PRC}
Z. Zhang and L.W. Chen,
\textit{Electric Dipole Polarizability in $^{208}\rm{Pb}$ as a Probe
of the Symmetry Energy and Neutron Matter around $\rho_0/3$},
\href{https://journals.aps.org/prc/abstract/10.1103/PhysRevC.92.031301}
{Phys. Rev. C \textbf{92}, 031301(R) (2015).}

\bibitem{Birkhan2017PRL}
J. Birkhan et al.,
\textit{Electric Dipole Polarizability of $^{48}\rm{Ca}$ and
Implications for the Neutron Skin},
\href{https://journals.aps.org/prl/abstract/10.1103/PhysRevLett.118.252501}
{Phys. Rev. Lett. \textbf{118}, 252501 (2017).}

\bibitem{Zhang2018PLB}
Z. Zhang, Y. Lim, J.W. Holt, and C.M. Ko,
\textit{Nuclear Dipole Polarizability from Mean-Field Modeling
Constrained by Chiral Effective Field Theory},
\href{https://www.sciencedirect.com/science/article/pii/S0370269317309899}
{Phys. Lett. B \textbf{777}, 73--79 (2018).}


\bibitem{Hao2026PRL}
Z.R. Hao et al.,
\textit{Precision Extraction of the Deuteron Electric Polarizability
via the Baldin Sum Rule with Full Low-Energy Coverage},
\href{https://journals.aps.org/prl/abstract/10.1103/xmmq-gnvq}
{Phys. Rev. Lett. \textbf{136}, 212502 (2026).}


\bibitem{Colonna2021PRC}
M. Colonna et al.,
\textit{Comparison of Heavy-Ion Transport Simulations: Mean-Field
Dynamics in a Box},
\href{https://journals.aps.org/prc/abstract/10.1103/PhysRevC.104.024603}
{Phys. Rev. C \textbf{104}, 024603 (2021).}

\bibitem{Xu2024PRC}
J. Xu et al.,
\textit{Comparing Pion Production in Transport Simulations of Heavy-Ion
Collisions at 270$A$ MeV under Controlled Conditions},
\href{https://journals.aps.org/prc/abstract/10.1103/PhysRevC.109.044609}
{Phys. Rev. C \textbf{109}, 044609 (2024).}

\bibitem{Zhao2026PRC}
X.-L. Zhao, X.-Y. Xie, Y. Li, and G.-L. Ma,
\textit{Sensitivity of Anisotropic Flow in Pb+Pb Collisions to the
Neutron Skin of $^{208}\rm{Pb}$ at Energies Available at the CERN
Large Hadron Collider},
\href{https://journals.aps.org/prc/accepted/10.1103/5qcl-8w8t}
{Phys. Rev. C, accepted (2026).}


\bibitem{Liu2022PLB}
L.-M. Liu, C.J. Zhang, J. Zhou, J. Xu, J. Jia, and G.X. Peng,
\textit{Probing Neutron-skin Thickness with Free Spectator Neutrons
in Ultracentral High-energy Isobaric Collisions},
\href{https://doi.org/10.1016/j.physletb.2022.137441}
{Phys. Lett. B \textbf{834}, 137441 (2022).}


\bibitem{Xu2022PRC}
H.-j. Xu, H. Li, Y. Zhou, X. Wang, J. Zhao, L.-W. Chen, and F. Wang,
\textit{Measuring Neutron Skin by Grazing Isobaric Collisions},
\href{https://journals.aps.org/prc/abstract/10.1103/PhysRevC.105.L011901}
{Phys. Rev. C \textbf{105}, L011901 (2022).}



\bibitem{VanderSchee2024PLB}
W. van der Schee, Y.-J. Lee, G. Nijs, and Y. Chen,
\textit{Hard Probes in Isobar Collisions as a Probe of the Neutron Skin},
\href{https://www.sciencedirect.com/science/article/pii/S0370269324005112}
{Phys. Lett. B \textbf{856}, 138953 (2024).}

\bibitem{Liu2023PLB}
L.-M. Liu, J. Xu, and G.-X. Peng,
\textit{Measuring Deformed Neutron Skin with Free Spectator Nucleons
in Relativistic Heavy-Ion Collisions},
\href{https://www.sciencedirect.com/science/article/pii/S0370269323000357}
{Phys. Lett. B \textbf{838}, 137701 (2023).}




\bibitem{Vitsos2026EPJC}
A. Vitsos, L.M.M. Soranzo, E.G.D. Nielsen, and Y. Zhou,
\textit{Characterizing the Neutron Skin of $^{48}\rm{Ca}$ through
Collective Flow at the CERN Large Hadron Collider},
\href{https://link.springer.com/article/10.1140/epjc/s10052-026-15751-8}
{Eur. Phys. J. C \textbf{86}, 536 (2026).}

\bibitem{Pihan2025}
G. Pihan, A. Monnai, B. Schenke, and C. Shen,
\textit{Neutron Skin from Conserved Charge Measurements at Collider
Experiments},
\href{https://arxiv.org/abs/2509.21644}
{arXiv:2509.21644 (2025).}

\bibitem{McLerran1994PRD}
L.~D. McLerran and R. Venugopalan,
\textit{Computing quark and gluon distribution functions for very large nuclei},
\href{https://doi.org/10.1103/PhysRevD.49.2233}
{Phys. Rev. D \textbf{49}, 2233 (1994)}.

\bibitem{McLerran1994PRD2}
L.~D. McLerran and R. Venugopalan,
\textit{Gluon distribution functions for very large nuclei at small transverse momentum},
\href{https://doi.org/10.1103/PhysRevD.49.3352}
{Phys. Rev. D \textbf{49}, 3352 (1994)}.

\bibitem{JalilianMarian1997PRD}
J. Jalilian-Marian, A. Kovner, L. McLerran, and H. Weigert,
\textit{Intrinsic glue distribution at very small $x$},
\href{https://doi.org/10.1103/PhysRevD.55.5414}
{Phys. Rev. D \textbf{55}, 5414 (1997)}.

\bibitem{Kovchegov1999PRD}
Y.~V. Kovchegov,
\textit{Small-$x$ $F_2$ structure function of a nucleus including multiple pomeron exchanges},
\href{https://doi.org/10.1103/PhysRevD.60.034008}
{Phys. Rev. D \textbf{60}, 034008 (1999)}.

\bibitem{Kovner1995PRD}
A. Kovner, L.~D. McLerran, and H. Weigert,
\textit{Gluon production from non-Abelian Weizs\"acker-Williams fields in nucleus-nucleus collisions},
\href{https://doi.org/10.1103/PhysRevD.52.6231}
{Phys. Rev. D \textbf{52}, 6231 (1995)}.

\bibitem{Krasnitz1999NPB}
A. Krasnitz and R. Venugopalan,
\textit{Nonperturbative computation of gluon minijet production in nuclear collisions at very high energies},
\href{https://doi.org/10.1016/S0550-3213(99)00366-1}
{Nucl. Phys. B \textbf{557}, 237 (1999)}.

\bibitem{Lappi2003PRC}
T. Lappi,
\textit{Production of gluons in the classical field model for heavy-ion collisions},
\href{https://doi.org/10.1103/PhysRevC.67.054903}
{Phys. Rev. C \textbf{67}, 054903 (2003)}.

\bibitem{Lappi2006NPA}
T. Lappi and L. McLerran,
\textit{Some features of the glasma},
\href{https://doi.org/10.1016/j.nuclphysa.2006.04.001}
{Nucl. Phys. A \textbf{772}, 200 (2006)}.

\bibitem{Gelis2010ARNPS}
F. Gelis, E. Iancu, J. Jalilian-Marian, and R. Venugopalan,
\textit{The Color Glass Condensate},
\href{https://doi.org/10.1146/annurev.nucl.010909.083629}
{Annu. Rev. Nucl. Part. Sci. \textbf{60}, 463 (2010)}.


\bibitem{Li2026CPL}
H. Li, L.-M. Liu, J. Chen, Y.-G. Ma, and C. Zhang,
\textit{Probing the Neutron-Skin Thickness Through $J/\psi$
Photoproduction in Ultra-Peripheral Collisions},
\href{https://cpl.iphy.ac.cn/article/doi/10.1088/0256-307X/43/5/050101}
{Chin. Phys. Lett. \textbf{43}, 050101 (2026).}

\bibitem{FRIB2022}
Lawrence Berkeley National Laboratory,
\textit{High Rigidity Spectrometer for FRIB},
\href{https://hrs.lbl.gov}
{HRS Project Website.}

\bibitem{NSAC2023}
Nuclear Science Advisory Committee,
\textit{A New Era of Discovery: The 2023 Long Range Plan for Nuclear Science},
\href{https://science.osti.gov/-/media/np/nsac/pdf/202310/NSAC-LRP-2023-v12.pdf}
{Nuclear Science Advisory Committee (2023).}

\bibitem{Noji2023NIMA}
S. Noji, R.G.T. Zegers, G.P.A. Berg, A.M. Amthor, T. Baumann,
D. Bazin, E.E. Burkhardt, M. Cortesi, J.C. DeKamp, M. Hausmann,
M. Portillo, D.H. Potterveld, B.M. Sherrill, A. Stolz,
O.B. Tarasov, and R.C. York,
\textit{Design of the High Rigidity Spectrometer at FRIB},
\href{https://doi.org/10.1016/j.nima.2022.167548}
{Nucl. Instrum. Methods Phys. Res. A \textbf{1045}, 167548 (2023).}

\bibitem{Ponnath2025NPA}
L. Ponnath et al. (R$^3$B Collaboration),
\textit{Precise Measurement of Nuclear Interaction Cross Sections
Towards Neutron-Skin Determination with R$^3$B},
\href{https://doi.org/10.1016/j.nuclphysa.2025.123022}
{Nucl. Phys. A \textbf{1056}, 123022 (2025).}


\bibitem{Hagen2016NatPhys}
G. Hagen, A. Ekstr\"om, C. Forss\'en, G.R. Jansen, W. Nazarewicz,
T. Papenbrock, K.A. Wendt, S. Bacca, N. Barnea, B. Carlsson, C. Drischler,
K. Hebeler, M. Hjorth-Jensen, M. Miorelli, G. Orlandini, A. Schwenk, and
J. Simonis,
\textit{Neutron and Weak-Charge Distributions of the $^{48}$Ca Nucleus},
\href{https://doi.org/10.1038/nphys3529}
{Nat. Phys. \textbf{12}, 186 (2016).}


\bibitem{Novario2023PRL}
S.J. Novario, D. Lonardoni, S. Gandolfi, and G. Hagen,
\textit{Trends of Neutron Skins and Radii of Mirror Nuclei from First
Principles},
\href{https://doi.org/10.1103/PhysRevLett.130.032501}
{Phys. Rev. Lett. \textbf{130}, 032501 (2023).}

\bibitem{Xu2024NSTabinitio}
X.-Y. Xu, S.-Q. Fan, Q. Yuan, et al.,
\textit{Progress in Ab Initio In-Medium Similarity Renormalization
Group and Coupled-Channel Method with Coupling to the Continuum},
\href{https://doi.org/10.1007/s41365-024-01585-0}
{Nucl. Sci. Tech. \textbf{35}, 215 (2024)}.

\bibitem{Yang2025CPL}
Y.-L. Yang and P.-W. Zhao,
\textit{Reconciling Light Nuclei and Nuclear Matter:
Relativistic Ab Initio Calculations},
\href{https://doi.org/10.1088/0256-307X/42/5/051201}
{Chin. Phys. Lett. \textbf{42}, 051201 (2025)}.


\bibitem{Furnstahl2015PRC}
R.J. Furnstahl, N. Klco, D.R. Phillips, and S. Wesolowski,
\textit{Quantifying Truncation Errors in Effective Field Theory},
\href{https://doi.org/10.1103/PhysRevC.92.024005}
{Phys. Rev. C \textbf{92}, 024005 (2015).}

\bibitem{Melendez2019PRC}
J.A. Melendez, R.J. Furnstahl, D.R. Phillips, M.T. Pratola, and S. Wesolowski,
\textit{Quantifying Correlated Truncation Errors in Effective Field Theory},
\href{https://doi.org/10.1103/PhysRevC.100.044001}
{Phys. Rev. C \textbf{100}, 044001 (2019).}

\bibitem{Drischler2020PRL}
C. Drischler, R.J. Furnstahl, J.A. Melendez, and D.R. Phillips,
\textit{How Well Do We Know the Neutron-Matter Equation of State at the
Densities Inside Neutron Stars? A Bayesian Approach with Correlated
Uncertainties},
\href{https://doi.org/10.1103/PhysRevLett.125.202702}
{Phys. Rev. Lett. \textbf{125}, 202702 (2020).}

\bibitem{Drischler2020PRC}
C. Drischler, J.A. Melendez, R.J. Furnstahl, and D.R. Phillips,
\textit{Quantifying Uncertainties and Correlations in the Nuclear-Matter
Equation of State},
\href{https://doi.org/10.1103/PhysRevC.102.054315}
{Phys. Rev. C \textbf{102}, 054315 (2020).}

\bibitem{LiXie2025PRC}
B.-A. Li and W.-J. Xie,
\textit{Evolution of in-medium baryon-baryon scattering cross sections
and stiffness of dense nuclear matter from Bayesian analyses of FOPI
proton-flow excitation functions},
\href{https://doi.org/10.1103/PhysRevC.111.054602}
{Phys. Rev. C \textbf{111}, 054602 (2025)}.

\bibitem{Wang2025CPC}
J.-M. Wang, X.-G. Deng, W.-J. Xie, B.-A. Li, and Y.-G. Ma,
\textit{Bayesian inference of nuclear incompressibility from collective
flow in mid-central Au+Au collisions at 400--1500 MeV/nucleon},
\href{https://doi.org/10.1088/1674-1137/adf4a1}
{Chin. Phys. C \textbf{49}, 124105 (2025)}.



\bibitem{Morfouace2019PLB}
P. Morfouace, C. Tsang, Y. Zhang, W.G. Lynch, M.B. Tsang, D.D.S. Coupland,
M. Youngs, Z. Chajecki, M.A. Famiano, T.K. Ghosh, G. Jhang, J. Lee, H. Liu,
A. Sanetullaev, R. Showalter, and J.R. Winkelbauer,
\textit{Constraining the Symmetry Energy with Heavy-Ion Collisions and
Bayesian Analyses},
\href{https://doi.org/10.1016/j.physletb.2019.135045}
{Phys. Lett. B \textbf{799}, 135045 (2019).}

\bibitem{Xu2020PRC}
J. Xu, W.-J. Xie, and B.-A. Li,
\textit{Bayesian Inference of Nuclear Symmetry Energy from Measured and
Imagined Neutron Skin Thickness in Sn Isotopes, $^{208}$Pb, and $^{48}$Ca},
\href{https://doi.org/10.1103/PhysRevC.102.044316}
{Phys. Rev. C \textbf{102}, 044316 (2020).}


\bibitem{Essick2021PRL}
R. Essick, I. Tews, P. Landry, and A. Schwenk,
\textit{Astrophysical Constraints on the Symmetry Energy and the Neutron Skin
of $^{208}$Pb with Minimal Modeling Assumptions},
\href{https://doi.org/10.1103/PhysRevLett.127.192701}
{Phys. Rev. Lett. \textbf{127}, 192701 (2021).}

\bibitem{Xu2021PRC}
J. Xu, Z. Zhang, and B.-A. Li,
\textit{Bayesian Uncertainty Quantification for Nuclear Matter
Incompressibility},
\href{https://doi.org/10.1103/PhysRevC.104.054324}
{Phys. Rev. C \textbf{104}, 054324 (2021).}

\bibitem{Qiu2024PLB}
M. Qiu, B.-J. Cai, L.-W. Chen, C.-X. Yuan, and Z. Zhang,
\textit{Bayesian Model Averaging for Nuclear Symmetry Energy from Effective
Proton-Neutron Chemical Potential Difference of Neutron-Rich Nuclei},
\href{https://doi.org/10.1016/j.physletb.2023.138435}
{Phys. Lett. B \textbf{849}, 138435 (2024).}

\bibitem{Neufcourt2018PRC}
L. Neufcourt, Y. Cao, W. Nazarewicz, and F. Viens,
\textit{Bayesian Approach to Model-Based Extrapolation of Nuclear
Observables},
\href{https://doi.org/10.1103/PhysRevC.98.034318}
{Phys. Rev. C \textbf{98}, 034318 (2018).}

\bibitem{Neufcourt2019PRL}
L. Neufcourt, Y. Cao, W. Nazarewicz, E. Olsen, and F. Viens,
\textit{Neutron Drip Line in the Ca Region from Bayesian Model Averaging},
\href{https://doi.org/10.1103/PhysRevLett.122.062502}
{Phys. Rev. Lett. \textbf{122}, 062502 (2019).}

\bibitem{Connell2021JPG}
M.A. Connell, I. Billig, and D.R. Phillips,
\textit{Does Bayesian Model Averaging Improve Polynomial Extrapolations?
Two Toy Problems as Tests},
\href{https://doi.org/10.1088/1361-6471/ac215a}
{J. Phys. G \textbf{48}, 104001 (2021).}

\bibitem{Kejzlar2023SciRep}
V. Kejzlar, L. Neufcourt, and W. Nazarewicz,
\textit{Local Bayesian Dirichlet Mixing of Imperfect Models},
\href{https://doi.org/10.1038/s41598-023-46568-0}
{Sci. Rep. \textbf{13}, 19600 (2023).}

\bibitem{Saito2024PRC}
Y. Saito, I. Dillmann, R. Kr\"ucken, M.R. Mumpower, and R. Surman,
\textit{Uncertainty Quantification of Mass Models Using Ensemble Bayesian
Model Averaging},
\href{https://doi.org/10.1103/PhysRevC.109.054301}
{Phys. Rev. C \textbf{109}, 054301 (2024).}

\bibitem{Boehnlein2022RMP}
A. Boehnlein, M. Diefenthaler, N. Sato, M. Schram, V. Ziegler, C. Fanelli,
M. Hjorth-Jensen, T. Horn, M.P. Kuchera, D. Lee, W. Nazarewicz, P. Ostroumov,
K. Orginos, P. Purohit, M.S. Smith, and L.-G. Pang,
\textit{Colloquium: Machine Learning in Nuclear Physics},
\href{https://doi.org/10.1103/RevModPhys.94.031003}
{Rev. Mod. Phys. \textbf{94}, 031003 (2022).}
\bibitem{Zhou2024PPNP}
K. Zhou, L. Wang, L.-G. Pang, and S. Shi,
\textit{Exploring QCD Matter in Extreme Conditions with Machine Learning},
\href{https://doi.org/10.1016/j.ppnp.2023.104084}
{Prog. Part. Nucl. Phys. \textbf{135}, 104084 (2024).}

\bibitem{He2023SCPMA}
W.-B. He, Q.-F. Li, Y.-G. Ma, Z.-M. Niu, J.-C. Pei, and Y.-X. Zhang,
\textit{Machine Learning in Nuclear Physics at Low and Intermediate
Energies},
\href{https://doi.org/10.1007/s11433-023-2116-0}
{Sci. China Phys. Mech. Astron. \textbf{66}, 282001 (2023).}

\bibitem{Konig2020PLB}
S. K\"onig, A. Ekstr\"om, K. Hebeler, D. Lee, and A. Schwenk,
\textit{Eigenvector Continuation as an Efficient and Accurate Emulator for
Uncertainty Quantification},
\href{https://doi.org/10.1016/j.physletb.2020.135814}
{Phys. Lett. B \textbf{810}, 135814 (2020).}

\bibitem{Duguet2024RMP}
T. Duguet, A. Ekstr\"om, R.J. Furnstahl, S. K\"onig, and D. Lee,
\textit{Eigenvector Continuation and Projection-Based Emulators},
\href{https://doi.org/10.1103/RevModPhys.96.031002}
{Rev. Mod. Phys. \textbf{96}, 031002 (2024).}

\bibitem{You2025NST}
H.-Q. You, X.-T. He, R.-H. Wu, S.-S. Zhang, J.-J. Li,
Q.-H. He, and H.-Q. Zhang,
\textit{Nuclear Deformation Effects on $\alpha$-Decay Half-Lives
with Empirical Formula and Machine Learning},
\href{https://doi.org/10.1007/s41365-025-01766-5}
{Nucl. Sci. Tech. \textbf{36}, 191 (2025)}.

\bibitem{Guo2025CPL}
S. Guo, L. Wang, K. Zhou, and G.-L. Ma,
\textit{Neural Unfolding of the Chiral Magnetic Effect in
Heavy-Ion Collisions},
\href{https://doi.org/10.1088/0256-307X/42/11/110101}
{Chin. Phys. Lett. \textbf{42}, 110101 (2025)}.

\bibitem{Zhang2025CPLML}
W. Zhang, Z. Zhang, J. Hu, B. Lu, J. Pang, and Q. Wang,
\textit{Machine Learning Unveils the Power Law of Finite-Volume
Energy Shifts},
\href{https://doi.org/10.1088/0256-307X/42/7/070202}
{Chin. Phys. Lett. \textbf{42}, 070202 (2025)}.

\bibitem{RWang2020PRR}
R. Wang, Y.-G. Ma, R. Wada, L.-W. Chen, W.-B. He, H.-L. Liu,
and K.-J. Sun,
\textit{Nuclear liquid-gas phase transition with machine learning},
\href{https://doi.org/10.1103/PhysRevResearch.2.043202}
{Phys. Rev. Research \textbf{2}, 043202 (2020).}

\bibitem{Wang2022PLBML}
Y. Wang, Z. Gao, H. L\"u, and Q. Li,
\textit{Decoding the nuclear symmetry energy event-by-event
in heavy-ion collisions with machine learning},
\href{https://doi.org/10.1016/j.physletb.2022.137508}
{Phys. Lett. B \textbf{835}, 137508 (2022).}

\bibitem{WangLi2023FrontPhys}
Y. Wang and Q. Li,
\textit{Machine learning transforms the inference of the nuclear
equation of state},
\href{https://doi.org/10.1007/s11467-023-1313-3}
{Front. Phys. \textbf{18}, 64402 (2023).}


\bibitem{Fujimoto2018PRD}
Y. Fujimoto, K. Fukushima, and K. Murase,
\textit{Methodology Study of Machine Learning for the Neutron Star Equation
of State},
\href{https://doi.org/10.1103/PhysRevD.98.023019}
{Phys. Rev. D \textbf{98}, 023019 (2018).}

\bibitem{Soma2022JCAP}
S. Soma, L. Wang, S. Shi, H. St\"ocker, and K. Zhou,
\textit{Neural Network Reconstruction of the Dense Matter Equation of State
from Neutron Star Observables},
\href{https://doi.org/10.1088/1475-7516/2022/08/071}
{J. Cosmol. Astropart. Phys. \textbf{08}, 071 (2022).}


\bibitem{Hebeler2013ApJ}
K.~Hebeler, J.~M.~Lattimer, C.~J.~Pethick, and A.~Schwenk,
Equation of state and neutron star properties constrained by nuclear physics and observation,
Astrophys. J. \textbf{773}, 11 (2013).
\href{https://doi.org/10.1088/0004-637X/773/1/11}
{https://doi.org/10.1088/0004-637X/773/1/11}

\bibitem{Ecker2023MNRAS}
C. Ecker and L. Rezzolla,
\textit{Impact of Large-Mass Constraints on the Properties of Neutron Stars},
\href{https://academic.oup.com/mnras/article/519/2/2615/6968615}
{Mon. Not. R. Astron. Soc. \textbf{519}, 2615--2622 (2023).}

\bibitem{Yue2026SciBull}
T.-G. Yue, Z. Zhang, and L.-W. Chen,
\textit{Evidence for Strong Isovector Nuclear Spin-Orbit Interaction},
\href{https://doi.org/10.1016/j.scib.2026.01.062}
{Sci. Bull. \textbf{71}, 1270 (2026).}

\bibitem{Qiu2026PLB}
M. Qiu, T.-G. Yue, Z. Zhang, and L.-W. Chen,
\textit{A Relativistic Mechanism for the Enhanced Isovector Spin-Orbit
Interaction Suggested by Parity-Violating Electron Scattering Experiments},
\href{https://doi.org/10.1016/j.physletb.2026.140648}
{Phys. Lett. B \textbf{879}, 140648 (2026).}

\bibitem{Yue2026Particles}
T.-G. Yue, Z. Zhang, and L.-W. Chen,
\textit{Effects of Isovector Spin-Orbit Interaction on the
Charge-Weak Form Factor Difference in $^{48}\rm{Ca}$,
$^{208}\rm{Pb}$, $^{90}\rm{Zr}$ and $^{62}\rm{Ni}$},
\href{https://doi.org/10.3390/particles9020054}
{Particles \textbf{9}, 54 (2026)}.

\bibitem{Zhao2025PRR}
T. Zhao, Z. Lin, B. Kumar, A.W. Steiner, and M. Prakash,
\textit{Characterizing the Nuclear Models Informed by PREX and CREX:
A View from Bayesian Inference},
\href{https://doi.org/10.1103/472x-9cxj}
{Phys. Rev. Research \textbf{7}, 043335 (2025).}

\bibitem{Kunjipurayil2025PRC}
A. Kunjipurayil, J. Piekarewicz, and M. Salinas,
\textit{Role of the Isovector Spin-Orbit Potential in Mitigating the
CREX-PREX Dilemma},
\href{https://doi.org/10.1103/tcy2-brmk}
{Phys. Rev. C \textbf{112}, 014310 (2025).}



\bibitem{RocaMaza2025PRL}
X. Roca-Maza and D.H. Jakubassa-Amundsen,
\textit{QED Corrections to the Parity-Violating Asymmetry in High-Energy
Electron-Nucleus Collisions},
\href{https://doi.org/10.1103/PhysRevLett.134.192501}
{Phys. Rev. Lett. \textbf{134}, 192501 (2025).}

\bibitem{Reed2026Comment}
B.T. Reed and C.J. Horowitz,
\textit{Comment on QED Corrections to the Parity Violating Asymmetry in
High-Energy Electron-Nucleus Scattering},
\href{https://arxiv.org/abs/2601.01615}
{arXiv:2601.01615 [nucl-th] (2026).}

\bibitem{Raduta2018PRC}
A.R. Raduta and F. Gulminelli,
\textit{Nuclear Skin and the Curvature of the Symmetry Energy},
\href{https://doi.org/10.1103/PhysRevC.97.064309}
{Phys. Rev. C \textbf{97}, 064309 (2018).}

\bibitem{Reed2024PRC}
B.T. Reed, F.J. Fattoyev, C.J. Horowitz, and J. Piekarewicz,
\textit{Density Dependence of the Symmetry Energy in the
Post-PREX-CREX Era},
\href{https://doi.org/10.1103/PhysRevC.109.035803}
{Phys. Rev. C \textbf{109}, 035803 (2024).}

\bibitem{Zhang2019JPG}
N.B. Zhang and B.A. Li,
\textit{Delineating Effects of Nuclear Symmetry Energy on the
Radii and Tidal Polarizabilities of Neutron Stars},
\href{https://doi.org/10.1088/1361-6471/aaef54}
{J. Phys. G \textbf{46}, 014002 (2019).}

\bibitem{Li2020PRC}
B.A. Li and M. Magno,
\textit{Curvature-Slope Correlation of Nuclear Symmetry Energy
and Its Imprints on the Crust-Core Transition, Radius, and
Tidal Deformability of Canonical Neutron Stars},
\href{https://doi.org/10.1103/PhysRevC.102.045807}
{Phys. Rev. C \textbf{102}, 045807 (2020).}

\bibitem{Krastev2019JPG}
P.G. Krastev and B.A. Li,
\textit{Imprints of the Nuclear Symmetry Energy on the Tidal
Deformability of Neutron Stars},
\href{https://doi.org/10.1088/1361-6471/ab1a7a}
{J. Phys. G \textbf{46}, 074001 (2019).}

\bibitem{Guan2025PRL}
Z.Y. Guan and Y.F. Niu,
\textit{Reconciliation between Neutron Skin Thickness from
PREX-2 Experiment and Neutron-Star Tidal Polarizability from
GW170817 Event: The Key Role of Symmetry Energy Curvature},
\href{https://doi.org/10.1103/6m4p-wmv3}
{Phys. Rev. Lett. \textbf{135}, 172701 (2025).}



\bibitem{Lehmann1954NC}H. Lehmann, \textit{\"Uber Eigenschaften von Ausbreitungsfunktionen und Renormierungskonstanten Quantisierter Felder}, \href{https://doi.org/10.1007/BF02783624}{Nuovo Cimento \textbf{11}, 342 (1954).}

\bibitem{Migdal1957JETP}A.B. Migdal, \textit{The Momentum Distribution of Interacting Fermi Particles}, \href{https://www.jetp.ras.ru/cgi-bin/e/index/e/5/2/p333?a=list}{Sov. Phys. JETP \textbf{5}, 333 (1957).}

\bibitem{Galitskii1958JETP}V.M. Galitskii, \textit{The Energy Spectrum of a Non-ideal Fermi Gas}, \href{https://www.jetp.ras.ru/cgi-bin/e/index/r/34/1/p151?a=list}{Sov. Phys. JETP \textbf{7}, 104 (1958).}

\bibitem{Abrikosov1963Book}A.A. Abrikosov, L.P. Gor'kov, and I.E. Dzyaloshinski, \textit{Methods of Quantum Field Theory in Statistical Physics}, \href{https://cds.cern.ch/record/107441}{Prentice-Hall, Englewood Cliffs, New Jersey (1963).}

\bibitem{Dickhoff2025Book}W.H. Dickhoff and D. Van Neck, \textit{Many-Body Theory Exposed!: Propagator Description of Quantum Mechanics in Many-Body Systems}, 3rd edn., \href{https://www.worldscientific.com/worldscibooks/10.1142/14178}{World Scientific Publishing Company, Singapore (2025).}

\bibitem{Koltun1972PRL}D.S. Koltun, \textit{Total Binding Energies of Nuclei, and Particle-Removal Experiments}, \href{https://journals.aps.org/prl/abstract/10.1103/PhysRevLett.28.182}{Phys. Rev. Lett. \textbf{28}, 182 (1972).}

\bibitem{Polls1995PPNP}A. Polls, A. Ramos, C.C. Gearhart, W.H. Dickhoff, and H. M\"uther, \textit{Short Range Correlations and Spectral Functions for Nuclear Matter and Finite Nuclei}, \href{https://doi.org/10.1016/0146-6410(95)00032-E}{Prog. Part. Nucl. Phys. \textbf{34}, 371 (1995).}

\bibitem{Soma2006PRC}V. Som\`a and P. Bo\.zek, \textit{Diagrammatic Calculation of Thermodynamical Quantities in Nuclear Matter}, \href{https://journals.aps.org/prc/abstract/10.1103/PhysRevC.74.045809}{Phys. Rev. C \textbf{74}, 045809 (2006).}

\bibitem{Mahzoon2017PRL}M.H. Mahzoon, M.C. Atkinson, R.J. Charity, and W.H. Dickhoff, \textit{Neutron Skin Thickness of $^{48}\rm{Ca}$ from a Nonlocal Dispersive Optical-Model Analysis}, \href{https://journals.aps.org/prl/abstract/10.1103/PhysRevLett.119.222503}{Phys. Rev. Lett. \textbf{119}, 222503 (2017).}

\bibitem{Atkinson2020PRC}M.C. Atkinson, M.H. Mahzoon, M.A. Keim, B.A. Bordelon, C.D. Pruitt, R.J. Charity, and W.H. Dickhoff, \textit{Dispersive Optical Model Analysis of $^{208}\rm{Pb}$ Generating a Neutron-Skin Prediction beyond the Mean Field}, \href{https://journals.aps.org/prc/abstract/10.1103/PhysRevC.101.044303}{Phys. Rev. C \textbf{101}, 044303 (2020).}



\bibitem{Hugenholtz1958Physica}N.M. Hugenholtz and L. Van Hove, \textit{A Theorem on the Single Particle Energy in a Fermi Gas with Interaction}, \href{https://doi.org/10.1016/S0031-8914(58)95281-9}{Physica \textbf{24}, 363 (1958).}

\bibitem{Xu2010PRC}C. Xu, B.A. Li, and L.W. Chen, \textit{Symmetry Energy, Its Density Slope, and Neutron-Proton Effective Mass Splitting at Normal Density Extracted from Global Nucleon Optical Potentials}, \href{https://journals.aps.org/prc/abstract/10.1103/PhysRevC.82.054607}{Phys. Rev. C \textbf{82}, 054607 (2010).}



\bibitem{Xu2011NPA}C. Xu, B.A. Li, L.W. Chen, and C.M. Ko, \textit{Analytical Relations between Nuclear Symmetry Energy and Single-Nucleon Potentials in Isospin Asymmetric Nuclear Matter}, \href{https://doi.org/10.1016/j.nuclphysa.2011.06.027}{Nucl. Phys. A \textbf{865}, 1 (2011).}

\bibitem{Chen2012PRC}R. Chen, B.J. Cai, L.W. Chen, B.A. Li, X.H. Li, and C. Xu, \textit{Single-Nucleon Potential Decomposition of the Nuclear Symmetry Energy}, \href{https://journals.aps.org/prc/abstract/10.1103/PhysRevC.85.024305}{Phys. Rev. C \textbf{85}, 024305 (2012).}

\bibitem{Li2015PLB}X.H. Li, W.J. Guo, B.A. Li, L.W. Chen, F.J. Fattoyev, and W.G. Newton, \textit{Neutron--Proton Effective Mass Splitting in Neutron-Rich Matter at Normal Density from Analyzing Nucleon--Nucleus Scattering Data within an Isospin Dependent Optical Model}, \href{https://doi.org/10.1016/j.physletb.2015.03.005}{Phys. Lett. B \textbf{743}, 408 (2015).}

\bibitem{Cai2019PRC}B.J. Cai and L.W. Chen, \textit{Relativistic Self-Energy Decomposition of Nuclear Symmetry Energy and Equation of State of Neutron Matter within QCD Sum Rules}, \href{https://journals.aps.org/prc/abstract/10.1103/PhysRevC.100.024303}{Phys. Rev. C \textbf{100}, 024303 (2019).}

\bibitem{Cai2018PRC}
B.-J. Cai and L.-W. Chen,
\textit{Neutron matter within QCD sum rules},
\href{https://doi.org/10.1103/PhysRevC.97.054322}
{Phys. Rev. C \textbf{97}, 054322 (2018).}



\bibitem{Cai2012PRC}
B.-J. Cai and L.-W. Chen,
\textit{Nuclear matter fourth-order symmetry energy in the relativistic
mean field models},
\href{https://doi.org/10.1103/PhysRevC.85.024302}
{Phys. Rev. C \textbf{85}, 024302 (2012)}.

\bibitem{Cai2012PLB}
B.-J. Cai and L.-W. Chen,
\textit{Lorentz covariant nucleon self-energy decomposition of the
nuclear symmetry energy},
\href{https://doi.org/10.1016/j.physletb.2012.03.058}
{Phys. Lett. B \textbf{711}, 104 (2012)}.

\bibitem{Cai2015PRC}
B.-J. Cai, F.~J. Fattoyev, B.-A. Li, and W.~G. Newton,
\textit{Critical density and impact of $\Delta(1232)$ resonance
formation in neutron stars},
\href{https://doi.org/10.1103/PhysRevC.92.015802}
{Phys. Rev. C \textbf{92}, 015802 (2015)}.

\bibitem{Cai2017NST}
B.-J. Cai and L.-W. Chen,
\textit{Constraints on the skewness coefficient of symmetric nuclear
matter within the nonlinear relativistic mean field model},
\href{https://doi.org/10.1007/s41365-017-0329-1}
{Nucl. Sci. Tech. \textbf{28}, 185 (2017)}.

\bibitem{Baym1961PR}G. Baym and L.P. Kadanoff, \textit{Conservation Laws and Correlation Functions}, \href{https://journals.aps.org/pr/abstract/10.1103/PhysRev.124.287}{Phys. Rev. \textbf{124}, 287 (1961).}

\bibitem{Baym1962PR}G. Baym, \textit{Self-Consistent Approximations in Many-Body Systems}, \href{https://journals.aps.org/pr/abstract/10.1103/PhysRev.127.1391}{Phys. Rev. \textbf{127}, 1391 (1962).}

\bibitem{Benhar1989NPA}
O. Benhar, A. Fabrocini, and S. Fantoni,
\textit{The nucleon spectral function in nuclear matter},
\href{https://doi.org/10.1016/0375-9474(89)90374-6}
{Nucl. Phys. A \textbf{505}, 267 (1989)}.

\bibitem{Benhar2008RMP}
O. Benhar, D. Day, and I. Sick,
\textit{Inclusive quasielastic electron--nucleus scattering},
\href{https://doi.org/10.1103/RevModPhys.80.189}
{Rev. Mod. Phys. \textbf{80}, 189 (2008)}.


\bibitem{Co2022PRC}G. Co', M. Anguiano, and A.M. Lallena, \textit{Effect of Short- and Long-Range Correlations on Neutron Skins of Various Neutron-Rich Doubly Magic Nuclei}, \href{https://journals.aps.org/prc/abstract/10.1103/PhysRevC.105.064316}{Phys. Rev. C \textbf{105}, 064316 (2022).}


\bibitem{Reinhard2022PRL}P.G. Reinhard, X. Roca-Maza, and W. Nazarewicz, \textit{Combined Theoretical Analysis of the Parity-Violating Asymmetry for $^{48}\rm{Ca}$ and $^{208}\rm{Pb}$}, \href{https://journals.aps.org/prl/abstract/10.1103/PhysRevLett.129.232501}{Phys. Rev. Lett. \textbf{129}, 232501 (2022).}



\bibitem{Yang2023PRC}S. Yang, R. Li, and C. Xu, \textit{$\alpha$ Clustering in Nuclei and Its Impact on the Nuclear Symmetry Energy}, \href{https://journals.aps.org/prc/abstract/10.1103/PhysRevC.108.L021303}{Phys. Rev. C \textbf{108}, L021303 (2023).}


\bibitem{LiP2025SRC}
P. Li, K.-J. Sun, B. Zhou, and G.-L. Ma,
\textit{Universal Imprinting of Short-Range Correlations in
Relativistic Heavy-Ion Collisions},
\href{https://arxiv.org/abs/2511.23293}
{arXiv:2511.23293 [nucl-th]}.


\bibitem{Freedman1974PRD}D.Z. Freedman, \textit{Coherent Effects of a Weak Neutral Current}, \href{https://journals.aps.org/prd/abstract/10.1103/PhysRevD.9.1389}{Phys. Rev. D \textbf{9}, 1389 (1974).}

\bibitem{Akimov2017Science}D. Akimov et al. (COHERENT Collaboration), \textit{Observation of Coherent Elastic Neutrino-Nucleus Scattering}, \href{https://www.science.org/doi/10.1126/science.aao0990}{Science \textbf{357}, 1123 (2017).}

\bibitem{Cadeddu2018PRL}M. Cadeddu, C. Giunti, Y.F. Li, and Y.Y. Zhang, \textit{Average CsI Neutron Density Distribution from COHERENT Data}, \href{https://journals.aps.org/prl/abstract/10.1103/PhysRevLett.120.072501}{Phys. Rev. Lett. \textbf{120}, 072501 (2018).}

\bibitem{Ciuffoli2018PRD}E. Ciuffoli, J. Evslin, Q. Fu, and J. Tang, \textit{Extracting Nuclear Form Factors with Coherent Neutrino Scattering}, \href{https://journals.aps.org/prd/abstract/10.1103/PhysRevD.97.113003}{Phys. Rev. D \textbf{97}, 113003 (2018).}

\bibitem{Payne2019PRC}C.G. Payne, S. Bacca, G. Hagen, W. Jiang, and T. Papenbrock, \textit{Coherent Elastic Neutrino-Nucleus Scattering on $^{40}\rm{Ar}$ from First Principles}, \href{https://journals.aps.org/prc/abstract/10.1103/PhysRevC.100.061304}{Phys. Rev. C \textbf{100}, 061304(R) (2019).}

\bibitem{Coloma2020JHEP}P. Coloma, I. Esteban, M.C. Gonzalez-Garcia, and J. Men\'endez, \textit{Determining the Nuclear Neutron Distribution from Coherent Elastic Neutrino-Nucleus Scattering: Current Results and Future Prospects}, \href{https://link.springer.com/article/10.1007/JHEP08%282020%29030}{J. High Energy Phys. \textbf{08}, 030 (2020).}

\bibitem{Huang2019PRD}X.R. Huang and L.W. Chen, \textit{Neutron Skin in CsI and Low-Energy Effective Weak Mixing Angle from COHERENT Data}, \href{https://journals.aps.org/prd/abstract/10.1103/PhysRevD.100.071301}{Phys. Rev. D \textbf{100}, 071301(R) (2019).}

\bibitem{Huang2022PRD}X.R. Huang and L.W. Chen, \textit{Supernova Neutrinos as a Precise Probe of Nuclear Neutron Skin}, \href{https://journals.aps.org/prd/abstract/10.1103/PhysRevD.106.123034}{Phys. Rev. D \textbf{106}, 123034 (2022).}

\bibitem{Reddy1998PRD}S. Reddy, M. Prakash, and J.M. Lattimer, \textit{Neutrino Interactions in Hot and Dense Matter}, \href{https://journals.aps.org/prd/abstract/10.1103/PhysRevD.58.013009}{Phys. Rev. D \textbf{58}, 013009 (1998).}

\bibitem{Burrows1998PRC}A. Burrows and R.F. Sawyer, \textit{Effects of Correlations on Neutrino Opacities in Nuclear Matter}, \href{https://journals.aps.org/prc/abstract/10.1103/PhysRevC.58.554}{Phys. Rev. C \textbf{58}, 554 (1998).}

\bibitem{Burrows2006NPA}A. Burrows, S. Reddy, and T.A. Thompson, \textit{Neutrino Opacities in Nuclear Matter}, \href{https://doi.org/10.1016/j.nuclphysa.2004.06.012}{Nucl. Phys. A \textbf{777}, 356 (2006).}

\bibitem{Roberts2012PRC}L.F. Roberts, S. Reddy, and G. Shen, \textit{Medium Modification of the Charged-Current Neutrino Opacity and Its Implications}, \href{https://journals.aps.org/prc/abstract/10.1103/PhysRevC.86.065803}{Phys. Rev. C \textbf{86}, 065803 (2012).}

\bibitem{Hutauruk2022PRC}P.T.P. Hutauruk, H. Gil, S.I. Nam, and C.H. Hyun, \textit{Effect of Nucleon Effective Mass and Symmetry Energy on the Neutrino Mean Free Path in a Neutron Star}, \href{https://journals.aps.org/prc/abstract/10.1103/PhysRevC.106.035802}{Phys. Rev. C \textbf{106}, 035802 (2022).}

\bibitem{Lin2023PRC}Z. Lin, A.W. Steiner, and J. Margueron, \textit{Uncertainty Quantification for Neutrino Opacities in Core-Collapse Supernovae and Neutron Star Mergers}, \href{https://journals.aps.org/prc/abstract/10.1103/PhysRevC.107.015804}{Phys. Rev. C \textbf{107}, 015804 (2023).}

\bibitem{Hoferichter2020PRD}
M. Hoferichter, J. Men{\'e}ndez, and A. Schwenk,
\textit{Coherent Elastic Neutrino--Nucleus Scattering:
EFT Analysis and Nuclear Responses},
\href{https://doi.org/10.1103/PhysRevD.102.074018}
{Phys. Rev. D \textbf{102}, 074018 (2020)}.

\bibitem{Zheng2014JCAP}
H. Zheng, Z. Zhang, and L.-W. Chen,
\textit{Form Factor Effects in the Direct Detection of
Isospin-Violating Dark Matter},
\href{https://doi.org/10.1088/1475-7516/2014/08/011}
{J. Cosmol. Astropart. Phys. \textbf{08}, 011 (2014)}.


\bibitem{Vietze2015PRD}
L. Vietze, P. Klos, J. Men{\'e}ndez, W.C. Haxton, and A. Schwenk,
\textit{Nuclear Structure Aspects of Spin-Independent
WIMP Scattering Off Xenon},
\href{https://doi.org/10.1103/PhysRevD.91.043520}
{Phys. Rev. D \textbf{91}, 043520 (2015)}.

\bibitem{Hoferichter2016PRD}
M. Hoferichter, P. Klos, J. Men{\'e}ndez, and A. Schwenk,
\textit{Analysis Strategies for General Spin-Independent
WIMP--Nucleus Scattering},
\href{https://doi.org/10.1103/PhysRevD.94.063505}
{Phys. Rev. D \textbf{94}, 063505 (2016)}.

\bibitem{Baudis2013PRD}
L. Baudis, G. Kessler, P. Klos, R.F. Lang, J. Men{\'e}ndez,
S. Reichard, and A. Schwenk,
\textit{Signatures of Dark Matter Scattering Inelastically
Off Nuclei},
\href{https://doi.org/10.1103/PhysRevD.88.115014}
{Phys. Rev. D \textbf{88}, 115014 (2013)}.

\bibitem{Gazda2017PRD}
D. Gazda, R. Catena, and C. Forss{\'e}n,
\textit{Ab Initio Nuclear Response Functions for Dark Matter Searches},
\href{https://doi.org/10.1103/PhysRevD.95.103011}
{Phys. Rev. D \textbf{95}, 103011 (2017)}.

\bibitem{Hu2022PRLDM}
B.-S. Hu, J. Padua-Arg{\"u}elles, S. Leutheusser, T. Miyagi,
S.R. Stroberg, and J.D. Holt,
\textit{Ab Initio Structure Factors for Spin-Dependent
Dark Matter Direct Detection},
\href{https://doi.org/10.1103/PhysRevLett.128.072502}
{Phys. Rev. Lett. \textbf{128}, 072502 (2022)}.

\bibitem{Hu2026FrontPhys}
B.-S. Hu,
\textit{Ab Initio Nuclear Theory for Heavy Nuclei and Its
Application to Dark Matter--Nucleus Scattering},
\href{https://doi.org/10.3389/fphy.2026.1880894}
{Front. Phys. \textbf{14}, 1880894 (2026)}.


\end{thebibliography}
\end{document}